\documentclass[letter, fontsize=10pt, twoside=true, numbers=noenddot]{kaobook}
\pdfoutput=1

\newtoggle{review}
\togglefalse{review} %

\ifxetexorluatex
    \usepackage{polyglossia}
    \setmainlanguage{english}
\else
    \usepackage[english]{babel} %
\fi
\usepackage[english=american]{csquotes}	%

\usepackage{blindtext}
\usepackage[style=philosophy-modern,latinemph=true,sorting=nyt]{kaobiblio}
\renewcommand{\formatmargincitation}[1]{%
    \sdcite{#1}%
}

\usepackage[framed=true]{kaotheorems}

\usepackage{kaorefs}

\usepackage[switch*,mathlines]{lineno}
\usepackage[nostamp]{draftwatermark}
\usepackage{import}

\newcommand{\Lagr}{\mathcal{L}}
\newcommand{\e}{\mathrm{E}}
\newcommand{\var}{\mathrm{Var}}

\newcommand*\linenomathpatch[1]{%
    \cspreto{#1}{\linenomath}%
    \cspreto{#1*}{\linenomath}%
    \csappto{end#1}{\endlinenomath}%
    \csappto{end#1*}{\endlinenomath}%
}
\newcommand*\linenomathpatchAMS[1]{%
    \cspreto{#1}{\linenomathAMS}%
    \cspreto{#1*}{\linenomathAMS}%
    \csappto{end#1}{\endlinenomath}%
    \csappto{end#1*}{\endlinenomath}%
}

\expandafter\ifx\linenomath\linenomathWithnumbers
    \let\linenomathAMS\linenomathWithnumbers
    \patchcmd\linenomathAMS{\advance\postdisplaypenalty\linenopenalty}{}{}{}
\else
    \let\linenomathAMS\linenomathNonumbers
\fi

\linenomathpatch{equation}
\linenomathpatchAMS{gather}
\linenomathpatchAMS{multline}
\linenomathpatchAMS{align}
\linenomathpatchAMS{alignat}
\linenomathpatchAMS{flalign}

\makeatletter
\patchcmd{\mmeasure@}{\measuring@true}{
    \measuring@true
    \ifnum-\linenopenaltypar>\interdisplaylinepenalty
    \advance\interdisplaylinepenalty-\linenopenalty
    \fi
}{}{}
\makeatother

\usepackage[color]{msyntax}

\makeatletter
\def\thanks#1{\protected@xdef\@thanks{\@thanks
        \protect\footnotetext{#1}}}
\makeatother

\usepackage{lettrine}
\usepackage{subcaption}
\usepackage{cancel}
\usepackage{ccicons}

\makeindex[columns=3, title=Index, intoc] 

\makeglossaries

\newcommand*{\glossfirstformat}[1]{\textit{#1}}

\newacronymstyle{myacro}
{%
    \GlsUseAcrEntryDispStyle{long-short}%
}%
{%
    \GlsUseAcrStyleDefs{long-short}%
}

\setacronymstyle{myacro}

\defglsentryfmt{%
    \ifglshaslong{\glslabel}{%
        \glsgenacfmt%
    }{%
        \ifglsused{\glslabel}{%
            \glsgenentryfmt%
        }{%
            \glossfirstformat{\glsgenentryfmt}%
        }%
    }%
}

\newglossaryentry{dissertation}{
    name=dissertation,
    description={a paricular type of write only document.}
}

\newglossaryentry{compression}{
    name=compression,
    description={Any operation which reduces the size in bits of a computational object.}
}

\newglossaryentry{lossy compression}{
    name=lossy compression,
    description={A \gls{compression} operation which removes information from the signal to save space.}
}

\newglossaryentry{lossless compression}{
    name=lossless compression,
    description={A \gls{compression} operation which preserves all information in the original signal.}
}

\newglossaryentry{deep learning}{
    name=deep learning,
    description={A machine learning technique that learns many layers of features jointly with a task objective.}
}

\newglossaryentry{JPEG}{
    name=JPEG,
    description={The Joint Photographic Experts Group, often referring to an image file or compression algorithm.}
}

\newglossaryentry{MPEG}{
    name=MPEG,
    description={The Motion Picture Experts Group, often referring to a compression algorithm.}
}

\newglossaryentry{first principles}{
    name=first principles,
    description={The underlying engineering decisions which motivate an algorithm.}
}

\newglossaryentry{metric tensor}{
    name=metric tensor,
    description={A tensor which relates a vector space and a co-vector space.}
}

\newglossaryentry{linear combination}{
    name=linear combination,
    description={A series of scalar multiplication and vector addition.}
}

\newglossaryentry{basis}{
    name=basis,
    plural=bases,
    description={A set of vectors which is linearly independent and spans a vector space.}
}

\newglossaryentry{multilinear map}{
    name=multilinear map,
    description={A mapping which is linear in exactly one argument.}
}

\newglossaryentry{linear map}{
    name=linear map,
    description={A mapping which preserves scalar multiplication and vector addition.}
}

\newglossaryentry{image}{
    name=image,
    description={A discrete 2D signal giving a sample value at integer positions $(x, y)$, the sample may be a scalar (grayscale) or a vector (color).}
}

\newglossaryentry{convolution}{
    name=convolution,
    description={The correlation of a signal and a kernel as the kernel is shifted across the signal.}
}

\newglossaryentry{cross-correlation}{
    name=cross-correlation,
    description={See \gls{convolution} (although they are technically different).}
}

\newglossaryentry{video}{
    name=video,
    description={A discrete 3D signal giving a sample value at integer positions $(x, y, t)$, the sample may be a scalar (grayscale) or a vector (color).}
}

\newglossaryentry{fourier transform}{
    name=Fourier transform,
    description={An integral transform defining an orthogonal basis for functions.}
}

\newglossaryentry{hadamard transform}{
    name=Hadamard transform,
    description={An approximation of the discrete cosine transform consisting of only 1s and -1s.}
}

\newglossaryentry{gabor transform}{
    name=Gabor transform,
    description={A special case of the \gls{stft} which uses a Gaussian filter to window the transform.}
}

\newglossaryentry{wavelet}{
    name=wavelet,
    description={A wave-like function with finite support.}
}

\newglossaryentry{wavelet transform}{
    name=wavelet transform,
    description={An integral transform using a set of wavelets as the basis, allows for multi-resolution and localization of frqeuencies in time.}
}

\newglossaryentry{multiresolution analysis}{
    name=multiresolution analysis,
    description={See \gls{wavelet transform}.}
}

\newglossaryentry{nyquist}{
    name=Nyquist Sampling Theorem,
    description={A signal with a maximum frequency $\zeta_m$ can be represented exactly by discrete samples with a sampling rate of at least $2\zeta_m$.}
}

\newglossaryentry{entropy}{
    name=entropy,
    description={The amount of information in a message, the amount of randomness in a system, the minimum number of bits required to encode a message.}
}

\newglossaryentry{communication system}{
    name=communication system,
    description={A system for conveying some message from one party to another. Consists of an information source, a transmitter, a signal, a source of noise corrupting the signal, a receiver, and a destination.}
}

\newglossaryentry{huffman coding}{
    name=Huffman coding,
    description={A method for producing optimal length codes for single symbols given the probabilities that each symbol will occur.}
}

\newglossaryentry{bayesian decision theory}{
    name=Bayesian decision theory,
    description={A method for making optimal decisions given perfect probability distributions describing possible events.}
}

\newglossaryentry{prior}{
    name=prior probability,
    description={The probability of an event occurring in the absence of any other information.}
}

\newglossaryentry{likelihood}{
    name=likelihood,
    description={The probability of an observation given that some event occurs.}
}

\newglossaryentry{posterior}{
    name=posterior probability,
    description={The probability of an event occurring given an observation.}
}

\newglossaryentry{evidence}{
    name=evidence,
    description={The probability of an observation.}
}

\newglossaryentry{decision boundary}{
    name=decision boundary,
    plural=decision boundaries,
    description={The manifold in space separating classification decisions.}
}

\newglossaryentry{linearly separable}{
    name=linearly separable,
    description={A decision which is able to be made using only the relative position with respect to a hyperplane.}
}

\newglossaryentry{gradient}{
    name=gradient,
    description={For a scalar valued function of a vector: the vector of partial derivatives of each component of the input with respect to the scalar output.}
}

\newglossaryentry{jacobian}{
    name=Jacobian,
    description={For a tensor valued function of a tensor: the tensor of partial derivatives of each component of the input with respect to each component of the output.}
}

\newglossaryentry{feature}{
    name=feature,
    description={An abstract or higher order representation of an image or a part of an image that is more suitable for input to a machine learning algorithm.}
}

\newglossaryentry{SVM}{
    name=SVM,
    description={Support Vector Machine, a linear model which separates examples using the maximum margin hyperplane.}
}

\newglossaryentry{image-to-image}{
    name=image-to-image,
    description={The machine learning problem which takes an image as input and produces an image as output.}
}

\newglossaryentry{semantic segmentation}{
    name=semantic segmentation,
    description={The machine learning problem which takes an image as input and produces a classification label for each pixel.}
}

\newglossaryentry{nash equilibrium}{
    name=Nash equilibrium,
    description={The state of a game where no player can obtain an advantage over any other player.}
}

\newglossaryentry{interlaced}{
    name=interlaced,
    description={A method for storing color images which stores color information in sequence, \eg, each pixel could consist of 24 bits with 8 bits for res, green, and blue.}
}

\newglossaryentry{planar}{
    name=planar,
    description={A method for storing color images which stores color information separately, \eg, the image may consist of all the red pixels followed by all the blue pixels, \etc.}
}

\newglossaryentry{luminance}{
    name=luminance,
    description={The brightness (``quantity'' of light) captured by a sensor.}
}

\newglossaryentry{chrominance}{
    name=chrominance,
    description={The color or hue of light captured by a sensor.}
}

\newglossaryentry{chroma subsampling}{
    name=chroma subsampling,
    description={The process for storing chrominance channels at a smaller resolution since human vision is less sensitive to changes in color information.}
}

\newglossaryentry{transform domain}{
    name=transform domain,
    description={A catch-all term for DCT coefficients, quantized JPEG data, or any other transformation of pixel data.}
}

\newglossaryentry{filter manifold}{
    name=filter manifold,
    description={A method for learning adaptable convolution kernels from a scalar input.}
}

\newglossaryentry{convolutional filter manifold}{
    name=convolutional filter manifold,
    description={A modification of the filter manifold which uses a spatial input.}
}

\newglossaryentry{mjpeg}{
    name=Motion JPEG,
    description={A video codec which stores each frame as a JPEG.}
}

\newglossaryentry{motion vectors}{
    name=motion vectors,
    description={Vector specifying the motion of video blocks.}
}

\newglossaryentry{motion compensation}{
    name=motion compensation,
    description={The process by which a video codec warps frames using estimated motion.}
}

\newglossaryentry{motion estimation}{
    name=motion estimation,
    description={The process by which a video encoder measures block motion.}
}

\newglossaryentry{error residual}{
    name=error residual,
    description={The difference between a motion compensated frame and the true frame.}
}

\newglossaryentry{slices}{
    name=slices,
    description={Regions of a video frame consisting of a whole number of macroblocks.}
}

\newglossaryentry{macroblocks}{
    name=macroblocks,
    description={Pixel blocks in a frame which are larger than the transform block size.}
}

\newglossaryentry{rate control}{
    name=rate control,
    description={Any method for tuning the bitrate of an image or video.}
}

\newacronym{dft}{DFT}{Discrete Fourier Transform}
\newacronym{dst}{DST}{Discrete Sine Transform}
\newacronym{dct}{DCT}{Discrete Cosine Transform}
\newacronym{stft}{STFT}{Short-Time Fourier Transform}
\newacronym{cwt}{CWT}{Continuous Wavelet Transform}
\newacronym{dwt}{DWT}{Discrete Wavelet Transform}
\newacronym{dtcwt}{DTCWT}{Dual Tree Complex Wavelet Transform}
\newacronym{cnn}{CNN}{Convolutional Neural Network}
\newacronym{mlp}{MLP}{Multilayer Perceptron}
\newacronym{gan}{GAN}{Generative Adversarial Network}
\newacronym{mcu}{MCU}{Minimum Coded Unit}
\newacronym{fcr}{FCR}{Frequency-Component Rearrangement}
\newacronym{jfif}{JFIF}{JPEG File Interchange Format}
\newacronym{exif}{EXIF}{Exchangeable Image File Format}
\newacronym{rrdb}{RRDB}{Residual-in-Residual Dense Block}

\makenomenclature

\begin{document}

\frontmatter

\iftoggle{review}{
    \renewcommand\linenumberfont{\normalfont\bfseries\color{RubineRed}\small}

    \linenumbers

    \begingroup
    \let\clearpage\relax
    \def\coltab{\rule{1.5em}{0.6em}}%
    \centering
    \begin{tabular}{llllll}
        \textcolor{BurntOrange}{\coltab}\ Max & \textcolor{SkyBlue}{\coltab}\ Abhinav & \textcolor{Goldenrod}{\coltab}\ Lillian & \textcolor{WildStrawberry}{\coltab}\ Shishira & \textcolor{SeaGreen}{\coltab}\ Namitha & \textcolor{Cerulean}{\coltab}\ Vatsal
    \end{tabular}

    \listoftodos
    \endgroup
    \clearpage

    \DraftwatermarkOptions{stamp=true, color={[gray]{0.9}}, scale=15}

    \newcommand{\smalltodo}[2][]{\todo[caption={#2}, #1]{\renewcommand{\baselinestretch}{0.5}\selectfont#2\par}}

    \newcommand{\mecomment}[2][]{\smalltodo[color=BurntOrange, #1]{#2}}
    \newcommand{\ascomment}[2][]{\smalltodo[color=SkyBlue, #1]{#2}}
    \newcommand{\lilcomment}[2][]{\smalltodo[color=Goldenrod, #1]{#2}}
    \newcommand{\smcomment}[2][]{\smalltodo[color=WildStrawberry, #1]{#2}}
    \newcommand{\npcomment}[2][]{\smalltodo[color=SeaGreen, #1]{#2}}
    \newcommand{\vacomment}[2][]{\smalltodo[color=Cerulean, #1]{#2}}

}{
    \newcommand{\mecomment}[1]{}
    \newcommand{\ascomment}[1]{}
    \newcommand{\lilcomment}[1]{}
    \newcommand{\smcomment}[1]{}
    \newcommand{\npcomment}[1]{}
    \newcommand{\vacomment}[1]{}
}

\chapter{Abstract}
\labch{abstract}

{
    \large
    \begin{tabular}{ll}
        \textbf{Title of Dissertation:}    & The First Principles of Deep Learning and Compression \\ \\
                                           & Max Ehrlich                                           \\
                                           & Doctor of Philosophy, 2022                            \\ \\
        \textbf{Dissertation Directed By:} & Professor Abhinav Shrivastava                         \\
                                           & Department of Computer Science                        \\ \\
                                           & Professor Larry S. Davis                              \\
                                           & Department of Computer Science
    \end{tabular}
}
\bigskip

{\setlength{\parindent}{0pt}
    The deep learning revolution incited by the 2012 Alexnet paper has been transformative for the field of computer vision. Many problems which were severely limited using classical solutions are now seeing unprecedented success. The rapid proliferation of deep learning methods has led to a sharp increase in their use in consumer and embedded applications. One consequence of consumer and embedded applications is lossy multimedia compression which is required to engineer the efficient storage and transmission of data in these real-world scenarios. As such, there has been increased interest in a deep learning solution for multimedia compression which would allow for higher compression ratios and increased visual quality.

    The deep learning approach to multimedia compression, so called Learned Multimedia Compression, involves computing a compressed representation of an image or video using a deep network for the encoder and the decoder. While these techniques have enjoyed impressive academic success, their industry adoption has been essentially non-existent. Classical compression techniques like JPEG and MPEG are too entrenched in modern computing to be easily replaced. This dissertation takes an orthogonal approach and leverages deep learning to improve the compression fidelity of these classical algorithms. This allows the incredible advances in deep learning to be used for multimedia compression without threatening the ubiquity of the classical methods.

    The key insight of this work is that methods which are motivated by first principles, \ie, the underlying engineering decisions that were made when the compression algorithms were developed, are more effective than general methods. By encoding prior knowledge into the design of the algorithm, the flexibility, performance, and/or accuracy are improved at the cost of generality. While this dissertation focuses on compression, the high level idea can be applied to many different problems with success.

    Four completed works in this area are reviewed. The first work, which is foundational, unifies the disjoint mathematical theories of compression and deep learning allowing deep networks to operate on compressed data directly. The second work shows how deep learning can be used to correct information loss in JPEG compression over a wide range of compression quality, a problem that is not readily solvable without a first principles approach. This allows images to be encoded at high compression ratios while still maintaining visual fidelity. The third work examines how deep learning based inferencing tasks, like classification, detection, and segmentation, behave in the presence of classical compression and how to mitigate performance loss. As in the previous work, this allows images to be compressed further but this time without accuracy loss on downstream learning tasks. Finally, these ideas are extended to video compression by developing an algorithm to correct video compression artifacts. By incorporating bitstream metadata and mimicking the decoding process with deep learning, the method produces more accurate results with higher throughput than general methods. This allows deep learning to improve the rate-distortion of classical MPEG codecs and competes with fully deep learning based codecs but with a much lower barrier-to-entry.
}

\titlehead{University of Maryland, College Park}
\subject{Submitted to the Faculty of the Graduate School}

\title[The First Principles of Deep Learning and Compression]{The First Principles of\\Deep Learning and Compression
    \begin{tikzpicture}[remember picture, overlay, opacity=0.15]
        \node[anchor=center, xshift=-2cm, yshift=-3cm, scale=3] at (current page.north east){%
            \pgfimage{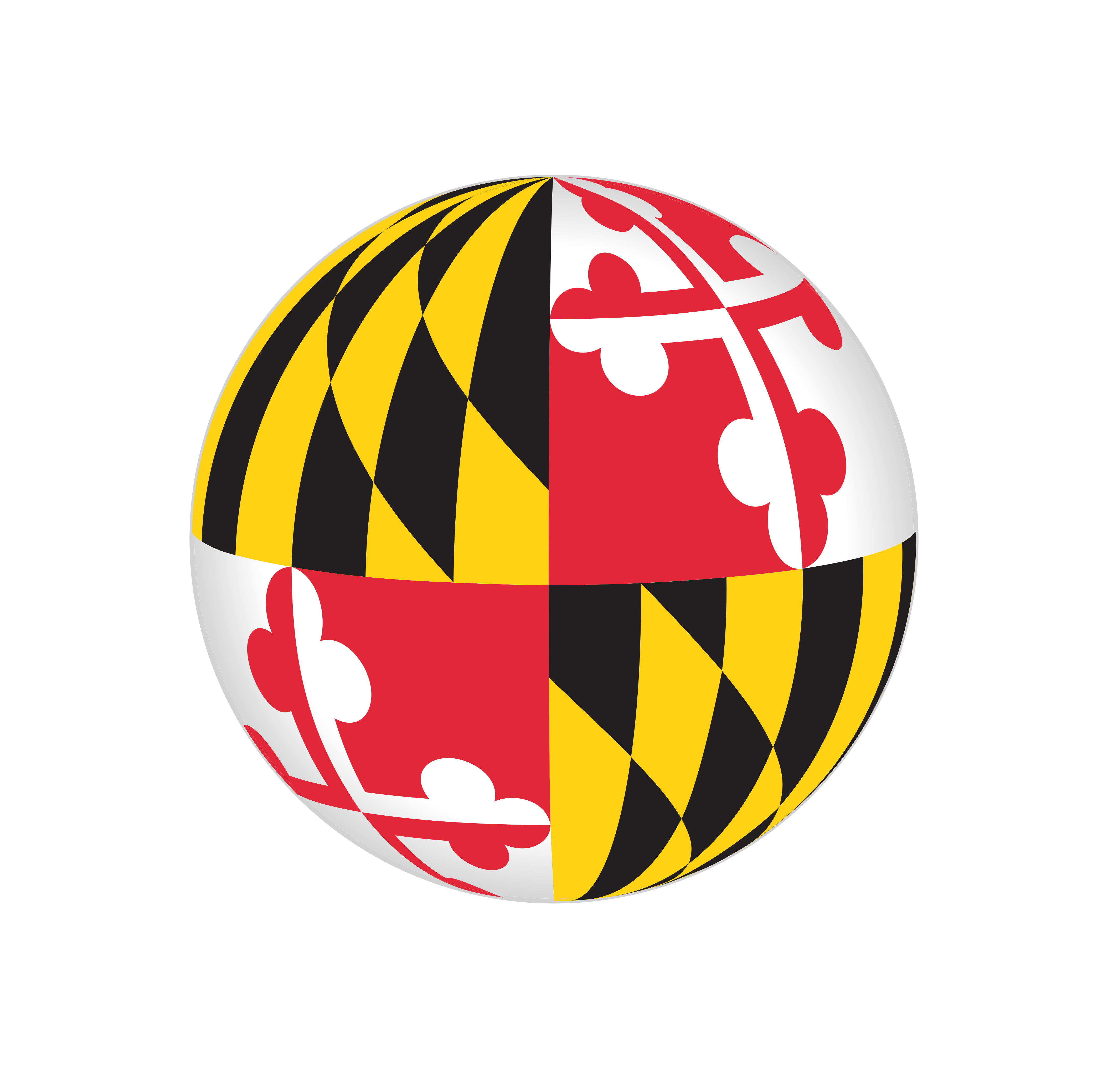}};
    \end{tikzpicture}
}
\subtitle{A Dissertation by}

\author[ME]{Max Ehrlich\\{\footnotesize\texttt{\href{mailto:maxehr@umiacs.umd.edu}{maxehr@umiacs.umd.edu}\qquad\href{https://maxehr.umiacs.io}{https://maxehr.umiacs.io}}}}

\thanks{\textbf{Advisors:} \href{https://www.cs.umd.edu/~abhinav/}{Professor Abhinav Shrivastava}, \href{https://lsd.umiacs.io/}{Professor Larry S. Davis}\newline\textbf{Dean's Representative:} \href{http://www.math.umd.edu/~czaja/}{Professor Wojciech Czaja}\newline\textbf{Examining Committee:} \href{http://users.umiacs.umd.edu/~ramani/}{Professor Ramani Duraiswami}, \href{https://www.cs.umd.edu/people/dmanocha}{Professor Dinesh Manocha}, \href{https://www.sri.com/bios/michael-a-isnardi/}{Dr. Michael A. Isnardi (SRI International)}, \href{http://luthuli.cs.uiuc.edu/~daf/}{Professor David A. Forsyth (UIUC)}}

\date{Wednesday March 30th 2022}

\publishers{For Partial Fulfillment of the Requirements of the Degree of\\Doctor of Philosophy in Computer Science}

\input{front/copyright}

\dedication{
	\noindent{\bfseries To my wife, \href{http://sujeong.kim}{Dr. Sujeong Kim}} \\
You supported me unwaveringly and unconditionally throughout this process and I am eternally grateful. \\

\bigskip
\noindent{\bfseries To my daughter Yena and my son Jaeo} \\
Knowing you will be the greatest privilege of my life.
}

\maketitle

\chapter*{Preface}
\addcontentsline{toc}{chapter}{Preface}
\labch{preface}

\lettrine{M}{ultimedia} compression\index{compression} is a critical feature of the modern internet \parencite{duggan2013photo}. Websites like Facebook, Instagram, and YouTube have increasingly coalesced around the sharing of images and video. Viewing and sharing such media are now prerequisite to modern internet interactions. When comparing media, we can create an approximate hierarchy with each successive level containing an order of magnitude more information. Text, which comprises a simple linguistic description. Images, which contain a full visualization of some scene. And videos, which contain a temporal evolution of the visualization of a scene. Naturally, as the amount of information contained in a particular medium increases, so too does the size of its digital representation.

Because of modern engineering constraints, it is not feasible to transmit image and video media in their native format (\eg, a 2D or 3D array of samples). As an example: a single frame of a 1080p video in a raw format, assuming single byte samples in three colors, would require around $1~\text{byte} \times 3~ \text{channels} \times 1920 \times 1080 = 6220800 ~\text{bytes} \approx 6\text{MB}$ to represent it natively. Extending this to 30 seconds of video at 30 frames per second would require $6220800 ~\text{bytes} \times 30 \text{s} \times 30~\text{fps} = 5598720000 ~\text{bytes} \approx 5\text{GB}$. We can observe that most videos are longer than 30 seconds and 4k videos are becoming more common. Transmitting these media over modern cellular or even wired connections would be quite difficult. A typical home internet connection bandwidth ranges from 10-50 Mbps. For the video example, this would take $5598720000 ~\text{bytes} \times 8 \div [10000000, 50000000]\text{bps} = [4478, 896]\text{s}$, \ie, anywhere from 15 minutes to $1.2$ hours for this short video. The situation is even worse on cellular connections where LTE upload speeds range from 2-5 Mbps \parencite{verizonwireless.com} (almost 3 hours for our example in the best case) and most users pay for a fixed amount of data.

To make the modern internet feasible, by reducing transmission and storage cost, we compress these media. For modern compression codecs, \gls{JPEG} \parencite{wallace1992jpeg} can reduce the 6MB image to around 25kB in size, and H.264 \parencite{richardson2011h} compression: the 5GB video to only a few megabytes depending on the spatial and temporal entropy. These impressive size reductions are a result of more than just entropy reducing operations: they also incur a loss of information. The removed information is designed to be as imperceptible as possible and is based on analyses of human visual perception. For images, we remove ``high spatial frequencies'' \parencite{ahmed1974discrete} or small spatial changes that would not ordinarily be noticed. For videos, we can take a further step to estimate motion between frames and encode only a description of the motion \parencite{le1991mpeg}. For modern codecs, the lossy effects are generally not noticed to laymen, and codec development continues to improve visual fidelity and reduce file sizes year-after-year.

Despite these amazing advances in compression, there are still problems. In many parts of the world, including in rural America, many people use metered internet connections \parencite{schmit2021exploring}. Under these connections they have a finite amount of data or pay for the data they use similar to most modern cell phone plans. For these people, participation in the modern internet is increasingly difficult. Not only are they expected to upload their own media, but they must view others' media in order to take part in the discourse on many websites. For this class of consumer, it is paramount that as minimal data as possible be used during any internet transmission, precluding most videos and some images from being accessible. The internet is historically unprecedented as both an entertainment medium and a repository of human knowledge, and benefits from maximal participation. Therefore, in order to reach these people more effectively, it is critical that further advances in multimedia compression be developed.

Meanwhile, \gls{deep learning}\index{deep learning} has revolutionized modern machine learning \parencite{krizhevsky2012imagenet, he2016deep,tan2019efficientnet}. In \gls{deep learning}, a model is trained to take an input in its native representation and learn a nonlinear mapping directly from it. This is in contrast to classical machine learning which depended on engineered features which were extracted prior to mappings being computed. By taking the native input representation, the deep model can be organized into many layers which function as their own feature extractors. Instead of engineered features, the best features to solve a given problem are learned jointly with the mapping function. The development of these techniques has enjoyed unprecedented success in all areas of machine learning, and these models are being rapidly deployed in consumer settings to solve problems which were once thought to be impossible for computers to solve.

Unsurprisingly, given the previous discussion, one area of interest for \gls{deep learning} applications is that of multimedia \gls{compression}. And also unsurprisingly, \gls{deep learning} has made amazing contributions here \parencite{toderici2015variable, balle2016end, toderici2017full, theis2017lossy, prakash2017semantic, stock2020and}. Deep models are able to compress both images and video significantly better than classical algorithms and with little loss of quality. Despite this, the classical algorithms stubbornly persist. \gls{JPEG} files are still ubiquitous and \gls{MPEG} standards continue to be developed and deployed in consumer application despite the amazing advances of machine learning. These algorithms, and their associated files, are simply too familiar and too ingrained in the code powering the modern internet to be easily replaced. Nevertheless, there is a plain socioeconomic need, as described previously, for deep learning in compression as there is for anything that reduces the size of images and videos.
\vspace{-1em}
\paragraph{In This Dissertation} I take an orthogonal approach to multimedia compression in deep learning. In my approach, I develop deep learning methods which work with the existing compression algorithms rather than replace them. In this way, our algorithms are easy to integrate into the modern internet as simple pre- or post-processing steps on images or videos. These classical compression algorithms are developed with a series of engineering decisions that determine how much and the nature of the information that is lost. I call these decisions ``\gls{first principles}''\index{first principles} and I develop machine learning algorithms that are explicitly aware of these decisions. I will show that this leads to a significant improvement in fidelity and/or flexibility of the solution. These advances have greatly improved the practicality of deploying machine learning solutions to solve compression problems, although their potential applications are widespread.
\vspace{-1em}
\paragraph{Organization Of This Document} This document is organized into three parts. In the first part, I will discuss, briefly, background knowledge that a reader should be equipped with in order to have a full understanding of this \gls{dissertation}. The next two parts discuss related works and my own contributions to image and video compression respectively. This \gls{dissertation} is written in a conversational style, and beyond this preface I will often refer to the reader as ``we''. This is indicating the ``we'', \ie, the reader and I, are discovering the knowledge together as the concepts in the dissertation are developed from prior work into completed topics. I strongly believe in the use of color for guidance. When I believe it will be helpful, I will use color in mathematics and figures to group related ideas. For complex mathematics specifically, I find this to be much clearer than braces alone especially for hinting from early in a derivation which parts of long equations are related and will eventually be grouped together or cancelled. When useful for clarifying an algorithm I have included code listings. These listings are written in something approximating python with pytorch \parencite{NEURIPS2019_9015} APIs where deep learning is required. These code samples are not guaranteed to run exactly as they are written.
\vspace{-1em}
\paragraph{What This Document Is}

\index{dissertation}
This document is, first and foremost, a \gls{dissertation}. This means that its primary purpose is to relay the unique contributions of the author over the course of about five years of research. The astute reader will notice, in the table of contents, section titles which are colored in {\color{Plum}\bfseries Plum}. These sections represent the unique contributions of my research program, \ie, papers which were published in the course of completing my Ph.D. These section titles are colored in the body of the document as well, so it is always easy for the reader to know if they are reading about background work or one of my contributions. Readers will, naturally, find these sections are the most detailed and well developed. In each of these chapters, I have included a dedicated section titled ``Limitations and Future Directions''. No scientific work is perfect and mine are no exception. I believe it is important to be up front about these limitations with a candid discussion along with guidance for future researchers in the field.
\vspace{-1em}
\paragraph{What This Document Is Not}

This document is not a textbook or survey of multimedia compression algorithms and their relationship with deep learning and readers should manage their expectations as such. For the purposes of imparting a full understanding of this dissertation's contributions to scientific discourse, there is a review of elementary concepts of mathematics, machine learning, and compression as well as an overview of related works and recent advances in machine learning. If, in the course of reading this dissertation, a reader gains any useful knowledge, this is welcome but entirely accidental.

\begingroup %

\setlength{\textheight}{230\hscale}

\etocstandarddisplaystyle %
\etocstandardlines %

\begingroup
\renewcommand{\baselinestretch}{1.0}
\tableofcontents %
\endgroup

\listoffigures %
\addcontentsline{toc}{chapter}{List of Figures}

\listoftables %
\addcontentsline{toc}{chapter}{List of Tables}

\renewcommand{\lstlistlistingname}{List of Code Listings}
\lstlistoflistings
\addcontentsline{toc}{chapter}{List of Code Listings}

\endgroup

\renewcommand{\nomname}{Notation}

\nomenclature{$a$}{A scalar (\ie, a member of a Field).}
\nomenclature{$F$}{A field of numbers.}
\nomenclature{$\bm{u}^i$}{A vector (\ie, a member of a Vector space) indexed by $i$.}
\nomenclature{$\bm{v}_j$}{A co-vector (\ie, a member of the dual of a Vector space) indexed by  $j$.}
\nomenclature{$V$}{A vector space.}
\nomenclature{$V^*$}{A co-vector space, the dual of $V$.}
\nomenclature{$M$}{A matrix in the classical sense, a 2D array of numbers, an order-2 tensor with unspecified type (type-(2, 0), (1, 1), or (0, 2)).}
\nomenclature{$S$}{A set, \ie, a collection of elements with no duplicates.}
\nomenclature{$\otimes$}{The tensor product.}
\nomenclature{$U^{abc\ldots}_{ijk\ldots}$}{A type-(m, n) Tensor indexed by $a,b,c,\ldots$ and $i,j,k,\ldots$, a member of the Tensor space generated by $V \otimes V \otimes \ldots \otimes V^* \otimes V^*$.}
\nomenclature{$\star$}{The discrete, valid convolution or cross-correlation operator.}
\nomenclature{$g_{ij}$}{A covariant \gls{metric tensor}.}
\nomenclature{$g^{ij}$}{A contravariant \gls{metric tensor}.}
\nomenclature{$\odot$}{The Hadamard (element-wise) product.}
\nomenclature{$\langle \bm{v}, \bm{u} \rangle$}{The inner product of $\bm{v}$ and $\bm{u}$.}
\nomenclature{$\nabla_x f(\bm{x})$}{The gradient of the scalar valued function of a vector $f()$ with respect to $x$.}
\nomenclature{$\times$}{The Cartesian product of two sets: the set of pairs of elements with one component from each set. Sometimes used to construct a vector (as in $\mathbb{R}^n$). Also used to denote the size (of an image or block, \eg, $8 \times 8$ pixel blocks.)}
\nomenclature{$\delta^i_j$}{The Kronecker delta which is 1 when $i = j$ and zero otherwise.}
\nomenclature{$H(x)$}{The Heaviside function which is 1 when $x > 0$ and 0 otherwise.}
\nomenclature{$\|\bm{x}\|_n$}{The $n$-norm of a vector $\bm{x}$.}
\nomenclature{$\|S\|$}{The cardinality of the set $S$.}
\nomenclature{$\land$}{The logical ``and'' operation which is true if both arguments are true and false otherwise.}
\nomenclature{$\lor$}{The logical ``or'' operation which is false if both arguments are false and true otherwise.}
\nomenclature{$i$}{The imaginary number such that $i^2 = -1$.}
\nomenclature{$\cdot$}{Scalar multiplication.}

\printnomenclature

\mainmatter %
\setchapterstyle{kao} %

\pagelayout{wide}
\addpart{Preliminaries}
\pagelayout{margin}

\setchapterpreamble[u]{\margintoc}
\chapter{Linear Algebra}
\labch{linear_algebra}

\lettrine{T}{o} begin the dissertation, we briefly review the fundamental ideas of linear algebra\index{linear algebra}. These concepts are extremely important for modeling in the high dimensional spaces used by deep learning, and indeed defining what a high dimensional space actually is and how it behaves. Generalizations of linear algebra, which we will cover in the next chapter, have a special relationship with the dissertation outside of this general importance: we will use these ideas to represent JPEG compression. Linear algebra also forms the basis of harmonic analysis which is central to lossy image compression.

\begin{warningbox}
    If you are familiar with the algebraic definitions of linear algebra, this chapter may seem somewhat hand-wavy. It is intended as a general introduction and we will generalize it later.
\end{warningbox}

\section{Scalars, Vectors, and Matrices}

All concepts in mathematics relate back to the foundational idea of the number. For our purposes, we will call a single number a \textit{scalar}\index{scalar}. Scalars will be denoted as a lower case letter in regular font: $a$.

We can ``stack'' several scalars in rows or columns to create \textit{vectors}\index{vector}. When we wish to call attention to a vector, we will use lowercase bold font. For example
\begin{align}
    \bm{b} = \begin{bmatrix}
                 b_0    \\
                 b_1    \\
                 \vdots \\
                 b_n
             \end{bmatrix}
\end{align}
is a vector made by stacking the $n$ scalars in a column.
\begin{align}
    \bm{c} = \begin{bmatrix}
                 c_0 & c_1 & \cdots & c_n
             \end{bmatrix}
\end{align}
is also a vector made by stacking $n$ scalars in a row. We will call $n$ the dimension of the vector. Note that in general $\bm{b} \neq \bm{c}$. We call $\bm{b}$ a vector and $\bm{c}$ a co-vector\index{co-vector}. The distinction will become important later. For now, we define the transpose\index{transpose} operation on a vector which transforms a vector into a co-vector or a co-vector into a vector
\begin{align}
    \bm{c}^T = \begin{bmatrix}
                   c_0    \\
                   c_1    \\
                   \vdots \\
                   c_n
               \end{bmatrix}
\end{align}

Given a scalar and a vector, we can multiply them to produce a new vector
\begin{align}
    \bm{d} = a\bm{c} \\
    = \begin{bmatrix}
          ac_0 & ac_1 & \cdots & ac_n
      \end{bmatrix}
\end{align}
where each component of $\bm{c}$ was multiplied by $a$, thus scaling the vector by $a$ hence the name scalar. We can also add vectors by summing their components to produce another vector. Given a set of $m$ vectors $V$ of dimension $n$.
\begin{align}
    \bm{e} = \sum_{\bm{v} \in V} v = \\
    = \begin{bmatrix}
          \sum_{\bm{v} \in V} v_0 & \sum_{\bm{v} \in V} v_1 & \cdots & \sum_{\bm{v} \in V} v_n
      \end{bmatrix}
\end{align}

We can now combine these operations to create one of the most fundamental ideas of linear algebra: the \gls{linear combination}\index{linear combination}. A linear combination is the sum of the product of some number of vectors and scalars and therefore produces a new vector. Let $S$ be a set of $m$ scalars
\begin{align}
    \bm{g} = \sum_{i=0}^m s_i\bm{v_i}
\end{align}
for $s_i \in S$ and $\bm{v_i} \in V$

Given two vectors, we can multiply them by computing their inner product\index{inner product} which produces a scalar
\begin{align}
    f = \langle\bm{b},\bm{c}\rangle = \sum_{i=0}^{n} b_ic_i
\end{align}
We define the $l_2$ norm\index{l2 norm} of a vector as
\begin{align}
    ||\bm{b}||_2 = \sqrt{\langle\bm{b},\bm{b}\rangle}
\end{align}
We call any vector $\bm{u}$ such that $||\bm{u}||_2 = 1$ a unit vector, noting that we normalize\index{normalize} a vector by computing the vector $\frac{\bm{b}}{||\bm{b}||_2}$. Any two vectors $\bm{v}$ and $\bm{w}$ such that
\begin{align}
    \langle\bm{v},\bm{w}\rangle = 0
\end{align}
are said to be perpendicular or orthogonal\index{orthogonal} to each other.

The formula for the $l_2$ norm
\begin{align}
    ||\bm{b}||_2 = \sqrt{\sum_{i=0}^n b_i^2}
\end{align}
implies a general formulation for an $l_n$ norm
\begin{align}
    ||\bm{b}||_n = \left(\sum_{i=0}^n b_i^n\right)^\frac{1}{n}
\end{align}
Another useful norm is the $l_1$ norm
\begin{align}
    ||\bm{b}||_1 = \left|\sum_{i=0}^n b_i\right|
\end{align}
Taking the limit as $n \rightarrow \infty$ gives the $l_\infty$ norm
\begin{align}
    ||\bm{b}||_\infty = \text{max}(b_i)
\end{align}

We can create two dimensional arrays of scalars which we call matrices\index{matrix} and which we denote with upper case normal font: $A$.
\begin{align}
    A = \begin{bmatrix}
            a_{11} & a_{12} & \cdots & a_{1n} \\
            a_{21} & a_{22} & \cdots & a_{2n} \\
            \vdots & \vdots & \ddots & \vdots \\
            a_{m1} & a_{m2} & \cdots & a_{mn}
        \end{bmatrix}
\end{align}
$A$ is said to be $m \times n$ dimensional.

We multiply a matrix and a vector by taking the \gls{linear combination} of each element of the vector with the corresponding column of the matrix
\begin{align}
    \bm{h} = A\bm{b} = \sum_{i=1}^n b_i\bm{A_i}
\end{align}
note that $b_i$ is the $i$th element of $\bm{b}$ which is a scalar and $\bm{A_i}$ is the $i$th column of $A$ which is a vector. Note also that the result $\bm{h}$ is a vector of dimension $m$. This equation implies that the number of columns in $A$ must match the number of rows in $\bm{b}$. We can extend this to matrix-matrix products, given an $n \times m$ matrix $B$
\begin{align}
    C = AB \\
    = \begin{bmatrix}
          \langle (A^1)^T,B_1\rangle & \langle (A^1)^T,B_2\rangle & \cdots & \langle (A^1)^T,B_m\rangle \\
          \langle (A^2)^T,B_1\rangle & \langle (A^2)^T,B_2\rangle & \cdots & \langle (A^2)^T,B_m\rangle \\
          \vdots                     & \vdots                     & \ddots & \vdots                     \\
          \langle (A^n)^T,B_1\rangle & \langle (A^n)^T,B_2\rangle & \cdots & \langle (A^n)^T,B_m\rangle
      \end{bmatrix}
\end{align}
where $A^j$ denotes the $j$th row\sidenote{We will use upper indices much more frequently than powers in this dissertation so it is advised to get familiar with the notation now. It will be left entirely to context to determine which we mean.} from $A$. In other words, each entry in $C$ is the inner product of the corresponding row of $A$ with the corresponding column of $B$. This construct implies a particular identity matrix\index{identity matrix}
\begin{align}
    I = \begin{bmatrix}
            1      & 0      & \cdots & 0      \\
            0      & 1      & \cdots & 0      \\
            \vdots & \vdots & \ddots & \vdots \\
            0      & 0      & \cdots & 1
        \end{bmatrix}
\end{align}

Matrices are important for representing linear maps on vectors. A linear map is any map which preserves vector addition and scalar multiplication. We can essentially ``store'' the coefficients of linear maps in matrices and use the action of matrix-vector multiplication to apply the map to a vector.

\section{Bases and Finite Dimensional Vector Spaces}

Given a set of vectors $B$, we define the Span\index{span} of $B$ as the set of all linear combinations of the vectors in $B$. Formally, given a set of scalars $S$
\begin{align}
    \text{span}(B) = \left\{ \sum_{i=1}^{|V|} \lambda_i \textbf{b}_i \middle|  \textbf{b}_i \in B, \lambda_i \in S \right\}
\end{align}
Given some arbitrary set of vectors $V$, we may wish to find $B$, \ie, a subset of vectors that spans $V$. If all elements of $B$ are linearly independent, we say that $B$ is a \gls{basis}\index{basis} of $V$. The \gls{basis} allows us to express any element of $V$ in terms of scalars of elements of $B$ and, in effect, defines $V$.

There may be many \glspl{basis} for the same set of vectors, so we may wish to change the basis and we may wish to define a particular basis as canonical. For example, consider the vector space $\mathbb{R}^3$. We often choose the following basis
\begin{align}
    \bm{e_0} = \begin{bmatrix}
                   1 \\ 0 \\ 0
               \end{bmatrix} \; \bm{e_1} = \begin{bmatrix}
                                               0 \\ 1 \\ 0
                                           \end{bmatrix} \; \bm{e_2} = \begin{bmatrix}
                                                                           0 \\ 0 \\ 1
                                                                       \end{bmatrix}
\end{align}
This canonical basis is desirable because the vectors are all orthonormal\index{orthonormal}, \ie, they are all orthogonal to each other and have magnitude of 1 which means there is no ``rotation'' or ``scaling'' of the coordinates. Moreover, this basis makes it extremely easy to express vectors in a familiar component notation. The vector
\begin{align}
    \bm{v} = \begin{bmatrix}
                 1 \\ 2 \\ 3
             \end{bmatrix}
\end{align}
is only expressed as such because we chose this canonical basis (implicitly) and defined $\bm{v}$ as
\begin{align}
    \bm{v} = 1\bm{e_0} + 2\bm{e_1} + 3 \bm{e_2}
\end{align}
If we wish to change the \gls{basis}\index{change of basis}, we first write the coordinates of the new basis ($B_1$) vectors in the old \gls{basis} ($B_0$) and then stack these vectors into a matrix $A$. We can then multiply any vector $\bm{v_0}$ written in terms of basis $B_0$ by $A$ to obtain the coordinates in terms of \gls{basis} $B_1$.
\begin{align}
    \bm{v_1} = A\bm{v_0}
\end{align}
We make the following notes about \glspl{basis} for $V$
\begin{enumerate}
    \item $V$ has a \gls{basis}
    \item All \glspl{basis} of $V$ have the same cardinality which is the dimension of $V$. If we write $\bm{v} \in V$ in coordinates and count the number of coordinates, that count will be the same as the number of basis vectors
\end{enumerate}
Note the implication of the last property, we can count the number of elements in the dimension of V, therefore, V is a finite dimensional vector space\index{finite dimensional vector space}.

\section{Infinite Dimensional Vector Spaces}
\labsec{la:infinite}

Infinite dimensional vector spaces will play an important role in the later analysis of compression, although the results of this analysis will eventually be discretized for use on a computer. In principle, infinite dimensional vector spaces behave in much the same way as finite dimensional ones. While a full treatment of this topic is beyond the scope of the dissertation, we will make some definitions in this chapter which will be expanded upon later.

Assume that $\bm{f}$ and $\bm{g}$ are members of an infinite dimensional vector space $V$. We can think about components of $\bm{f}$ and $\bm{g}$ as being indexed by any real number instead of a finite number of natural numbers. For example we might have $\bm{f}(2) = 4$ for the second component and $\bm{f}(-12.5) = -156.25$ for the negative-twelve-point-five-th component. In other words, $\bm{f}$ and $\bm{g}$ are functions, and these functions are vectors in a vector space.

With that established, our next goal should be to produce a \gls{basis} for these functions. After all, being able to express a function as the coefficients of some \gls{basis} should have myriad uses especially if we do not know exactly the form of the function we wish to express. We will develop this basis later in the dissertation but for now we can define two important concepts: orthogonality of functions and normality of functions.

Recall that two vectors were said to be orthogonal If they point at right angles to each other, \ie, their inner product is zero. To determine orthogonality we need an inner product for functions. We make the following definition
\begin{align}
    \left\langle f(x), g(x) \right\rangle = \int_{-\infty}^\infty f(x)g(x) dx
\end{align}
which is exactly the same as the inner product formula in finite dimensions with the sum expanded to an integral. As usual, if $\left\langle f(x), g(x) \right\rangle = 0$ then the functions are orthogonal.

Next we need to define normality of a function. Recall that a vector was said to be normal if its length is 1. So we need a way of defining the ``length'' of a function. We make the following definition
\begin{align}
    \|f(x)\|_2 = \sqrt{\int_{-\infty}^\infty f^2(x) dx}
\end{align}
Once again, this is the same as the discrete formula using only an integral and if $\|f(x)\| = 1$ then the vector is a normal vector.

Given the tools to determine if a set of functions are orthonormal we can now develop what is essentially a canonical \gls{basis} for functions. This discussion will be continued in \nrefch{frequency_analysis}.

\section{Abstractions}
\labsec{la:abs}

While the geometric interpretations provide useful intuitions, there is a limit to how far we can take them mathematically. We conclude by briefly introducing the abstract forms of the ideas in this chapter.

A field\index{field} $F$ is a set on which addition and multiplication are defined. Specifically we define
\begin{align}
    + : F \times F \rightarrow F \\
    \cdot : F \times F \rightarrow F
\end{align}
and stipulate that if they meet the following criteria for $a,b,c \in F$:
\begin{description}
    \item[Associativity] addition and multiplication are associative: $a + (b + c) = (a + b) + c$ and $a \cdot (b \cdot c) = (a \cdot b) \cdot c$
    \item[Commutivity] addition and multiplication are commutative: $a + b = b + a$ and $a \cdot b = b \cdot a$
    \item[Identity] Two different elements 0 and 1 exist that satisfy additive and multiplicative identity respectively: $a + 0 = a$, $a \cdot 1 = a$
    \item[Inverse] There exists an additive inverse $-a$ and a multiplicative inverse $a^{-1}$ such that $a + (-a) = 0$ and $a \cdot a^{-1} = 1$
    \item[Distributivity] Multiplication and addition distribute according to $a \cdot (b + c) = (a \cdot b) + (a \cdot c)$
\end{description}
then $F$ is a field.

We define a vector space\index{vector space} $V$ over the field $F$ in a similar way. We have two operations
\begin{align}
    + : V \times V \rightarrow V \\
    \cdot : F \times V \rightarrow V
\end{align}
and we call $V$ a vector space, elements of $V$ vectors\index{vector}, and elements of $F$ scalars\index{scalar} if for $\bm{u}, \bm{v}, \bm{w} \in V$ and $a, b \in F$,
\begin{description}
    \item[Associativity] addition is associative: $\bm{u} + (\bm{v} + \bm{w}) = (\bm{u} + \bm{v}) + \bm{w}$
    \item[Commutivity] addition is commutative: $\bm{u} + \bm{v} = \bm{v} + \bm{u}$
    \item[Identity and Inverse] two elements 0 and $-\bm{v}$ exist such that $\bm{v} + 0 = \bm{v}$ and $\bm{v} + (-\bm{v}) = 0$
    \item[Compatibility of Multiplication] Scalar and field multiplication are compatible: $a \cdot (b \cdot \bm{v}) = (a \cdot b) \cdot \bm{v}$
    \item[Scalar Multiplication Identity] Multiplication with the scalar identity: $1 \cdot \bm{v} = \bm{v}$
    \item[Distributivity] Scalar multiplication is distributive with respect to both vector and field addition: $a \cdot (\bm{u} + \bm{v}) = a \cdot \bm{u} + a \cdot \bm{v}$ and $(a + b) \cdot \bm{v} = a \cdot \bm{v} + b \cdot \bm{v}$
\end{description}
Note that we made no mention of coordinates or numbers, we only defined sets and operations along with their behavior.

With these definitions we can form \glspl{linear combination}\index{linear combination} for $\bm{w}, \bm{v_0} \ldots \bm{v_N} \in V$ and $a_0 \ldots a_N \in F$
\begin{align}
    w = \sum_{i=0}^N a_i\cdot\bm{v_i}
\end{align}
\setchapterpreamble[u]{\margintoc}
\chapter{Multilinear Algebra}
\labch{multilinear_algebra}

\lettrine{T}{he} previous chapter developed vectors\index{vector} and matrices\index{matrix} where vectors are a primary ``mathematical object'' in a high-dimensional space and a matrix represents a map which can transform that object. In a sense, this discussion feels unfinished. We had scalars which were zero dimensional, vectors which were one dimensional, and matrices which were two dimensional. Why stop there?

In this chapter we develop the extremely high level ideas of multilinear algebra which generalizes linear algebra to higher dimensional objects. This is a large and complex topic which we only need a small piece of for understanding this dissertation, in fact this entire chapter may be closer to the first lecture in a semester long graduate course. This chapter will immediately obsolete the matrix and vector notation we introduced in the previous chapter for reasons which will be explained in the first section.

The primary goal of multilinear algebra is to study \glspl{multilinear map}. Recall that \glspl{linear map} are maps which preserve vector addition and scalar multiplication. More formally, we call $A: V \rightarrow V$ a \gls{linear map} on the vector space $V$ over the field $F$ if, for $v, u \in V$ and $a \in F$
\begin{itemize}
    \item $A(v + u) = A(v) + A(u)$
    \item $A(a \cdot v) = a \cdot A(v)$
\end{itemize}
A bilinear map is an extension of this concept to two arguments where the map is linear in each argument. We call $B : V \times V \rightarrow V$ a bilinear map with vector space $V$ and field $F$ for $v_0, v_1, u_1, u_1 \in V$ and $a \in F$ if
\begin{itemize}
    \item $B(v_0 + v_1, u_0) = B(v_0, u_0) + B(v_1, u_0)$ and $B(v_0, u_0 + u_1) = B(v_0, u_0) + B(v_0, u_1)$
    \item $B(a \cdot v_0, u_0) = B(v_0, a \cdot u_0) = a \cdot B(v_0, u_0)$
\end{itemize}
Continuing this until its natural end, we call a \gls{multilinear map} a function of multiple arguments which is linear in each one\sidenote{One important thing to note at this point is that we can only tweak \textbf{one} argument at a time, we cannot, for example, compute $B(v_0 + v_1, u_0 + u_1)$ and expect a linear result.}. $M: V \times V \times \cdots \times V \rightarrow V$ is a multilinear map with $v_{0 \cdots N}, u \in V$ and $a \in F$ if
\begin{itemize}
    \item $M(v_0 + u, v_1, \cdots, v_N) = M(v_0, v_1, \cdots, v_N) + M(u, v_1, \cdots, v_N)$ and $M(v_0, v_1 + u, \cdots, v_N) = M(v_0, v_1, \cdots, v_N) + M(v_0, u, \cdots, v_N)$, \etc
    \item $M(a \cdot v_0, v_1, \cdots, v_N) = M(v_0, a \cdot v_1, \cdots, v_N) \cdots M(v_0, v_1, \cdots, a \cdot v_N) = a\cdot M(v_1, v_1, \cdots, v_N)$
\end{itemize}
We will represent \glspl{multilinear map} using higher order objects called tensors\index{tensor}. Perhaps surprisingly, the practical use of these concepts in the dissertation will still only be linear or bilinear maps, however, we leverage multilinear algebra by working on tensor inputs and outputs which serve as a natural representation for images \vs the vectors that are traditionally used in these maps.

\section{Tensors}

Traditionally in computer science we think of tensors as multidimensional arrays of numbers. Despite the protests of many physicists and mathematicians, this is a perfectly reasonable definition of a tensor. For example, we might have a 3- or 4- or 5D array of numbers and call this a tensor. In the mathematical sense, a tensor is a representation of a \gls{multilinear map}. We will denote tensors and tensor spaces with uppercase math font: $T$.

Recall the concepts of vectors\index{vector} and co-vectors\index{co-vector}. We will refer to vector spaces as $V$ and co-vector spaces as $V^*$. It is important to keep in mind that although these spaces are related they are not the same. These vectors and co-vectors will, in a sense, be the primitives that we use to construct tensors. We will index vectors using upper indices and co-vectors using lower indices.

All tensors have a type, which is the primary way we will refer to them. Some texts refer to a tensor rank, we do not use this convention because it is ambiguous. Rank has other meanings in linear algebra and tensor rank does not explain the composition of the tensor in terms of vectors and co-vectors. If we absolutely have to refer to the sum of the number of vector and co-vector spaces we will call this the order of the tensor, though this situation will be extremely rare. We will say that vectors are type-(1, 0) tensors and co-vectors are type-(0, 1) tensors. Matrices can then be type-(2,0) tensors, type-(1, 1) tensors, or type-(0, 2) tensors. The distinction between type is important. Since matrices and vectors now have concrete definitions as tensors, this obsoletes our earlier notation which drew a distinction between them. From this point on, all non-scalars will be written in tensor notation.

\section{Tensor Products and Einstein Notation}

We construct arbitrary tensors using products of vectors and co-vectors. To do this we define the tensor product of two tensors. We will build up to this by revisiting some concepts from linear algebra. Given two vectors $v, u$ in some vector space $V$ on a field $F$, we defined the inner product as
\begin{align}
    \sum_{i=0}^N v_i u_i = a
\end{align}
where $a \in F$. Given a matrix (a type-(1, 1) tensor) $M$ we can compute the matrix-vector product as
\begin{align}
    \sum_{i=0}^N M_i x_i = w
\end{align}
for $w \in V$. Similarly we compute the matrix-matrix product given another matrix $N$ as
\begin{align}
    \sum_{i=0}^N M_iN^i = O
\end{align}
These expressions can by simplified using Einstein notation \sidecite{einstein1923grundlage}. In Einstein notation, repeated indices that appear as upper and lower indices are assumed to be summed out, allowing us to remove the summations from the previous equations. For example, the matrix-matrix product is now simply
\begin{align}
    M^j_iN^i_k = O^j_k
\end{align}
where the non-summed indices are added in for clarity. This is extremely important when working with general tensors because the expressions are quite verbose with summation notation. We will make heavy use of Einstein notation in this dissertation so it is important to understand it now.

Given two arbitrary tensors we can now define the generic tensor product
\begin{align}
    T \otimes U = T_{l_0, l_1, \cdots, l_N}^{u_0, u_1, \cdots, u_N} U_{l'_0, l'_1 \cdots, l'_N}^{u'_0, u'_1, \cdots, u'_N} = V_{l_0, \cdots, l_N, l'_0, \cdots, l'_N}^{u_0,\cdots, u_N, u'_0 \cdots, u'_N}
\end{align}
Of course we are free to form other useful products for tensors. For example, given a type-(2, 3) tensor $P$ and a type-(4, 2) tensor $Q$ we could compute the type-(4, 3) tensor $R$ as
\begin{align}
    P_{ij}^{kml}Q^{ij}_{abcd} = R^{kml}_{abcd}
\end{align}
where we have summed out the $i,j$ indices.

To construct a tensor from vectors and co-vectors we can use this tensor product. Consider the vectors $u, v \in V$ and the co-vectors $p, q, r \in V^*$. We can construct a type-(3, 2) tensor from these by computing
\begin{align}
    u^i v^j p_k q_l r_m = T^{ij}_{klm}
    \labeq{mla:tp}
\end{align}

In many situations it will be useful for us to raise or lower indices (sometimes called index juggling\index{index juggling}). In other words, given a tensor $T_{ij}$ we may want to construct $T^i_j$ or $T^{ij}$. These tensors are related to $T_{ij}$ but they are not the same. We can accomplish this by multiplying $T$ by the covariant or contravariant \gls{metric tensor}\index{metric tensor} which relates to vector and co-vector spaces. These tensors are defined such that
\begin{align}
    g^{ik}g_{kj} = \delta^i_j
\end{align}
where $\delta$ is the Kronecker delta
\begin{align}
    \delta^i_j = \begin{cases}
                     0 & i \neq j \\
                     1 & i = j
                 \end{cases}
\end{align}
a generalization of the identity matrix from linear algebra, $g^{ij}$ is the contravariant metric (for converting co-vectors to vectors) and $g_{ij}$ is the covariant metric (for converting vectors to co-vectors).

For various reasons we will consider a general derivation of the \glspl{metric tensor} to be beyond the scope of this \gls{dissertation}, and in fact we will always be using tensors defined with respect to the canonical basis which has a metric of $\delta$. This means we can freely raise and lower indices without considering the metric.

\section{Tensor Spaces}

If we needed to start with vectors every time we wanted to build a tensor it would quickly become unsustainable. Instead, we need a way to refer to tensor spaces, or sets of tensors. This is sometimes referred to as the intrinsic definition of a tensor. We again use the tensor product but this time we use vector and co-vector spaces. Recalling the type-(2, 3) tensor $T$ which we constructed from vectors and co-vectors, we can define $T$ directly as
\begin{align}
    T \in V \otimes V \otimes V^* \otimes V^* \otimes V^*
\end{align}
in other words, $V \otimes V \otimes V^* \otimes V^* \otimes V^*$ defines a space of tensors. This space contains all tensors which can be constructed by the tensor product of $V$ twice and $V^*$ three times. In other words, all tensors which can be built from an equation like \refeq{mla:tp} but with any $u ,v \in V$ and $p, q, r \in V^*$.

For a generic tensor T, we say that it is of type-(p, q) for
\begin{align}
    T \in \underbrace{V \otimes \cdots \otimes V}_{\text{$p$ times}} \otimes \underbrace{V^* \otimes \cdots \otimes V^*}_{\text{$q$ times}}
\end{align}
This will be the primary convention that we use to define tensors in the rest of this dissertation. Note that this mimics some of the definitions from \nrefsec{la:abs} in that we no longer have need of coordinates, we only deal with arbitrary vector spaces, co-vector spaces, and the tensor products of their members which is why this is called the intrinsic definition. We close by noting that although we have only used $V$ and $V^*$, in general, the vector spaces defining a tensor can be different provided that the spaces are defined over the same field.

\section{Linear Pixel Manipulations}

\begin{marginfigure}
    \centering
    \includegraphics{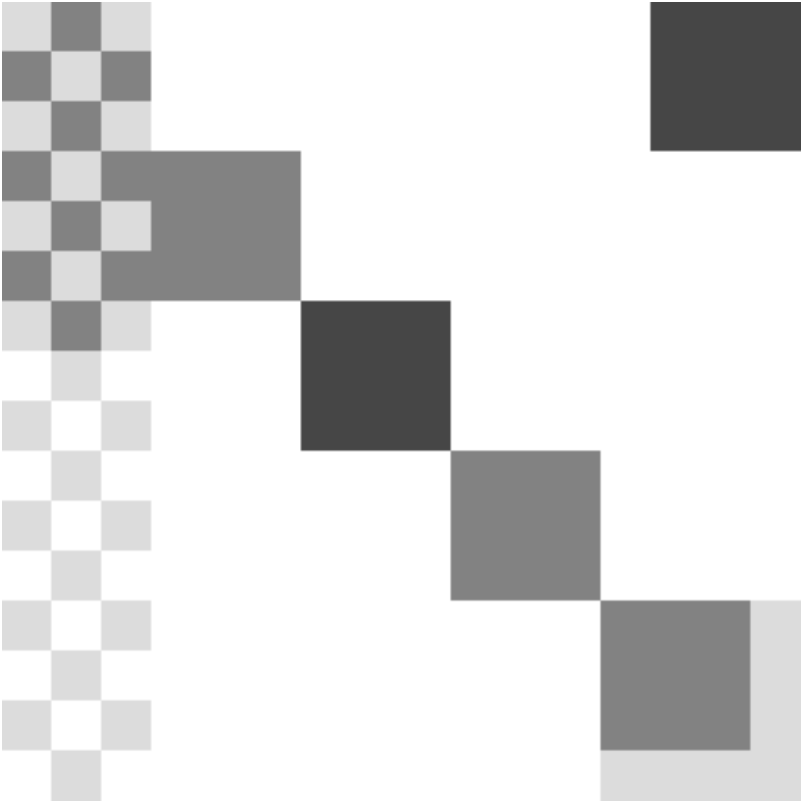}
    \caption[Grayscale Example Image]{\textbf{Grayscale Example Image.}}
    \labfig{mla:probe_gray}
\end{marginfigure}
With the boring theory out of the way we can look at an interesting practical application of tensors: linear pixel manipulations\index{linear pixel manipulations}. By representing an \gls{image} as a tensor we can compute many complex transformations of the image using other tensors. Some of these are not traditionally thought of as being ``linear'' when we restrict our thinking to two-dimensional matrices as linear maps that transform images through matrix multiplication. Instead of thinking of images as ``collections of vectors'' we treat the image as one object: a higher order tensor and then we consequently define the linear map on this object in even higher dimensions.

More formally, we will deal with planar images. The image may have any number of channels but it always has two spatial dimensions. So a grayscale image would be a type-(0, 2) tensor. A traditional color image would be a type-(0, 3) tensor. In most cases, even for color images it will suffice to define linear maps as type-(2, 2) tensors which transform the spatial dimensions while preserving the channel dimension.

\begin{marginfigure}
    \centering
    \includegraphics{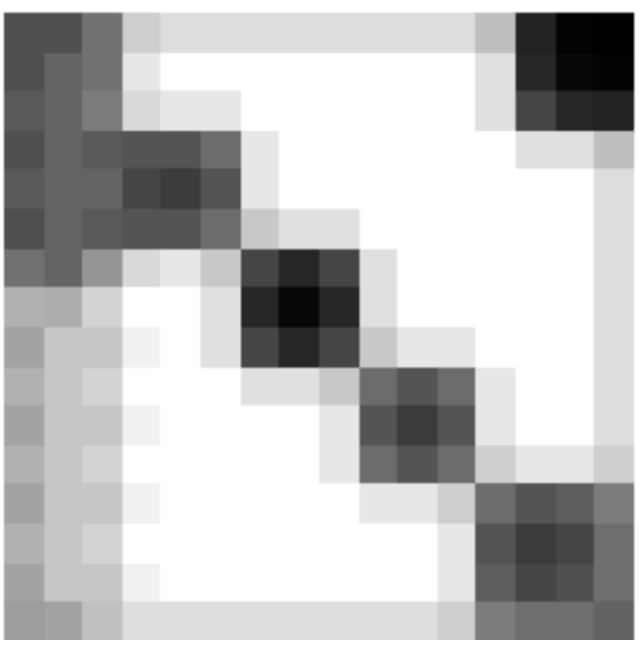}
    \caption[Grayscale Gaussian Smoothing]{\textbf{Grayscale Gaussian Smoothing.}}
    \labfig{mla:smooth_gray}
\end{marginfigure}
We begin with a simple example. Consider the example image in \reffig{mla:probe_gray}. We can represent this grayscale image as a type-(0, 2) tensor $I \in H^* \otimes W^*$. One simple linear manipulation we can perform on this image is Gaussian smoothing in a $3 \times 3$ window. We can represent this linear map as a type-(2, 2) tensor
\begin{align}
    G : H^* \otimes W^* \rightarrow H^* \otimes W^* \\
    G \in H \otimes W \otimes H^* \otimes W^*       \\
    G^{ij}_{uv} = \begin{cases}
                      0.5   & i = u \land j = v                      \\
                      0.125 & i = u \land (j = v - 1 \lor j = v + 1) \\
                      0.125 & (i = u - 1 \lor i = u + 1) \land j = v \\
                      0     & \text{otherwise}
                  \end{cases}
\end{align}

\begin{marginfigure}
    \centering
    \includegraphics{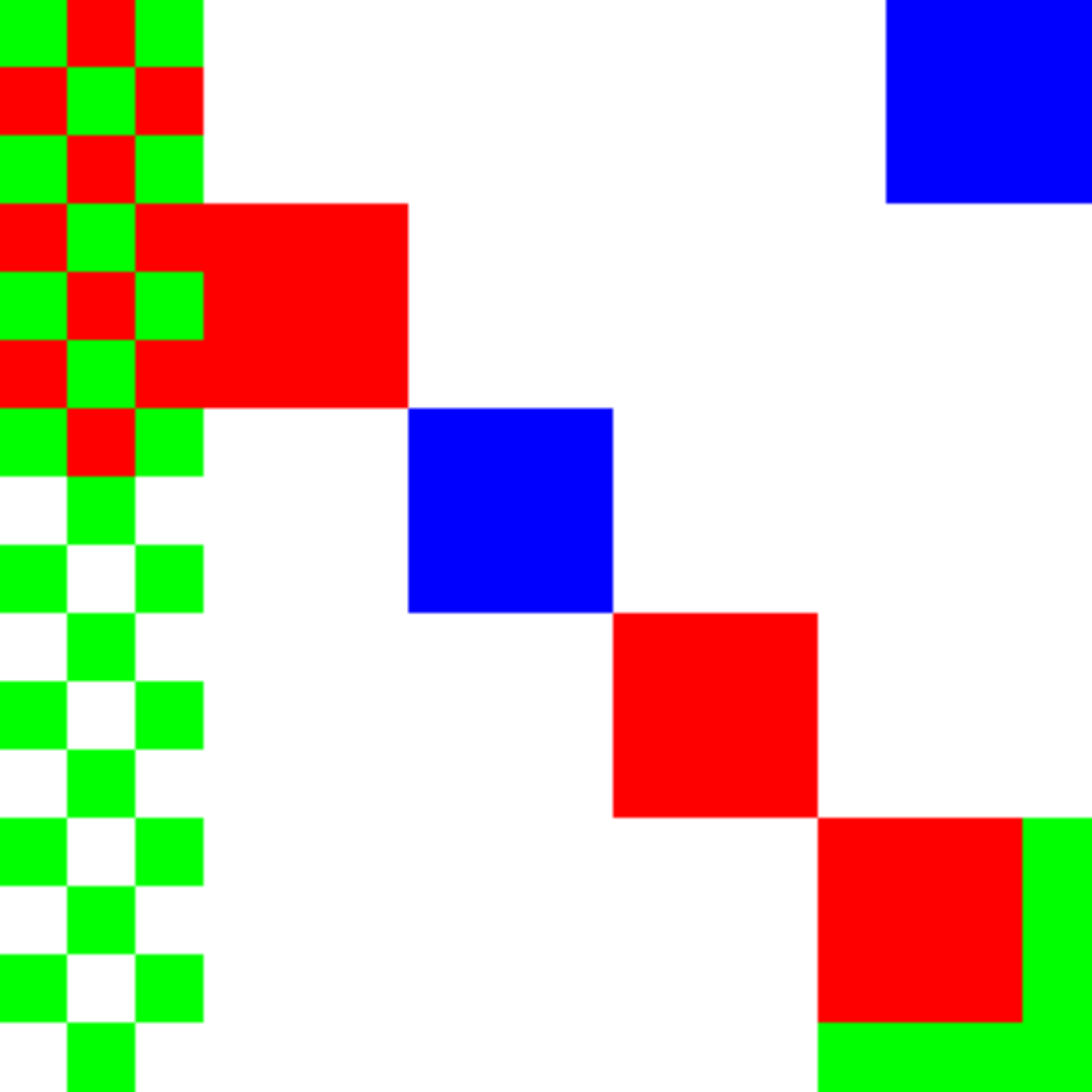}
    \caption[Color Example Image]{\textbf{Color Example Image.}}
    \labfig{mla:probe}
\end{marginfigure}
From the first equation, we can see that $G$ is a linear map on type-(0, 2) tensors that transforms them into type-(0, 2) tensors. From the second equation we see that $G$ is a type-(2, 2) tensor (this is a consequence of the first equation). The third equation defines the form of $G$ for arbitrary indices $i,j,u,v$. In this case, $i, j$ index the input pixel and $u, v$ index the output pixel, the value stored at the index is the coefficient of the pixel. So that is $0.5$ when the indices are equal and $0.125$ for any neighboring pixels, all other pixel have a zero coefficient. We apply this linear map by computing
\begin{align}
    I'_{uv} = G^{ij}_{uv}I_{ij}
\end{align}
The result of this computation is show in \reffig{mla:smooth_gray}.

Next we can consider a color image. The color version of the example images is shown in \reffig{mla:probe}. Converting this color image to grayscale is a linear manipulation. We represent the color image as $I \in P^* \otimes H^* \otimes W^*$. We then define the following linear map
\begin{align}
    Y : P^* \otimes H^* \otimes W^* \rightarrow H^* \otimes W^* \\
    Y \in P \otimes H \otimes W \otimes H^* \otimes W^*         \\
    Y^{pij}_{uv} = \begin{cases}
                       0.299 & p = 0 \\
                       0.587 & p = 1 \\
                       0.114 & p = 2
                   \end{cases}
\end{align}
which comes directly from the grayscale conversion equation
\begin{align}
    Y = 0.299 R + 0.587 G + 0.114 B
\end{align}
\begin{marginfigure}[-5\baselineskip]
    \centering
    \includegraphics{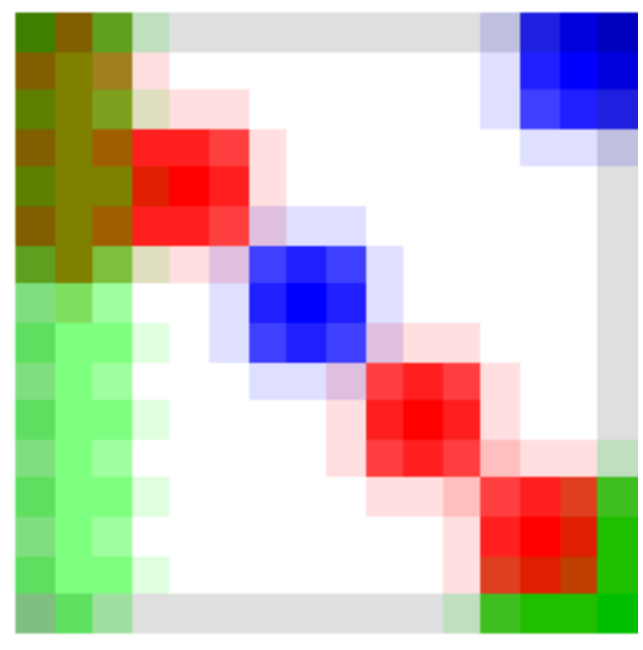}
    \caption[Color Smoothing]{\textbf{Color Smoothing.}}
    \labfig{mla:color_smooth}
\end{marginfigure}
We apply this map as
\begin{align}
    I'_{uv} = Y^{pij}_{uv}I_{pij}
\end{align}

If we apply this map to the example image we get the same image as \reffig{mla:probe_gray}. Interestingly, we can apply $G$ to this color image as well and it will perform correct smoothing on the color image (\reffig{mla:color_smooth}). In this case we would be computing
\begin{align}
    I'_{puv} = G^{ij}_{uv}I_{pij}
\end{align}
and since $G \in H \otimes W \otimes H^* \otimes W^*$, the channel dimension of $I$, $P^*$, is preserved.

\begin{marginfigure}
    \centering
    \includegraphics{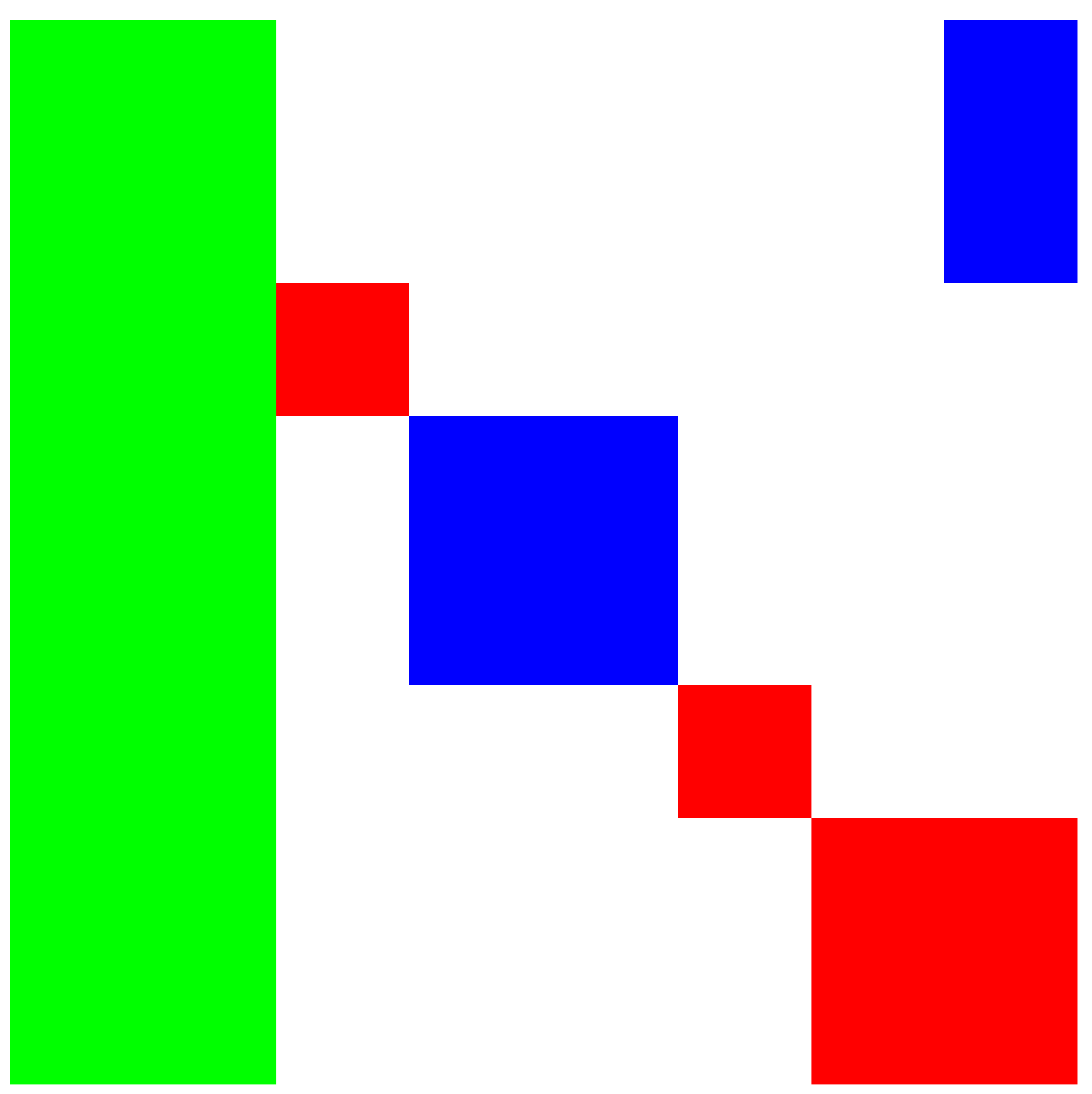}
    \caption[Color Downsampling]{\textbf{Color Downsampling.}}
    \labfig{mla:downsample}
\end{marginfigure}
Let's try something more interesting: resampling. We can define nearest neighbor up- and downsampling as linear maps. This works for both color and grayscale images, the computation is the same and the tensor will be type-(2, 2). For downsampling by a factor of 2 we define the following linear map
\begin{align}
    H : H^* \otimes W^* \rightarrow H'^* \otimes W'^* \\
    H \in H \otimes W \otimes H'^* \otimes W'^*       \\
    H^{ij}_{uv} = \begin{cases}
                      1 & i = 2u \land j = 2v \\
                      0 & \text{otherwise}
                  \end{cases}
\end{align}
where $H'$ and $W'$ are vector spaces with half the dimension of $H$ and $W$. We can define upsampling in a similar way. For upsampling by a factor of 2 we define the following linear map
\begin{align}
    D : H'^* \otimes W'^* \rightarrow H^* \otimes W^* \\
    H \in H' \otimes W' \otimes H^* \otimes W^*       \\
    H^{ij}_{uv} = \begin{cases}
                      1 & i = \lfloor u / 2 \rfloor \land j =  \lfloor v / 2 \rfloor \\
                      0 & \text{otherwise}
                  \end{cases}
\end{align}
We apply these maps by computing
\begin{align}
    I'_{uv} = H^{ij}_{uv}I_{ij} \\
    I_{uv} = D^{ij}_{uv}I'_{ij}
\end{align}
for grayscale images and
\begin{align}
    I'_{puv} = H^{ij}_{uv}I_{pij} \\
    I_{puv} = D^{ij}_{uv}I'_{pij}
\end{align}
for color images. The result for the color image is shown in \reffig{mla:downsample}.

Taking this further, we can define any \gls{convolution}\index{convolution} or \gls{cross-correlation}\index{cross-correlation} using tensors. This is reasonable since we know that convolutions are linear operations although we do not always see them written out as \glspl{linear map}. We consider a general convolution kernel $K$ with any shape. We will denote the shape of the kernel as the tuple $S = (s_0, s_1)$. Then we define the following linear map
\begin{align}
    C : H^* \otimes W^* \rightarrow H^* \otimes W^* \\
    C \in H \otimes W \otimes H^* \otimes W^*       \\
    C^{ij}_{uv} = \begin{cases}
                      K_{u - i + s_0, v - j + s_1} & u - s_0 \leq i \leq u + s_0 \land v - s_1 \leq j \leq v + s_1 \\
                      0                            & \text{otherwise}
                  \end{cases}
\end{align}
note that this does not consider a mapping between channels like we would use in a convolutional network (this is simple enough to add in though). We apply this to grayscale or color images as
\begin{align}
    I'_{uv} = C^{ij}_{uv}I_{ij} \\
    I'_{puv} = C^{ij}_{uv}I_{pij}
\end{align}

As a taste of what's to come, let's try something more surprising. It may sound surprising but breaking an \gls{image} into evenly sized blocks is a linear operation, and we can derive a tensor which represents this map. We will first define two new co-vector spaces, the block dimensions $M^*$ and $N^*$\sidenote{For example if we wanted even $8 \times 8$ blocks we might write these as $\mathbb{R}^{8*}$ although I do not like this notation}. We will also define the spaces $X^*$ and $Y^*$ with dimension equal to the dimension of $H^*$ and $W^*$ divided by the block size (\ie, the number of blocks that can fit in the image). Then we can define a type-(2, 4) tensor, the \gls{linear map}
\begin{align}
    B : H^* \otimes W^* \rightarrow X^* \otimes Y^* \otimes M^* \otimes N^* \\
    B \in H \otimes W \otimes X^* \otimes Y^* \otimes M^* \otimes N^*       \\
    B^{ij}_{xymn} = \begin{cases}
                        1 & \text{pixel $h, w$ belongs in block $x, y$ at offset $m, n$} \\
                        0 & \text{otherwise}
                    \end{cases}
\end{align}
which may seem like kind of a let down but this is the canonical form we will use later in the dissertation. A more satisfying and programmer-oriented definition might be
\begin{align}
    B^{ij}_{xymn} = \begin{cases}
                        1 & x \cdot \text{dim}(M) + m = i \land y \cdot \text{dim}(N) + n = j \\
                        0 & \text{otherwise}
                    \end{cases}
\end{align}
We apply the map as
\begin{align}
    I'_{pxymn} = B^{ij}_{xymn}I_{pij}
\end{align}

\begin{marginfigure}[-10\baselineskip]
    \centering
    \includegraphics{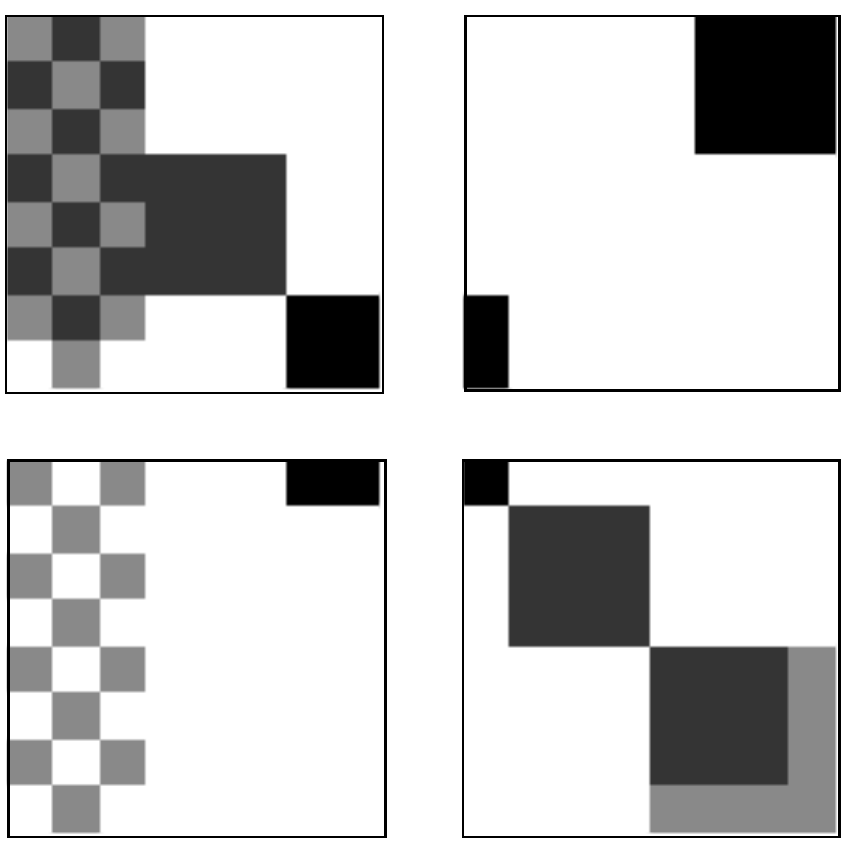}
    \caption[Block Linear Map Example]{\textbf{Block Linear Map Example.} The blocks are arranged spatially but note that in tensor form there are separate indices for the block position and the 2D offset into each block.}
    \labfig{mla:blocks}
\end{marginfigure}
Since this one might be a little confusing, consider a concrete example with the example image in \reffig{mla:probe_gray}. This is a $16 \times 16$ image and we want to break it into $8 \times 8$ blocks, so there will be four total blocks in a $2 \times 2$ grid (\reffig{mla:blocks}). In this case, $\text{dim}(M) = \text{dim}(N) = 8$ and $ \text{dim}(X) = \text{dim}(Y) = 2$. So after applying $B$ to the $16 \times 16$ image we would get a tensor of shape $2 \times 2 \times 8 \times 8$ giving the spatial arrangement of the $8 \times 8$ blocks.

While this was a fun exercise the actual practical application of this idea is fairly limited since the tensors must be on the order of the image size. A critical component of the dissertation is that we can actually represent all of JPEG as a linear map. This is extremely powerful because linear maps are well studied phenomena, so expressing something as complex as JPEG as a single linear map gives us myriad tools for further analysis and manipulation.

\setchapterpreamble[u]{\margintoc}
\chapter{Harmonic Analysis}
\labch{frequency_analysis}

\lettrine{H}{armonic} analysis\index{harmonic analysis} is an invaluable tool for mathematics and engineering that enables some of the most important technologies in existence today.  In \nrefsec{la:infinite} we touched briefly on the concept of infinite dimensional vector spaces and we noted that the vector space of functions of real variables is one such space. In this chapter we expand upon this idea and introduce the Fourier transform\index{fourier transform} and harmonic analysis. The ideas we present in this chapter will be fundamental guiding principles behind \gls{image} and \gls{video} compression.

Fourier was interested in solving the heat equation\index{heat equation} which describes the temperature of an ideal length of wire over space and time. The equation defines a function $u(t,x)$ as a partial differential equation with conditions:
\begin{align}
    \frac{\partial }{\partial t} u(t, x) = \frac{\partial^2}{\partial x^2}u(t,x) \\
    u(0, x) = f(x)                                                               \\
    u(t, 0) = 0                                                                  \\
    u(t, 1) = 0
\end{align}
for $t \geq 0$ and $x \in [0, 1]$.

Of critical importance to us is the second equation which relates the form of $u(t, x)$ at time $t = 0$ to some arbitrary function of space. Fourier showed that, since the other conditions of the heat equation yield a harmonic function, one composed of simple waves, $f(x)$ must be able to be decomposed as such \sidenote{It is interesting to note that although this result is one of the most influential results in all of engineering, it was given negative reviews at the time Fourier published it.}. While we will not go into the full derivation that Fourier used, or even touch on the modern understanding of the transform, we will show how this implies an orthonormal basis for functions which allows us to express them as a sum of coefficients of simple waves.

\begin{kaobox}[frametitle=Note]
    There are many different ways to think about the fourier transform. Fourier was thinking in terms of the heat equations, many people like to envision a ``machine'' that isolates frequencies. I prefer the model which is motivated by linear algebra and that is what I discuss in this chapter although all views on the subject are equally correct and interesting.
\end{kaobox}

\section{The Fourier Transform}
\labsec{fourier}

Recall our definitions for the $l_2$ norm and inner product of functions
\begin{align}
    \|f(x)\|_2 = \sqrt{\int_{-\infty}^\infty f^2(x)\, dx} \\
    \langle f(x), g(x) \rangle = \int_{-\infty}^\infty f(x)g(x)\, dx
\end{align}
given these tools we can try to find something resembling a canonical \gls{basis}\index{canonical basis} for functions. We would like a canonical \gls{basis} to be a set of functions that is orthonormal\index{orthonormal}, \ie, a set of functions which are all of unit length and which are all orthogonal to each other.

Consider the functions $\sin(x)$ and $\cos(x)$. We can show easily that these functions are orthogonal to each other by solving
\begin{align}
    \langle \sin(x), \cos(x) \rangle = \int_{-\infty}^\infty \sin(x)\cos(x)\, dx
\end{align}
We start by restricting the domain to $[-\pi,\pi]$ since the sine and cosine functions are periodic.
\begin{align}
    = \int_{-\pi}^\pi \sin(x)\cos(x)\, dx
\end{align}
Then we use substitution to solve the integral. Let $u = \cos(x)$ and $du = -sin(x)\,dx$
\begin{align}
    \int_{-\pi}^\pi \sin(x)\cos(x)\, dx = \int u\, -du \\
    = -\int u\,du                                      \\
    = - \frac{u^2}{2} + C
\end{align}
substituting and evaluating the result gives
\begin{align}
    - \frac{u^2}{2} + C = -\frac{\cos^2(x)}{2} + C     \\
    \left. -\frac{\cos^2(x)}{2} + C \right|_{-\pi}^\pi \\
    = -\frac{\cos^2(-\pi)}{2} + \frac{\cos^2(\pi)}{2} = 0
\end{align}
so sine and cosine are indeed orthogonal. To check if they are normal we compute
\begin{align}
    \int_{-\pi}^\pi \cos^2(x)\,dx \\
    \int_{-\pi}^\pi \sin^2(x)\,dx
\end{align}
We can solve the first integral with the trigonometric identity
\begin{align}
    \cos^2(x) = \frac{1 + \cos(2x)}{2}
\end{align}
substituting gives
\begin{align}
    \int_{-\pi}^\pi  \frac{1 + \cos(2x)}{2}\,dx    \\
    = \frac{1}{2} \int_{-\pi}^\pi 1 + \cos(2x)\,dx \\
    = \frac{1}{2} \left(\int_{-\pi}^\pi\,dx + \int_{-\pi}^\pi \cos(2x)\,dx\right)
\end{align}
\begin{align}
    = \left. \frac{1}{2} \left(x + \frac{\sin(2x)}{2}\right) \right|_{-\pi}^\pi     \\
    = \left. \frac{x}{2} + \frac{\sin{2x}}{4} \right|_{-\pi}^\pi                    \\
    = \frac{\pi}{2} + \frac{\sin{2\pi}}{4} - \frac{-\pi}{2} - \frac{\sin{-2\pi}}{4} \\
    = \pi + \frac{\sin{2\pi}}{4} - \frac{\sin{-2\pi}}{4}                            \\
    = \pi
\end{align}
We get the same result for sine, so the functions are not normal but they can be easily made normal by dividing by $\pi$. Therefore, sine and cosine seem like ideal candidates provided we can produce an infinite set from these two.

In order to have a basis for the infinite dimensional space of functions we need an infinitely large set of basis vectors. Without further elaboration, the \gls{fourier transform} defines this set as
\begin{align}
    \{\sin(-2\pi x \zeta), \cos(-2\pi x \zeta ) | \zeta \in \mathbb{R}\}
\end{align}
or simply
\begin{align}
    \{e^{-2\pi i x \zeta}    | \zeta \in \mathbb{R}\}
\end{align}
Note that this is an uncountable infinite set of vectors, which is what we needed, and we call $\zeta$ the frequency. The actual integral transform\index{integral transform} is then
\begin{align}
    F(\zeta) = \int_{-\infty}^\infty f(x) e^{-2 \pi i x \zeta}\, dx
\end{align}
Note that, as we described for the norm and inner product of functions, this is simply generalizing the expression for a \gls{linear combination} of a finite dimensional vector with its basis vectors.

As useful as this result is, it is not readily applicable to computation as is the case with many concepts dealing with infinity. We can, however, define the \gls{dft}\index{discrete fourier transform} as the following type-(1, 1) tensor
\begin{align}
    F \in \mathbb{C}^N \otimes \mathbb{C}^{N*} \\
    F_{mn} = \frac{1}{\sqrt{N}}e^{-2\pi i \frac{mn}{N}}
    \labeq{ha:dft}
\end{align}
$F$ is a linear map $F: \mathbb{C}^N \rightarrow \mathbb{C}^N$ acting on complex vectors of dimension $N$. Note that $F$ is symmetric, \ie, $F^i_j = F^j_i$. For practical applications, this matrix multiply would be prohibitively expensive so we use the fast Fourier transform to recursively memoize the transform result reducing the number of computations to $O(N\log(N))$. We do not describe this algorithm in detail here.

There are some other transforms which are related to the \gls{dft} and are useful. Specifically a major downside to the \gls{dft} is the dependence on complex numbers. For many discrete applications, real numbers would work fine. This motivates the \gls{dst}\index{discrete sine transform} \sidecite{jain1976loeve, kekre1978comparative} and the \gls{dct}\index{discrete cosine transform} \sidecite{ahmed1974discrete}.

These transforms can be thought of as taking only the imaginary (sine) or real (cosine) part of the \gls{dft}. We can get away with this on discrete samples by assuming that the signal, outside of the region we sampled, is an odd or even function. We are free to do this since we do not care at all about what the function actually looks like outside where we sampled so it does not need to be accurate.

For our purposes, the \gls{dct} will play an outsize role since it is central to our later discussion of \gls{JPEG}. The \gls{dst} will come up briefly in video coding, however. The \gls{dct} can be defined differently depending on how boundary conditions are handled. We will not detail all of these, but the two important ones for us are the DCT-II, which we will call ``the \gls{dct}``, and is defined in two dimensions as
\begin{align}
    D^i_j = \frac{1}{\sqrt{2N}}C(i)C(j)\sum_{x=1}^N\sum_{y=1}^N \cos\left[\frac{(2x+1)i\pi}{2N}\right]\cos\left[\frac{(2y+1)j\pi}{2N}\right] \\
    C(u) = \begin{cases}
               \frac{1}{\sqrt{2}} & u = 0    \\
               1                  & u \neq 0
           \end{cases}
\end{align}
and the DCT-III, which we will call ``the inverse \gls{dct}'', and is defined in two dimensions as
\begin{align}
    (D^{-1})^x_y = \frac{1}{\sqrt{2N}}\sum_{i=1}^N\sum_{j=1}^NC(i)C(j)\cos\left[\frac{(2x+1)i\pi}{2N}\right]\cos\left[\frac{(2y+1)j\pi}{2N}\right]
\end{align}
In both cases, $C(u)$ is a scale factor which makes the transform orthonormal. As in the \gls{dft}, these are both \glspl{linear map}, this time with $D : \mathbb{R}^N \rightarrow \mathbb{R}^N$ and $D^{-1} : \mathbb{R}^N \rightarrow \mathbb{R}^N$ and are type-(1, 1) tensors.

We note here an important theorem which will be useful for us later in the dissertation

\begin{theorem}[The DCT Least Squares Approximation Theorem]
    \labthm{dctlqa2}
    Given a set of $N$ samples of a signal ${X}$, let ${Y}$ be the DCT coefficients of ${X}$. Then for $1 \leq m \leq N$ the approximation of $X$ given by
    \begin{align}
        p_m(t) = \frac{1}{\sqrt{N}}y_0 + \sqrt{\frac{2}{N}}\sum_{k=1}^m y_k \cos\left(\frac{k(2t+1)\pi}{2N}\right)
        \labeq{ha:dct1d}
    \end{align}
    minimizes the least-squared error
    \begin{align}
        e_m = \sum_{i=1}^N(p_m(i) - x_i)^2
    \end{align}
\end{theorem}
\begin{proof}
    First consider that since \refeq{ha:dct1d} represents the Discrete Cosine Transform, which is a Linear map, we can write rewrite it as
    \begin{equation}
        D^T_my = x
    \end{equation}
    where $D_m$ is formed from the first $m$ rows of the DCT matrix, $y$ is a row vector of the DCT coefficients, and $x$ is a row vector of the original samples.

    To solve for the least squares solution, we use the the normal equations, that is we solve
    \begin{equation}
        D_mD^T_my = D_mx
    \end{equation}
    and since the DCT is an orthonormal transformation, the rows of $D_m$ are orthonormal, so $D_mD^T_m = I$. Therefore
    \begin{equation}
        y = D_mx
    \end{equation}
    Since there is no contradiction, the least squares solution must use the first $m$ DCT coefficients.

\end{proof}

A related transform to the ``trigonometric'' transforms is the \gls{hadamard transform} or Walsh-Hadamard transform\index{Hadamard transform}. The \gls{hadamard transform} defines the transformation matrix recursively as
\begin{align}
    H_0 = 1 \\
    H_m = \begin{bmatrix}
              H_{m - 1} & H_{m - 1}  \\
              H_{m - 1} & -H_{m - 1}
          \end{bmatrix}
\end{align}
The obvious advantage of this transform is that it contains only $-1$ and $1$ entries, so it can be computed quite efficiently without even multiplication operations (only sign changes are needed).

\section{The Gabor Transform}

While the \gls{fourier transform} is useful for telling us what frequencies make up a given signal, it cannot tell us when those frequencies occur. It considers all the samples we have and tells us which frequencies explain all the samples. In some cases, it would be useful to know both which frequencies occur and where they occur. For example, if we are examining seismic data, it may be important to know when high frequency vibrations occurred to predict the time of a future earthquake. With a Fourier transform, we would only know that there were high frequency vibrations.

We can accomplish this in a naive way with a \gls{stft}\index{short-time fourier transform}. The high level idea is extremely simple. The input signal is broken up into smaller blocks of time and the Fourier transform is computed on each block separately. Then, for each block of time we can see which frequencies are available, and we can adjust the block size to increase the time resolution.

The \gls{gabor transform} is an interesting twist on this idea. Instead of a hard window, we use a soft window by convolving the \gls{fourier transform} with a Gaussian kernel. In a continuous representation this is
\begin{align}
    G(\tau, \zeta) = \int_{-\infty}^\infty f(t) e^{-\pi(t - \tau)^2}e^{-2 \pi i t \zeta} \; dt
\end{align}
yielding amplitude results with time offsets $\tau$ as well as frequencies $\zeta$ \sidenote{in contrast to the Fourier result which is amplitude \vs frequency with no time component}. While this yields a smooth windowed response in time, it still suffers from what we call the uncertainty principle which all STFTs are subject to. That is, the larger the time window, the worse the localization is, and the smaller the time window, the more constrained we are in the frequencies we can represent. Put another way, time-resolution and frequency-resolution are inverses: can only have one and not both.

To see this result, consider the \gls{dft} matrix given in \refeq{ha:dft}. This matrix has a finite number of frequencies that it can represent because of the discrete representation. The high frequency represents each sample in a single period. If we restrict the size of the \gls{dft} to windows, as in the \gls{gabor transform}, we reduce the size of this matrix and therefore we reduce the number of frequencies we can represent. Conversely, if we allow the size of the window to increase without bound, so as to get the best frequency resolution, we will eventually end up with a window size that is the length of the original signal and therefore is equivalent to the standard \gls{fourier transform} that has no temporal component at all. As we will see in the next section, this uncertainty principle extends to more sophisticated methods and is a fundamental limitation of harmonic analysis.

\begin{figure}[t]
    \centering
    \includegraphics{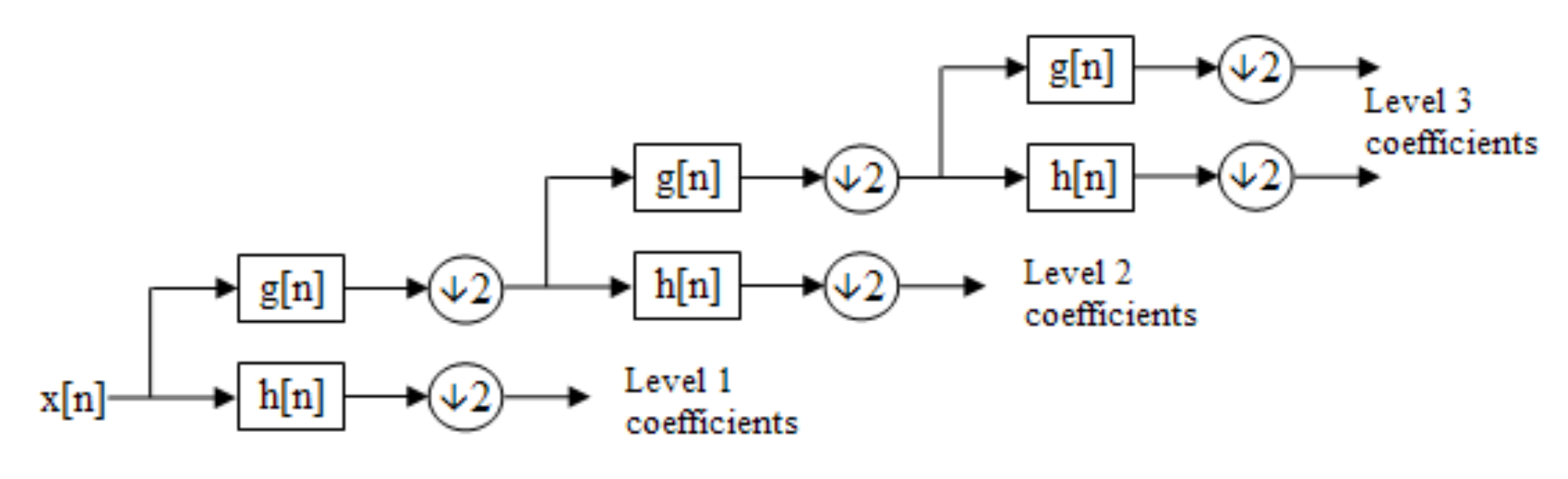}
    \caption[Discrete Wavelet Transform]{\textbf{Discrete Wavelet Transform.} The DWT repeats the sampling process recursively on the low frequency band.}
    \labfig{ha:dwt}
\end{figure}

\vspace{1em}
\section{Wavelet Transforms}

\begin{marginfigure}[2\baselineskip]
    \centering
    \includegraphics{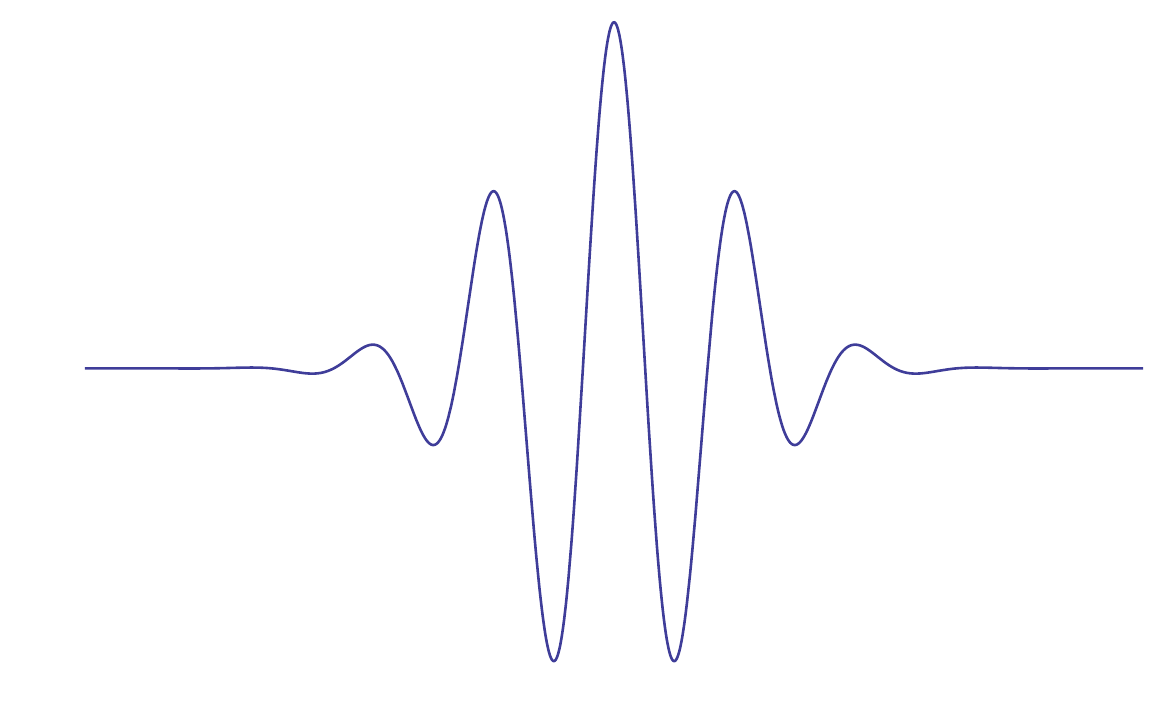}
    \caption[Morlet Wavelet]{\textbf{Morlet Wavelet.} The Morlet wavelet illustrates the high amplitude in the center of the wave with decreasing amplitude moving to the sides. \textit{Image credit: Wikipedia.}}
    \labfig{ha:morlet}
\end{marginfigure}

\Gls{wavelet}\index{wavelet} transforms \index{wavelet transforms} extend the concept of the \gls{stft} to what, at the time of writing, can be considered its natural end. Instead of using sine and cosine bases, the \gls{wavelet transform} defines other functions which have "finite support". In other words, they have a high amplitude at time $t = 0$ with the amplitude gradually decreasing as $t$ moves away from 0 (this is shown in \reffig{ha:morlet} with the Morlet wavelet). As in the \gls{stft}, this measures a local response to the \gls{wavelet}. Then, as in the \gls{gabor transform}, we can slide the \gls{wavelet} around by shifting it along the input signal to compute local responses at different times.

The key improvement of wavelet transforms is that they include a term which controls the frequency of the wave. This allows for a full bank of frequencies to be computed at each time representing the response of the signal to wavelets of increasing frequency. Note that because of the uncertainty principal, this generates a tree-like structure. For a given time $t$, there may be multiple high frequency wavelet responses for a single low frequency wavelet (\reffig{ha:uncertainty}). As in the last section, the more precisely we wish to describe the constituent frequencies in a signal the less precisely we can localize them in time. Unlike the last section, however, we can still localize the high frequencies well even if we cannot localize the low frequencies, with a \gls{stft}, our localization capability is defined entirely by the block size (or Gaussian standard deviation for the \gls{gabor transform}). Since we examine the same signal at multiple scales, or resolutions, we call this \gls{multiresolution analysis}\index{multiresolution analysis}.

\begin{marginfigure}
    \centering
    \includegraphics{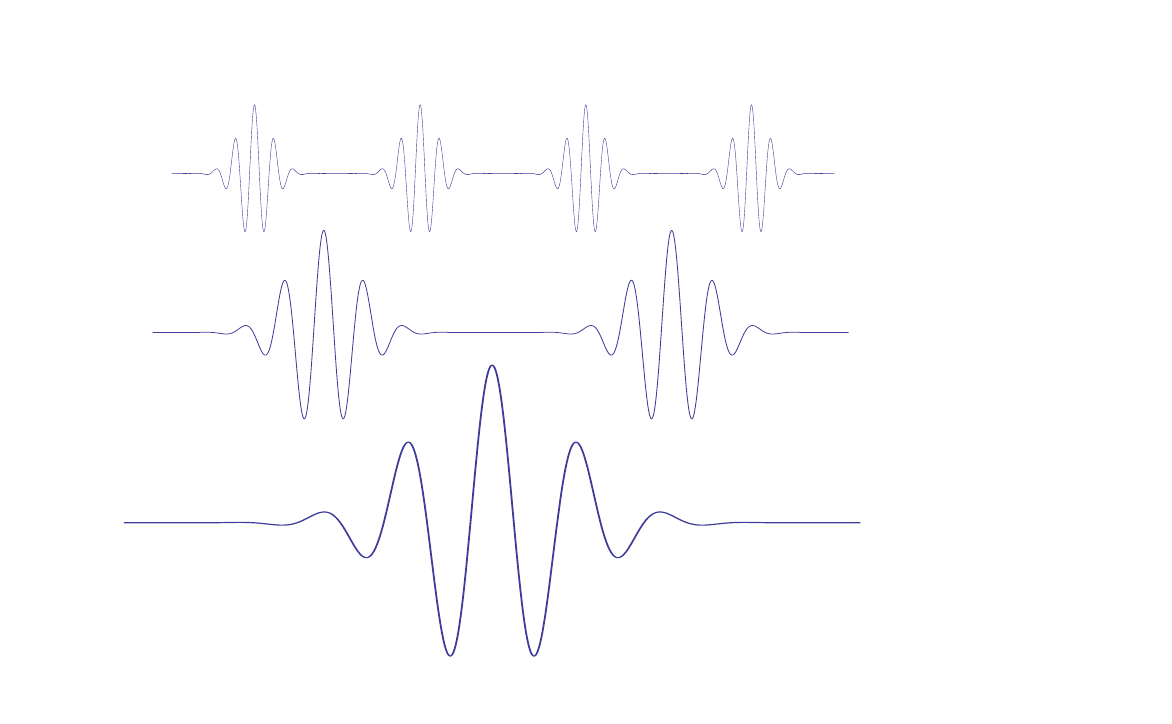}
    \caption[Wavelet Uncertainty]{\textbf{Wavelet Uncertainty.} The low frequency wavelet has poor time resolution, in other words, we cannot tell as exactly the time where that frequency occured as we can with the high frequency wavelets. \textit{Image credit: wikipedia.}}
    \labfig{ha:uncertainty}
\end{marginfigure}

Formally, we define a mother wavelet\index{mother wavelet} $\psi(t)$ which we can then shift and scale as desired. This yields a basis for the space of functions, just as with the fourier transform, given by the following set
\begin{align}
    W = \left\{ \psi_{\zeta, s} (t) \; \middle|\; \zeta \in \mathbb{R}, s \in \mathbb{R}, \psi_{\zeta, s}(t) = \frac{1}{\sqrt{\zeta}}\psi\left(\frac{t - s}{\zeta}\right) \right\}
\end{align}
where $\zeta$ determines the frequency (or scale) of the wavelet and $s$ determines the shift. We then compute the integral transform
\begin{align}
    T(\zeta, s) = \int_{-\infty}^\infty f(t)\psi_{\zeta, s}(t)\;dt
    \labeq{ha:wavtran}
\end{align}
for a function of time (a signal) $f(t)$. Just as with the \gls{fourier transform} this is simply a linear combination of the signal with each of the basis entries, but we have generalized from the Fourier basis ($e^{-2\pi i t \zeta}$) to the more general $\psi(f, s)$. In the rest of this section, we will discuss how to apply the wavelet transform to discrete signals and how certain important wavelets are defined.

\subsection{Continuous and Discrete Wavelet Transforms}

As with the \gls{fourier transform}, in order to use these tools on real signals, we must discretize them for execution on a computer. There are several ways we can do this, the first one we will discuss is the \gls{cwt}\index{continuous wavelet transform} which, despite the name, is not exactly continuous. To define this we simply assume that the signal $f(t)$ is finite and discretely sampled, and we rewrite the integral of \refeq{ha:wavtran} as a sum
\begin{align}
    T(\zeta, s) = \sum_k f_k \psi_{\zeta,s}(k)
\end{align}
then we stipulate that the wavelet function have finite support\index{finite support}, in other words, we assume that it is zero outside of a certain range so we can represent it with a finite number of samples. We can then define the wavelet transform using \gls{convolution}. We define the kernel
\begin{align}
    \widetilde{\psi}_{\zeta, t} = \frac{1}{\zeta}\psi\left(\frac{\zeta T - t}{\zeta}\right)
\end{align}
for a wavelet with support $T$ and we compute
\begin{align}
    T_{st}^m = f_t \star \widetilde{\psi}_{\zeta,m + \zeta T}
\end{align}

The \gls{dwt}\index{discrete wavelet transform} takes this idea further. The idea is that instead of dealing with the wavelet basis change equations directly, we can simply express the transform as a series of high pass/low pass filters which coarsely discretize the scale. We first construct \gls{convolution} kernels for a high pass and low pass filter $g$ and $h$ and compute the \glspl{convolution}
\begin{align}
    y_{\text{low}} = f \star g \\
    y_{\text{high}} = f \star h
\end{align}
By definition, these filters pass half the frequencies they are given as input. Therefore, by the \gls{nyquist}\index{nyquist sampling theorem}, we can also discard half the samples of each result without losing information. We represent this with a downsampling by two operation ($\downarrow$)
\begin{align}
    y_{\text{low}} = (f \star g) \downarrow 2 \\
    y_{\text{high}} = (f \star h) \downarrow 2
\end{align}
This process is repeated recursively on $y_\text{low}$ while $y_\text{high}$ is retained as an output. This yields a tree structure (\reffig{ha:dwt}).

We briefly mention a newer technique here, the \gls{dtcwt} \sidecite{selesnick2005dual}\index{dual tree complex wavelet transform}. This is a complex \gls{wavelet transform} which is inspired by real cosine and imaginary sine components of the \gls{fourier transform}. The main advantage of this transform is shift invariance, \ie, a shift in the input signal yields the same transform coefficients. While the theory of the \gls{dtcwt} is quite involved, the algorithm is simple assuming suitable wavelets exist. As in the \gls{dwt}, high and low pass filters are applied with the results decimated, only this time there are two wavelets producing two trees (\reffig{ha:dtcwt}). The results of one tree are treated as the real part of a complex output and the results of the other tree are used as the imaginary part.

All of these methods require suitable definitions of the $\psi(t)$ function. While the natural instinct is to choose orthogonal \glspl{wavelet}, biorthogonal wavelets\index{biorthogonal}, which relax the orthogonal constraint as long as the transform is still invertible, have also been shown to work well and have more flexibility in their design. Note that the definition of a basis does not require orthogonality. Common choices for $\psi(t)$ include the Haar wavelets\index{Haar wavelet} (discussed next), the Morlet wavelet\index{Morlet wavelet} which is related to the Gabor transform, and the Daubechies wavelets\index{Daubechies wavelet} among others. While most tasks will work fine with the simplistic Haar wavelets, knowing the properties of each wavelet to pick the ideal one for a given task can make a difference.

\begin{figure}[t]
    \centering
    \includegraphics{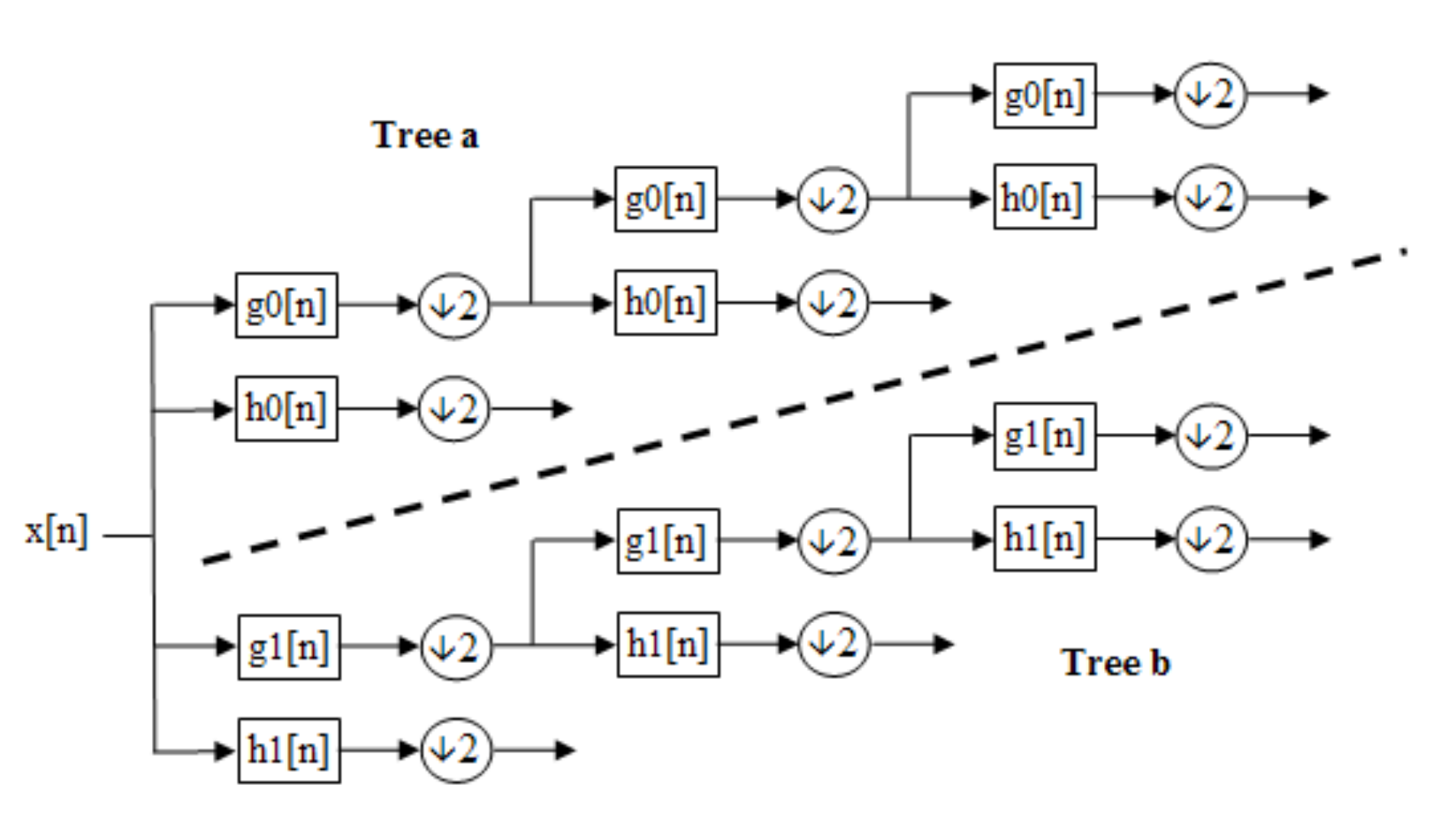}
    \caption[Dual Tree Complex Wavelet Transform]{\textbf{Dual Tree Complex Wavelet Transform.} The DTCWT is computed in the same way as the DWT but with two trees.}
    \labfig{ha:dtcwt}
\end{figure}

\subsection{Haar Wavelets}

\begin{marginfigure}
    \centering
    \includegraphics{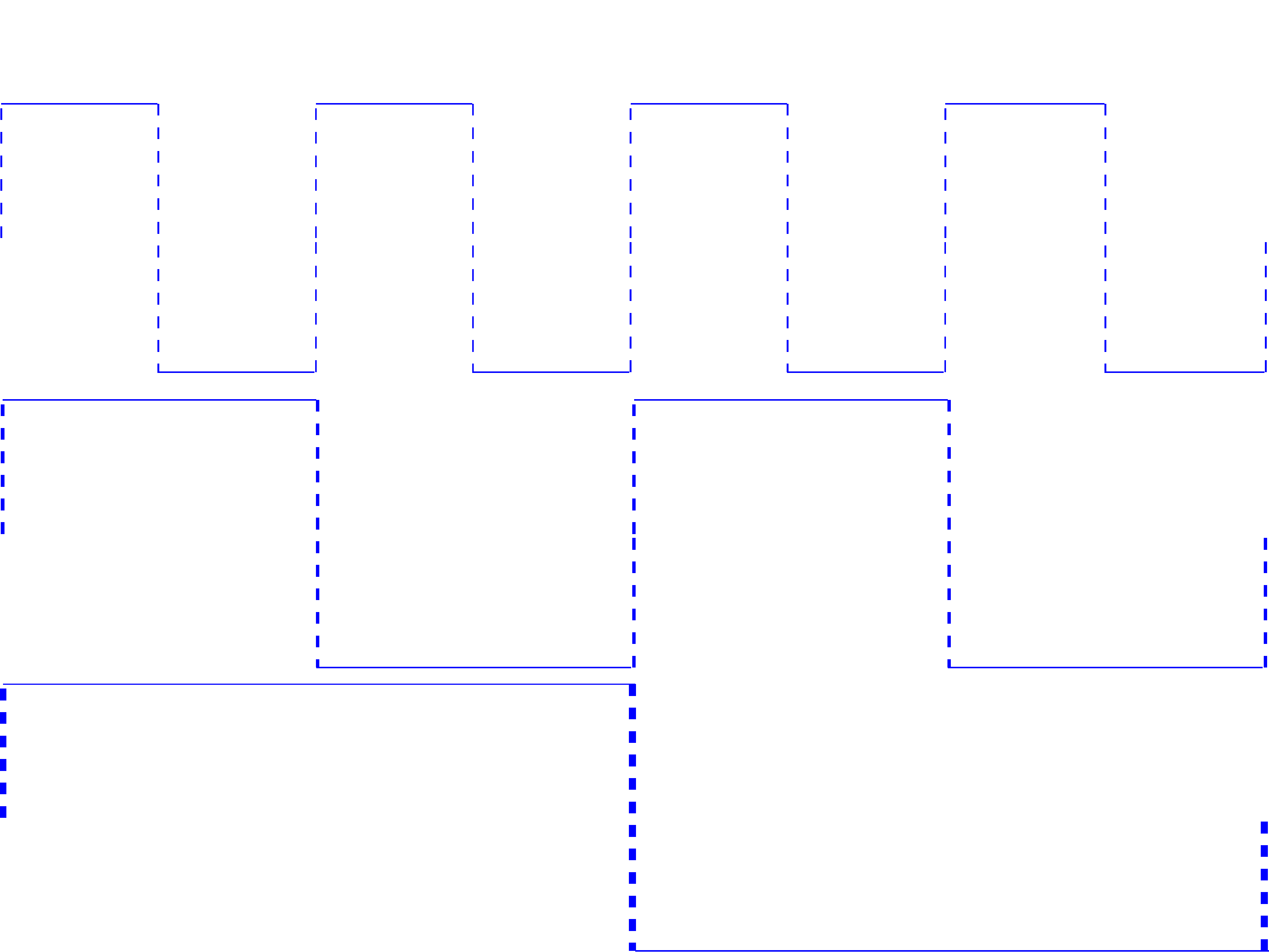}
    \caption[Haar Wavelet]{\textbf{Haar Wavelet.} Frequency increases vertically, time increase to the right.}
    \labfig{ha:haar}
\end{marginfigure}

The Haar wavelet\index{Haar wavelet} is one of the most simple and popular choices for $\psi(t)$. It is defined as
\begin{align}
    \psi(t) = \begin{cases}
                  1  & 0 \leq t \leq \frac{1}{2}  \\
                  -1 & \frac{1}{2}  \leq t \leq 1 \\
                  0  & \text{otherwise}
              \end{cases}
\end{align}
The Haar wavelet transform is simple to implement and computationally efficient leading to its widespread use. The wavelets have compact support and are orthogonal making the Haar transform effective for conducting localized frequency analysis, in fact they were the first attempt at a basis for multiresolution analysis. The 1D Haar wavelet is plotted in \reffig{ha:haar} for three frequencies and several shifts per frequency. Note that the time axis (horizontal) spans from 0 to 1. The Haar wavelet has very compact support, outside the support region, which naturally shrinks with increasing frequency, the value of the wavelet is zero, so any samples outside the considered region contribute no information to the frequency response.

In the 1D transform the wavelet was measuring differences along the time axis to measure the frequency response. In 2D, we must consider differences on two axes including the diagonal (both axes simultaneously). \reffig{ha:haar_dwt} shows an example of this for a single level DWT. Note that each of the four frequency bands, called LL, HL, LH, HH, are stored at half the width and height leading to the $4 \times 4$ arrangement on the left hand side. In this case, the top-left is the LL band, the top right is the LH band, the bottom left is the HL band, and the bottom right is the HH band. Note the different features that each band responds to: the HL and LH bands respond to horizontal and vertical structures respectively and the HH band respond to diagonal structures.

\begin{figure}[t]
    \centering
    \includegraphics{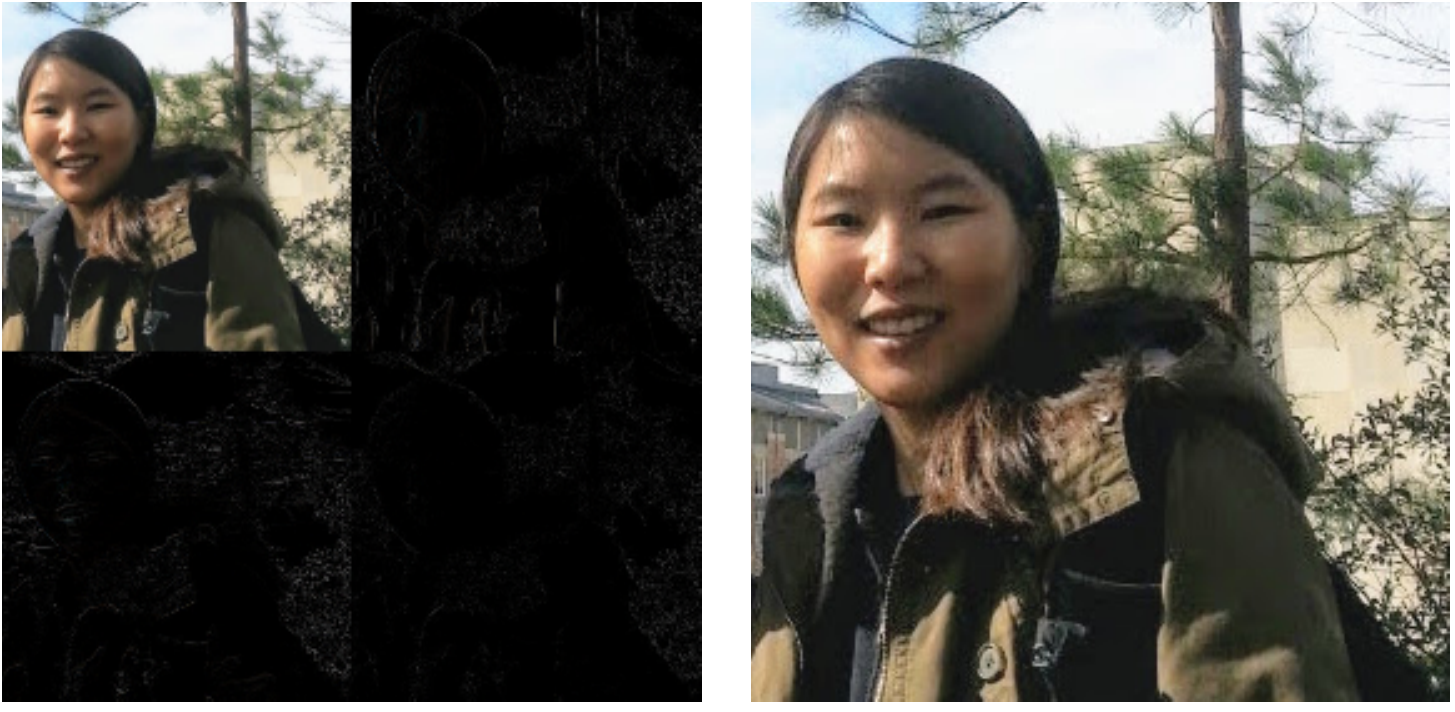}
    \caption[DWT Using Haar Wavelets]{\textbf{DWT Using Haar Wavelets.} The left image is the single level DWT of the right image. Note that each filtered image is stored at half the resolution in the width and height so each of the four filtered images can be arranged in the same shape as the original image.}
    \labfig{ha:haar_dwt}
\end{figure}

While the Haar transform's simplicity and effectiveness allow for widespread use there may be more suitable wavelets for a given task. The Daubechies wavelets\index{Daubechies wavelet} \sidecite{daubechies1992ten} in particular have come into common use as they were designed based on the analysis of Ingrid Daubachies who made numerous contributions to multiresolution analysis. For example Daubachies showed that if the number of vanishing moments\index{vanishing moments} is $N$, then the support of the wavelet is at least $2N - 1$. Vanishing moments, which relate the wavelet to a polynomial, can be of critical importance in choosing a wavelet if there is some understanding of the function to be analysed. Generally, a wavelet with $N$ vanishing moments is orthogonal to a polynomial of degree $N - 1$.

In this section we covered only the most basic ideas of multiresolution analysis as it does not factor into the work of this dissertation. However, the \gls{wavelet transform}, which was a critical part of the last decade of signal processing, is now making its way rapidly into deep learning applications \sidecite{liu2018multi, bruna2013invariant, zhao2022wavelet} \sidenote{among many others including currently unpublished work.} so knowledge of these techniques will rapidly become important for the computer vision researchers.
\setchapterpreamble[u]{\margintoc}
\chapter{Entropy and Information}
\labch{entropy}

\lettrine{I}{nformation} theory\index{information theory} marked a major advancement in the understanding of communication. Claude Shannon's 1943 paper ``A Mathematical Theory of Communication'' was rare in that it both introduced the field of information theory and then systematically solved all major problems within it, essentially an entire field in one paper. Importantly for us, Shannon's formulations for measuring the information contained in a message gave rise to lossless compression\index{lossless compression} algorithms which are still used to this day. In this chapter, we review the high level ideas of information theory, specifically \gls{entropy}\index{entropy}, and how these ideas were used to develop compression algorithms.

The overall goal of information theory~\sidecite{shannon1948mathematical} is to measure the amount of information contained in a signal. The signal can be discrete (\eg, words) or continuous (\eg, television, sound, \etc). Shannon was responding to a recent development in communication: modulation. These techniques were rudimentary \gls{lossy compression} methods which introduced noise into the messages in exchange for reducing their size (similar to \gls{JPEG} and \gls{MPEG} as we will see later). Exactly how much noise was introduced and the limits of the system with respect to how much noise would make the message unintelligible was a mystery. As expected this was preventing the full and effective use of these technologies, since operators would either introduce too much distortion and be left with an unintelligible message or introduce too little noise and be faced with transmission delay.

\section{Shannon Entropy}

Mathematically, we are free to make any choice to define a measure of information. In other words, any monotonic function of the number of possible messages since all are equally likely. However, Shannon chooses to define information on a log scale since it has some useful properties
\begin{itemize}
    \item Many practical properties vary with the logarithm. For example, two wires have double the bandwidth of one wire.
    \item It makes the math considerably easier since logarithms have nice properties around addition, multiplication, differentiation, \etc
\end{itemize}
therefore, we define\sidenote{Note that I am choosing these words carefully. We are deciding to measure information in this way and developing a field around that decision rather than measuring some natural property of the world like a physicist might.} the ``amount of information'' $I$ as
\begin{align}
    I \propto \log(M)
\end{align}
for some message $M$. For logarithm base 2, we will call the unit of information "bits". Since this is our measure of information, we can also measure the information capacity of a channel as
\begin{align}
    C = \lim_{t \rightarrow \infty}\frac{\log(N(t))}{t}
\end{align}
where $N(t)$ messages can be transmitted in time $t$.

\begin{figure}[t]
    \centering
    \includegraphics{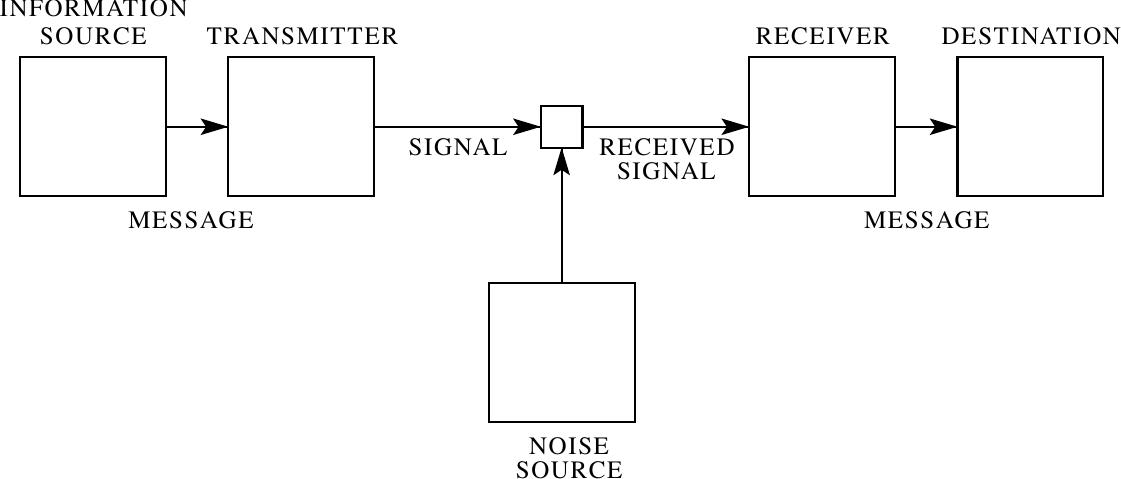}
    \caption[The General Communication System]{\textbf{The General Communication System.} One of Shannon's most important contributions was the idea that any communication system can be divided into parts and developed separately. \textit{Image credit: Claude Shannon \parencite{shannon1948mathematical}.}}
    \labfig{ent:comms}
\end{figure}

Before we continue, however, we touch on one of Shannon's most influential contributions. That is the general definition of a \gls{communication system}, given in \reffig{ent:comms}. Shannon showed that any \gls{communication system} consists of the same fundamental parts. Even systems such as telegraphy and color television which seem very different from each other are fundamentally the same. This model drives much of Shannon's analysis of information content.

Since the communication system must be designed to support any possible message, we must take a probabilistic approach to describing the generation of messages by the information source. In other words, for a discrete communication system, the information source will generate messages by producing discrete symbols one at a time. The generation of a given symbol is determined based on the past symbols and we can therefore compute a probability for each symbol.

As an example of this consider the English language. Given a set of letters: "FIRE BA" we can say that the letter "D" is highly likely to be the next letter. This is a Markov process and while incredibly complicated to produce for real scenarios, Markov modeling would allow us to produce probabilities for each symbol. The important point here is that since we are fairly certain about "D", a "D" being generated has low information and therefore requires less space to transmit. Something unexpected like an "X" would have high information content. So we can represent expected or frequent results with fewer bits.

Another example, assume I wish to communicate the weather in Seattle, and I know that there is a 100\% chance of rain in Seattle. This information can be transmitted with zero bits, since there is no need to communicate anything. Suppose that I instead wish to communicate the weather in College Park where it rains roughly 50\% of the time, then I would require the same amount of bits to transmit raining or sunny.

So now we have established an intuitive idea of the information content of a message. That is, we are measuring how ``expected'' or ``surprising'' or ``random'' a message appears. Given a set of symbols with probabilities $p_i$ for the $i$th symbol, we define the entropy $H$ as
\begin{align}
    H = -\sum_{i=1}^N p_i\log p_i
\end{align}
This measure has some important properties
\begin{itemize}
    \item $H = 0$ if and only if all of the $p_i$ are zero except for one, in other words, there is only one symbol and it always occurs (like the Seattle example). This means there is no entropy.
    \item $H$ is maximized when all $p_i$ are the same ($\frac{1}{N}$), since this is the most uncertain situation (like in the College Park example).
\end{itemize}

At this point we have developed information theory to the barest minimum extent in order to define entropy of a discrete channel. We are not taking into account noise or continuous signals, all of which are discussed at length in Shannon's paper along with much more thorough derivations. We have already touched on the idea that low entropy symbols can be represented with fewer bits. In the next two sections we will develop algorithms for computing these representations. These methods are examples of \gls{lossless compression} where all information in the original message is preserved.

\section{Huffman Coding}

\gls{huffman coding} \sidecite{huffman1952method} is a method for producing optimal length codes for symbols based on their probability of occurrence. It was the first method for finding optimal codes (Shannon presented a method which was not guaranteed to be optimal) and it is still in heavy use at the time of writing by image and video codecs 70 years after its invention.

\begin{marginfigure}
    \centering
    \includegraphics{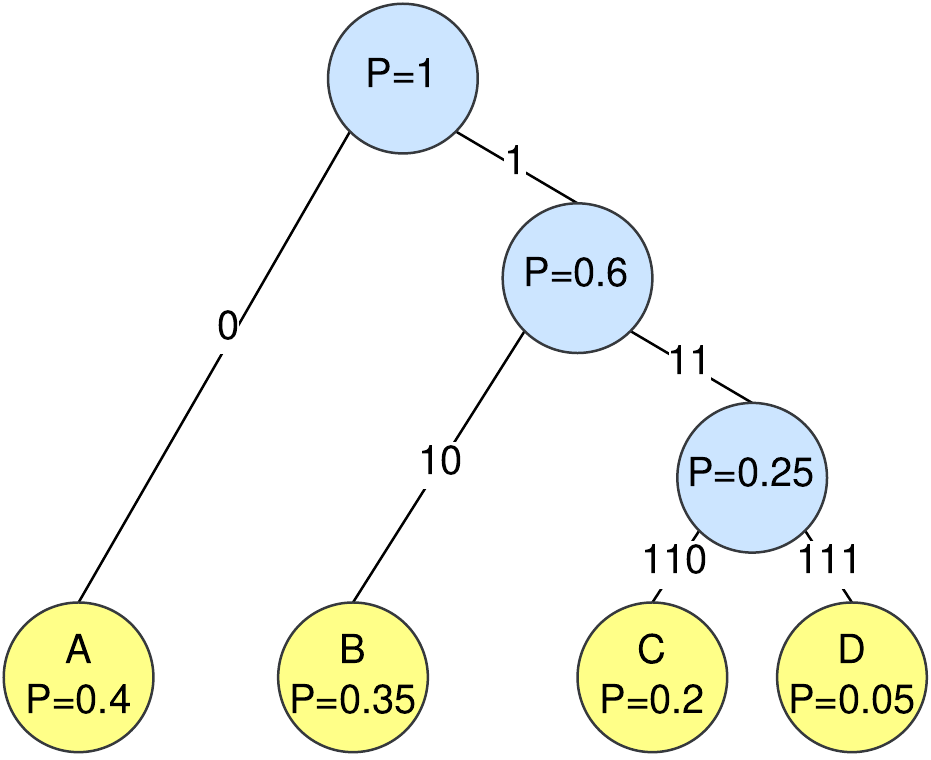}
    \caption[Huffman Tree Example]{\textbf{Huffman Tree Example.} The following tree structure assigned the smallest length sequence to the most probable symbol and the longest length sequence to the least probable.}
    \labfig{ent:huffman}
\end{marginfigure}

\gls{huffman coding} requires a set of symbols and their probabilities of occurrence as input. Then, given a message as a sequence of symbols, the algorithm produces the minimum length code that uniquely conveys the message. This requires assigning the shortest codes to the most probable symbols and the longest codes to the least probable symbols.

We do this using a binary tree. Start with a leaf node for each symbol that stores the probability of that symbol and insert them into a priority queue. Then, at each step, remove the two nodes with the lowest probability and merge them into an internal node with probability equal to the sum of the probabilities of these nodes. Then insert this new node into the priority queue and repeat until the queue has only one node on it. This node is the root of the tree. The process is a simple greedy algorithm. Approximate code is given in Listing \ref{lst:ent:buildhuff}.

\begin{lstlisting}[caption={Building a Huffman Tree.},language=Python,label={lst:ent:buildhuff}]
def build_tree(symbols: List[Tuple(float, str)]) -> Node:
    leaves = [(s[0], Node(s[0], s[1], None, None)) for s in symbols]
    p = heapq.heapify(leaves)
    
    while len(p) > 1:
        l = heapq.heappop(p)
        r = heapq.heappop(p)

        n = Node(l.probability + r.probability, None, l, r)
        heapq.heappush((n.probability, n))
        
    return p[0]
\end{lstlisting}

To encode, for each symbol traverse the tree from the root tracking the series of left and right child's used in the traversal. Add a 0 to the symbol for a left and a 1 for a right. When the correct leaf node is reached, the resulting string of 0s and 1s encodes the symbol. To decode, simply read each bit at a time and traverse the tree (right or left) based on the bit value. When a leaf node is encountered, emit that symbol and return to the root of the tree.

Let's consider a simple example. Suppose we are given a simple four letter alphabet with symbols $M = \{A, B, C, D\}$. These four symbols are known to occur with probabilities $P = \{p_A = 0.4, p_B = 0.35, p_C = 0.2, p_D = 0.05\}$. Since we have four symbols, the default encoding would be 2 bits per symbol, $\{A = 00, B = 01, C = 10, D = 11\}$. However computing the entropy of the set $P$ gives
\begin{align}
    H(P) = -\sum_{p \in P} p \log p                                   \\
    = -0.4 \log(0.4) - 0.35\log(0.35) - 0.2\log(0.2) - 0.05\log(0.05) \\
    = 0.529 + 0.530 + 0.464 + 0.216                                   \\
    = 1.74
\end{align}
so approximately 1.74 bits, meaning that the default encoding of 2 bits wastes 0.26 bits per symbol on average.

\begin{figure}[t]
    \centering
    \includegraphics{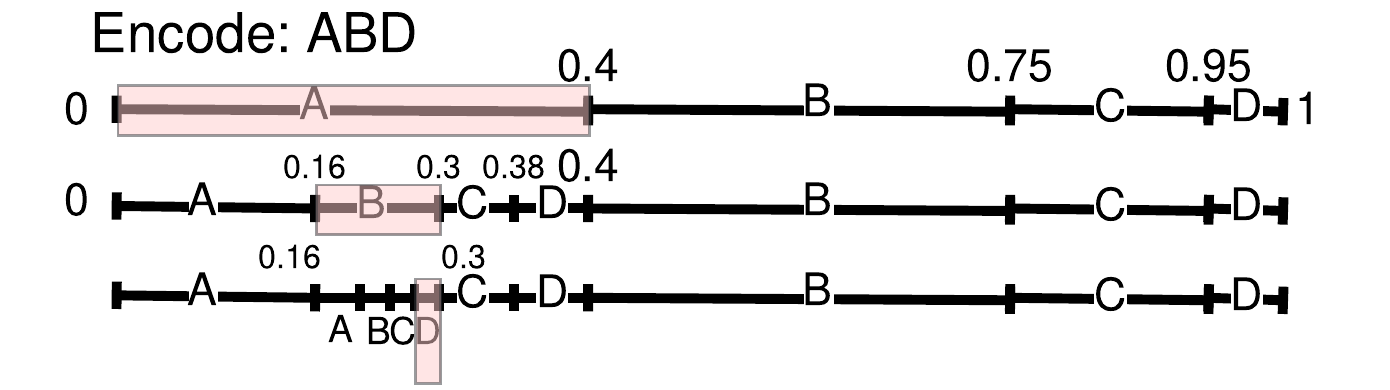}
    \caption[Arithmetic Coding Example]{\textbf{Arithmetic Coding Example.} Using the same alphabet and probabilities as the last section, we encode ABD into the range $[0.29, 0.3)$.}
    \labfig{ent:arith}
\end{figure}

We construct a Huffman tree for the above set in \reffig{ent:huffman}. This gives the following variable length codes $\{A = 0, B = 10, C = 110, D = 111\}$ obtained by traversing the tree for each symbol. Note that although there are some symbols which now require 3 bits to encode, these are the least probable symbols and the most probable symbol, $A$, requires only 1 bit. If we compute the average size of a symbol with these codes we actually have 1.85 bits/symbol on average so we are still above the limit in terms of entropy. This is because symbols cannot occupy a fraction of a bit.

\section{Arithmetic Coding}

Although Huffman codes were optimal in terms of the number of bits to encode single symbols, we saw that Huffman coding was not able to reach the theoretical minimum number of bits defined by the entropy of the set. By computing an encoding for an entire message rather than one symbol at a time we can overcome this limitation. This is the motivation behind arithmetic coding, which stores an entire message into an arbitrary number $q$ such that $0 \leq q < 1$.

Once again the algorithm is given a set of symbols and their probabilities. The encoder starts with the interval $[0, 1)$ and divides the interval into sub-intervals for each symbol. The algorithm picks the interval which corresponds to the current symbol and proceeds to the symbol. When all symbols are consumed, the resulting interval uniquely identifies the message, and since the intervals are unique we only need to transmit a single element of the final interval to identify the message\sidenote{Specifically, enough bits such that any fraction beginning with the transmitted number falls into the desired interval.}. To decode we can follow the same process, but this time we are given the number $q$. At each step we construct the same intervals and simply check which one the given number falls into, emitting that symbol at each step. This does require either a special terminating symbol or a known message length to stop. The algorithm is shockingly simple and highly effective.

An example encoding is shown in \reffig{ent:arith}. In that example, we encode the message ``ABD'' following the same alphabet and probabilities we used for the Huffman coding example. We start by dividing $[0, 1)$ into proportional parts for each symbol, we find that the first symbol is $A$ so we choose the interval from $[0, 0.4)$. Next we divide that interval into proportional parts and since the next symbol is $B$, we choose $[0.16, 0.3)$ since $0.16 = 0.4 \times 0.4$ and $0.3 = 0.16 + (0.4 \times 0.35)$. The final symbol is $D$ so we choose the interval from $[0.29, 0.3)$ and transmit (arbitrarily) 0.295. Again, decoding follows a similar process. We are given the number 0.295 as input and we divide up the interval $[0, 1)$, finding that this falls into $[0, 0.4)$, we emit $A$. Then we find that 0.295 falls into $[0.16, 0.3)$ and we emit $B$. Finally, we find that 0.295 falls into $[0.29, 0.3)$ and we emit $D$, having decoded the message ``ABD''.

While it may seem remarkable that a message can be transmitted in a single number, the algorithm does have faults. Again, the message must fit into a discrete number of bits, which can reduce the efficiency compared to the theoretical maximum. Furthermore, we are assuming that we have an accurate probability model of the symbol frequencies. This may not be possible to obtain exactly, and in fact, we may not even want global symbol probabilities. Since we are encoding a message, the most efficient encoding of that message would model the probabilities of symbols in that message only (\eg, $\frac{1}{3}$ for $A,B,D$ and $0$ for $C$ in our example). However, this requires transmitting the probability model which may remove any gains in efficiency from the coding. In general, these are still open problems and while we can obtain ``optimal'' codes with respect to some specific definition of optimal, the theoretical entropy limit that Shannon's work gives us remains elusive.
\setchapterpreamble[u]{\margintoc}
\chapter{Machine Learning and Deep Learning}
\labch{dl}

\lettrine{M}{achine} learning\index{machine learning} is rapidly revolutionizing the way that people interact with computers. This is largely driven by the explosive proliferation of \glspl{cnn}\index{convolutional neural networks} \parencite{lecun1990handwritten} since they were shown to be computationally viable for large problems in 2012 \parencite{krizhevsky2012imagenet}. Although machine learning seems commonplace today, this was not the case ten years ago (at the time of writing) and there were many who believed that machine learning would never achieve widespread success.

While this \gls{dissertation} is centered on compression as an application, it is first and foremost a contribution to machine learning for computer vision. In this chapter, we develop a high-level understanding of machine learning concepts which relate to the rest of the dissertation. This discussion is grounded in Bayesian decision theory which is often overlooked in machine learning discourse. Otherwise, the focus is on computer vision methods rather than general methods.

\begin{kaobox}[frametitle=Note]
    Some of the material in this chapter is based on the book Pattern Classification \parencite{hart2000pattern} which I strongly recommend to interested readers for more in-depth information.
\end{kaobox}

\section{Bayesian Decision Theory}

\Gls{bayesian decision theory}\index{bayesian decision theory} tells us the best possible decision we can make about data even if we know exactly the underlying generating distributions. In a sense this can be thought of as a best case scenario because in real life we do not know the underlying distributions so we must either approximate them or approximate decision criteria directly. The classic example of this proceeds as follows. Dockworker Dave is observing fish as they are unloaded from boats. His task is to sort the fish into bins, one for sea bass which we will denote as $c_b$ and one for salmon which we will denote as $c_s$. The fish come out of the boat randomly. In the absence of any other information (such as identifying markers), how can he develop a strategy to sort them with minimal errors?

Let's give Dave some knowledge to help. Since the fish are coming off the boat in a random order, we must describe the occurrence of each fish probabilistically. Assume that Dave knows how many fish were caught of each type, then he knows the \gls{prior}\index{prior probability} $P(c_b)$ and $P(c_s)$. For example if $P(c_b) = 0.7$ and $P(c_s) = 0.3$ then Dave should classify \textit{all of the fish} as bass and he will have 70\% accuracy. Of course this will entail him dumping all fish into the bass bin which is a bit odd considering that he knows there are two types of fish. Nevertheless this strategy will attain the lowest error given what Dave knows.

We can give Dave some more information to help him. Dave's daughter Wendy studies fish and she informs him that the color can be used to differentiate bass from salmon although it is not a perfect indicator (see \reffig{dl:fish}). In this case we would say that there is a continuous random variable $x$ which yields conditional probabilities $P(x | c_b)$ which is the probability of each color value for sea bass and $P(x | c_s)$ for salmon. We call this the \gls{likelihood} of the color given the type of fish and we will call the color of the fish a feature\index{feature}.

\begin{marginfigure}
    \centering
    \includegraphics{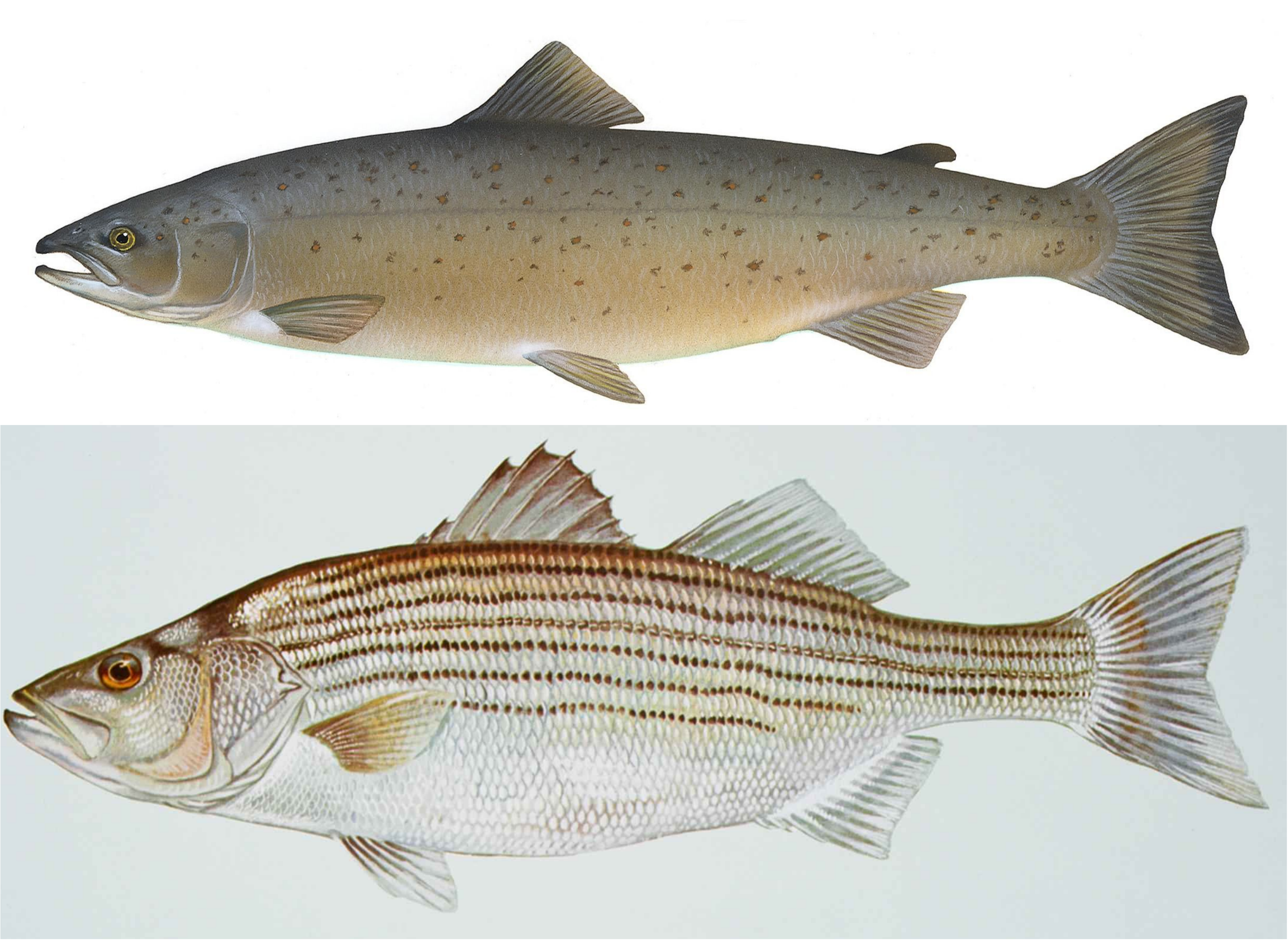}
    \caption[Salmon \vs Sea bass]{\textbf{Salmon \vs Sea bass.} Top: Salmon, Bottom: Sea bass. The two fish have different colors.}
    \labfig{dl:fish}
\end{marginfigure}

How does Dave use this information? In order to make a decision given color, we want to compute $\mathcolor{BurntOrange}{P(c_s | x)}$ and $\mathcolor{BurntOrange}{P(c_b | x)}$, which we call \gls{posterior}\index{posterior probabilities}, and take the larger probability, but we only have $\mathcolor{JungleGreen}{P(c_b), P(c_s), P(x | c_b), P(x | c_s)}$. We also know that there is a joint distribution for each class $\mathcolor{Plum}{P(c_{s,b}, x)}$ which is the probability of a fish being class $s$ or $b$ and having color $x$ that relates these quantities. From probability theory, we can write this in terms of the conditional
\begin{align}
    \mathcolor{Plum}{P(c_{s, b}, x)} = \mathcolor{BurntOrange}{P(c_{s, b} | x)}\mathcolor{Plum}{p(x)} = \mathcolor{JungleGreen}{P(x | c_{s, b})P(c_{s, b})}
\end{align}
this is the definition of conditional probability. Rearranging to group the quantities that we know gives
\begin{align}
    \mathcolor{BurntOrange}{P(c_{s, b} | x)} = \frac{\mathcolor{JungleGreen}{P(x | c_{s, b})P(c_{s, b})}}{\mathcolor{Plum}{P(x)}}
    \labeq{dl:br}
\end{align}
which is known as Bayes rule\index{Bayes rule}. This allows us to compute the class probability given some measurement as long as we have the known likelihood and prior probabilities. We have another unknown term, the \gls{evidence} term, in \refeq{dl:br}, $\mathcolor{Plum}{P(x)}$, which is the probability of any fish having the measured color: in general we do not need this term. The Bayes decision rule\index{Bayes decision rule} is
\begin{align}
    c = \begin{cases}
            c_s & \mathcolor{BurntOrange}{P(c_s | x)} > \mathcolor{BurntOrange}{P(c_b | x)} \\
            c_b & \mathcolor{BurntOrange}{P(c_b | x)} > \mathcolor{BurntOrange}{P(c_s | x)}
        \end{cases}
\end{align}
expanding one of these inequalities gives
\begin{align}
    \frac{\mathcolor{JungleGreen}{P(x | c_{s})P(c_{s})}}{\mathcolor{Plum}{P(x)}} > \frac{\mathcolor{JungleGreen}{P(x | c_{b})P(c_b)}}{\mathcolor{Plum}{P(x)}}                     \\
    = \frac{\mathcolor{JungleGreen}{P(x | c_{s})P(c_{s})}}{\cancel{\mathcolor{Plum}{P(x)}}} > \frac{\mathcolor{JungleGreen}{P(x | c_{b})P(c_b)}}{\cancel{\mathcolor{Plum}{P(x)}}} \\
    = \mathcolor{JungleGreen}{P(x | c_{s})P(c_{s})} > \mathcolor{JungleGreen}{P(x | c_{b})P(c_b)}
\end{align}
in terms of only known quantities. This is good because the evidence term is often hard to measure.

So now Dave can use his knowledge of the prior probabilities and Wendy's color probabilities and multiply them to produce the probability of sea bass or salmon, binning the fish based on whichever is more probable. This seems like a perfectly reasonable idea, but what kinds of errors will Dave make? Let's compute the probability of Dave's error
\begin{align}
    P(\text{error} | x) = \begin{cases}
                              P(c_b | x) & c = c_s \\
                              P(c_s | x) & c = c_b
                          \end{cases}
\end{align}
In other words, the error rate will be the probability of the other class. To compute the average error rate, we marginalize $x$ from the joint distribution
\begin{align}
    P(\text{error}) = \int_{-\infty}^\infty P(\text{error}, x)\;dx \\
    = \int_{-\infty}^\infty P(\text{error} | x)P(x)\;dx
\end{align}
We cannot control, or really even measure, $P(x)$ but we can control $P(\text{error} | x)$ by making it as small as possible. And the only way to accomplish that is by picking the higher probability for $P(c_{s, b} | x)$ as our classification choice, thus proving the optimality of the Bayesian decision.

So now we have a way of making the best possible classification decisions. Given prior probabilities of the different classes and likelihoods of each feature given each class, we can then compute the posterior probabilities and pick the higher one. This guarantees the minimum error: we cannot achieve lower error than this. However we now have a new problem: how do we produce these probabilities for real problems? In general, we can not, and we will have to approximate the distributions leading to an even high error rate. In this sense, the Bayesian decision can be thought of as a theoretical lower limit for the error rate. Even if we know everything, because of the probabilistic nature of decision problems, we will not make the right choice for every input.

This sets up a theoretical dichotomy. Do we approximate the underlying prior and likelihood distributions which generated the data and then make Bayesian decisions based on our observations? Or instead can we simply compute the boundary between the posterior distributions as a function of the observation that makes a decision directly? Either way, these two questions are the entire purpose of machine learning. Given some data, sampled from unknown distributions, how do we compute approximations which match the true distributions or decision boundaries as closely as possible.

\section{Perceptrons and Multilayer Perceptrons}

One simple way of learning \glspl{decision boundary}\index{decision boundary} is the perceptron\index{perceptron} \sidecite{rosenblatt1957perceptron}. The perceptron defines a simple linear model for making a binary decision between two classes (although it can be extended to more complex scenarios). Given a vector of weights $\bm{w}$, and an input feature vector $\bm{x}$, the perceptron makes the following decision
\begin{align}
    f(\bm{x}) = \begin{cases}
                    1 & \langle \bm{w}, \bm{x} \rangle > 0 \\
                    0 & \text{otherwise}
                \end{cases}
\end{align}
or simply
\begin{align}
    f(\bm{x}) = H(\langle \bm{w}, \bm{x} \rangle)
\end{align}
where $H()$ is the Heaviside function\index{Heaviside function}, for classes 1 and 0. The decision boundary in this case is a linear function of $\bm{x}$. The task then is to compute a suitable $\bm{w}$ given some data.

Starting from a randomly initialized $\bm{w}_0$ and some set of training data $\bm{x}_i$ with labels $y_i$ the learning algorithm first computes the decision on $\bm{x}_i$.
\begin{align}
    \widehat{y}_i = f(\bm{x_i}) = \langle \bm{w}_0, \bm{x} \rangle
\end{align}
which may be incorrect. The algorithm then updates the weights as
\begin{align}
    \bm{w}_1 = \bm{w}_0 + (y_i - \widehat{y}_i)\bm{x}_i
\end{align}
This process is repeated for all pairs $(\bm{x_i}, y_i)$ until some predefined stopping criterion is met. In the case that all $\bm{x}_i$ are \gls{linearly separable} with respect to $y_i$, this stopping criterion may be convergence, but this is almost never the case in real life.

\begin{marginfigure}
    \centering
    \includegraphics{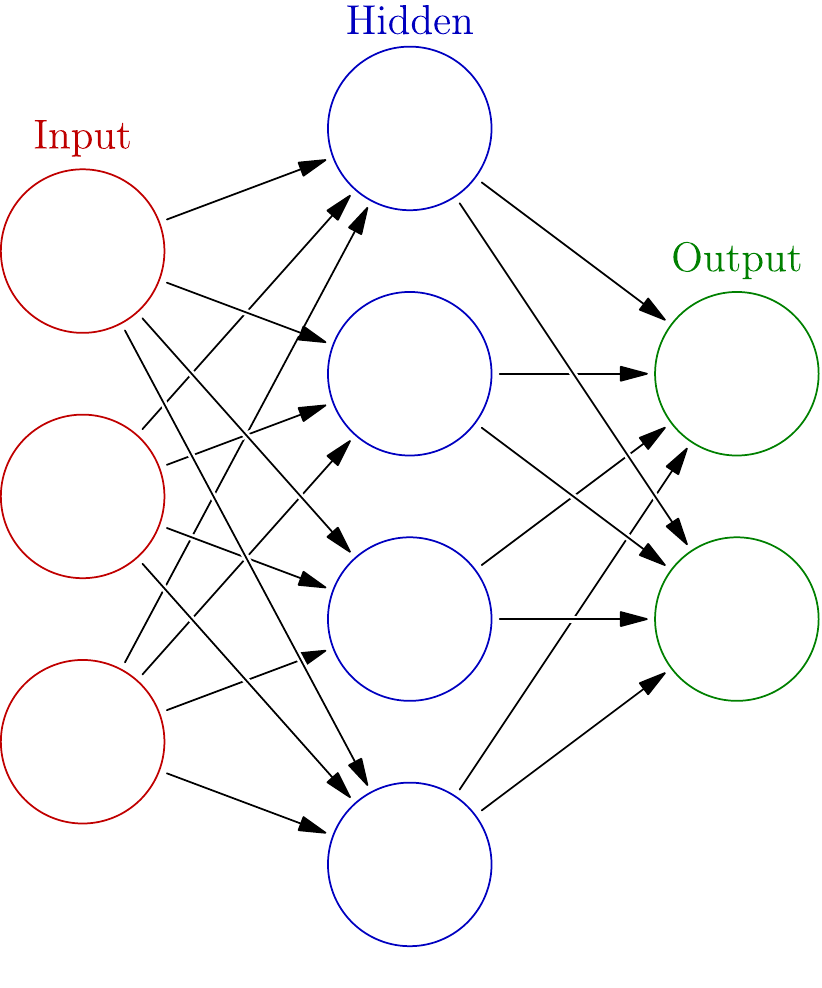}
    \caption[Multilayer Perceptron]{\textbf{Multilayer Perceptron.} The multilayer perceptron organizes groups of perceptrons into layers separated by non-linearities. In this case each circle represents a perceptron. The first and last layers are termed the input and output layers respectively; any layers in between are termed hidden layers. \textit{Image credit: wikipedia.}}
    \labfig{dl:mlp}
\end{marginfigure}

To model real scenarios, a more complex model is needed: one that can model non-linear relationships. We can extend the perceptron to model these more complex scenarios by building a \gls{mlp}\index{multilayer perceptron} (MLP). The \gls{mlp} stacks layers of perceptrons separated by non-linearity (\reffig{dl:mlp}). More formally, for layer weights $W_l$ (a matrix), input $\bm{x}$, and nonlinearity $\sigma()$, a \gls{mlp} can be defined as
\begin{align}
    f(\bm{x}) = W_N \sigma(\ldots \sigma(W_1\sigma(W_0 \bm{x})))
\end{align}
for an MLP with $N$ layers. We call the first layer (weights $W_0$) the input layer, the last layer (weights $W_N$) the output layer, and the intermediate layers (weights $W_1,\ldots,W_{N - 1}$) the hidden layers. In practice we will also define a loss function $l()$ which takes the network output and the true classification and tell use how wrong it was. Importantly this function needs to be scalar valued
\begin{align}
    e(W) = l(y, f(\bm(x); W))
\end{align}
describing the error for some set of weights $W$.

Training this model requires some tricks. We use an algorithm called backpropagation\index{backpropagation} \sidecite{lecun1990handwritten}. If we observe the form of $l()$, we can see that it is a scalar valued function of a vector. This means that we can compute the \gls{gradient} of the output with respect to the input
\begin{align}
    \bm{\nabla}_W l(y, f(\bm(x); W)) = \begin{bmatrix}
                                           \frac{\partial}{\partial w^0_{00}}  l(y, f(\bm(x); W)) \\
                                           \frac{\partial}{\partial w^0_{10}} l(y, f(\bm(x); W))  \\
                                           \vdots                                                 \\
                                           \frac{\partial}{\partial w^l_{ij}} l(y, f(\bm(x); W))  \\
                                           \vdots                                                 \\
                                           \frac{\partial}{\partial w^L_{MN}} l(y, f(\bm(x); W))
                                       \end{bmatrix}
\end{align}
for $L$ layers and weights of size $MN$, which tells us ``in what direction and by how much'' we would need to change the network in order to classify $\bm{x}$ correctly. We can compute these quantities using the chain rule. For each layer we compute the \gls{jacobian} $\frac{\partial W^L}{W^{L - 1}}$ (since these are vector valued functions) with respect to the previous layer and continue until we have differentiated every layer.
\begin{align}
    \bm{\nabla}_{W^0} l =  \bm{\nabla}_{W^N} l \odot \frac{\partial{W^N}}{{\partial W^{N - 1}}} \odot \frac{\partial{W^{N - 1}}}{{\partial W^{N - 2}}} \cdots \frac{\partial{W^1}}{{\partial W^0}}
    \labeq{dl:backprop}
\end{align}
which gives updates for the weights in each layer.

\section{Image Features}

In order to apply any of these models to images, we need some way of representing images as the input vectors $x$ to the functions in the previous section. While we could simply flatten the images into vectors, this may cause issues with the learning process. Small perturbations of the input pixels can cause large changes in their actual values. Also pixels themselves can vary considerably in appearance yet still represent the same class. These issues impact the separability of the problem, and create extremely complex \glspl{decision boundary} that are difficult if not impossible to model without arbitrarily deep networks.

A more successful strategy would be to compute some higher order representation of the images which we can show is more meaningful. Although we may explore ideas like extracting numerical shape descriptions or color conversions, there are some abstract representations which have been shown to be effective. We will explore two of these in this section: Histogram of Oriented Gradients (HOG)\index{histogram of oriented gradients} and the Scale-Invariant Feature Transform (SIFT)\index{scale-invariant feature transform}. Both of these techniques transform an image into a series of vectors, which we call \glspl{feature}, that can then be input to a machine learning model.

\subsection{Histogram of Oriented Gradients}

The Histogram of Oriented Gradients \sidecite{dalal2005histograms} captures shape and orientation of objects using a local descriptor. Often the image will be contrast normalized in blocks before the histogram of gradients is computed on each pixel in small cells. The descriptor for each cell is the concatenation of the histograms for all of the pixels in the cell.

To compute the gradient of an image, it can simply be convolved with a gradient kernel. There are many such kernels but
\begin{align}
    h = \begin{bmatrix}
            -1 & 0 & 1
        \end{bmatrix} \\
    v = \begin{bmatrix}
            -1 \\ 0 \\ 1
        \end{bmatrix}
\end{align}
are the popular choices for computing horizontal and vertical gradients respectively. After the gradient is computed it can be binned per cell (usually $8 \times 8$ pixel cells) to compute the histogram. These histogram cells are then normalized with respect to larger blocks (usually $16 \times 16$ blocks) to further increase invariance to image transformations. This gives a descriptor for each cell which can be input to any classifier. For example, Dalal and Triggs used the HoG feature for pedestrian detection an \gls{SVM} classifier\sidenote{We do not cover SVMs here}.

\begin{marginfigure}
    \centering
    \includegraphics{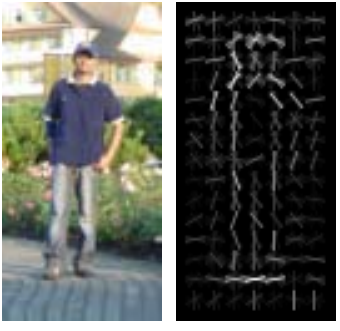}
    \caption[HoG Features]{\textbf{HoG Features.} The left shows an example image and the right shows HoG features which classify as ``person''. The HoG features are shown as the weighted orientation based on the histogram of the cell and classification confidence. \textit{Image credit: \cite{dalal2005histograms}.}}
    \labfig{dl:hog}
\end{marginfigure}
The result at the time was quite impressive and HoG features came into widespread use. HoG features are ``dense'' in the sense that every block in the image is covered in some sense which means that the model is given a strong prior on the local shapes present in the image. This can be seen visually in \reffig{dl:hog}. In the figure we see a man in the example image. The HoG features visualized on the right show outlines of the important shapes in each region. This visualization is produced by drawing tangent lines for each orientation in the histogram and weighting the lines by the histogram values on each cell. The lines are then further weighted by the SVM confidence to show which lines are contributing to the human classification. We see strong responses on the feet, shoulders, and head meaning that the model considers these unique identifiers of people that are not present in other objects.

\subsection{Scale-Invariant Feature Transform}

One of the most popular and powerful image features is the Scale-Invariant Feature Transform\index{scale-invariant feature transform} \sidecite{lowe1999object}. Like HoG features, SIFT features capture a local description of shape using orientation. Unlike HoG, the primary purpose of SIFT was to find scale-invariant \index{scale-invariant} keypoints\index{keypoint} which are unique locations that appear the same under scale changes. These points can be used for object matching. Since the points should be rotation and scale invariant, a query object should be able to be located even if it is subject to complex deformations.

\begin{marginfigure}
    \centering
    \includegraphics{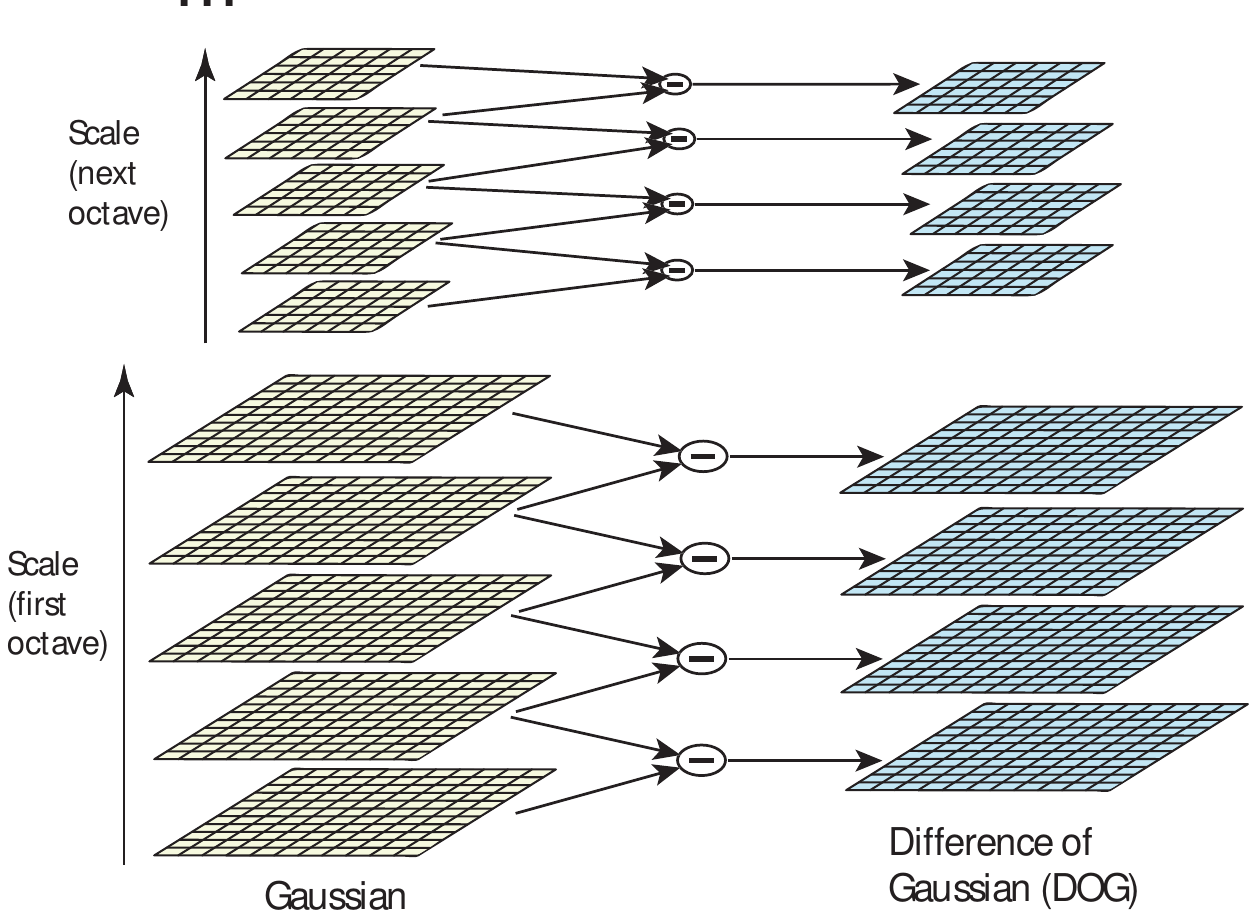}
    \caption[Difference of Gaussians]{\textbf{Difference of Gaussians.} The difference of Gaussian's scale space computes gaussian blurs of increasing strength to the input image. The blurred images are then subtracted from each other. Points which survive this process are scale invariant. \textit{Image credit: \cite{lowe1999object}.}}
    \labfig{dl:ss}
\end{marginfigure}
To compute the scale space, SIFT uses a difference of Gaussian's (DoG)\index{difference of gaussians}. The image is computed at different scales by Gaussian blurring the image successively, then the difference between neighboring blurred images is taken (\reffig{dl:ss}). When this is done over many scales, any points which survive for the entire stack of DoG images are considered scale-invariant since they are clearly localized across scales. These points are pixel localized by applying non-maximum suppression and then sub-pixel localized by computing a second order Taylor expansion on the pixel which can produce a zero point in between pixel boundaries.

For each keypoint, the rotation invariant descriptor is computed. The gradient magnitude $m(x, y)$ and orientation $o(x, y)$ are computed as
\begin{align}
    m(x, y) = \sqrt{(P_{x + 1, y} - P_{x - 1, y})^2 + (P_{x, y + 1} - P_{x, y - 1})^2 } \\
    o(x, y) = \arctan\frac{P_{x, y + 1} - P_{x, y - 1}}{P_{x + 1, y} - P_{x - 1, y}}
\end{align}
for image $P$. This is computed in a $3 \times 3$ neighborhood around the keypoint and then a histogram is computed. The orientation with the highest bin is assigned to the keypoint. To further improve the invariance, these descriptors are compiled in a $4 \times 4$ grid into an 8 bin histogram. The resulting 128 dimensional descriptor resulting from the concatenation of these histograms is assigned as the keypoint descriptor. This descriptor can be normalized to improve invariance to lighting changes.

SIFT features were the de facto standard in image features for many years. During the end of the classical/feature based learning era, there was a particular shift towards dense SIFT features. This step simply forgoes the keypoint detection steps and computes a descriptor for each pixel. This is useful for tasks like semantic segmentation that require per-pixel labels but it can also be used as a rotation invariant base for more general tasks.

\begin{figure}[t]
    \centering
    \includegraphics[width=\textwidth]{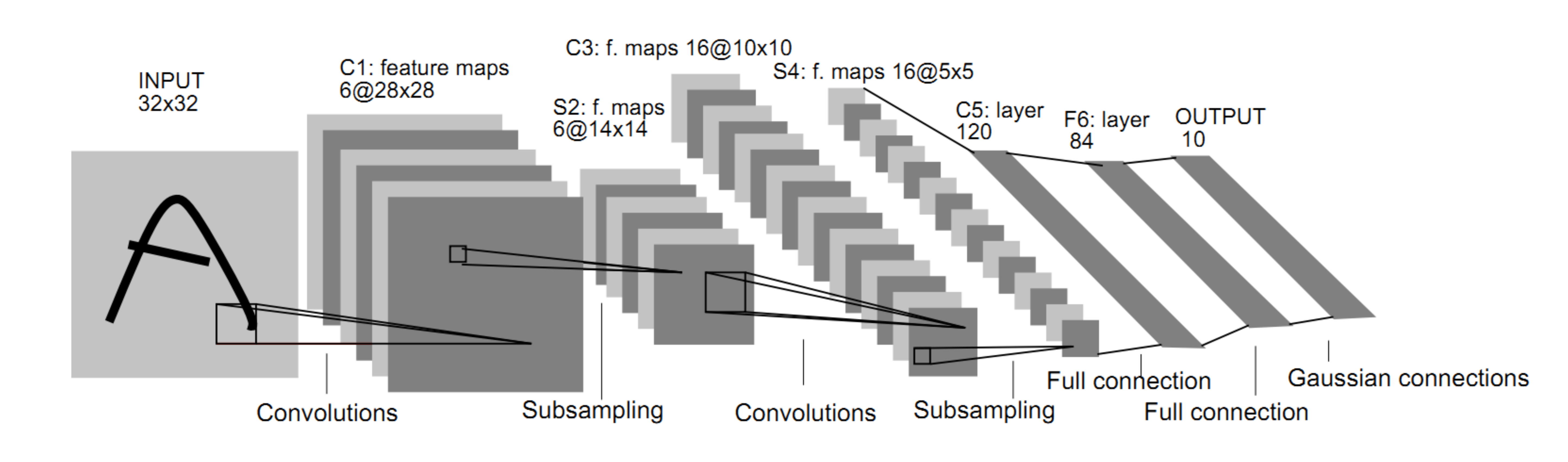}
    \caption[Convolutional Neural Network]{\textbf{Convolutional Neural Network.} Diagram shows the LeNet-5 architecture. The model takes pixels as input and computes feature maps using successive convolutions, non-linearities, and subsampling layers. For classification, the network terminates in a MLP. This allows the classifier and the feature extractors to be trained jointly using backpropagation. \textit{Image credit: \cite{lecun1998gradient}}}
    \labfig{dl:cnn}
\end{figure}

\section{Convolutional Networks and Deep Learning}

Feature engineering is a complex process. The two algorithms we described in the last section are non-trivial to understand, much less to develop on one's own. Furthermore, it is not clear if a given feature is suitable to any particular task, we only have vague motivation and intuition to guide us. The fundamental contribution of \gls{deep learning}\index{deep learning} was that the best features for a given problem can be learned along with the classifier using only pixels as input. This replaced the tedious feature engineering process with something much more powerful and much simpler to develop.

Deep learning is powered by the \gls{cnn}\index{convolutional neural network} \sidecite{lecun1990handwritten}. These ideas had been around for some time but it was not until Alexnet \sidecite{krizhevsky2012imagenet} that deep enough and complex enough networks were shown to be computationally viable with a GPU implementation. This quickly revolutionized machine learning with entire scientific careers dedicated to feature engineering becoming obsolete in a short time frame.

The \gls{cnn} itself is not particularly complex. It is a \gls{mlp} with the matrix multiplications replaced with convolutions, formally
\begin{align}
    f(\bm{x}) = W_N \star \sigma(\ldots \sigma(W_1 \star  \sigma(W_0 \star \bm{x})))
\end{align}
The advantage of this is that the weights can be small kernels, usually $3 \times 3$ instead of the large matrices required to process an image with a \gls{mlp} (these matrices would need to be the same width and height as the image). This already made \glspl{cnn} much more efficient than \glspl{mlp} even without the GPU implementation. Furthermore, many seemingly complex image transformations can be computed with convolutions which is why we say that the convolutional network computes learned feature representations. These non-linear feature extractors replace the hand designed feature extractors of classical machine learning.

One of the more influential and yet simple architectures is shown in \reffig{dl:cnn}, LeNet-5 \sidecite{lecun1998gradient}. Many \gls{cnn} variants can be described by the components in that figure. The convolutional layers are paired with non-linearity and subsampling layers. The subsampling layers are usually some kind of pooling (max pooling or average pooling) which helps aggregate feature information spatially. The actual classification decision is made using a MLP once the feature maps have been reduced to a sufficiently small and abstract representation. For non-linearity currently ReLU is the most popular choice which I like to define in terms of the Heaviside function
\begin{align}
    R(x) = H(x)x
\end{align}
but which most people like to write as
\begin{align}
    R(x) = \begin{cases}
               x & x > 0    \\
               0 & x \leq 0
           \end{cases}
\end{align}

Why \glspl{cnn} work as well as they do remains somewhat of a mystery, but like much of machine learning, we can get an idea using intuition. As we have already discussed, the hand designed features of classical machine learning may not have been the best for a given task. The learned features of a convolutional network are likely more suited since they are customized to the task. Images are a discrete sampling of a 2D signal, and nearby pixels are often highly correlated or anti-correlated (in terms of edges). \glspl{cnn} can pick up on these correlations because they use a translation-invariant learned convolution which is moved across the image spatially in a sliding window. Finally, since convolutional networks are highly efficient, they can be made deeper and wider to learn more complex mappings.

In this dissertation we will be exclusively exploring convolutional neural network architectures. While there have been some major advancements to \glspl{cnn} which we touch on in the rest of this chapter and throughout the dissertation, it is worth noting that the \glspl{cnn} of today are largely the same as those used by the pioneers of deep learning.

\section{Residual Networks}
\labsec{resnets}

Residual networks \sidecite{he2016deep}\index{residual networks} were a major advancement in the design of convolutional networks. Instead of learning a mapping $y = f(x)$ like a traditional network, the residual network defines a mapping $y = f(x) + x$. This, along with some other notable architectural changes, makes the residual network highly effective. The precise reason why this helps so much is still debated however it likely makes ``gradient flow'' easier (gradient flow was also explored by the VGG \sidecite{simonyan2014very} and Inception \cite{szegedy2015going} architectures). Examining \refeq{dl:backprop}, we can spot a potential problem. As the depth of the network increases, carrying the gradient from the loss through all \glspl{jacobian} to the earliest layer may be difficult. The gradient tends to shrink as we move backwards through the layers, we call this problem vanishing gradient\index{vanishing gradient}. Residual learning is likely solving this problem by allowing a shortcut connection around some of the convolution layers which carries a stronger gradient signal to the early layers.

\begin{figure}[t]
    \centering
    \includegraphics{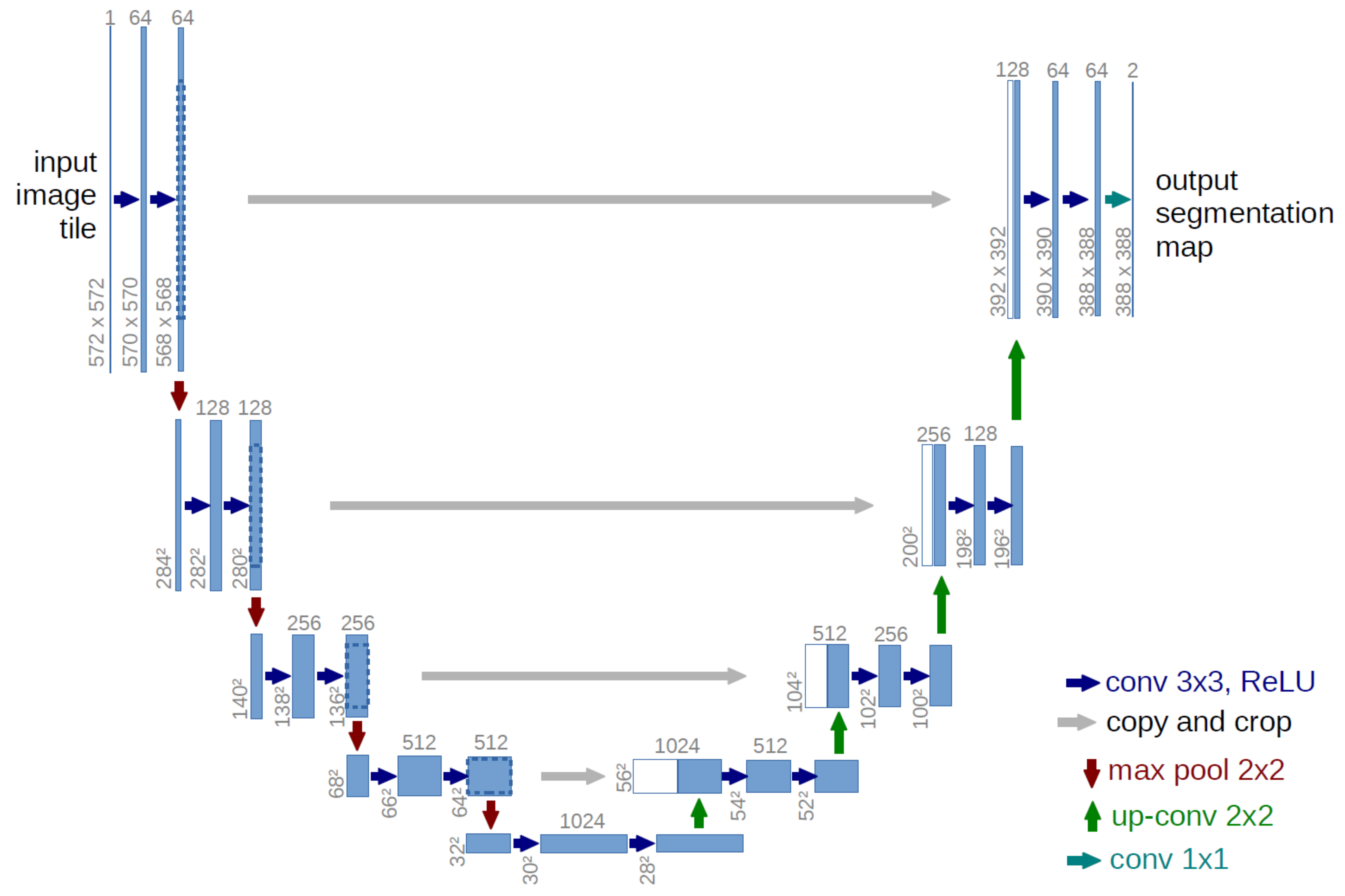}
    \caption[U-Net]{\textbf{U-Net}. U-Nets arrange convolutional layers in a U-shape of decreasing and then increasing size. Skip connections allow for better gradient flow to early layers. \textit{Image credit: \cite{ronneberger2015u}.}}
    \labfig{dl:unet}
\end{figure}
\begin{marginfigure}
    \centering
    \includegraphics{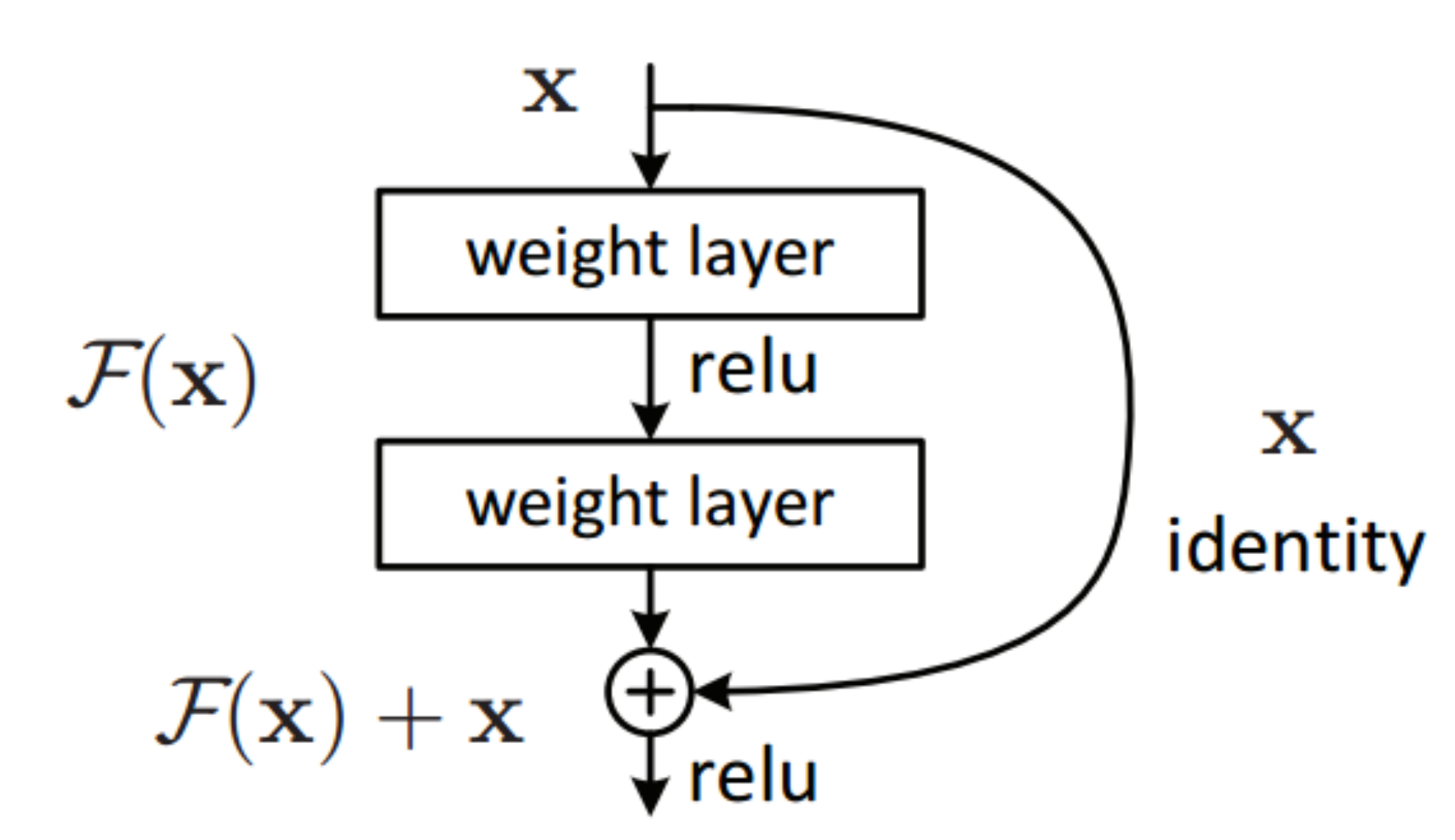}
    \caption[Residual Block]{\textbf{Residual Block.} Each residual block consists of two convolutions with a ReLU non-linearity and a batch normalization layer. \textit{Image credit: \cite{he2016deep}.}}
    \labfig{dl:resblock}
\end{marginfigure}
The actual design of the network is based on a so called ``residual block''\index{residual block} pictured in \reffig{dl:resblock}. Each block has two weight layers with batch normalization \sidecite{ioffe2015batch} separated by a ReLU non-linearity. The addition of batch normalization is thought to simplify the learning process even further for the weight layers by removing lower order statsitics of mean and variance. For each batch, the layer tracks the running mean $\mu$ and variance $\sigma^2$ and computes
\begin{align}
    \text{BN}(x) = \gamma \frac{x - \mu}{\sigma} + \beta
\end{align}
for learned $\gamma$ and $\beta$. The block includes the hallmark residual connection short-circuiting the two weight layers.

The entire network architecture stacks the residual blocks using strided convolutions to perform learned downsampling instead of using pooling. The network terminates with a ``global average pooling'' layer which performs spatial averaging over each channel of the output to produce a small vector suitable for input to a \gls{mlp} like prior network designs.

\section{U-Nets}

It is worth noting that we are, of course, not limited to only classification problems. The U-Net\index{u-net} \sidecite{ronneberger2015u} architecture is suitable for problems which require a spatial output like \gls{image-to-image} translation and \gls{semantic segmentation}. In this dissertation, we will almost exclusively be dealing with \gls{image-to-image} problems although the architectures we discuss later will differ greatly from the U-Nets. Similar to residual networks, U-Nets were a major advancement in these spatial tasks. And also like residual networks, the major contribution was likely in gradient flow.

U-Nets define the network in two distinct parts: the encoder and the decoder, the schematic is shown in \reffig{dl:unet}. The encoder is much like a traditional convolutional network. There are alternating convolutions and non-linearities with downsampling. The decoder is the reverse process, taking the compact representation from the encoder and using upsampling operations to compute a result which has the same dimensions as the input image. The major design feature of this is the skip connections\index{skip connections}. These connections take feature maps from the encoder and concatenate them with the feature representations of the same size in the decoder which allows a strong gradient signal to flow to the early layers avoiding the vanishing gradient problem.

The U-Net was revolutionary at the time for its results on the extremely difficult semantic segmentation problem. However, the U-Net would quickly become widely used for any spatial task, and is still used quite frequently. Pix2Pix \sidecite{isola2017image} for example was based entirely on the U-Net. While U-Nets were highly influential on all image-to-image problems, we will employ very different architectures later in the dissertation, and indeed very few works in compression actually use U-Nets. This is because there are other ways to deal with vanishing gradients (like residual blocks and their derivatives) and the downsampling operations in U-Nets tend to remove fine details which we want to preserve in restoration tasks.

\section{Generative Adversarial Networks}

\Glspl{gan}\index{generative adversarial networks} \sidecite{goodfellow2014generative} will be relied upon heavily in the methods we detail in the dissertation. \glspl{gan} were a truly revolutionary moment in the generation of images using \glspl{cnn}. Prior error based methods, called autoencoders\sidenote{Although this is an abuse of the term, technically an autoencoder should generate the exact image it is given as input and nothing else.}, produced very poor results even for simple datasets like MNIST \sidecite{lecun1998mnist}.

\begin{marginfigure}
    \centering
    \includegraphics{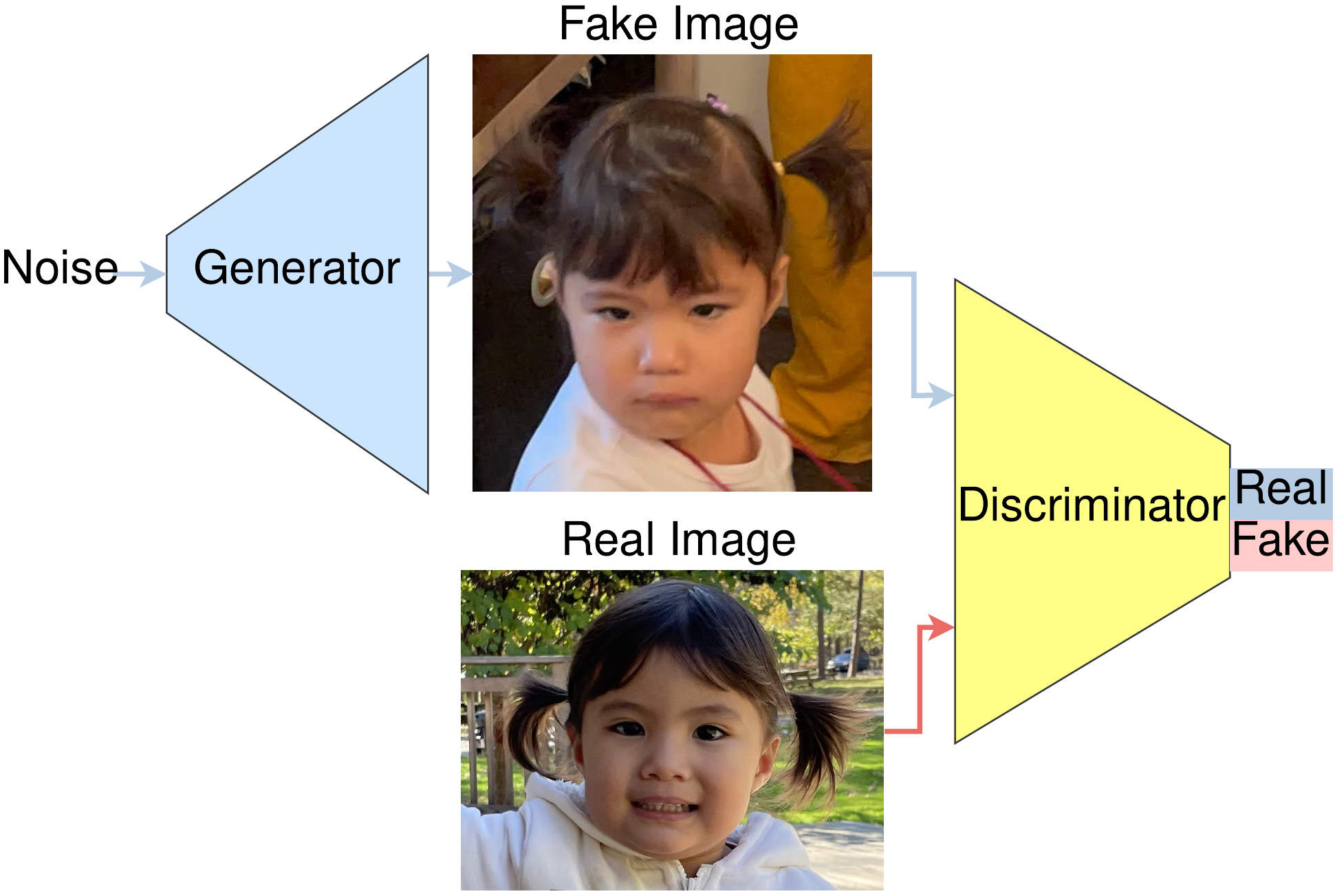}
    \caption[GAN Procedure]{\textbf{GAN Procedure.} The generator creates an image from random noise and provides it to the discriminator along with real images. The discriminator must identify which images are real and which are fake.}
    \labfig{dl:gan}
\end{marginfigure}
The many variants of the \glspl{gan} would change this dramatically using an ingenious and fairly simple idea. The \gls{gan} methods sets up an adversarial game with two networks. One network, the generator\index{generator}, generates images, and another, the discriminator\index{discriminator}, tries to identify which images are real and which are fake. The generator is rewarded for fooling the discriminator into classifying its images as real and penalized for getting caught. Conversely, the discriminator is rewarded for correctly identifying fake images and penalized for incorrectly classifying them. Training (theoretically) ends when the two networks achieve a \gls{nash equilibrium} \sidecite{nash1950equilibrium, nash1951non, heusel2017gans}. This procedure is shown in \reffig{dl:gan}.

We train this pair of networks using standard cross entropy classification loss. The only difference is that we reverse the labels when training the generator since we want it to fool the discriminator. This is sometimes call the minimax loss. Given real samples $x$, noise vectors $z$, discriminator $D()$, and generator $G()$, we define the loss
\begin{align}
    l(x, z) = \log(D(x)) + \log(1 - D(G(z)))
\end{align}
and we train the discriminator to maximize $l()$ while training the generator to minimize $l()$. In other words
\begin{align}
    \underset{G}{\text{min}} \underset{D}{\text{max}} \e_{x \in \text{real}}[\log(D(x))] + \e_{z \in \text{noise}}[\log(1 - \log(D(G(Z))))]
\end{align}

As these two networks play their game over the course of training, the discriminator will start to identify more and more fake images. The increasing loss on the generator will cause it to generate more realistic images. Since identifying fake images is relatively easy for a \gls{cnn}, by the end of training, the generator will be producing extremely realistic images in order to continue to fool the discriminator.  In practice the Nash equilibrium is hard to achieve and we simply stop training GANs after a certain number of steps. \Glspl{gan} also chronically diverge since it is hard for the \gls{gan} to recover from a situation where the discriminator has a large advantage over the generator.

\section{Recap}

To recap, we have reviewed machine learning from the ground up. We built the ideas of machine learning on a foundation of how to make decisions in the presence of perfect information. We then developed the perceptron and its extension, the multilayer perceptron which is the progenitor of modern deep learning. We discussed hand engineered features and why they were necessary and finally developed deep learning as a replacement for these features. We then reviewed some of the most important ideas of deep learning including convolutional networks, residual learning, U-Nets, and GANs. This concludes the foundational knowledge which is required to fully understand the original research developed in the remainder of this dissertation.
\pagelayout{wide}
\addpart{Image Compression}
\pagelayout{margin}

\setchapterpreamble[u]{\margintoc}
\chapter{JPEG Compression}
\labch{jpeg}

\lettrine{J}{PEG}\index{JPEG} has been a driving force for internet media since its standardization in 1992 \parencite{wallace1992jpeg}. The principal idea in \gls{JPEG} compression is to identify which details of an image are the least likely to be noticed if they are missing. These details can then be replaced with lower \gls{entropy} versions. By removing information, there is a significant size reduction over methods which perform entropy coding alone. This is called \gls{lossy compression}\index{lossy compression}, since information is lost in the encoding process.

The lost information is, in general, not recoverable. Usually this is not a major issue, as the JPEG algorithm was designed to remove unnoticed details. However, there are situations where the information loss is noticeable in the form of unpleasant artifacts (\reffig{jpeg:info}). This is particularly true when a JPEG image is saved multiple times, which causes repeated application of the lossy process. A significant portion of the dissertation is devoted to using machine learning to approximate the lost information.

\begin{marginfigure}
    \centering
    \includegraphics{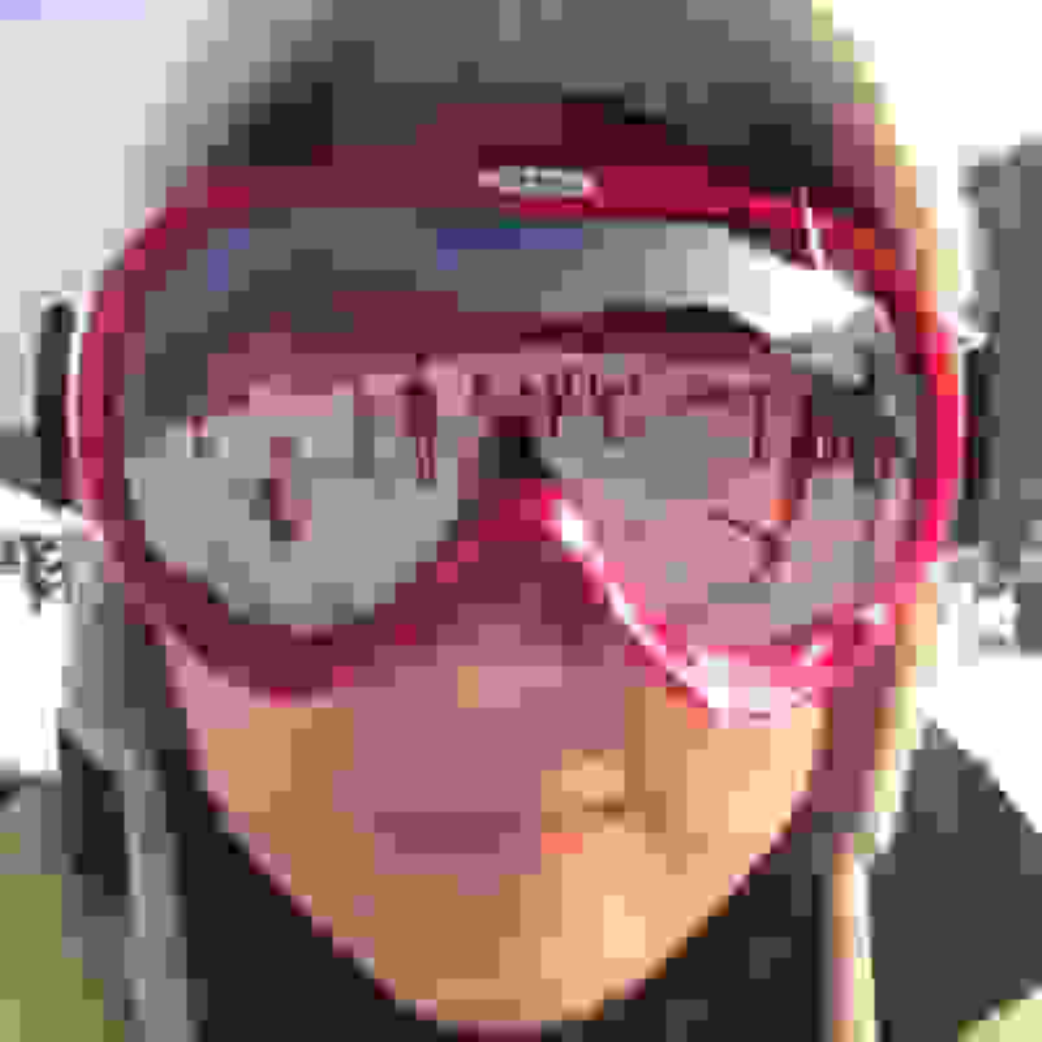}
    \caption[JPEG Information Loss]{\textbf{JPEG Information Loss.} This image suffers from extreme degradations caused by JPEG compression. Zoom in on this image, it probably has fewer details than you think it does.}
    \labfig{jpeg:info}
\end{marginfigure}

A common source of consumer confusion with JPEG is in the name itself. JPEG refers to three things simultaneously
\begin{description}
    \item[The JPEG Algorithm] The algorithm for compressing images.
    \item[JPEG Files] The disk file format for storing JPEG compressed data and its associated metadata. This is actually either a \gls{jfif}\index{JFIF} file or an \gls{exif}\index{EXIF} file.
    \item[The Joint Photographic Experts Group] The working group that maintains the JPEG standard.
\end{description}

This chapter is devoted to giving the reader an understanding of JPEG compression which is sufficient to motivate the first-principles that we use in developing the algorithms later in the dissertation. We will review the function of JPEG compression and decompression step-by-step and we will discuss the extremely important view of JPEG as a linear map\index{linear map}. We will also briefly discuss other image compression algorithms.

\section{The JPEG Algorithm}

We now present the JPEG algorithm step-by-step. Where the standard is ambiguous we defer to the Independent JPEG Group's libjpeg software \sidecite{libjpeg}. This software is widely considered standard in the industry, although there are other implementations of JPEG. We start by describing the compression process and then conclude with the decompression process, which is largely the inverse. Throughout the description we will place emphasis on which parts of the standard are motivated by human perception and which steps involve loss of information.

\subsection{Compression}

JPEG compression starts with an RGB image usually in \gls{interlaced} (RGB24) format. This image is then converted to the YCbCr \gls{planar} format, however, this is not the more common ITU-R BT.601 \sidecite{ITURBT601} format, which produces values in [16, 235] for $Y$ and [16, 240] for $C_b$, $C_r$. Instead, this format uses the full range of byte values ([0, 255]). The color conversion uses the following three equations
\begin{align}
    Y = \mathcolor{BrickRed}{2.99R} + \mathcolor{ProcessBlue}{0.587B} + \mathcolor{JungleGreen}{0.114G}                                       \\
    \mathcolor{YellowGreen}{C_b} = 128 - \mathcolor{BrickRed}{0.168736R} - \mathcolor{ProcessBlue}{0.331264B} + \mathcolor{JungleGreen}{0.5G} \\
    \mathcolor{RubineRed}{C_r} = 128 + \mathcolor{BrickRed}{0.5R} - \mathcolor{ProcessBlue}{0.418688B} - \mathcolor{JungleGreen}{0.081312G}
\end{align}
This color conversion is designed to better represent human perception of the image which treats changes in \gls{luminance} (the Y channel) with more weight than \gls{chrominance} (the $C_b$ and $C_r$ channels). Therefore, the $C_b$ and $C_r$ channels can have more information removed with less of an effect on the overall image.

One operation in particular which removes additional information from the color channels is \gls{chroma subsampling}. \Gls{chroma subsampling} describes a $4 \times 2$ block of pixels and is represented as a triple, \eg, $\mathcolor{JungleGreen}{4}:\mathcolor{BurntOrange}{2}:\mathcolor{Plum}{0}$. The $\mathcolor{JungleGreen}{4}$ represents the number of luma samples per row. The $\mathcolor{BurntOrange}{2}$ represents the number of chroma samples in the 1st row. The $\mathcolor{Plum}{0}$ represents the number of chroma samples which change in the second row. So in this example, there are $\mathcolor{JungleGreen}{4}$ luma samples in each row, $\mathcolor{BurntOrange}{2}$ chroma samples in the first row, and \textcolor{Plum}{none} of them change in the second row, meaning that the chroma channels should be stored at half the width and height of the luma channel. Another example is $\mathcolor{JungleGreen}{4}:\mathcolor{BurntOrange}{2}:\mathcolor{Plum}{2}$ which indicates that the $\mathcolor{BurntOrange}{2}$ chroma samples in the first row \textcolor{Plum}{both} change in the second row, so the chroma channels are stored with half the width but the same height as the luma channel\sidenote{This is an archaic and confusing notation but unfortunately it is still used.}.

Before we remove information we need to pad the image. JPEG is based on $8 \times 8$ blocks so at the least the image needs to be padded to a multiple of 8 in the width and height. If \gls{chroma subsampling}\index{chroma subsampling} is used, this needs to be taken into account during padding and the image may need to be padded to a multiple of 16 or more in the width, height, or both. This defines the \gls{mcu}\index{minimum coded unit}, \ie, the minimum size block which can be encoded using the given settings. The padding in this case is always done on the bottom and right edges of the image and repeats the final sample as the padding value. With the image padded, the chroma channels can be subsampled.

Next comes the main feature of the JPEG algorithm, the \gls{dct}\index{dct} on non-overlapping $8 \times 8$ blocks. Before computing this, the pixels are centered by subtracting 128. The \gls{dct} is applied using
\begin{align}
    D_{ij} = \frac{1}{4}C(i)C(j)\sum_{x=0}^7\sum_{y=0}^7 P_{xy} \cos\left[\frac{(2x+1)i\pi}{16}\right]\cos\left[\frac{(2y+1)j\pi}{16}\right] \\
    C(u) = \begin{cases}
               \frac{1}{\sqrt{2}} & u = 0    \\
               1                  & u \neq 0
           \end{cases}
\end{align}
for $8 \times 8$ block of pixels $P$. This accomplishes two goals. First, it concentrates the energy of each block into the top left corner. Second, it serves as a frequency transform which allows us to remove frequencies which we believe viewers will be less likely to notice.

The \gls{dct}\index{dct} coefficients are then quantized by dividing by a quantization matrix\index{quantization matrix}. This is an $8 \times 8$ matrix of coefficients which reduce the magnitude of the \gls{dct} coefficients. Since humans tend not to notice missing high spatial frequencies, the quantization matrices generally target these. However, most encoders compute the quantization matrix from a scalar quality factor which is easier for users to comprehend, and as this quality decreases, the quantization matrix removes lower and lower frequencies. After quantization, the result is truncated to an integer. This removes information in the fractional part and permits the result to be stored in an integer which takes up less space. In a sense this is the first ``compression'' operation. The entire operation is given by
\begin{align}
    Y'_{ij} = \left\lfloor\frac{Y_{ij}}{(Q_y)_{ij}}\right\rceil         \\
    (C_b')_{ij} = \left\lfloor\frac{(C_b)_{ij}}{(Q_c)_{ij}}\right\rceil \\
    (C_r')_{ij} = \left\lfloor\frac{(C_r)_{ij}}{(Q_c)_{ij}}\right\rceil
\end{align}
for \gls{luminance} quantization matrix $Q_y$ and chrominance quantization matrix $Q_c$. The color channels are often quantized more coarsely as human vision is less sensitive to color data. Note that since we truncate, any fractional part after division is irrevocably lost, the resulting coefficient can only be approximated from the integer part. Any coefficient which is less than zero after division is set to zero and cannot be recovered even approximately. Other than chroma subsampling, this is the only source of loss in JPEG compression. In order to decode the image, $Q_y$ and $Q_c$ are both stored in the JPEG file.

\begin{marginfigure}
    \centering
    \includegraphics{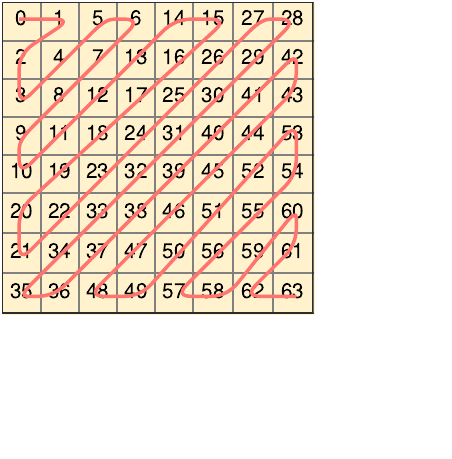}
    \caption[Zig-Zag Order]{\textbf{Zig-Zag Order.} This ordering is intended to put low frequencies in the beginning and high frequencies at the end.}
    \labfig{jpeg:zigzag}
\end{marginfigure}

These quantized coefficients are then vectorized in a zig-zag order (\reffig{jpeg:zigzag}) which is designed to put low frequencies in the beginning of the 64 dimensional vectors and high frequencies at the end. This is because the next step is to run-length code\index{run-length code} this vector. Since the quantization process was more likely to zero out high frequency coefficients, this concentrates the zeros at the end of the vector and leads to more effective run-length coding. This is the second ``compression'' operation.

The final run-length coded vectors are then entropy coded. This can use either Huffman coding \sidecite{huffman1952method} or arithmetic coding \sidecite{rissanen1979arithmetic}. With a significant amount of redundant or unnoticeable information removed, these entropy coding operations are extremely efficient and yield a significant space reduction over the uncompressed image.

\subsection{Decompression}

The decompression algorithm is largely the reverse operation. After undoing entropy coding we have the quantized coefficients. These are element-wise multiplied by the quantization matrices to compute the approximated coefficients
\begin{align}
    \widehat{Y}_{ij} = Y'_{i, j}(Q_y)_{ij}         \\
    (\widehat{C_b})_{ij} = (C_b')_{i, j}(Q_c)_{ij} \\
    (\widehat{C_r})_{ij} = (C_r)'_{ij}(Q_c)_{ij}
\end{align}

We can then compute the inverse \gls{dct}\index{dct} of the approximated coefficients
\begin{align}
    P_{xy} = \frac{1}{4}\sum_{i=0}^7\sum_{j=0}^7C(i)C(j)\widehat{D}_{ij}\cos\left[\frac{(2x+1)i\pi}{16}\right]\cos\left[\frac{(2y+1)j\pi}{16}\right]
\end{align}
and uncenter the spatial domain result by adding 128. The color channels are interpolated to remove chroma subsampling. We then remove any padding that was added, and convert the image back to the RGB24 color space
\begin{align}
    \mathcolor{BrickRed}{R} = Y + \mathcolor{RubineRed}{1.402(C_r - 128)}                                                      \\
    \mathcolor{JungleGreen}{G} = Y - \mathcolor{YellowGreen}{0.344136(C_b - 128)} - \mathcolor{RubineRed}{0.714136(C_r - 128)} \\
    \mathcolor{ProcessBlue}{B} = Y + \mathcolor{YellowGreen}{1.772(C_b - 128)}
\end{align}
and the image is ready for display.

There are three important things to take away from this discussion. First, other than \gls{chroma subsampling}, which is optional, the only lossy operation is the truncation during the quantization step. This is a fairly simple operation considering the \gls{dct} coefficients but it creates complex patterns in the spatial domain. Next, the blocks are non-overlapping, so for each block there is no dependence on pixels outside of the block. Finally, each pixel in the block depends on all of the coefficients in the block. Conversely, each coefficient in the block also depends on all of the pixels in the block. We will exploit this property later in the dissertation.

\section{The Multilinear JPEG Representation}
\labsec{multilinear_jpeg}

In what is perhaps a surprising result, the steps of the JPEG transform are easily linearizable \sidecite{smith1994fast}, a property that was explored significantly in the 1990s \sidecite{chang1992video, shen1995inner, natarajan1995fast, smith1993algorithms}. Indeed, outside of entropy coding, the only non-linear step in compression is the truncation that occurs during quantization, and all the steps of decompression are linear. Furthermore, when we process JPEG images, we are either dealing with the decompression process, or we are in full control over the compression process and it is therefore our choice if and when we truncate. We would only need to do this if we were saving the result as a JPEG. We now develop the steps of the JPEG algorithm into linear maps and compose them into a single linear map that models compression and a single linear map which models decompression.

Without loss of generality, consider a single channel (grayscale) image. We model this image as the type-(0, 2) tensor $I \in H^* \otimes W^*$. Note that although we are essentially dealing with real numbers, we have intentionally left $H^*$ and $W^*$ as arbitrary co-vector spaces because there is no reason to define them concretely for our purposes. We will, however, make the stipulation that they are defined with respect to a standard orthonormal basis so that we can freely convert between the co-vector and vector spaces without the use of a metric tensor. Note that the following equations are written in Einstein notation; see \nrefch{multilinear_algebra} if this is unfamiliar.

Our first task is to break this image into $8 \times 8$ blocks. We define the linear map
\begin{align}
    B : H^* \otimes W^* \rightarrow X^* \otimes Y^* \otimes M^* \otimes N^* \\
    B \in H \otimes W \otimes X^* \otimes Y^* \otimes M^* \otimes N^*       \\
    B^{hw}_{xymn} = \begin{cases}
                        1 & \text{pixel}\;h, w\;\text{belongs in block}\;x, y\;\text{at offset}\;m, n \\
                        0 & \text{otherwise}
                    \end{cases}
\end{align}
where $B$ is a type-(2, 4) tensor defining a linear map on type-(0, 2) tensors. The result of this map will be a tensor with $8 \times 8$ blocks indexed by $x, y$ and 2D offsets for each block indexed by $m, n$. Although this definition is fairly abstract, it can be computed fairly easily using modular arithmetic, although it does need to be recomputed for each image.

Next we compute the DCT of each block,. We define the following linear map
\begin{align}
    D : M^* \otimes N^* \rightarrow A^* \otimes B^*                                                                                          \\
    D \in M \otimes N \otimes A^* \otimes B^*                                                                                                \\
    D_{\alpha\beta}^{mn} = \frac{1}{4}C(\alpha)C(\beta)\cos\left(\frac{(2m+1)\alpha\pi}{16}\right)\cos\left(\frac{(2n+1)\beta\pi}{16}\right) \\
    C(u) = \begin{cases}
               \frac{1}{\sqrt{2}} & u = 0    \\
               1                  & u \neq 0
           \end{cases}
\end{align}
The equation for $D$ should look familiar by now. $D$ is a type-(2, 2) tensor defining a linear map on type-(0, 2) tensors. The $m,n$ block offset indices in the input tensor will index spatial frequency after applying this map.

Next we linearize the coefficients \sidenote{Note that we're doing things slightly out of order but ultimately the order does not matter here and doing it this way simplifies the form of the next tensor}. We define the following linear map
\begin{align}
    Z : A^* \otimes B^* \rightarrow \Gamma^* \\
    Z \in A \otimes B \otimes \Gamma^*       \\
    Z^{\alpha\beta}_\gamma = \begin{cases}
                                 1 & \alpha, \beta\;\text{is at}\;\gamma\;\text{under zigzag ordering} \\
                                 0 & \text{otherwise}
                             \end{cases}
\end{align}
This is a type-(2, 1) tensor defining a linear map on type-(0, 2) tensors. It flattens the $8 \times 8$ blocks into 64 dimensional vectors. In other words, the $\alpha, \beta$ indices indicate which indexed spatial frequency will be indexed with a single $k$ after applying this transformation. This tensor depends on the zigzag ordering and can simply be hard coded.

Finally, we divide by the quantization matrix. We still need to scale the coefficients even though we are not rounding them. We define the linear map
\begin{align}
    S : \Gamma^* \rightarrow K^* \\
    S \in \Gamma \otimes K^*     \\
    S^\gamma_k = \frac{1}{q_k}
\end{align}
where $q_k$ is the $k$th entry in the quantization matrix for the JPEG image. This is a type-(1, 1) tensor defining a linear map on co-vectors.

In order to define both compression and decompression, we need only one more linear map, scaling by the quantization matrix
\begin{align}
    \widetilde{S} : K^* \rightarrow \Gamma^* \\
    \widetilde{S} \in K \otimes \Gamma^*     \\
    \widetilde{S}^k_\gamma = q_k
\end{align}
for the same definition of $q_k$ as above.

We now have the fairly simple task of assembling these steps into single tensors. We say this is simple because all of the operations are linear maps and therefore are readily composable. We define
\begin{align}
    J : H^* \otimes W^* \rightarrow X^* \otimes Y^* \otimes K^* \\
    J \in H \otimes W \otimes X^* \otimes Y^* \otimes K^*       \\
    J^{hw}_{xyk} = B^{hw}_{xymn}D^{mn}_{\alpha\beta}Z^{\alpha\beta}_\gamma S^\gamma_k
\end{align}
for compression and
\begin{align}
    \widetilde{J} : X^* \otimes Y^* \otimes K^* \rightarrow H^* \otimes W^* \\
    \widetilde{J} \in X \otimes Y \otimes K \otimes H^* \otimes W^*         \\
    \widetilde{J}^{xyk}_{hw} = B^{xymn}_{hw}D^{\alpha\beta}_{mn}Z^\gamma_{\alpha\beta}\widetilde{S}^k_\gamma
\end{align}
for decompression.

It is difficult to express how powerful this result is and how easily it is achieved using rudimentary concepts from multilinear algebra. What seems like a fairly complex algorithm, and indeed is, when thought of as an operation on a matrix, reduces to a simple linear map when we model the inputs and intermediate steps as tensors. Equipped with this linear map, we can and will model complex phenomena on compressed \gls{JPEG}\index{JPEG} data directly without needing to decompress it.

\section{Other Image Compression Algorithms}

The astute reader will have noticed early on that we, quite confidently, are in a part labeled ``image compression'' and yet we are only discussing \gls{JPEG}. There are other image compression algorithms, so a natural question is ``why are we not discussing those?''

For myriad reasons, there are really no interesting problems to study for other compression algorithms. PNG \sidecite{rfc2083}, for example, is widely used. But this is lossless compression, so there are no artifacts to overcome. GIF \sidecite{gif} is also lossless, although many GIF services quantize colors into a palette to save more space, which is a potential problem that could be interesting to work on. More modern formats are based on video compression and while they are lossy, they are simply unused. BPG \sidecite{bpg} is the most promising of these, and therefore the least used, but there is also HEIC/HEIF \sidecite{heif} which is currently being unsuccessfully pushed by Apple on the iPhone.

Probably the most interesting of these algorithms is JPEG 2000 \sidecite{skodras2001jpeg}, which is lossy and widely used in digital cinema \sidecite{swartz2004understanding}, although it was completely ignored by consumers. This codec is interesting because it would require us to update our theory to take into account the discrete wavelet transform that JPEG 2000 uses in place of the DCT. However, the use of this transform imposes a major practical problem as well: JPEG 2000 images look good even at low bitrates because the wavelet transform is so effective, so they may not require correction.

Instead, we will focus our energy where it can have the most impact: by exclusively studying JPEG. Even at the time of writing, 30 years after standardization, JPEG is the most commonly used image file format. It is easy to use, familiar to consumers, and has become the backbone of the internet, making it incredibly resilient to any challenger, no matter how much better the compression or quality of the images. At the same time, JPEG does suffer from some extreme artifacts in many conditions. It is this combination of visible quality loss and widespread use that makes JPEG ideal for further study.
\setchapterpreamble[u]{\margintoc}
\chapter{\color{Plum}JPEG Domain Residual Learning}
\labch{jdr}

\lettrine{N}{ow} we develop a general method for performing residual network~(\cite{he2016deep}, \nrefsec{resnets})\index{residual learning} learning and inference on \gls{JPEG}\index{JPEG} data directly, \ie, without the need for decompressing the images. This method was published separately in the proceedings of the International Conference on Computer Vision~\parencite{ehrlich2019deep}.

\begin{warningbox}
    This chapter is extremely math-heavy and dry. It is strongly recommended to review the background math outlined in the first chapters of the dissertation for a complete understanding of the material. This chapter may serve as a powerful sleep aid: do not operate heavy machinery while reading this chapter.
\end{warningbox}

Compared to processing data in the pixel domain, directly working on JPEG data has several advantages. First, we observe that most images on the internet are compressed, including deep learning datasets such as ImageNet~\parencite{deng2009imagenet}. Next we observe that JPEG\index{JPEG} images, being compressed, are naturally smaller and more sparse than their uncompressed counterparts. These are all desirable properties for memory- and compute- hungry \gls{deep learning} algorithms.

The primary goal of the method presented in this section is to be as close as possible to the pixel domain result. In other words, given a learned pixel domain mapping $H: {I} \rightarrow {\mathbb{R}^c}$ mapping images to class probabilities for ${c}$ classes\sidenote{We frame this discussion in terms of classification but it applies equally well to any problem type}, we want to define a mapping $\Delta(H)$ such that $\left|H({I_n}) - \Delta(H)(\text{DCT}({I_n}))\right|$ is minimized for any ${I_n} \in {I}$. We can accomplish this goal analytically and we will develop the theory in the coming sections, including a discussion of why it is (likely) not possible to generate a mathematically exact $\Delta(H)$ function\sidenote{\ie, a $\Delta$ function such that for all $i$, $\left|H(i) - \Delta(H)(\text{DCT}(i))\right| = 0$} and what guarantees are available on the deviation.

Recall that a residual network requires several components to operate
\begin{description}
    \item[Convolution] The primary learned linear mapping between feature maps at each layer. Each ``residual block'' contains two of these operations.
    \item[Batch Normalization] Produces normalized features for the convolutions; this is thought to ease the learning process by removing unnecessary statistics from the input features which can be represented exactly \sidecite{ioffe2015batch}.
    \item[Global Average Pooling] An innovation of the ResNet. When the convolution layers are exhausted, the features are averaged channel-wise to produce a vector suitable for input to a fully connected layer
    \item[ReLU] The non-linearity of the ``residual block'', this allows the network to learn complex mapping.
\end{description}
Our task will be to derive transform domain versions of each of these operations.

\begin{fpbox}
    \begin{itemize}
        \item JPEG is easily linearized, convolutions are linear, composing them expressed a learned convolution exactly in the JPEG domain.
        \item Other components of the residual network can be expressed analytically in the JPEG domain.
        \item ReLU can be approximated with a bilinear map.
    \end{itemize}
\end{fpbox}

\section{New Architectures}
\labsec{newarch}

Before discussing the proposed technique, we will first make a detour to review two popular methods of JPEG\index{JPEG} and DCT\index{DCT} deep learning. These methods are all new architectures which enable effective processing in the \gls{transform domain}\index{transform domain} but which do not attempt to replicate any pixel domain result. These methods have some advantages over both the method presented in the rest of this chapter and when compared to pixel-domain networks. For example, both methods show good task accuracy with faster processing. There are some notable disadvantages, however. In particular, these methods are not suitable for situations when a pixel domain network already exists and its results need to be replicated on \glspl{JPEG}\index{JPEG}.

These ideas were inspired by the ``do nothing'' approach published in both NeurIPS and an ICLR workshop~\sidecite{gueguen2018faster}. In this approach, the transform coefficients are passed into a mostly unmodified ResNet for classification. The authors postulate that with the higher-level representation of the \gls{dct}, fewer layers are required to achieve similar accuracy and therefore the network will be faster. Indeed the authors show this is true empirically. However, despite the intuition, this paper's evaluation leaves much to be desired, and it is unclear what the contribution of the \gls{dct} is to the result. Meanwhile, it is well known in \gls{JPEG} artifact correction literature (discussed later in the dissertation), where ``dual-domain'' methods are commonplace, that providing \gls{dct} coefficients to a network is not successful without considerable effort, a result seeming at odds with Gueguen \etal.

Instead, the following methods are inspired by unique attributes of \gls{dct} coefficients. Namely: that each coefficient is a function of \textit{all} pixels in a block, that a block of pixels is only correctly represented by \textit{all} DCT coefficients, and that the DCT coefficients are orthogonal and arranged in a grid simply for convenience. This last point is critical. One of the reasons that small convolutional kernels work well on pixels is that nearby pixels are usually correlated in some way, so translation invariant features are readily learned. If that convolution is instead applied to coefficients, this is then a mapping on arbitrary orthogonal measurements which are \textit{intentionally decorrelated}, leaving little hope for success.

\subsection{Frequency-Component Rearrangement}

This is treated by Lo~\etal~\sidecite{lo2019exploring} by simply rearranging the frequencies into the channel dimension before processing (\reffig{fcr}), yielding a feature map which is $\frac{1}{8}$th the width and height and with 64 channels (per input channel). Note how this allows a convolution to capture the information contained in the DCT. Since the convolution operation used in deep learning maps \textit{all} channels in the input to \textit{each} channel in the output, every coefficient plays a role in the resulting map and therefore complete block information is captured. They call this method \gls{fcr}\index{frequency component rearrangement}

\begin{marginfigure}
    \centering
    \includegraphics{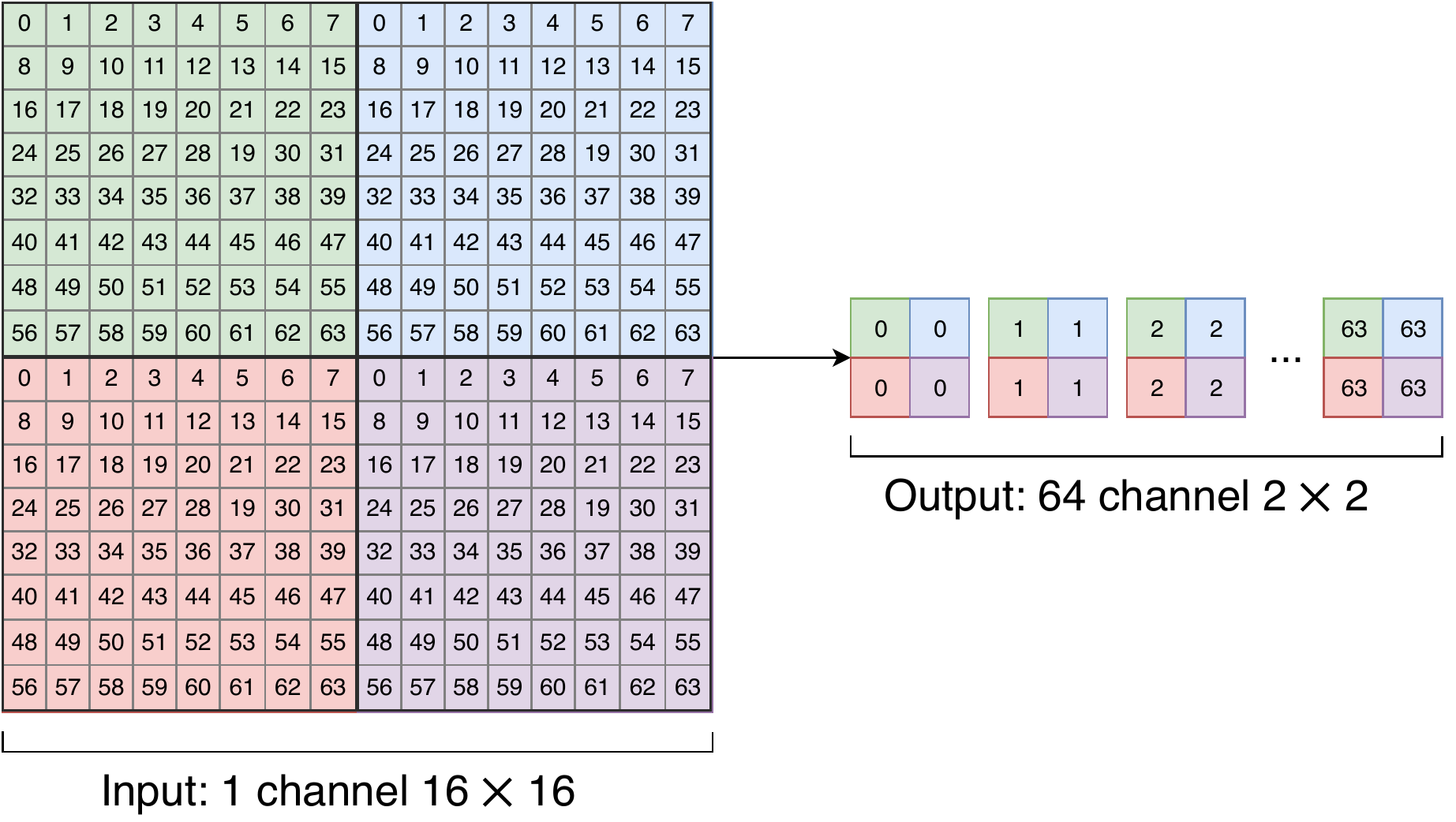}
    \caption[Frequency Component Rearrangement]{Frequency component rearrangement.}
    \labfig{fcr}
\end{marginfigure}

Lo~\etal use their network for semantic segmentation of road features quite successfully. At the time of publication, their method was both fast and accurate.

\subsection{Strided Convolutions}

A similar solution proposed by Deguere~\etal~\sidecite{deguerre2019fast} uses strided convolutions instead of \gls{fcr}. Specifically, this method uses an $8 \times 8$ stride-8 convolution such that each DCT block is processed in isolation. Note again how this makes good use of the coefficients: the $8 \times 8$ convolution ensures that every coefficient plays a role in the resulting mapping, and the stride-8 ensures that there is no leakage of information across blocks. Once these ``block representations'' are computed, the resulting feature map is again $\frac{1}{8}$th the width and height (now with a variable number of features) . Deguere~\etal use this method for object detection in the DCT domain and again performed admirably at the time of publication.
\section{Exact Operations}

In the previous section we discussed novel architectures that equip \glspl{cnn}\index{convolutional neural networks} with the ability to process data in the transform domain\index{transform domain}. While this is useful and important, it requires training a new \gls{cnn} from scratch and has no particular relationship to the underlying pixels that the \gls{cnn} is processing. Since \glspl{cnn} were designed to process pixel domain data, and the \gls{dct} is a transform of pixel data, a natural question is whether a method can be formulated that is capable of processing transform domain data and which has some mathematical guarantee or relationship to the underlying pixel domain model. We now develop just such a method.

\subsection{JPEG Domain Convolutions}
\labsec{jdc}

Recall from \nrefsec{multilinear_jpeg} that the JPEG transform can be linearized and written as linear maps on tensor inputs and that this analysis yields the following linear maps:
\begin{align}
    \mathcolor{BurntOrange}{J} : \mathcolor{JungleGreen}{H^*} \otimes \mathcolor{JungleGreen}{W^*} \rightarrow \mathcolor{Plum}{X^*} \otimes \mathcolor{Plum}{Y^*} \otimes \mathcolor{Plum}{K^*}
\end{align}
for compression of an image represented by $\mathcolor{JungleGreen}{I} \in \mathcolor{JungleGreen}{H^*} \otimes \mathcolor{JungleGreen}{W^*}$ to transform coefficients $\mathcolor{Plum}{F} \in \mathcolor{Plum}{X^*} \otimes \mathcolor{Plum}{Y^*} \otimes \mathcolor{Plum}{K^*}$, and
\begin{align}
    \mathcolor{BurntOrange}{\widetilde{J}} : \mathcolor{Plum}{X^*} \otimes \mathcolor{Plum}{Y^*} \otimes \mathcolor{Plum}{K^*} \rightarrow \mathcolor{JungleGreen}{H^*} \otimes \mathcolor{JungleGreen}{W^*} \\
\end{align}
to decompress. We proceed by considering only single channel images. We will add in channels and batch dimensions later since they have no bearing on the derivation.

We know that convolutions are linear maps, therefore, deriving a JPEG domain convolution is fairly simple. Assume that $\mathcolor{BurntOrange}{C} : \mathcolor{JungleGreen}{H^*} \otimes \mathcolor{JungleGreen}{W^*} \rightarrow \mathcolor{JungleGreen}{H^*} \otimes \mathcolor{JungleGreen}{W^*}$ is a linear map representing an arbitrary convolution. This convolution would be applied to an image $I$ in the pixel domain by computing
\begin{align}
    \mathcolor{JungleGreen}{I'_{h'w'}} = \mathcolor{BurntOrange}{C^{hw}_{h'w'}}\mathcolor{JungleGreen}{I_{hw}}
\end{align}
Given transform coefficients $\mathcolor{Plum}{F} \in \mathcolor{Plum}{X^*} \otimes \mathcolor{Plum}{Y^*} \otimes \mathcolor{Plum}{K^*}$ for $\mathcolor{JungleGreen}{I}$, we can derive $\mathcolor{JungleGreen}{I}$ as
\begin{align}
    \mathcolor{JungleGreen}{I_{hw}} = \mathcolor{BurntOrange}{\widetilde{J}^{xyk}_{hw}}\mathcolor{Plum}{F_{xyk}}
\end{align}
Similarly, we can derive transform coefficients $\mathcolor{Plum}{F'}$ for $\mathcolor{JungleGreen}{I'}$ by applying $\mathcolor{BurntOrange}{J}$
\begin{align}
    \mathcolor{Plum}{F'_{x'y'k'}} = \mathcolor{BurntOrange}{J^{h'w'}_{x'y'k'}}\mathcolor{JungleGreen}{I'_{h'w'}}
\end{align}
Substituting these two expressions yields
\begin{align}
    \mathcolor{JungleGreen}{I'_{h'w'}} = \mathcolor{BurntOrange}{C^{hw}_{h'w'}}\mathcolor{BurntOrange}{\widetilde{J}^{xyk}_{hw}}\mathcolor{Plum}{F_{xyk}} \\
    \mathcolor{Plum}{F'_{x'y'k'}} = \mathcolor{BurntOrange}{J^{h'w'}_{x'y'k'}}\mathcolor{BurntOrange}{C^{hw}_{h'w'}}\mathcolor{BurntOrange}{\widetilde{J}^{xyk}_{hw}}\mathcolor{Plum}{F_{xyk}}
\end{align}
And we make the following definition
\begin{align}
    \mathcolor{Plum}{F'_{x'y'k'}} = \left[\mathcolor{BurntOrange}{J^{h'w'}_{x'y'k'}}\mathcolor{BurntOrange}{C^{hw}_{h'w'}}\mathcolor{BurntOrange}{\widetilde{J}^{xyk}_{hw}}\right]\mathcolor{Plum}{F_{xyk}} \\
    \mathcolor{BurntOrange}{\Xi^{xyk}_{x'y'k'}} = \mathcolor{BurntOrange}{J^{h'w'}_{x'y'k'}}\mathcolor{BurntOrange}{C^{hw}_{h'w'}}\mathcolor{BurntOrange}{\widetilde{J}^{xyk}_{hw}}
\end{align}
giving a simple expression for computing $\mathcolor{BurntOrange}{\Xi} : \mathcolor{Plum}{X^*} \otimes \mathcolor{Plum}{Y^*} \otimes \mathcolor{Plum}{K^*} \rightarrow \mathcolor{Plum}{X^*} \otimes \mathcolor{Plum}{Y^*} \otimes \mathcolor{Plum}{K^*}$, a convolution in the compressed domain, given a convolution in the pixel domain. It is important to note that this is not a simple notational trick. Because ${J}, {C},$ and ${\widetilde{J}}$ are linear maps, the resulting ${\Xi}$ performs all three operations in a single step and is significantly faster than performing them separately\sidenote{Much in the same way that linear functions $f({x}) = 5{x}$ and $g({x}) = 2{x}$ can be combined into $(f \circ g)({x}) = 10{x}$ which has only a single multiply \vs the two multiplies of separately applying $g$ and then $f$}.

With the mathematics satisfied, we now turn to the development of an efficient algorithm for computing ${\Xi}$. After all, the convolution ${C}$ is usually represented as a simple $3 \times 3$ matrix of numbers\sidenote{Although other sizes and shapes are possible}. However our derivation is expressed in terms of an $\text{dim}(H) \times \text{dim}(W) \times \text{dim}(H) \times \text{dim}(W)$ (2, 2)-tensor.

One way to understand ${C}$ is as a look-up table of coefficients. For example, if we index ${C}$ as \texttt{C[5, 7]}, we are given a tensor of coefficients for \textit{every pixel} in the input representing its contribution to the $(5,7)$ pixel in the output. Naturally, many of these coefficients are 0. In fact, the only non-zero pixels are those from $(2, 4)$ to $(8, 10)$. Similarly, if we index ${C}$ as \texttt{C[:, :, 5, 7]} we can see the contribution of pixel $(5, 7)$ in the input to every output pixel (which again is mostly zero). This implies a naive algorithm: Exploding Convolutions (Listing \ref{lst:expconv}) where the entire (2,2)-tensor is iterated and the correct coefficients are copied from the convolution kernel. The resulting map is then composed with ${J}$ and ${\widetilde{J}}$ to produce the transform domain map.

\begin{lstlisting}[caption={Exploding Convolutions (Naive)},language=Python,label={lst:expconv}]
def explode_convolution(shape: Tuple[int, int], conv: Tensor, J: Tensor, J_tilde: Tensor) -> Tensor:
    size = (conv.shape[0] // 2, conv.shape[1] // 2)
    shape = (shape[0] + size[0] * 2, shape[1] + size[1] * 2)

    c = torch.zeros((shape[0], shape[1], shape[0], shape[1]))
    for i in range(shape[0]):
        for j in range(shape[1]):            
            for u in range(shape[0]):
                for v in range(shape[1]):
                    hrange = (u - size[0], u + size[0])
                    vrange = (v - size[1], v + size[1])
            
                    if hrange[0] <= i <= hrange[1] and vrange[0] <= j <= vrange[1]:
                        x = u - i + size[0] 
                        y = v - j + size[1]
                        c[i, j, u, v] = conv[x, y]

    xi = torch.einsum("h'w'x'y'k',hwh'w',xykhw->xykx'y'k'", [J,c,J_tilde])
return xi
\end{lstlisting}

Although this algorithm is simple, it comes with some notable disadvantages. First, it is slow. Iterating over the entire (2,2)-tensor is time consuming even for a small image. Second, it is difficult to parallelize without domain knowledge of low-level programming. In other words, a CUDA kernel (or similar construct) would need to be produced to efficiently implement this algorithm. A better algorithm would be readily and efficiently programmed in a high level deep learning library like PyTorch~\sidecite{NEURIPS2019_9015}.

Examine the tensor ${\widetilde{J}}$ and note that
\begin{align}
    {\widetilde{J}} \in \mathcolor{Plum}{X} \otimes \mathcolor{Plum}{Y} \otimes \mathcolor{Plum}{K} \otimes \mathcolor{JungleGreen}{H^*} \otimes \mathcolor{JungleGreen}{W^*}
\end{align}
Recall that our model of single channel images uses ${I} \in \mathcolor{JungleGreen}{H^*} \otimes \mathcolor{JungleGreen}{W^*}$, therefore, the last two dimensions of ${\widetilde{J}}$ \textit{are} a single channel image and we can model ${\widetilde{J}}$ as a batch of single channel images by reshaping it to fold $\mathcolor{Plum}{X}, \mathcolor{Plum}{Y}, \mathcolor{Plum}{K}$ into a single dimension $\mathcolor{Plum}{N}$, giving
\begin{align}
    {\widehat{J}} \in \mathcolor{Plum}{N} \otimes \mathcolor{JungleGreen}{H^*} \otimes \mathcolor{JungleGreen}{W^*}
\end{align}
We are then free to convolve ${\widehat{J}}$ with the kernel\sidenote{Note that the definition of ${C}$ has changed slightly and is now kernel} ${C}$:
\begin{align}
    {\widehat{C}} = {C} \star {\widehat{J}}
\end{align}
and then reshape ${\widehat{C}}$ giving
\begin{align}
    {\widetilde{C}} \in \mathcolor{Plum}{X} \otimes \mathcolor{Plum}{Y} \otimes \mathcolor{Plum}{K} \otimes \mathcolor{JungleGreen}{H^*} \otimes \mathcolor{JungleGreen}{W^*}
\end{align}
Note that the shape of ${\widetilde{C}}$ and ${\widetilde{J}}$ are the same, all we have done here is compose the convolution kernel ${C}$ into the decompression operation ${\widetilde{J}}$. Next, we compose ${\widetilde{C}}$ and ${J}$
\begin{align}
    {\Xi}^{xyk}_{x'y'k'} = {\widetilde{C}}^{xyk}_{hw}{J}^{hw}_{x'y'k'}
\end{align}
to compute ${\Xi}$.

\begin{lstlisting}[caption={Exploding Convolutions (Fast)},language=Python,label={lst:expconv_fast}]
def explode_convolution(J_tilde: Tensor, J: Tensor, C: Tensor) -> Tensor:
    J_hat = J_tilde.flatten(0, 2)
    c_hat = torch.nn.functional.conv2d(J_hat, C)
    c_tilde = c_tilde.view_as(J_tilde)
    xi = torch.einsum("xykhw,x'y'k'hw->xykx'y'k'",[C_tilde, J])
    return xi
\end{lstlisting}

This algorithm (Listing \ref{lst:expconv_fast}) is simple to code in machine learning libraries. Here, it takes up only six lines of code and involves no loops. Furthermore, since this algorithm depends only on reshaping, convolution, and einsum, it can take advantage of the built-in optimizations that these libraries include resulting from years of research into these algorithms \sidecite{daniel2018opt, chetlur2014cudnn}. It is also worth noting that autograd algorithms used by these libraries will work as expected for this algorithm, \ie, it is straightforward to optimize ${C}$ with respect to some objective when ${\Xi}$ is used to transform the input feature maps.

Extending this to batches of multi-channel images is straightforward. First, we define the convolution ${C}$ as ${C} : {P^*} \otimes {H^*} \otimes {W^*} \rightarrow {P'^*} \otimes {H^*} \otimes {W^*}$ adding the input and output plane dimensions ${P}, {P'}$ and noting that ${C}$ lacks any batch dimension since the same operation is applied to each image in the batch. Next, we simply define ${\Xi}$ as
\begin{align}
    {\Xi}^{pxyk}_{p'x'y'k'} = {J}^{h'w'}_{x'y'k'}{C}^{phw}_{p'h'w'}{\widetilde{J}}^{xyk}_{hw}
\end{align}
where the ${J}, {\widetilde{J}}$ tensors have not changed. This simply adds the plane dimensions ${P}, {P'}$ to to ${\Xi}$. This map is applied to transform coefficients ${F} \in {N^*} \otimes {P^*} \otimes {X^*} \otimes {Y^*} \otimes {K^*}$ as
\begin{align}
    {F'}_{np'x'y'k'} = {\Xi}^{pxyk}_{p'x'y'k'}{F}_{npxyk}
\end{align}
where the batch dimension ${N}$ is preserved. With the exception of some extra indices, this does not change the algorithm in Listing \ref{lst:expconv_fast}.

\subsection{Batch Normalization}
\labsec{jdbn}

Batch normalization~\sidecite{ioffe2015batch}\index{batch normalization} is a commonly used technique which ensures each layer receives normalized feature maps. For a single channel feature map ${I} \in {H^*} \otimes {W^*}$ batch normalization uses the sample mean $\e[{I}]$ and variance $\var[{I}]$ along with learnable affine parameters ${\gamma}, {\beta}$. These parameters are then applied as
\begin{align}
    \text{BN}({I}) = {\gamma}\frac{{I} - \e[{I}]}{\sqrt{\var[{I}]}} + {\beta}
\end{align}
The batch statistics are used to update running statistics which are applied at inference time instead of the sample statistics. This equation has a simple closed-form expression in the \gls{transform domain}\index{transform domain}.

We start with the mean and variance. Recall from \nrefsec{fourier} the definition of the 2D Discrete Cosine Transform over ${N} \times {N}$ blocks
\begin{align}
    D({i},{j}) = \frac{1}{\sqrt{2{N}}}C({i})C({j})\sum_{{x}=0}^{{N-1}}\sum_{{y}=0}^{{N-1}} {I}({x}, {y})\cos\left(\frac{(2{x}+1){i}\pi}{2{N}}\right)\cos\left(\frac{(2{y}+1){j}\pi}{2{N}}\right) \\
    C({k}) = \begin{cases}
                 \frac{1}{\sqrt{2}} & {k} = 1    \\
                 1                  & {k} \neq 0
             \end{cases}
\end{align}
Let us compute an expression for the (0, 0) coefficient
\begin{align}
    D(0,0) = \frac{1}{\sqrt{2{N}}}C(0)C(0)\sum_{{x}=0}^{{N-1}}\sum_{{y}=0}^{{N-1}} {I}({x}, {y})\cos\left(\frac{(2{x}+1)0\pi}{2{N}}\right)\cos\left(\frac{(2{y}+1)0\pi}{2{N}}\right) \\
    = \frac{1}{2\sqrt{2{N}}}\sum_{{x}=0}^{{N-1}}\sum_{{y}=0}^{{N-1}} {I}({x}, {y}) \cos(0)\cos(0)                                                                                    \\
    = \frac{1}{2\sqrt{2{N}}}\sum_{{x}=0}^{{N-1}}\sum_{{y}=0}^{{N-1}} {I}({x}, {y})
\end{align}
We further assume $8 \times 8$ blocks as used by JPEG
\begin{align}
    = \frac{1}{2\sqrt{2\cdot8}}\sum_{{x}=0}^7\sum_{{y}=0}^7 {I}({x}, {y}) \\
    = \frac{1}{8}\sum_{x=0}^7\sum_{y=0}^7 {I}({x}, {y})                   \\
\end{align}
Since
\begin{align}
    \e[{I}] = \frac{1}{64}\sum_{{x}=0}^7\sum_{{y}=0}^7 {I}({x}, {y})
\end{align}
we have
\begin{align}
    \e[{I}] = \frac{1}{8}D(0,0)
\end{align}
yielding a simple expression for the sample mean of a block given DCT coefficients. Note that this is extremely efficient compared to computing the mean on the feature maps directly: it requires one read operation and one multiply operation per block \vs 64 reads, 63 sums, and one multiply~\sidenote{If there are multiple coefficient blocks (as is common) their means will need to be combined.}.

To compute the variance we use the following theorem
\begin{theorem}[The DCT Mean-Variance Theorem]
    Given a set of samples of a signal ${X}$ such that $E[{X}] = 0$, let ${Y}$ be the DCT coefficients of ${X}$. Then
    \begin{align}
        \var[{X}] = \e[{Y}^2]
    \end{align}
\end{theorem}
\vspace{-2em}
\begin{proof}
    Start by considering $\var[{X}]$, we write this as
    \begin{align}
        \var[{X}] = \e[{X}^2] - \e[{X}]^2
    \end{align}
    We are given $\e[{X}] = 0$, so we simplify this to
    \begin{align}
        \var[{X}] = \e[{X}^2]
    \end{align}
    Next, we use the DCT linear map ${D}: {M^*} \otimes {N^*} \rightarrow {A^*} \otimes {B^*}$ where the vector spaces ${M}$ and ${N}$ indicate the block dimensions and ${A}, {B}$ indicate spatial frequencies. Then:
    \begin{align}
        {X}_{mn} = {D}^{\alpha\beta}_{mn}{Y}_{\alpha\beta}
    \end{align}
    and
    \begin{align}
        E[{X}_{mn}^2] = E[(D^{\alpha\beta}_{mn}Y_{\alpha\beta})^2]
    \end{align}
    Expanding the squared term gives
    \begin{align}
        E[{X}_{mn}{X}_{mn}] = E[{D}^{\alpha\beta}_{mn}{Y}_{\alpha\beta}{D}^{\alpha\beta}_{mn}{Y}_{\alpha\beta}]
    \end{align}
    And expanding the expectation gives
    \begin{align}
        \frac{1}{{|M||N|}} {X}_{mn}{X}_{mn} =\frac{1}{{|A||B|}} {D}^{\alpha\beta}_{mn}{Y}_{\alpha\beta}{D}^{\alpha\beta}_{mn}{Y}_{\alpha\beta}
    \end{align}
    Note that $\frac{1}{{|M||N|}} = \frac{1}{{|A||B|}}$ so we cancel giving
    \begin{align}
        {X}_{mn}{X}_{mn} = {Y}_{\alpha\beta}{D}^{\alpha\beta}_{mn}{Y}_{\alpha\beta}{D}^{\alpha\beta}_{mn}
    \end{align}
    Rearranging the right-hand side gives
    \begin{align}
        {X}_{mn}{X}_{mn} = {D}^{\alpha\beta}_{mn}{D}^{\alpha\beta}_{mn}{Y}_{\alpha\beta}{Y}_{\alpha\beta}
    \end{align}
    Since the tensors ${D}$ are defined with respect to a standard orthonormal basis, we can freely raise and lower their indices (their metric tensor is identity). Lowering $\alpha, \beta$ and raising $m,n$  on one of the ${D}$ tensors gives:
    \begin{align}
        {X}_{mn}{X}_{mn} = {D}^{\alpha\beta}_{mn}{D}^{mn}_{\alpha\beta}{Y}_{\alpha\beta}{Y}_{\alpha\beta}
    \end{align}
    Since ${D}^{\alpha\beta}_{mn}{D}^{mn}_{\alpha\beta} = 1$ we have
    \begin{align}
        {X}_{mn}{X}_{mn} = {Y}_{\alpha\beta}{Y}_{\alpha\beta} \\
        = {X}_{mn}^2 = {Y}_{\alpha\beta}^2
    \end{align}
    Substituting gives
    \begin{align}
        \var[{X}] = \e[{X}^2] = E[{Y}^2]
    \end{align}
\end{proof}

Therefore, it is sufficient to compute the mean of the squared DCT coefficients to get the variance of the underlying pixels. This is no faster or slower than the pixel domain algorithm.

Next, we move on to the affine parameters ${\gamma}, {\beta}$. Applying ${\gamma}$ is easy: since the transform we are using is linear, multiplying by a scalar can happen before or after the transform, \ie,
\begin{align}
    {J}({\gamma} {I}) = {\gamma} {J}({I})
\end{align}
so we can simply multiply the transform coefficients by ${\gamma}$. Applying ${\beta}$ is also straightforward, since adding the scalar ${\beta}$ would raise the mean by ${\beta}$, we can add ${\beta}$ to only the (0,0) coefficient. This yields a simple closed-form algorithm for computing batch normalization.

\begin{lstlisting}[caption={Transform Domain Batch Norm},language=Python,label={lst:tdbatchnorm}]
def batch_norm(F: Tensor, gamma: float, beta: float) -> Tensor:
    mu = F[0, 0]
    F[0, 0] = 0
    var = torch.mean(F**2)
    
    F *= gamma / torch.sqrt(var)
    F[0, 0] = beta
    return F
\end{lstlisting}

Note that the algorithm in Listing \ref{lst:tdbatchnorm} assumes each sample is a single $8 \times 8$ block. If this is not the case, then the algorithm can be easily adjusted to compute combined mean and variance over several blocks (and multiple channels)\sidenote{Depending on the batch norm implementation, it may be necessary to apply Bessel correction to the variance computation as well.}.

\subsection{Global Average Pooling}
\labsec{jdgac}

\begin{marginfigure}
    \centering
    \includegraphics{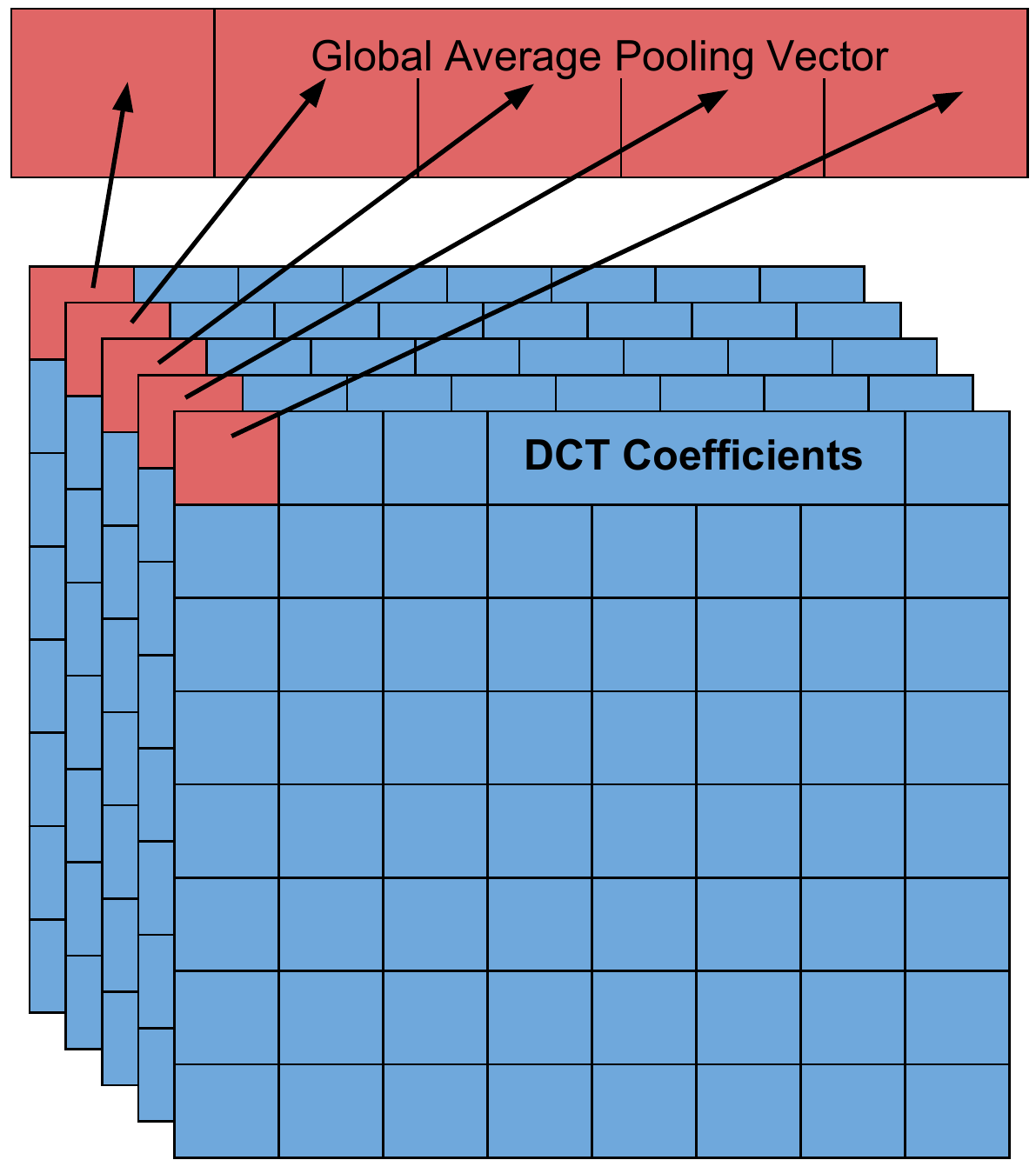}
    \caption[Transform Domain Global Average Pooling]{Illustration of transform domain global average pooling.}
    \labfig{globalavgpooling}
\end{marginfigure}

Global average pooling reduces feature maps to a single scalar per channel. In other words, spatial information is averaged "globally". Given the discussion in the previous section, this is extremely simple to compute in the transform domain. As the (0, 0) coefficient is proportional to the mean of each block, we can simply read off these coefficients and scale them to produce the global average pooling vector (\reffig{globalavgpooling}). This is significantly faster than the pixel domain algorithm. Note that this is exactly the result that the pixel domain algorithm would have generated, so from this point forward we no longer need to worry about operations in the transform domain (\ie, the fully-connected layers do not need modification).
\section{ReLU}
\labsec{jdrelu}

Having defined the exact operations, we now turn to a missing and critical component of residual networks: ReLU \sidecite{nair2010rectified, fukushima1982neocognitron}. Note that we have dedicated an entire section to what is a relatively simple operation in the pixel domain. ReLU is defined as
\begin{align}
    R(x) = \begin{cases}
               x & x \geq 0 \\
               0 & x < 0
           \end{cases}
\end{align}
The previous section made use of mathematical properties of the \gls{JPEG} transform in order to derive closed form solutions for \gls{transform domain} operations. Since ReLU is necessarily non-linear, we will have no such luck with that approach. In fact, not only is ReLU non-linear, it is piecewise linear depending on the pixel domain value, information which we do not have access to in the transform domain. Instead, we will develop an approximation technique for ReLU that works in the transform domain and is tunable giving an accuracy-speed trade-off.

We compute this approximation by partially decoding each block of coefficients. This is still fast since only a subset of coefficients are required and since the result of the approximation is in the pixel domain we can freely compute ReLU on it. Recall the DCT Least Squares Approximation Theorem proven in \nrefsec{fourier}.
\begin{theorem}[The DCT Least Squares Approximation Theorem]
    \labthm{dctlqa2}
    Given a set of $N$ samples of a signal ${X}$ let ${Y}$ be the DCT coefficients of ${X}$. Then for $1 \leq m \leq N$ the approximation of $X$ given by
    \begin{align}
        p_m(t) = \frac{1}{\sqrt{N}}y_0 + \sqrt{\frac{2}{N}}\sum_{k=1}^m y_k \cos\left(\frac{k(2t+1)\pi}{2N}\right)
    \end{align}
    minimizes the least-squared error
    \begin{align}
        e_m = \sum_{i=1}^N(p_m(i) - x_i)^2
    \end{align}
\end{theorem}
\refthm{dctlqa2} guides us in choosing the lowest $m$ frequencies when we decode (rather than some arbitrary set) in order to constrain the error of the approximation. For a 2D DCT, we use all frequencies $(i, j)$ such that $i + j \leq m$ yielding 15 frequencies. The threshold $m$ is freely tunable to the problem and we will examine its effect later.

Although we now have a reasonable algorithm for computing ReLU from transform coefficients, we are left with two major problems. The first is that although our approximation was motivated by a least-squares minimization, it is not guaranteed to reproduce any of the original samples. Since ReLU preserves positive samples (only zeroing negative samples) it would be nice if at least those were preserved. The second is that our network expects transform coefficients as input but the ReLU we have computed is in the spatial domain. It would be expensive to have to convert the result back to transform coefficients before continuing our computation.

Consider for a moment the nature of our first problem. Suppose we have a sample with value $0.7$. After taking the DCT and computing the least-squares approximation with a subset of coefficients, the value of this sample is changed to $0.5$. We can observe that although the least-squares approximation is incorrect, it is still \textit{positive}. In other words, the reconstruction has not changed the sign of the sample so it will not be zeroed by ReLU. The more coefficients we use the more likely it is that these reconstructions are sign-preserving\sidenote{This is true for other piecewise function intervals as well. The technique described here is general.} since the high frequencies contribute less to the accuracy of the result (otherwise they would not be a least-squares minimization). In this sense we can observe that it is easier to preserve the sign than the exact pixel value.

Therefore, rather than compute ReLU on this approximation, we can instead compute a mask and apply that mask. We reformulate ReLU as follows
\begin{align}
    R(x) = H(x)x \\
    H(x) = \begin{cases}
               1 & x \geq 0 \\
               0 & x < 0
           \end{cases}
\end{align}
where $H(x)$ is the Heaviside step function which we treat as a mask. If we compute $H(p_m)$ on the approximation $p_m$, and multiply the result by the original samples $x$, we will have masked the negative samples while preserving the positive ones. We call this technique Approximated Spatial Masking (ASM)\index{approximated spatial masking}. See \reffig{asm} for a visual example of this algorithm.

\begin{marginfigure}
    \centering
    \includegraphics{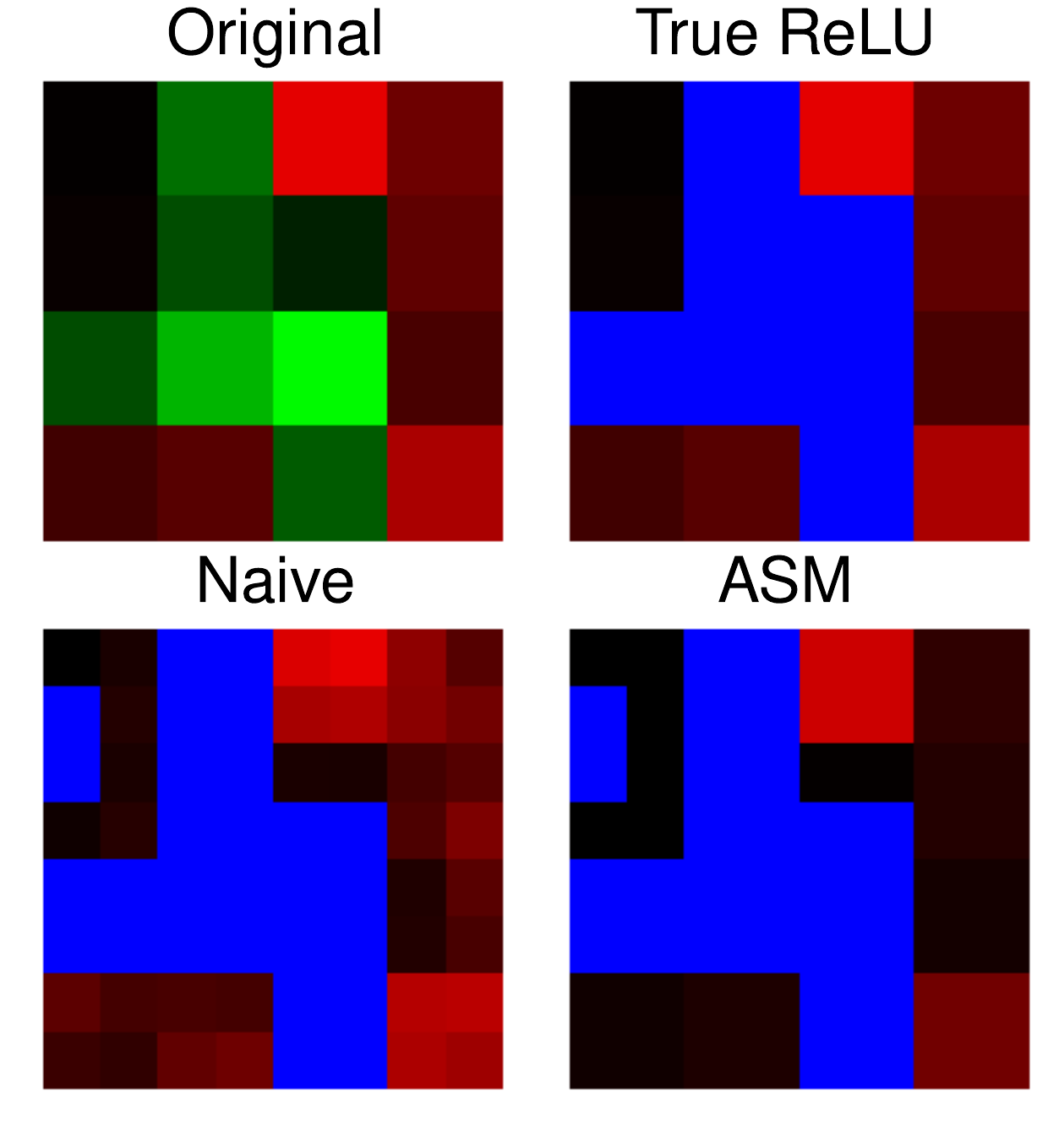}
    \caption[ReLU Approximation Example]{\textbf{ReLU Approximation Example.} Green pixels are negative, red pixels are positive, blue pixels are exactly zero. The top-left shows the original image. The top-right is the true ReLU. The bottom-left shows a naive approximation using only the least squares approximation. Note that while negative pixels are zeroed, very few positive pixels have the correct value and there are mask errors resulting from the approximation. The bottom-right image shows the ASM technique. Note that while there are still mask errors, positive pixel values are preserved.}
    \labfig{asm}
\end{marginfigure}

The only problem left to solve is that our original samples are in the transform domain and the mask is in the pixel domain. To simplify the following discussion, we consider only DCT blocks here (extending to the full transform is trivial). We can solve this using our multilinear model of the JPEG transform. Given transform coefficients $\mathcolor{JungleGreen}{F} \in \mathcolor{JungleGreen}{A^*} \otimes \mathcolor{JungleGreen}{B^*}$, a spatial domain mask $\mathcolor{Plum}{G} \in \mathcolor{Plum}{M^*} \otimes \mathcolor{Plum}{N^*}$, and the masked result $\mathcolor{JungleGreen}{F'} \in \mathcolor{JungleGreen}{A^*} \otimes \mathcolor{JungleGreen}{B^*}$, consider the steps such an algorithm would perform
\begin{enumerate}
    \item Take the inverse DCT of $\mathcolor{JungleGreen}{F}$ to give $\mathcolor{Plum}{I} \in \mathcolor{Plum}{M^*} \otimes \mathcolor{Plum}{N^*}$
    \item Pixelwise multiply the mask $\mathcolor{Plum}{G}$ and $\mathcolor{Plum}{I}$ to give $\mathcolor{Plum}{I'}$
    \item Take the DCT of $\mathcolor{Plum}{I'}$ to give the masked result $\mathcolor{JungleGreen}{F'}$
\end{enumerate}
All of these steps are linear or bilinear
\begin{align}
    \mathcolor{Plum}{I_{mn}} = \mathcolor{BurntOrange}{D^{\alpha\beta}_{mn}}\mathcolor{JungleGreen}{F_{\alpha\beta}} \\
    \mathcolor{Plum}{I'_{mn}} = \mathcolor{Plum}{G_{mn}}\mathcolor{Plum}{I_{mn}}                                     \\
    \mathcolor{JungleGreen}{F'_{\alpha'\beta'}} = \mathcolor{BurntOrange}{D^{mn}_{\alpha'\beta'}}\mathcolor{Plum}{I'_{mn}}
\end{align}
Substituting, we have
\begin{align}
    \mathcolor{JungleGreen}{F'_{\alpha'\beta'}} = \mathcolor{BurntOrange}{D^{mn}_{\alpha'\beta'}}\mathcolor{Plum}{G_{mn}}\mathcolor{Plum}{I_{mn}}                  \\
    = \mathcolor{BurntOrange}{D^{mn}_{\alpha'\beta'}}\mathcolor{Plum}{G_{mn}}\mathcolor{BurntOrange}{D^{\alpha\beta}_{mn}}\mathcolor{JungleGreen}{F_{\alpha\beta}} \\
    = \mathcolor{Plum}{G_{mn}}\mathcolor{BurntOrange}{D^{mn}_{\alpha'\beta'}}\mathcolor{BurntOrange}{D^{\alpha\beta}_{mn}}\mathcolor{JungleGreen}{F_{\alpha\beta}}
\end{align}
And we make the following definition (after raising some indices to preserve dimensions)
\begin{align}
    \mathcolor{JungleGreen}{F'_{\alpha'\beta'}} = \mathcolor{Plum}{G_{mn}}\left[\mathcolor{BurntOrange}{D^{mn}_{\alpha'\beta'}}\mathcolor{BurntOrange}{D^{\alpha\beta m n}}\right]\mathcolor{JungleGreen}{F_{\alpha\beta}} \\
    \mathcolor{BurntOrange}{\Psi^{\alpha\beta m n}_{\alpha'\beta'}} = \mathcolor{BurntOrange}{D^{mn}_{\alpha'\beta'}}\mathcolor{BurntOrange}{D^{\alpha\beta m n}}
\end{align}
giving the bilinear map $\mathcolor{BurntOrange}{\Psi} : \mathcolor{Plum}{M^*} \otimes \mathcolor{Plum}{N^*} \times \mathcolor{JungleGreen}{A^*} \otimes \mathcolor{JungleGreen}{B^*} \rightarrow \mathcolor{JungleGreen}{A^*} \otimes \mathcolor{JungleGreen}{B^*}$. This map can be computed once and reused. We can use this map along with our approximate mask and original DCT coefficients to produce a highly accurate ReLU approximation with few coefficients.
\section{Recap}

Before continuing to empirical concerns, we briefly recap the theoretical discussion in the previous sections. Residual networks consist of four basic operations: \textbf{Convolution}, \textbf{Batch Normalization}, \textbf{Global Average Pooling}, and \textbf{ReLU}.

In \nrefsec{jdc} we found that JPEG domain convolutions can be expressed as
\begin{align}
    \mathcolor{BurntOrange}{\Xi^{pxyk}_{p'x'y'k'}} = \mathcolor{BurntOrange}{\widetilde{J}^{h'w'}_{x'y'k'}C^{phw}_{p'h'w'}J^{xyk}_{hw}}
\end{align}
and in Listing \ref{lst:expconv_fast} we developed a fast algorithm for computing this.

In \nrefsec{jdbn} we developed a closed form solution for JPEG domain batch normalization. We found that
\begin{align}
    \e[I] =\frac{1}{8}D(0,0) \\
    \var[I] = \e[D^2] \;\text{iff}\; D(0,0) = 0
\end{align}
and that we can apply $\beta$ by adding it to $D(0, 0)$ and we can apply $\gamma$ as we would to a spatial domain input (by multiplying it by each coefficient).

In \nrefsec{jdgac} we found that global average pooling in the JPEG domain is as simple as computing $\frac{1}{8}D(0, 0)$ from each channel. We also noted that since this is equivalent to the spatial domain mean, there is no need to derive the fully-connected layers.

Finally, in \nrefsec{jdrelu} we developed an approximation technique for ReLU where we use a subset of coefficients to decode each block and compute and approximate $H(x)$ on each block where $H()$ is the Heaviside step function producing a mask $\mathcolor{Plum}{G_{mn}}$. Then we apply this mask to the original coefficients $\mathcolor{JungleGreen}{F_{\alpha\beta}}$ using
\begin{align}
    \mathcolor{JungleGreen}{F'_{\alpha'\beta'}} = \mathcolor{Plum}{G_{mn}}\mathcolor{BurntOrange}{\Psi^{\alpha\beta mn}_{\alpha'\beta'}}\mathcolor{JungleGreen}{F_{\alpha\beta}}
\end{align}
This concluded our theoretical derivations.
\begin{kaobox}[frametitle=Model Conversion]
    One important thing to note is that at no time did we stipulate that the convolution weights or batch norm affine parameters need to be learned from scratch. Indeed, this method can take any such values, random or learned, and produce JPEG domain operations. Therefore, we can use the method to \textit{convert pre-trained models} to operate in the JPEG domain. This idea has some powerful implications and we will examine it's trade-offs in the empirical analysis.
\end{kaobox}
\section{Empirical Analysis}

We now turn out attention to an empirical evaluation of the algorithm. After all, the discussion in the previous sections was highly theoretical and altogether divorced from practical concerns. A natural question at this point is: ``How well does this actually work?''

We will start by creating a toy network. This small network will be used in the experiments in this section to evaluate and benchmark the technique.
\begin{figure}[t]
    \centering
    \includegraphics{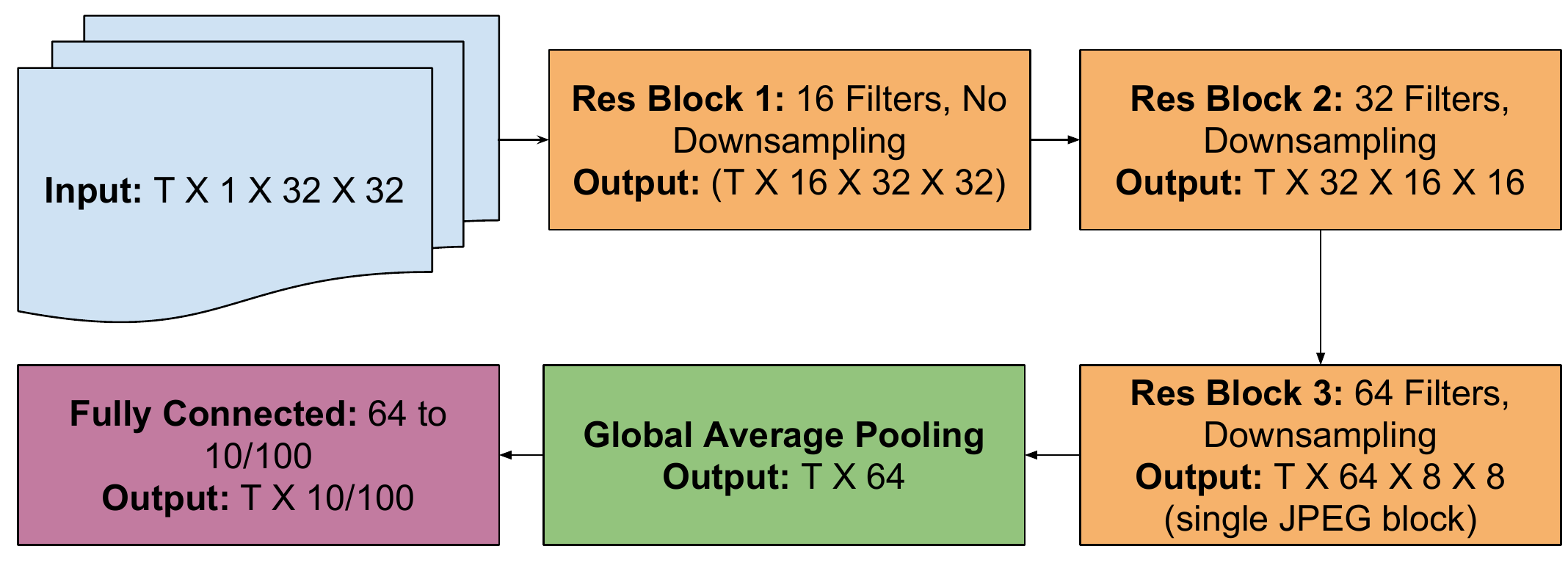}
    \caption[Toy Network Architecture]{\textbf{Toy Network Architecture.} Note that by the final ResBlock, the image is reduced to $8 \times 8$ which is a single block of coefficients. This simplifies the global average pooling layer.}
\end{figure}
This toy architecture consists of three residual blocks followed by global average pooling and a single fully connected layer. Although this is a simple architecture, it will more than suffice for our benchmarks of MNIST~\sidecite{lecun1998mnist} and CIFAR 10/100~\sidecite{krizhevsky2009learning}. The inputs will always be $32 \times 32$ images to ensure an even number of JPEG blocks\sidenote{MNIST inputs are zero padded with two pixels on each side}. We consider two versions of this network, one which processes images in the spatial domain (\ie, a traditional ResNet) and one which we have applied algorithm on to allow it to process JPEG transform coefficients.

For those unconvinced by mathematics (or maybe suspicious of the ability to implement the math in PyTorch), we first examine whether our derivations were correct at all. This is straightforward: we simply use an exact ReLU, taking all 15 frequencies for the JPEGified version of the toy network. For more meaningful accuracies, the network is trained until convergence in the pixel domain and the weights are then converted. Since our other operations are supposed to be "exact", this should yield the same accuracy as a pixel domain network to within some small floating point error, which is confirmed by the result in \reftab{model_conversion}.

\begin{margintable}
    \caption[Model Conversion Accuracies]{\textbf{Model Conversion Accuracies.} Note that the deviation is small between the spatial domain and JPEG domain network.}
    \labtab{model_conversion}
    \resizebox{\textwidth}{!}{
        \begin{tabular}{@{}rlll}
            \toprule
            Dataset   & Spatial & JPEG  & Deviation \\
            \midrule
            MNIST     & 0.988   & 0.988 & 2.999e-06 \\
            CIFAR-10  & 0.725   & 0.725 & 9e-06     \\
            CIFAR-100 & 0.385   & 0.385 & 1e-06     \\
            \bottomrule
        \end{tabular}
    }
\end{margintable}

Next we examine the accuracy of the ReLU approximation. Since this is not a true ReLU, we expect there to be some effect on overall network accuracy when fewer frequencies are used. However, it is still a non-linearity which should enable the network to learn effective mappings. We consider ReLU accuracy from three perspectives
\begin{description}
    \item[Absolute Error] How accurate is our ASM approximation compared with an naive approximation?
    \item[Conversion Error] If we convert pre-trained weights, how much does frequency effect the final accuracy result?
    \item[Training Error] If we train a network from scratch using the ReLU approximation, how much does frequency affect the final accuracy result?\sidenote{Assuming the same number of frequencies are used for training and inference.}
\end{description}
We show results to this effect in \reffig{relu_accuracy}.
\begin{figure*}[t]
    \centering
    \includegraphics[width=0.49\textwidth]{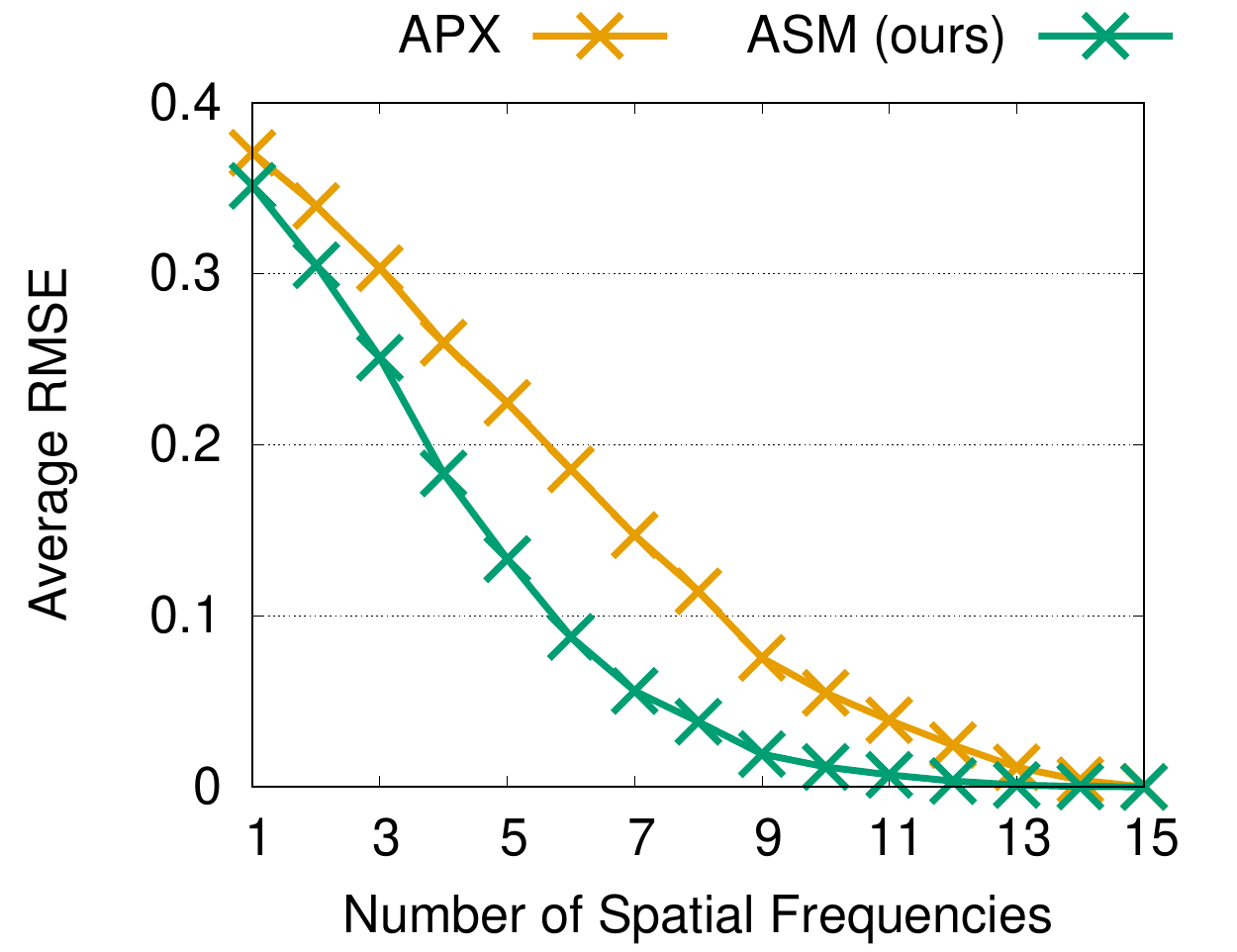} \hfill \includegraphics[width=0.49\textwidth]{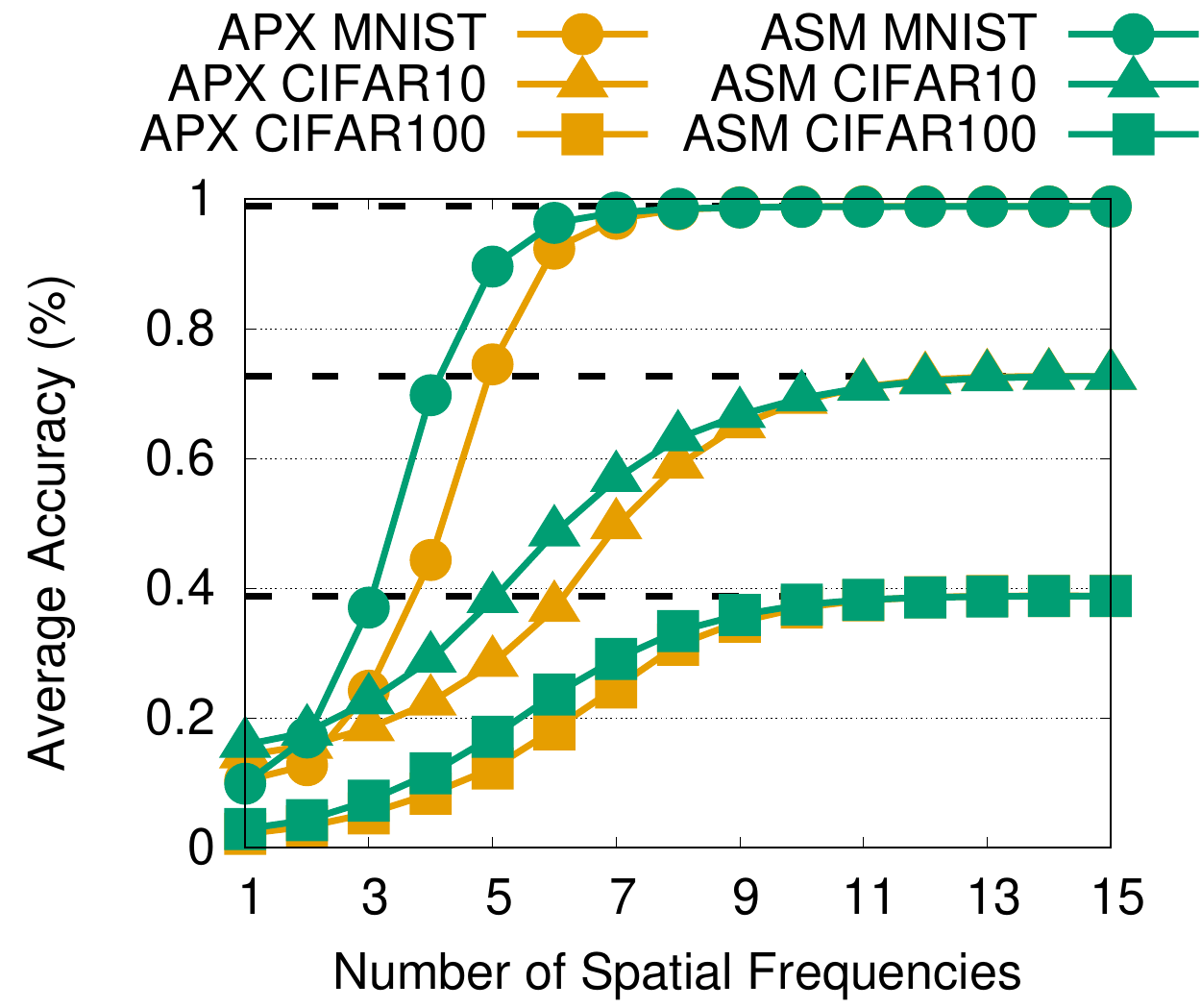} \hfill \includegraphics[width=0.49\textwidth]{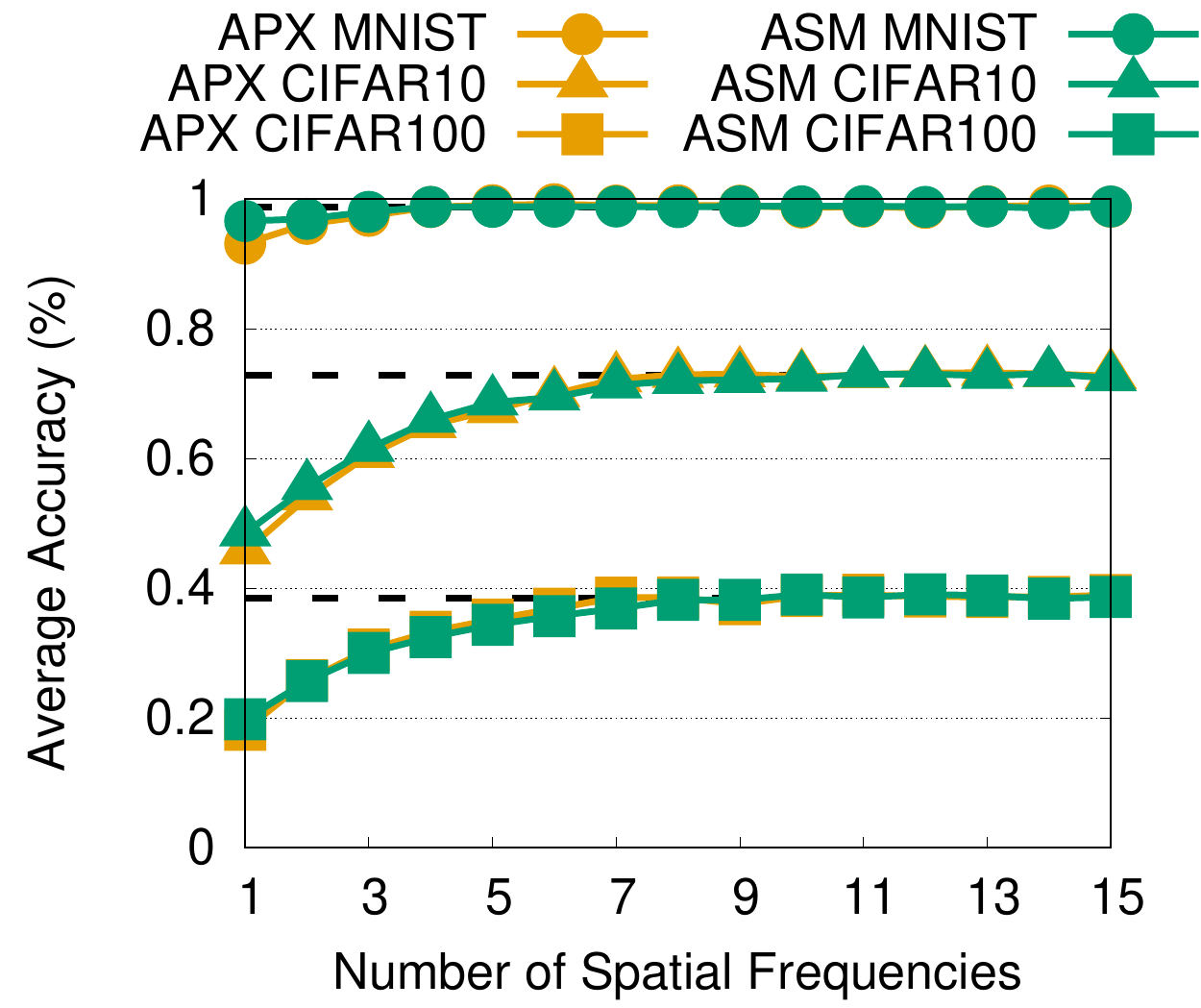}
    \caption[ReLU Approximation Accuracy]{\textbf{ReLU Approximation Accuracy.} Left: RMSE error. Middle: Model accuracy after model conversion. Right: Model accuracy when re-training from scratch. Note that APX denotes he naive ReLU approximation. Dotted lines represent spatial domain accuracy.}
    \labfig{relu_accuracy}
\end{figure*}
The left graph shows the absolute error of the ReLU approximation. For this experiment, 10 million $8 \times 8$ blocks are generated by upsampling random $4 \times 4$ pixel blocks. We then measure RMSE between the true block and the approximated block. Note that compared to the naive approximation, the ASM method we developed has lower error throughout and the error drops faster. In the middle graph, we show model conversion error. We train 100 models from random weights in the pixel domain and then apply our algorithm to convert the weights, and measure the resulting classification accuracy. Again we see that the ASM method has better performance. In the final graph, we train networks from random weights using our JPEG domain algorithm. Interestingly, this performs significantly better than model conversion indicating that the weights have learned to adapt to the ReLU approximation.

\begin{marginfigure}
    \centering
    \includegraphics{figures/throughput.pdf}
    \caption[Throughput Comparison]{\textbf{Throughput Comparison.} We compare JPEG domain and spatial domain training and inference.}
    \labfig{jdr_throughput}
\end{marginfigure}

The final result we show is throughput. In general, the method developed here should be fast if for no other reason than the JPEG images do not need to be decompressed before being processed. In \reffig{jdr_throughput} we compare throughput for training and testing in the JPEG domain \vs in the spatial domain. As expected, inference is significantly faster in the JPEG domain. Curiously, however, training is only slightly faster. This is caused by the more complex update rule for autograd to compute through the ReLU approximation and the JPEG domain conversion for the convolutions.

\section{Limitations and Future Directions}

The astute reader will have noticed by now a major limitation with this work: memory usage. Recall that compressed domain convolutions are formed by convolving the kernel $\mathcolor{BurntOrange}{C}$, a $\text{dim}(\mathcolor{BurntOrange}{P}) \times \text{dim}(\mathcolor{BurntOrange}{P'}) \times 3 \times 3$ matrix, with the JPEG decompression tensor $\mathcolor{BurntOrange}{\widetilde{J}} \in \mathcolor{Plum}{X} \otimes \mathcolor{Plum}{Y} \otimes \mathcolor{Plum}{K} \otimes \mathcolor{JungleGreen}{H^*} \otimes \mathcolor{JungleGreen}{W^*}$ and then applying the JPEG compression tensor $\mathcolor{BurntOrange}{J} \in \mathcolor{Plum}{X^*} \otimes \mathcolor{Plum}{Y^*} \otimes \mathcolor{Plum}{K^*} \otimes \mathcolor{JungleGreen}{H} \otimes \mathcolor{JungleGreen}{W}$. This yields a the type (3, 3) tensor $\mathcolor{BurntOrange}{\Xi} \in \mathcolor{BurntOrange}{P'} \otimes \mathcolor{Plum}{X} \otimes \mathcolor{Plum}{Y} \otimes \mathcolor{Plum}{K} \otimes \mathcolor{BurntOrange}{P} \otimes \mathcolor{Plum}{X^*} \otimes \mathcolor{Plum}{Y^*} \otimes \mathcolor{Plum}{K^*}$.

Observe the size of this tensor. For an image of size $\text{dim}(\mathcolor{JungleGreen}{H}) \times \text{dim}(\mathcolor{JungleGreen}{W})$ it is in $O((\text{dim}(\mathcolor{JungleGreen}{H}) \times \text{dim}(\mathcolor{JungleGreen}{W}))^2)$. In other words we have taken a small constant size weight and expanded it to be on the order of the image size squared. This is perhaps the primary direction for future work. The massive size of this tensor entirely prevents the method from being useful for anything beyond the toy network and small image datasets presented in the previous section. While a constant size kernel could be created using tiling (each convolution depends on at most the blocks one outside of the ``currently processed'' block), this would still be significantly larger than the small kernel used by spatial domain networks. By restricting the convolution to a single block, an $\text{dim}(\mathcolor{BurntOrange}{P'}) \times \text{dim}(\mathcolor{BurntOrange}{P}) \times 8 \times 8 \times 8 \times 8$ kernel could be created with an approximate result which would significantly improve the situation. It is left to future work to determine the practicality of these ideas and what their effect on network accuracy is.

Our ReLU formulation is currently an approximation. As we studied in the previous section, this approximation does impact the overall network accuracy even when retraining. It would be nice if an exact ReLU could be formulated to avoid this issue. It is currently unknown if this is possible.

While on the topic of ReLU, software support for our method is currently quite lacking. In essence, many of our memory and speed savings come from the sparse nature of JPEG compressed data. Zero elements could take up no memory and contribute no operations to the compute graph, but this depends on adequate software support for sparse operations which is currently missing from libraries like PyTorch. Specifically, support for sparse \texttt{einsum} would need to be added. This is perhaps the low-hanging fruit that would immediately reduce the memory footprint while further increasing the speed the algorithm.
\setchapterpreamble[u]{\margintoc}
\chapter{Improving JPEG Compression}
\labch{dl-jpeg}

\lettrine{W}{ith} a good understanding of \gls{JPEG}\index{JPEG} compression and how it relates to deep learning, we turn to a survey of methods which improve JPEG compression. These methods are essentially specializations of image enhancement. So sister problems in this domain are super-resolution, denoising, deraining, \etc Notably, we will not be considering new deep learning based codecs which are beyond the scope this this dissertation. These methods are reviewed briefly in Appendix \ref{app:fulldl}, however. We focus on historical methods which made significant advancements in the understanding of JPEG artifact correction and present them roughly in publication order although they are grouped into sections by their high level ideas.

Before discussing the \gls{deep learning} techniques we first mention two classical methods for correction of JPEG artifacts. The first method uses a ``pointwise shape-adaptive DCT''\index{shape-adaptive discrete cosine transform} (SA-DCT) \sidecite{foi2006pointwise}. The SA-DCT can be thought of as a generalization of the block DCT used by JPEG to account for block sizes of varying shape. Foi \etal model JPEG compression artifacts as Gaussian noise with zero mean and compute $\sigma^2$ using an empirically developed formula on the quantization matrix. For each point in the image, the technique computes a DCT kernel that best fits the underlying data (hence shape-adaptive). This filter is then used to estimate the Gaussian noise term for enhancement. The next method \sidecite{yang2000blocking} use a generalized lapped biorthogonal transform\index{generalize lapped biorthogonal transform} (GLBT) \sidecite{tran1998generalized}. In this technique, the JPEG DCT coefficients are modeled as an intermediate output of the GLBT and the remaining filters in the method are designed to remove blocking artifacts. Prior to deep learning, these techniques were the most successful at removing JPEG artifacts.

\begin{warningbox}
    This chapter is mostly a history lesson. Skip to the last section if you want a TL;DR.
\end{warningbox}

\section{Pixel Domain Techniques}

We begin our discussion with the straightforward ``pixel domain'' techniques. These networks function as traditional convolutional networks. They use pixels and input and output either the corrected network or its residual. The first such technique was the ARCNN\index{arcnn} \sidecite{dong2015compression} later followed up by Fast ARCNN \sidecite{yu2016deep}. These networks followed a traditional encoder-decoder architecture and are based off of the contemporary SRCNN \sidecite{dong2014learning}.

ARCNN is tiny by modern standards with four convolutional layers. The first is a $9 \times 9$ layer with 64 channels, next a $7 \times 7$ with 32 channels, then a $1 \times 1$ with 16 channels and finally a $5 \times 5$ decoder with 1 channel (for grayscale only). The authors of ARCNN claim that each layer is designed for a specific purpose but there is no deep supervision on the layers and they are trained end-to-end so it is unlikely that they learn a particular task.

Fast ARCNN changes this architecture to an ``hourglass'' shape, essentially a U-Net \sidecite{ronneberger2015u} without skip connections which was common at the time. The architecture uses strided convolutions for the downsampling operations. Since the size of the feature maps is reduced, the architecture processes images faster, hence the name. This does reduce the overall reconstruction accuracy, however.

The L4/L8 networks \sidecite{svoboda2016compression} introduce two major new ideas to artifact correction. The first is the idea of residual learning, where the network is encouraged to learn only the difference between the input image and the true reconstruction. In other words, the reconstructed image $X_r$ is expressed as
\begin{align}
    X_r = X_c + f(X_c)
\end{align}
for compressed image $X_c$ and learned network $f()$. The second contribution is that of an edge preserving loss. The authors rightly observe that prior networks, due to their regression only losses, have blurry edges. They solve this by using Sobel filters to compute the partial first derivatives of the reconstructed image and computing loss on these filtered images which focuses the network on edge reconstructions. As expected the L4/L8 architectures have four and eight layers respectively and otherwise do not differ significantly from ARCNN.

CAS-CNN \sidecite{cavigelli2017cas} build on the previous idea by employing a significantly more complex architecture. This architecture contains skip connections, not unlike a U-Net, and upgrades the traditional regression loss to use multiple scales. These scales are computed using deep supervision of the downsampled feature maps and make a fairly significant improvement to the overall accuracy. This is likely helped by the skip connections in the U-Net architecture.

We now jump to the MWCNN \sidecite{liu2018multi} which is a major difference in architecture. MWCNN is a fascinating method for general image restoration which was applied directly to JPEG artifacts at the time of publication (along with other problems). The key idea is to replace the pooling layers in a traditional CNN with a discrete wavelet transform. Recall that a discrete wavelet transform computes band-pass filters which restrict each output to half the frequency range of the input. By the nyquist sampling theorem, we can then discard half the samples without losing any information. MWCNN exploits this by using the DWT in place of a pooling operation, stacking the resulting four frequency sub-bands in the channel dimension without any significant loss of information. The original image can then be reconstructed by using the inverse wavelet transform on the feature maps after traditional convolutional layers. Otherwise the architecture resembles U-Net. The use of this clever signal processing trick allows MWCNN to achieve remarkable results on a number of restoration tasks including JPEG artifact correction.

Honorable mention at this point goes to DPW-SDNet \sidecite{chen2018dpw}. This could be considered a dual-domain method although we take a somewhat stricter definition of domain so instead we list it here with MWCNN. The main contribution of DPW-SDNet was to include two networks, one which processes the image in the pixel domain and another which processes it after a single level DWT.

Another method from 2018, S-Net \sidecite{zheng2018s}, introduces a scalable network. This is based on the apt observation that more quantization requires a deeper network and ``more work'' to restore. Their architecture is, therefore, scalable either based on the amount of degradation applied to the image or constraints on the compute budget of the hardware. This was an important contribution toward the practical use of artifact correction and remains an under-explored idea.

Two works by Galteri \etal \sidecite{galteri2017deep, galteri2019deep} introduce GANs to the problem of artifact correction. As we observed in the discussion of L4/L8 and CAS-CNN, regression losses produce a blurry result. This is both because of the CNN's inherent bias towards error minimization, something which is easiest to accomplish with a low-frequency reconstruction, and because of JPEG's tendency to destroy high frequency details in the first place. Although L4/L8/CAS-CNN make progress on this problem with specialized losses, they had obvious limitations which Galteri \etal overcome with a GAN loss. This generates significantly more realistic reconstructions, although there is no attempt at an ``accurate'' reconstruction with good numerical results\sidenote{Which, in my opinion, is completely acceptable.}. The 2019 version of this work even includes a rudimentary attempt at a ``universal'' architecture which can operate independently of quality setting although it accomplishes this with an ensemble.

The final technique we discuss in this section is RDN \sidecite{zhang2020residual}. This represents a departure from the more traditional U-Net style networks we have been discussing. Instead, RDN is based on ESRGAN \sidecite{wang2018esrgan} and its RRDB layers. These layers are an enhanced version of the traditional residual layer \sidecite{he2016deep} with more residual connections. Just as these layers were a huge improvement for super-resolution, they are a huge improvement for artifact correction.

\section{Dual-Domain Techniques}

Dual domain\index{dual domain} techniques are the result of an attempt to inject some low level JPEG data into the learning process. The high level idea is to process the input in both the spatial (pixel) domain and the frequency (DCT) domain. This is done with two separate networks and their result is fused. This way if there is some information that either domain does not capture, it can potentially be exploited by the other domain. The technique was introduced with a sparse-coding method \sidecite{liu2015data} that we will examine in the next section.

On the deep learning side the idea is first addressed with DDCN \sidecite{guo2016building}. The idea is very straightforward. There are two separate encoders, one for the pixel domain and one for the DCT domain. The output of both networks is processed by a third aggregation network which decodes to a residual that is added to the input image.

DMCNN \sidecite{zhang2018dmcnn} extends this idea in two ways. The first is with a multiscale loss on the pixel branch as we saw in L4/L8. The next is with a DCT Rectifier which constrains the magnitude of the DCT residual based on the possible values that the true coefficients could take. Recall the formula for quantization
\begin{align}
    Y'_{ij} = \left\lfloor\frac{Y_{ij}}{(Q_y)_{ij}}\right\rceil
\end{align}
shown here for the $Y$ channel only. The approximated coefficient is then
\begin{align}
    \widehat{Y}_{ij} = \left\lfloor\frac{Y_{ij}}{(Q_y)_{ij}}\right\rceil (Q_y)_{ij}
\end{align}
Dividing by $(Q_y)_{ij}$ gives
\begin{align}
    \frac{\widehat{Y}_{ij}}{(Q_y)_{ij}} = \left\lfloor\frac{Y_{ij}}{(Q_y)_{ij}}\right\rceil
\end{align}
We can now expand this as an inequality since the rounded value must range from $[-\frac{1}{2}, \frac{1}{2}]$ around the rounding result
\begin{align}
    \frac{\widehat{Y}_{ij}}{(Q_y)_{ij}} - \frac{1}{2} \leq \frac{Y_{ij}}{(Q_y)_{ij}} \leq \frac{\widehat{Y}_{ij}}{(Q_y)_{ij}} + \frac{1}{2}
\end{align}
multiplying by $(Q_y)_{ij}$ yields our desired constraint on $Y_{ij}$
\begin{align}
    \widehat{Y}_{ij}- \frac{(Q_y)_{ij}}{2} \leq Y_{ij} \leq \widehat{Y}_{ij} + \frac{(Q_y)_{ij}}{2}
\end{align}
Since the artifact correction network is trying to compute $Y_{ij}$ from $\widehat{Y}_{ij}$, this constraint helps reduce the space of possible solutions.

The next major innovation in dual-domain methods is IDCN \sidecite{zheng2019implicit}. The major advantage of IDCN is that it is designed for color images and uses ``variance maps'' to account for the differences in statistics between the channels. Their dual-domain formulation is also of interest. They introduce a dual-domain layer which is ``implicit'', similar to our result in \nrefsec{multilinear_jpeg}, the DCT transform can be composed such that the DCT result and the pixel result are computed simultaneously.

Finally Jin \etal \sidecite{jin2020dual} extend the dual domain concept to process frequency bands in different paths. This is based on two observations: firstly, some artifacts are restricted to particular frequency bands and secondly, as we have said many times, accurate high-frequency reconstructions are difficult. By separating out the frequency bands for separate processing, the network is able to focus on restoring those particular frequencies as well as freeing up model capacity for artifacts which occur only in the considered frequency bands of the branch.

\section{Sparse-Coding Methods}

Sparse coding\index{sparse coding} is a dictionary learning method. A series of representative examples are learned which (we hope) form an ``over complete'' basis for our solution space \sidenote{I do not believe that an ``over complete basis'' is an actual concept in linear algebra. I assume the developers of this method are referring to a frame}. Because the input is no longer uniquely determined by the basis, we also try to enforce sparsity such that the members of the basis are as sparse as possible. We do not cover sparse coding in more detail in this dissertation.

Sparse coding was introduced to artifact correction by Li \etal \sidecite{liu2015data} where they also introduced dual domain learning. The idea is straightforward: learn sparse codes in pixel space and DCT space and fuse the results.

D3 \sidecite{wang2016d3} makes an interesting extension to Li \etal. They formulate the problem in a ``feed forward'' manner. In other words, sparse coding is used first on the DCT coefficients and then the result of that is fed into anther sparse coding module in the pixel domain. Both stages are supervised with loss functions similar to neural networks.

The final sparse coding method we consider, DCSC \sidecite{fu2019jpeg} is pixel domain only. However, they incorporate a simple convolutional network into their architecture such that the sparse codes are computed on CNN features. This gives a sort of ``best case'' scenario where the powerful convolutional features can be exploited by the sparse coding method. As a bonus, their method uses a single model for all quality settings, although they do not train in the general case and only target qualities 10 and 20.

\begin{table*}[t]
    \caption[Summary of JPEG Artifact Correction Methods]{\textbf{Summary of JPEG Artifact Correction Methods.} The methods are all listed with their technique (CNN or Sparse Coding) and whether they incorporate dual domain information or not. This table is not exhaustive. Methods are sorted by year.}
    \resizebox{1.5\textwidth}{!}{
        \begin{tabular}{@{}llllll@{}}
            \toprule
            Year & Method                     & Citation                      & Technique     & Dual Domain                         & Note                                \\
            \midrule
            2015 & ARCNN                      & \cite{dong2014learning}       & CNN           & \textcolor{BrickRed}{$\times$}      &                                     \\
                 & Data driven sparsity ...   & \cite{liu2015data}            & Sparse Coding & \textcolor{JungleGreen}{\checkmark} &                                     \\
            2016 & L4/L8                      & \cite{svoboda2016compression} & CNN           & \textcolor{BrickRed}{$\times$}      &                                     \\
                 & DDCN                       & \cite{guo2016building}        & CNN           & \textcolor{JungleGreen}{\checkmark} &                                     \\
                 & D3                         & \cite{wang2016d3}             & Sparse Coding & \textcolor{JungleGreen}{\checkmark} &                                     \\
            2017 & CAS-CNN                    & \cite{cavigelli2017cas}       & CNN           & \textcolor{BrickRed}{$\times$}      &                                     \\
                 & Deep Genarative ...        & \cite{galteri2017deep}        & CNN           & \textcolor{BrickRed}{$\times$}      & GAN, Color                          \\
            2018 & MWCNN                      & \cite{liu2018multi}           & CNN           & \textcolor{BrickRed}{$\times$}      & Uses DWT instead of pooling         \\
                 & DPW-SDNet                  & \cite{chen2018dpw}            & CNN           & \textcolor{BrickRed}{$\times$}      & Dual wavelet and pixel domain       \\
                 & S-Net                      & \cite{zheng2018s}             & CNN           & \textcolor{BrickRed}{$\times$}      & Scalable                            \\
                 & DMCNN                      & \cite{zhang2018dmcnn}         & CNN           & \textcolor{JungleGreen}{\checkmark} & DCT Rectifier                       \\
            2019 & Deep Generative ...        & \cite{galteri2019deep}        & CNN           & \textcolor{BrickRed}{$\times$}      & GAN, Universal with ensemble, Color \\
                 & IDCN                       & \cite{zheng2019implicit}      & CNN           & \textcolor{JungleGreen}{\checkmark} & Implicit DCT Layer, Color           \\
                 & DCSC                       & \cite{fu2019jpeg}             & Sparse Coding & \textcolor{BrickRed}{$\times$}      & Uses CNN Features                   \\
            2020 & RDN                        & \cite{zhang2020residual}      & CNN           & \textcolor{BrickRed}{$\times$}      & Uses RRDB                           \\
                 & Dual stream multi path ... & \cite{jin2020dual}            & CNN           & \textcolor{JungleGreen}{\checkmark} &                                     \\
            \bottomrule
        \end{tabular}
    }
    \labtab{jpeg-dl:summary}
\end{table*}

\section{Summary and Open Problems}

We summarize all the methods discussed in this chapter in \reftab{jpeg-dl:summary}. There are some interesting things we can take away from this discussion. For example, it seems that dual-domain methods work well and they are continually revisited. Deeper networks have also naturally been successful but the switch to RRDB layers by RDN was particularly interesting. More complex techniques like wavelet based or sparse coding based methods are underutilized and may be more complex than is needed with the advances of vanilla neural networks.

One noteworthy takeaway is that while there are pixel domain techniques and dual-domain techniques, there is not a single DCT domain only technique. A careful examination of ablation studies in the dual-domain papers explains this: their DCT branches do not perform well on their own. Somehow, the DCT branch is capturing new information that the pixel branch does not, but not enough to carry out restoration on its own. This is likely caused by the DCT being a set of coefficients for orthogonal basis functions rather than a single correlated signal like pixels. We will consider this an open problem as we move into the next section.

Also somewhat surprising is that although many methods recognized that JPEG artifact correction struggles to restore high frequencies with regression losses, only one author thought to use a GAN for correction. This is at least partially because of the community's incessant focus on benchmark results as the criterion for publication. GAN restoration does not perform well on the benchmarks. We consider this an open problem as well.

Another oddity: very few of the methods explicitly treat color images. This is odd on its own but even more so when we consider that JPEG explicitly handles chrominance differently than luminance by compressing it more aggressively and downsampling it. Also, there is spatial correlation between luminance and chrominance, which could and should be exploited in the reconstruction. Only the works of Galteri \etal and IDCN explicitly handle color data. This is another open problem.

Finally, and crucially, there are very few ``universal'' or ``quality blinded'' techniques. In fact, the only ones discussed in this section were Galteri \etal \sidecite{galteri2019deep} and DCSC \sidecite{fu2019jpeg}. All other networks in this section trained a different network for each quality setting they consider, a practice which is not sustainable in real deployments. Although solutions to this problem have been cropping up at the time of writing \sidecite{kim2019pseudo, zini2019deep, kim2020agarnet, jiang2021towards} we again consider this to be an open problem. In the next chapter, we will develop a method that addresses these problems.

\setchapterpreamble[u]{\margintoc}
\chapter{{\color{Plum}Quantization Guided JPEG Artifact Correction}}
\labch{qgac}

\lettrine{I}{n} the previous chapter we discussed several methods for using deep networks to improve \gls{JPEG} compression. These techniques augment \gls{JPEG} with significantly better rate-distortion while allowing users to still produce and share their familiar \gls{JPEG} files\sidenote[][15\baselineskip]{With the added benefit that users without special software can still view the files albeit at lower quality.}. These networks, however, come with three major disadvantages that have so far made them purely academic successes.

First and foremost, these methods are so called ``quality aware'' methods in which all training and testing data is compressed at a single quality level. This yields a single model per \gls{JPEG} quality, which is undesirable for several reasons. Recall that quality is an integer in $[0, 100]$, thus potentially requiring 101 different models to be trained. Although it is likely that the models may generalize to nearby qualities (we will examine this somewhat in \nrefsec{qgac-generalize}), at the very least this still requires the training and deployment of \textit{more than one} model, something which is still considered expensive for most institutions. Furthermore, when these models are deployed, they will be given arbitrary \gls{JPEG} files to correct and the \gls{jfif} file format \textit{does not store quality} leaving a real system with no reliable method to choose a model. This problem could be solved with an auxiliary model that regresses image to quality~\sidecite[2\baselineskip]{galteri2019deep} but this still requires training and deploying an ensemble and now an additional model to pick the quality.

The next, and perhaps more peculiar, problem with these methods is that they are grayscale only. In other words these models only work on the \gls{luminance} channel of the compressed images. While this does align well with human perception, humans can certainly perceive color degradations (see \reffig{qgac-overview} to perceive this yourself). There is an implicit assumption that \gls{luminance} models could be applied channel-wise to YCbCr or RGB images however we find that this does not hold well in practice as we show in \nrefsec{qgac-comparisons}.

Lastly these methods are hyper-focused on error metrics. While this has proven to be a reliable way to improve rate-distortion, it generally does not translate to improved perceptual quality producing blurry edges and an overall lack of texture. To improve perceptual quality, more complex techniques are required.

In this chapter we develop a technique which addresses all three of these major problems. Our method leverages low-level \gls{JPEG} information to condition a single network on quantization data stored in the \gls{jfif} file, allowing one network to achieve good results on a wide range of qualities. Our network treats color channels as first class citizens and takes concrete steps to correct them effectively, keeping in mind that \gls{JPEG} compression treats color and luminance differently applying more compression to the color channels. Finally we develop texture restoring and GAN losses that are designed to produce a visually pleasing result especially at low qualities. This method was published separately in the proceedings of the European Conference on Computer Vision \sidecite{ehrlich2020quantization}.

\begin{fpbox}
    \begin{itemize}
        \item Conditioning the network on the quantization matrix allows it to correct at many different qualities using information available to a real system
        \item Explicitly modeling color degradation improves performance on color images
        \item Formulating \gls{dct} domain regression allows the network to leverage quantization data more effectively
        \item GAN loss functions for high frequency restoration
    \end{itemize}
\end{fpbox}

\section{Overview}

The method we develop in this chapter consists of several parts, all of which operate together to produce the final result. We will develop this method from the bottom up, starting with the individual building blocks of the method and then describing how they are connected. At a high level, our network operates in several stages, these are illustrated in \reffig{qgac-overview}.

\begin{figure*}[t]
    \centering
    \includegraphics{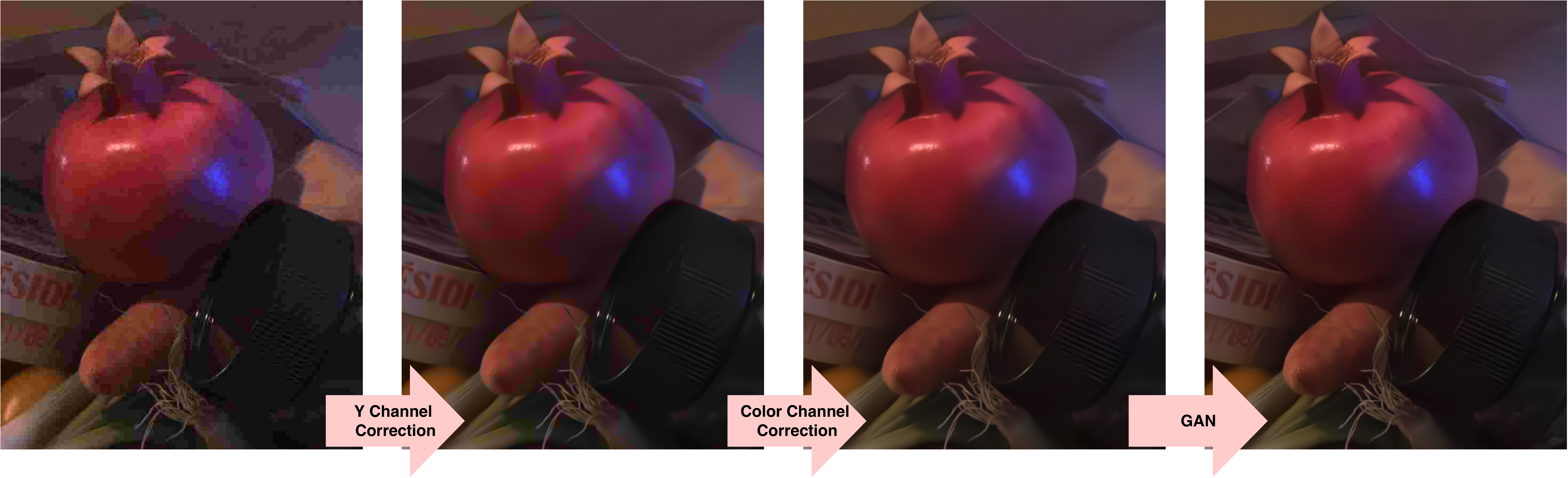}
    \caption[Overview]{\textbf{Overview.} The network first restores the Y channel, then the color channels, then applies GAN correction.}
    \labfig{qgac-overview}
\end{figure*}

Our network first corrects the luminance (Y) channel of the image. The luminance channel has less aggressive compression applied to it and serves as a base for further correction. Our network then moves on to correcting the color channels. As these channels are further compressed, they lack fine detail and structure that may have been present in the luminance channel especially after correction. Therefore, we provide the corrected luminance channel along with the degraded color channel to the color correction network to give it additional information.

Throughout the network, we condition carefully selected layers on the JPEG quantization matrix. Recall that this $8 \times 8$ matrix describes how much rounding was applied to each DCT coefficient. Because this is directly describing a phenomenon in the frequency domain, our entire network processes the DCT coefficients of the input only: no pixels are used, and the network produces DCT coefficients as output. This is in stark contrast to other methods which use only pixel or both pixels and coefficients and depends on new developments in DCT domain networks. We use methods described in \nrefsec{newarch} to correctly process these data. Before these methods were developed, DCT domain networks had objectively inferior performance to pixel and dual-domain networks.

Our training likewise proceeds in stages. After training the network to produce only luminance coefficients using regression, we then add the color network in and train it again using regression. This way, the color network is always getting a high quality luminance result to condition its own correction on. After the luminance and chrominance networks are trained, we then fine-tune the entire network using GAN and texture losses. This adds significant detail to the result while preventing it from quickly deviating and diverging.

\section{Convolutional Filter Manifolds}

One potential limitation of traditional convolutional networks is that they learn only a single mapping from input features ($F_i$) to output features, in other words
\begin{align}
    h(F_i) = \sigma(W \star F_i)
\end{align}
for non-linearity $\sigma$ and learned weight $W$. While this is sufficient for many use cases, it can be limiting in others.

Specifically in our case, we would like to specialize the learned filters for different quantization matrices, in other words, the learned weight $W$ should be a function of the quantization matrix $Q$. One simple way to do this is to tile $Q$ to match the shape of $F$ and concatenate the two
\begin{align}
    h(F_i, Q) = \sigma(W \star \left[F_i\;Q]\right])
\end{align}
however, this yields a linear mapping between $F_i$ and $Q$ limits the learned relationship between $F_i$ and $Q$.

Instead, we can use a \gls{filter manifold}\index{filter manifold}~\sidecite{kang2016crowd}, sometimes called a kernel predictor\index{kernel predictor}~\sidecite{mildenhall2018burst}. The goal of the filter manifold is to predict a convolutional kernel given a \textit{scalar} side input, i.e.,
\begin{align}
    h(F_i, s) = \sigma(W(s) \star F_i)
\end{align}
for $s \in \mathbb{R}$, so now the weight $W$ is a non-linear function of $s$. Kang \etal choose a small \gls{mlp}\index{multilayer perceptron} for $W$
\begin{align}
    h(F_i, s) = \sigma(W(s) \star F_i) \\
    W(s) = \sigma(W_2(\sigma(W_1s)))
\end{align}
allowing the network to learn a non-linear relationship between the side data and $F_i$ along with the learned mapping between $F_i$ and the network output.

However, our side input $Q$ is not a scalar, it is an $8 \times 8$ matrix. Using a \gls{mlp} for this input would be computationally expensive, so we propose a simple extension, termed \glspl{convolutional filter manifold}\index{convolutional filter manifold}(CFM), to replace $W()$ with a small convolutional network. We additionally learn a bias term along with the weight
\begin{align}
    h(F_i, Q) = \sigma(W(Q) \star F_i + b(Q)) \\
    b(Q) = W_b \star F_q(Q)                   \\
    W(Q) = W_w \star F_q(Q)                   \\
    F_q(Q) = \sigma(W_2 \star (\sigma(W_1 \star Q)))
\end{align}
This formulation allows us to learn parameterized weights representing the complex relationship between the \gls{JPEG} \gls{dct} features and the quantization matrix and can be thought of as generating a ``quantization invariant'' representation for the network to operate on. This is the primary contribution which allows the network to model degradations from many different quality levels. In \nrefsec{qgac-blocks} we will describe primitive layers which make use of this formulation and in \nrefsec{qgac-net} we will describe where these layers are placed in the overall network structure in order to maximize their effectiveness. In \nrefsec{qgac-cfm-res}, we will explore some interesting properties of these layers.

\section{Primitive Layers}
\labsec{qgac-blocks}

The network we develop in this chapter is dependent on several ``primitive layers'' or basic operations which we will use to build the network. In this section, we describe them in detail. The first is the \gls{rrdb}\index{residual-in-residual dense block} layer~\sidecite{wang2018esrgan} first developed for super-resolution. This layer consists of three ``Dense Blocks'' in a residual sequence. Each of these ``Dense Blocks'' consists of five convolution-relu layers with skip connections between each layer forming an enhanced version of the standard residual block\index{residual block}. See \reffig{qgac-rrdb} for a schematic depiction of this layer. We make only one change to the \gls{rrdb} used in ESRGAN by replacing the Leaky ReLU~\sidecite{maas2013rectifier} with Parametric ReLU~\sidecite{he2015delving}.

\begin{marginfigure}
    \centering
    \includegraphics{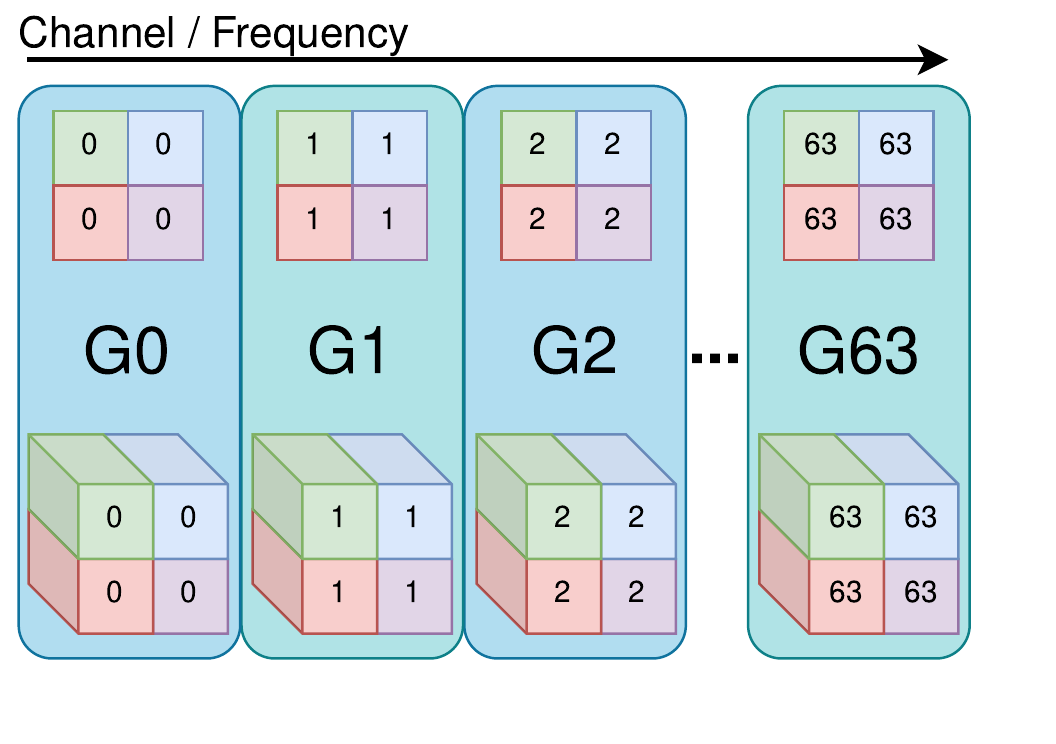}
    \caption[FCR With Grouped Convolutions]{\textbf{FCR With Grouped Convolutions.} Each frequency component is processed in isolation with its own convolution weights. We implement this using a grouped convolution with 64 groups.}
    \labfig{qgac-fcr}
\end{marginfigure}
In \nrefsec{newarch} we discussed recent advances in convolutional networks that can take advantage of the unique characteristics of \gls{dct} coefficients. We employ both of these layers in our network. The first is frequency-component rearrangement\index{frequency component rearrangement} where the \gls{dct} coefficients for each block are arranged in the channel dimension yielding 64 channels and $\frac{1}{8}$th the width and height of the input. We take the additional step of using grouped convolutions with 64 groups to ensure that each frequency is processed in isolation. See \reffig{fcr} for the frequency rearrangement and \reffig{qgac-fcr} for an illustration of the grouped convolution. We insert these layers into the RRDB described above. This paradigm allows our network to focus on enhancing individual frequency bands more effectively.

However, many frequency bands are entirely zeroed out by the compression process. Completely relying on the grouped convolution would be destined for failure because if a frequency band is set to zero, no amount of convolutional layers can change its value (it will either remain zero or be set to the layer biases). Therefore, we need a layer which is also capable of looking at multiple frequency bands, and for this we choose the $8 \times 8$ stride-8 layer. This layer produces a representation of each \gls{dct} block by considering all the frequency bands in the block at once. Since the stride is set to 8, the representation does not include information from nearby blocks. Information from nearby blocks is incorporated by processing the block representations with \gls{rrdb} layers. Since these layers are considering the \gls{dct} coefficients of the entire block, we make the additional step to use CFMs instead of regular convolutions to equip the layers with quantization information, thus generating the ``quantization invariant'' block representation. This is shown in \reffig{qgac-8by8}. For our applications, the input to the CFM is the $8 \times 8$ quantization matrix. This is processed with a convolutional network to produce the weight and bias. Note that the weight layer has $\text{in\_channels} \times \text{out\_channels}$ channels and is a transposed $8 \times 8$ convolution. The result is reshaped to an $\text{out\_channels} \times \text{in\_channels} \times 8 \times 8$ convolution kernel.

\begin{figure}[t]
    \centering
    \includegraphics{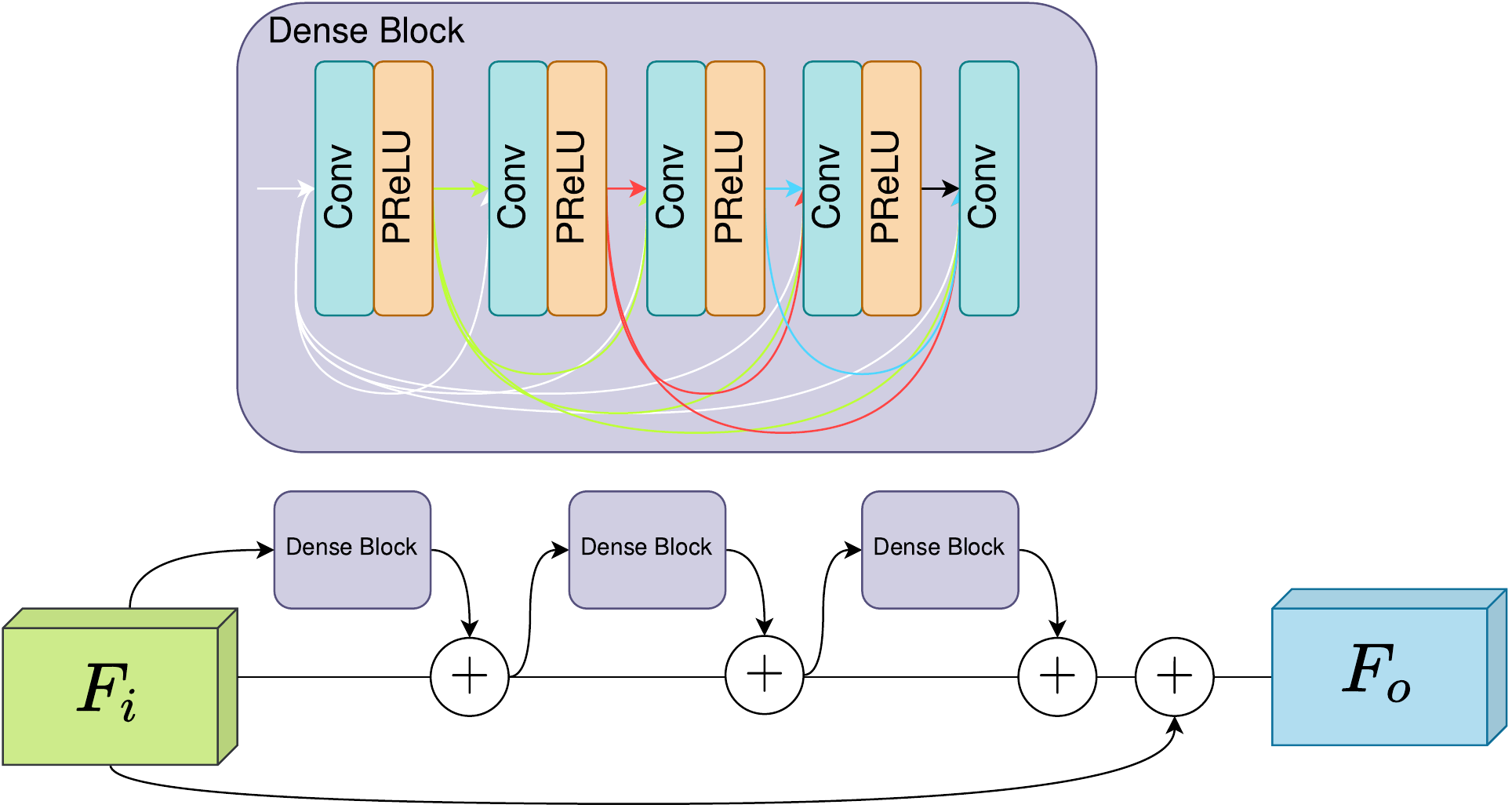}
    \caption[RRDB]{\textbf{RRDB} Layer shown with input feature map $F_i$ and output feature map $F_o$. Note that we change the original RRDB layer by adding PReLU layers.}
    \labfig{qgac-rrdb}
\end{figure}

\section{Full Networks}
\labsec{qgac-net}

\begin{figure}[t]
    \centering
    \includegraphics{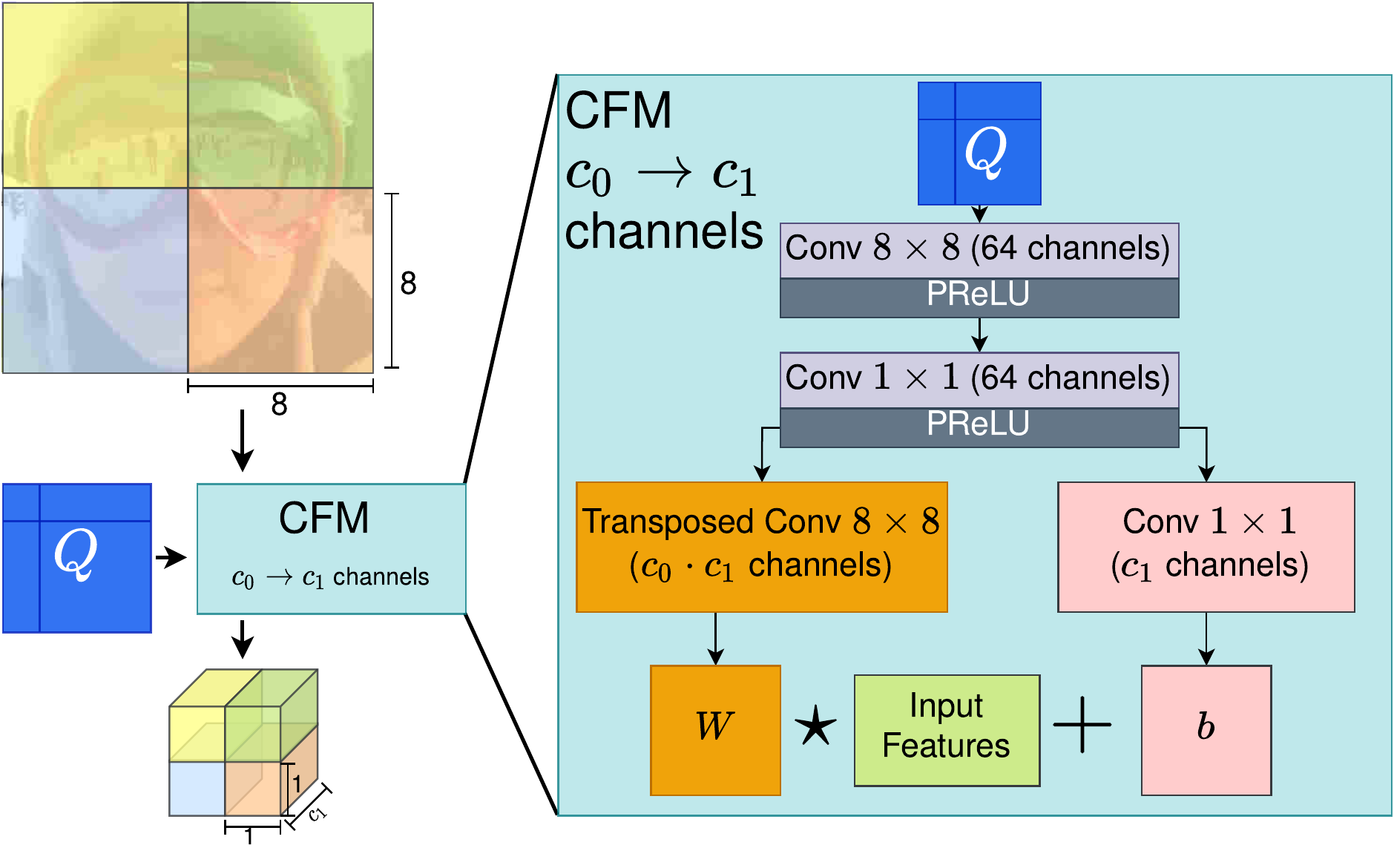}
    \caption[$8 \times 8$ stride-8 CFM]{\textbf{$\bm{8 \times 8}$ stride-8 CFM.} Note that the numbers in parenthesis denote the number of channels. The CFM layer computes a weight and bias from the quantization matrix using a small convolutional network. The result of this network is reshaped as either a weight or bias.}
    \labfig{qgac-8by8}
\end{figure}

With the primitive layers defined, we now show how to build those layers into the networks and subnetworks our method uses for correction of JPEG artifacts. Recall that our method first corrects the grayscale channel and then uses that result to aid correction of the chroma channels. Therefore we start by describing the grayscale correction network. After that we will describe the color correction network.

The grayscale correction network, shown in \reffig{qgac-nets} left, consists of four subnetworks which work in series to produce the final correction: two blocknets, a frequencynet, and a fusion network which we describe next. The blocknet (\reffig{qgac-subs} left) uses the $8 \times 8$ stride-8 CFM layers described in the previous section. It computes block representations and then processes the representations with stacked RRDB layers before decoding the block representations with a transposed CFM layer. Between the two blocknets we place a frequencynet (\reffig{qgac-subs} middle). This uses the FCR grouped convolutions to enhance frequency bands in isolation. The frequencies are first rearranged before being processed with RRDB layers. The result is then rearranged to restore the frequencies to the spatial dimensions. The intermediate results from all of the subnetworks are then passed to a fusion layer (\reffig{qgac-subs} right). The primary purpose of this is to strengthen the gradient received by the early layers which would be prone to gradient vanishing otherwise \sidecite{hochreiter2001gradient}.

The color correction network (\reffig{qgac-nets} right) borrows the main ideas from the blocknet in the grayscale correction network. We assume that inputs are 4:2:0 chroma subsampled, which means they must be upsampled by a factor of two in each dimension to match the grayscale resolution. We use the block representation of the color channels and use a $4 \times 4$ stride-2 layer to do the upsampling. The result is concatenated channelwise with the block representation of the restored Y-channel before being processed further and finally decoded. In both the grayscale and color networks, we treat network outputs as residuals which are added to the degraded input coefficients.

\section{Loss Functions}
\labsec{qgac:loss}

\begin{figure*}[t]
    \centering
    \includegraphics{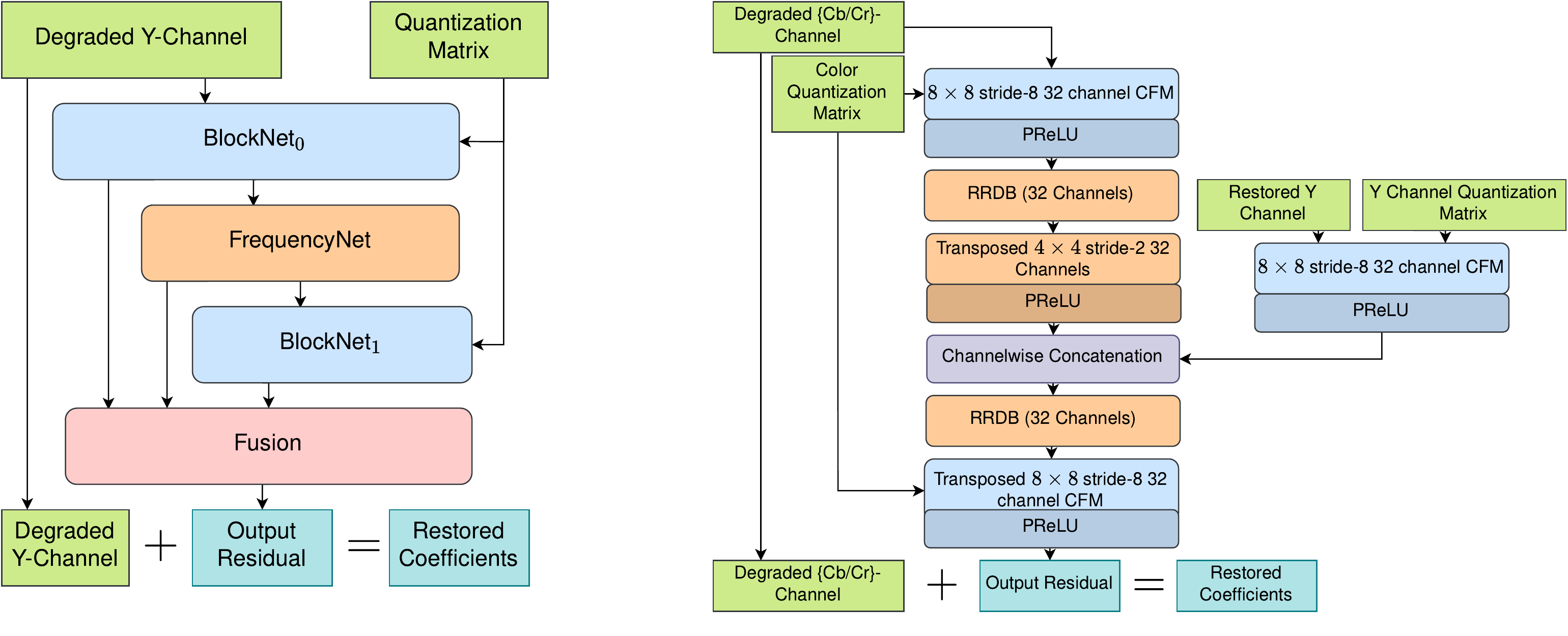}
    \caption[Restoration Networks]{\textbf{Restoration Networks.} Left: Y-Channel Network. Right: Color Channel Network. Note the skip connections around each of the subnetworks in the Y-Channel Network which promotes gradient flow to these early layers.}
    \labfig{qgac-nets}
\end{figure*}

\begin{figure*}[t]
    \centering
    \includegraphics{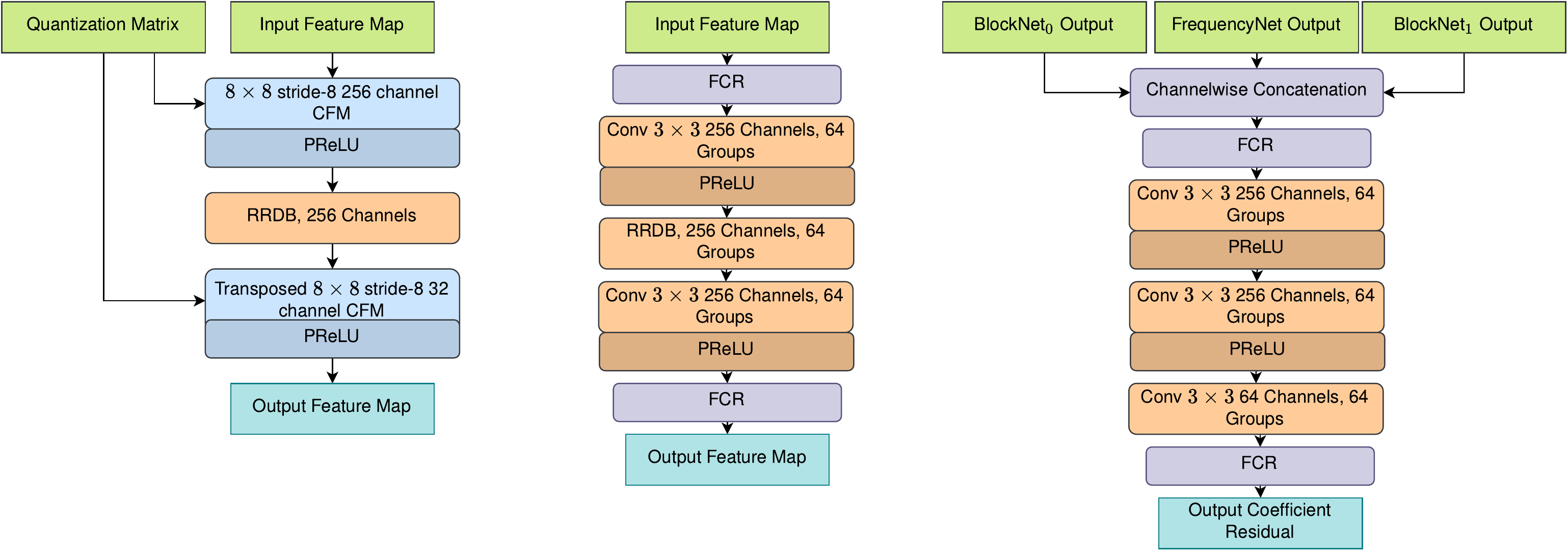}
    \caption[Subnetworks]{\textbf{Subnetworks.} Left: BlockNet, Center: FrequencyNet, Right: Fusion.}
    \labfig{qgac-subs}
\end{figure*}

A well documented problem with image-to-image translation is that of a blurry result. Intuitively, since the network is told to optimize $l_1$ or $l_2$ distance between the input and output, the easiest way to accomplish its goal is to produce a sort of ``averaging''. The human perception of this averaging is as a low-frequency image which lacks fine details. This is exacerbated by compression which intentionally removes high frequency details. In this sense, a simple error based loss function is, in essence, asking the network to solve the wrong problem. What we really want the network to do is restore high frequencies.

Nevertheless, an error based loss is useful for correcting hard block boundaries that JPEG creates as well as for preventing divergence with more complex losses. Therefore, we pre-train the grayscale and color networks using $l_1$ and Structural Similarity (SSIM)\index{structural similarity} \sidecite[-2\baselineskip]{wang2004image} losses to ensure that they start from a reasonable location when we fine-tune with the more interesting loss functions\sidenote{Note that this is a visual improvement on its own, however it is nothing compared to the result from the GAN and texture losses.}. We denote this loss function as
\begin{align}
    \Lagr_R(X_u, X_r) = ||X_u - X_r||_1 - \lambda\text{SSIM}(X_u, X_r)
    \labeq{loss_r}
\end{align}
for restored image $X_r$ and uncompressed image (\ie, a version of $X_r$ which was never compressed) $X_u$ and $\lambda$ is a balancing hyperparameter.

With the color and grayscale networks trained for regression we now move on to GAN \sidecite{goodfellow2014generative}\index{generative adversarial network} and texture losses. GANs were originally introduced purely for generating realistic images. The algorithm pits a generator network against discriminator network where the generator's goal is to produce an image which is realistic enough to fool the discriminator and the discriminator's goal is to discover which images were generated by the generator. In this way, the two networks are adversaries in a game and by rewarding them for doing well, the generator can create more realistic images. For our purposes, we use a GAN to hallucinate plausible high frequency details, edges, and textures onto the compressed images.

For this we employ the relativistic average GAN loss \sidecite{jolicoeur2018relativistic}. This loss function tweaks the original GAN definition to encourage the generator to produce images which appear ``more realistic than the average fake data'' and is generally more stable than a vanilla GAN. For our purposes, we redefine ``fake'' as the restored image $X_r$ and ``real'' as the uncompressed image $X_u$. We then define the loss as
\begin{align}
    \Lagr_\text{RA}(X_u, X_r) = \log(L(X_u)) - \log(1 - L(X_r)) \\
    L(x) = \begin{cases}
               \sigma(D(x) - \e_{x_r \in \text{Restored}}[D(x_r)])     & x \text{is uncompressed} \\
               \sigma(D(x) - \e_{x_u \in \text{Uncompressed}}[D(x_u)]) & x \text{is restored}
           \end{cases}
\end{align}
For discriminator $D()$ and sigmoid $\sigma()$. We base the discriminator $D()$ on DCGAN \sidecite{radford2015unsupervised}, its architecture is shown in \reffig{qgac-disc}. All convolutional layers use spectral normalization \sidecite{miyato2018spectral}. Note that we provide both the compressed as well as the uncompressed/restored version of the image and discriminator decisions are made on a per-JPEG block basis.

\begin{marginfigure}
    \centering
    \includegraphics{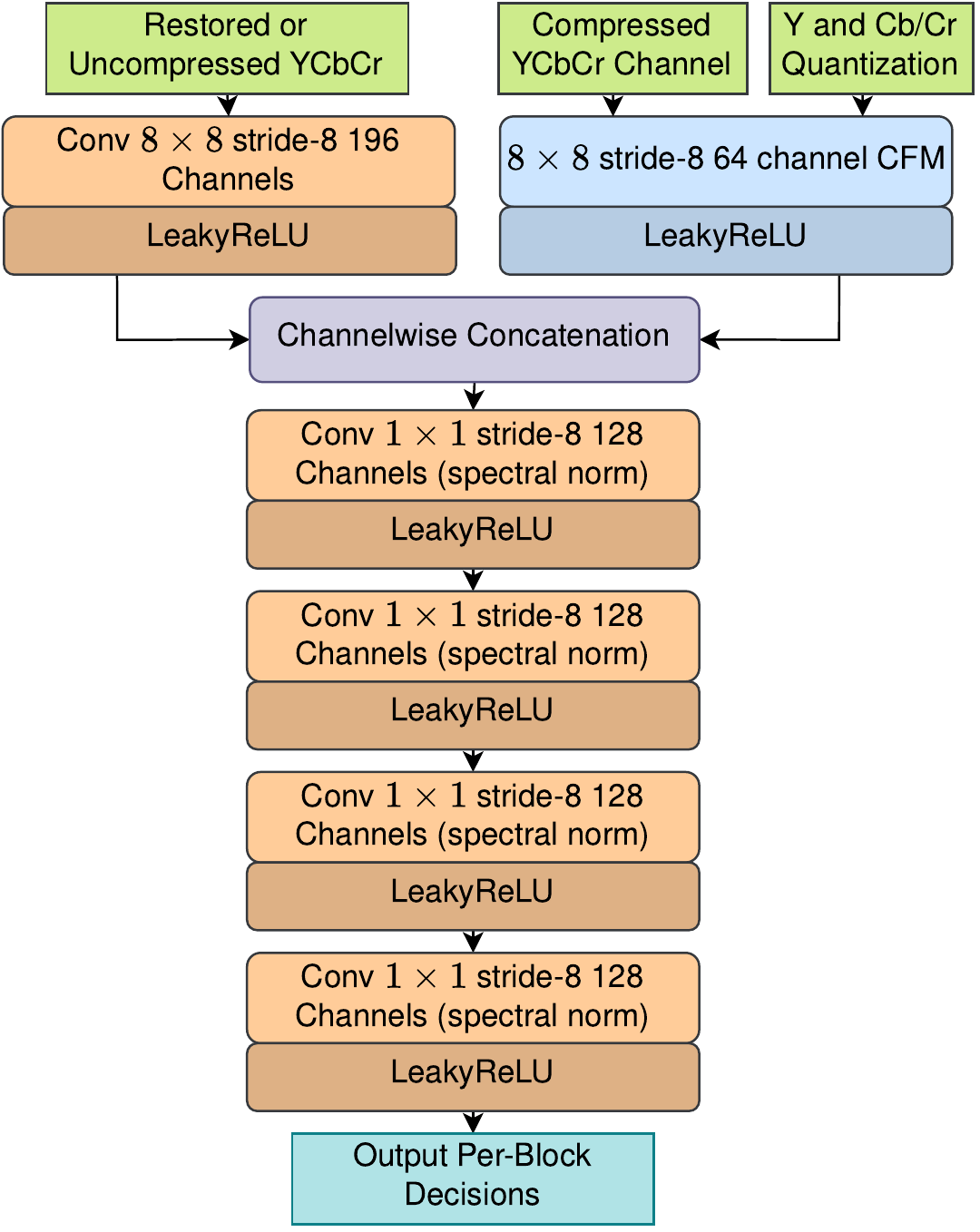}
    \caption[GAN Discriminator]{\textbf{GAN Discriminator.} Note that the discriminator makes decisions for each JPEG block.}
    \labfig{qgac-disc}
\end{marginfigure}
While the GAN is a useful tool for generating realistic corrections, the general notion of real or fake only provides so much information. In practice, GAN losses for image-to-image translation are often coupled with ``perceptual losses'' \sidecite{johnson2016perceptual}. More specifically, these losses use an ImageNet \sidecite{deng2009imagenet} trained VGG network \sidecite{simonyan2014very}. The intuition is that this auxiliary network measures a semantic similarity between the input image and the desired target since this network was trained for classification. By encouraging semantic similarity, a more realistic result can be achieved since the images appear to fall into the same class.

While this is useful for general image-to-image translation we find an alternative approach is more useful for compression. Since compression destroys high frequency details, like textures, the more these details can be recovered or sufficiently hallucinated, the more realistic the reconstruction. Therefore, we use a VGG network trained on the MINC \sidecite{bell2015material} dataset for material classification. The main idea here is that if a restored and uncompressed image have similar logits for a material classification task, they would likely be classified as the same material and therefore have realistic textures. We denote this loss function as
\begin{align}
    \Lagr_\text{t}(X_u, X_r) = ||\text{MINC}_{5,3}(X_u) - \text{MINC}_{5,3}(X_r)||_1
\end{align}
where $\text{MINC}_{5,3}$ indicates layer 5 convolution 4 from the MINC trained VGG.

This yields the complete GAN loss
\begin{align}
    \Lagr_{\text{GAN}}(X_u, X_r) = \Lagr_t(X_u, X_r) + \gamma\Lagr_{\text{RA}}(X_u, X_r) + \nu||X_u - X_r||_1
    \labeq{loss_g}
\end{align}
for balancing hyperparameters $\gamma, \nu$. Note that the $l_1$ loss makes another appearance here to prevent the GAN from diverging.

\section{Limitations and Future Directions}

Although this work represents a major step forward in the usability of JPEG artifact correction methods, there are still some major problems to be solved. First and foremost is the double compression problem. Because QGAC parameterizes itself only on the quantization matrix of the file it is correcting, it has no way of knowing if this image was recompressed. For example a real-life company which will not be named directly and with a complex image processing pipeline decompresses and recompresses each JPEG it receives multiple times. Realizing that this would lead to significant degradation, this company recompresses its images at quality 100, mitigating most quality loss. However, QGAC will treat this as a quality 100 JPEG and perform essentially no restoration on it: it knows no better. In effect the image processing pipeline has lied to QGAC about the nature of the compression. This was partially addressed by AGARNET \sidecite{kim2020agarnet} which allows for a spatially varying ``Q-map'', essentially per-pixel quality, to be used as an auxiliary input, however generating the Q-map is not straightforward.

Then there is the related problem: the JPEG degraded image may not be stored as a JPEG at all. It is fairly common to transcode JPEG files to PNGs where they can be stored without further degradation. QGAC, of course, cannot operate on PNGs because they do not contain quantization information. This was addressed by FBCNN \sidecite{jiang2021towards} which trains a network to predict quality level from pixels (along with the restored output) thus implicitly parameterizing the network on quality and allowing it to take any kind of image as input.

There is the longstanding problem of high frequency reconstructions. It seems that there are currently two paradigms in restoration: low-frequency but accurate reconstructions and high frequency but inaccurate reconstructions (\eg GAN reconstruction which looks nice but has little relationship to the ground truth image). A ``holy grail'' of reconstruction work would be allow accurate reconstructions in the high frequencies. This is partially addressed later in the dissertation with the scale-space loss of Metabit.

Another important direction to consider for practical usage of artifact correction is runtime, memory, and hardware concerns. The end goal is to put these methods into the hands of users who may be on smartphones or laptops but little attention has been paid to this so far. Current techniques often require datacenter machines with powerful, often multiple, GPUs in order to run in a timely manner (or at all). More attention to practical, efficient formulations and to quantized integer models or specialized hardware is important to the widespread dissemination of this technology.

\setchapterpreamble[u]{\margintoc}
\chapter{{\color{Plum}Task-Targeted Artifact Correction}}
\labch{ttac}

\lettrine{T}{hus} far we have considered artifact correction as a tool for presenting attractive images to a user. In other words, where a compressed image contains certain artifacts, we want to suppress those artifacts so that the user can view something closer to the uncompressed image. We noted that this was a difficult task to accomplish for some time because artifact correction methods were trained on a ``per-quality'' basis with a different model for each quality and we proceeded to develop a method for correction of JPEG artifacts that is ``quality-blind'', \ie, only a single model is trained for all JPEG qualities.

We now consider a slightly different question: what if the images are intended for machine consumption and not human consumption? How does this change the problem, if at all, and how do machine learning algorithms respond to JPEG compression? In this contribution of the dissertation, we develop a flexible method of overcoming the accuracy loss caused by JPEG compression on common computer vision models. This includes both a study of how JPEG compression affects these models and the examination of different methods for mitigation of the accuracy loss. This method was published separately in the MELEX workshop in the proceedings of the International Conference on Computer Vision \sidecite{ehrlich2021analyzing}.

The method presented in this chapter trains an artifact correction network to target a specific computer vision task. This has significant advantages over off-the-shelf techniques which we examine in \nrefsec{ttac-trans}. Namely, the method is transferable between models. In other words, once trained to assist a particular model, it is general enough to assist other models. Similarly, it can be trained to assist multiple tasks simultaneously without a significant penalty on its effectiveness. We call this method Task-Targeted Artifact Correction (TTAC)\index{task-targeted artifact correction}.

\begin{fpbox}
    \begin{itemize}
        \item JPEG degrades task performance. Leveraging explicit JPEG correction can mitigate the problem
        \item Supervise the JPEG correction method using differences between the uncompressed and corrected images
        \item Task trained correction networks are generalizable to many downstream tasks
    \end{itemize}
\end{fpbox}

\section{Standard JPEG Compression Mitigation Techniques}

Before moving on we briefly review other techniques which are commonly thought to mitigate JPEG artifacts.

\paragraph{Supervised Fine-Tuning/Data Augmentation} The simplest possible scheme, JPEG compressed inputs are mixed in during training as a form of data augmentation. The goal here is to train the network to expect JPEG compressed inputs and map them correctly. While this idea works, often well, it has several disadvantages. The first is that it sacrifices accuracy on clean images. So the result of the network is no longer ``at a theoretical maximum'' because it has, in some sense, expended capacity modeling JPEG compressed inputs. Additionally, this method requires ground-truth labels which can be expensive to obtain.

\paragraph{Off-the-Shelf Artifact Correction} Another exceedingly simple method: simply apply an artifact correction network to JPEG compressed inputs. Since the artifact correction method is reducing error with respect to the clean image, intuition states that this should help the performance of a downstream task (and indeed it does). Moreover, this technique could be employed practically with the development of QGAC which does not require knowledge of the JPEG quality. This technique also has the advantage of not requiring any training at all and indeed keeping the clean accuracy intact is a selling point of the method. However, this is an ``all-or-nothing'' approach in that there is no way to tune it when it does not work.

\paragraph{Stability Training} This technique \sidecite{zheng2016improving} is more interesting than the last two ideas and involves logit matching between the network output on clean and perturbed (in this case JPEG compressed) images. In this case, the stability loss is defined as
\begin{align}
    \Lagr_\text{stability}(x, x') = ||f(x) - f(x')||_2
\end{align}
where $f(x)$ is some neural network and $x'$ is the perturbed version of $x$. This objective is then minimized along with the primary task objective during training. While this technique does encourage robustness and is self-supervised it inherits several drawbacks from the supervised method. The task network now needs to expend capacity to model the compressed mapping and performance on clean images is sacrificed.

\section{Artifact Correction for Computer Vision Tasks}

\begin{marginfigure}
    \centering
    \includegraphics{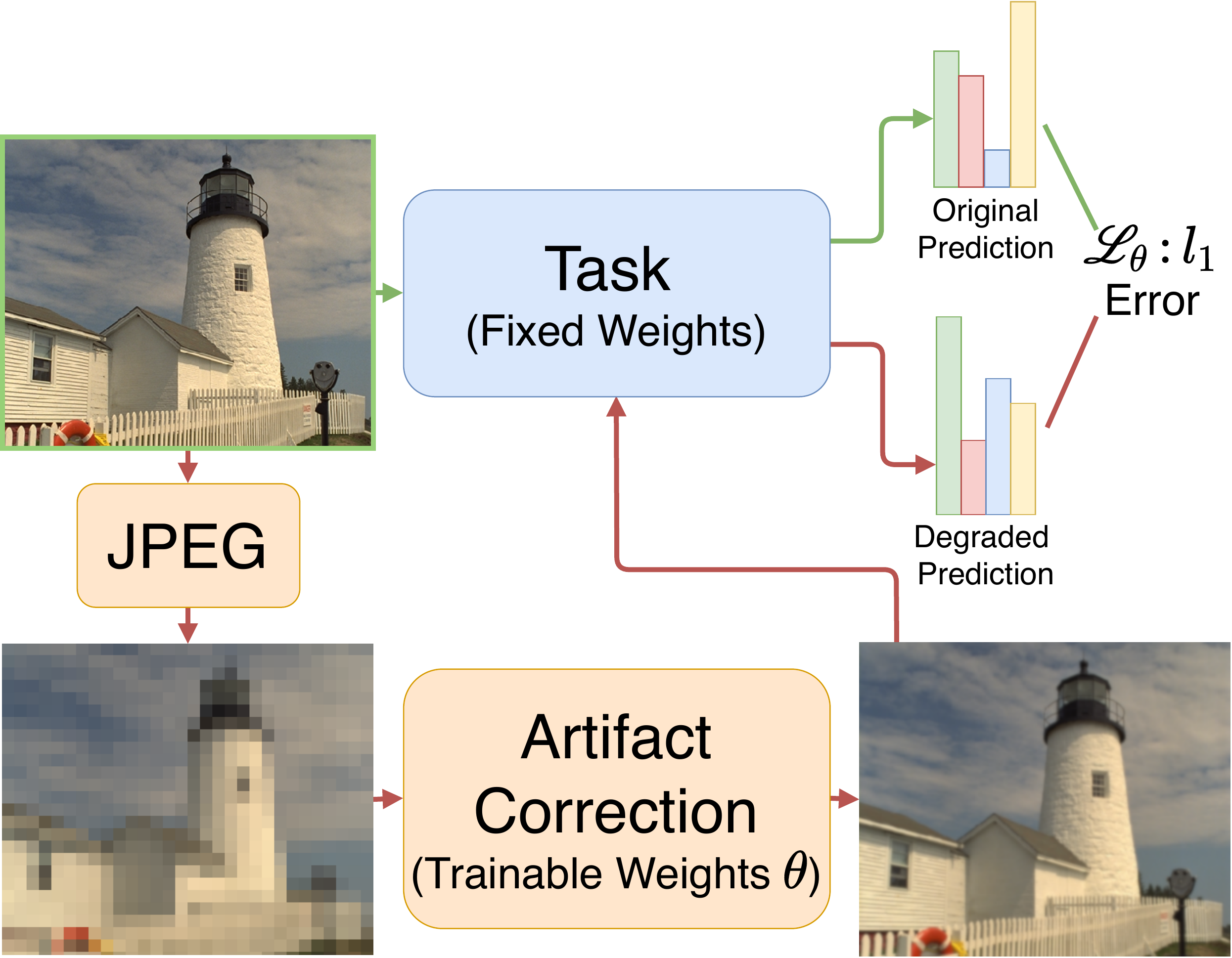}
    \caption[Task-Targeted Artifact Correction]{\textbf{Task-Targeted Artifact Correction.} The logit difference from the task network between clean and artifact-corrected versions of the same image is used to train the artifact correction network.}
    \labfig{ttac-explainer}
\end{marginfigure}
The algorithm we propose in this chapter targets an artifact correction network to a particular task. In all cases we will use QGAC from \nrefch{qgac} for the artifact correction network. Starting from pre-trained weights, we fine-tune the artifact correction network using logit error from the task network between clean inputs and compressed inputs.

Formally, given a task $t()$, and artifact correction network $q()$, we minimize
\begin{align}
    \Lagr_\theta(B) = ||t(b) - t(q(\text{JPEG}_q(B);\theta))||_1
\end{align}
where $\text{JPEG}_q$ denotes JPEG compression at quality $q$. Note that the parameters, $\theta$, that we optimize belong to $q$. The task network is unchanged during this process. See \reffig{ttac-explainer} for a visual depiction of this process.

While the intuition behind this process is simple there are several details that need to be accounted for. First consider that we are not training the artifact correction network based on any decision by the task network, \eg, classification or detection. Instead, we are matching the actual logit values. These are vectors of real numbers and are much finer grained than the actual decision which may be binary. In effect we are rewarding the artifact correction network for inducing the same perception of an input image in the task network. Note that since there is no hard decision required for training the method is entirely self-supervised. Only the logit values, which are independent of any ground truth, are considered during the training process.

This differs from stability training in several key ways. First, it does not modify the task network, so performance on clean images is unchanged and the task network is free to expend its entire capacity learning the relationship between clean data and the output. Next, since the correction task is given to an auxiliary network, this network can be reused for other tasks. As we examine in \nrefsec{ttac-trans}, this works surprisingly well allowing the artifact correction network to be trained using a lightweight task and reused for more complex tasks. To summarize, task-targeted artifact correction takes the advantages of all prior techniques with none of the disadvantages and adds in transferability as a bonus.

\begin{figure*}[t]
    \centering
    \includegraphics[width=0.5\textwidth]{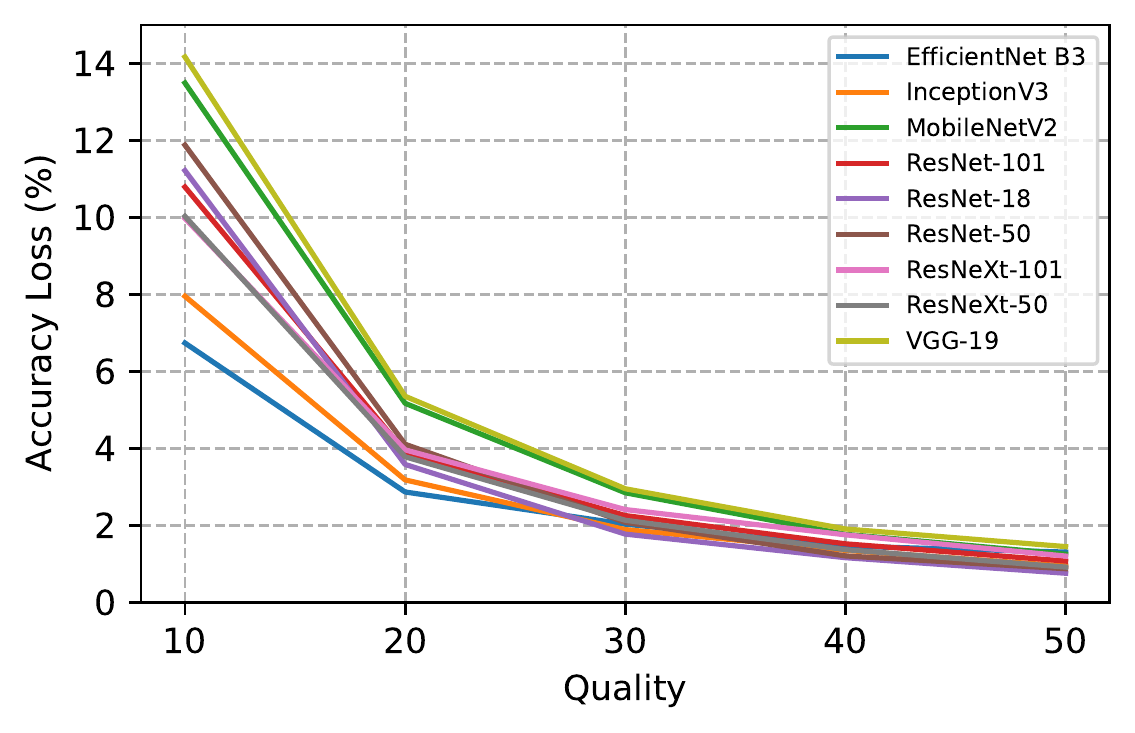}\hfill\includegraphics[width=0.5\textwidth]{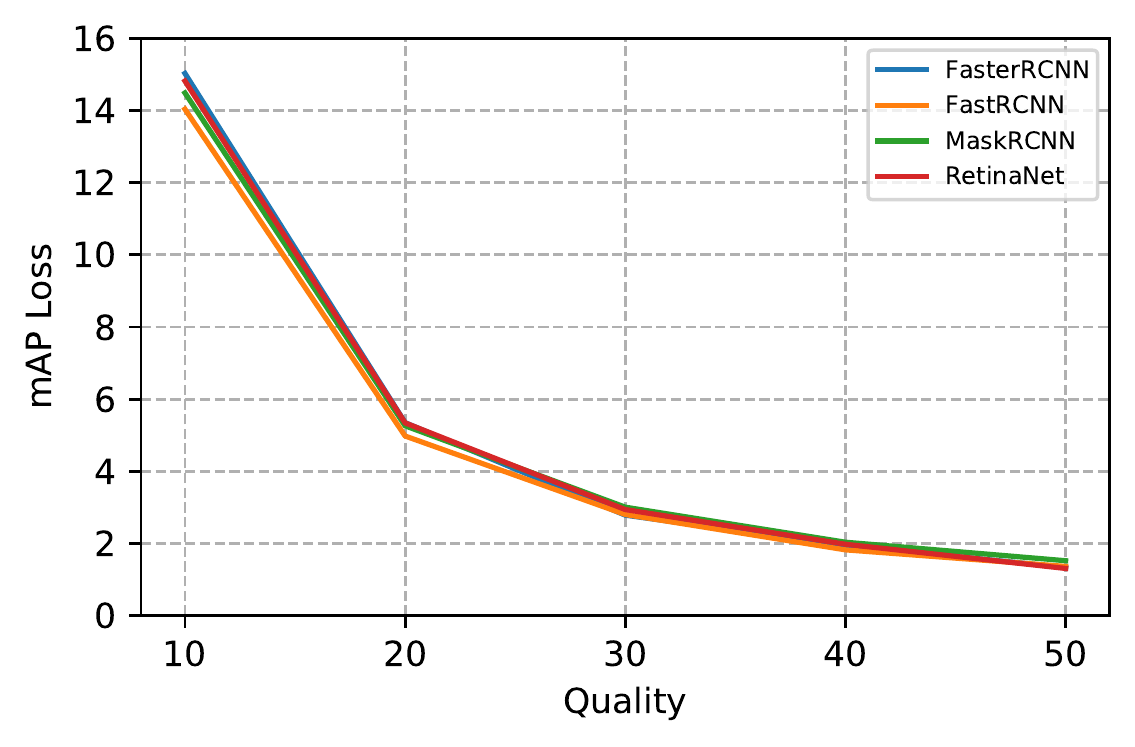}\hfill\includegraphics[width=0.5\textwidth]{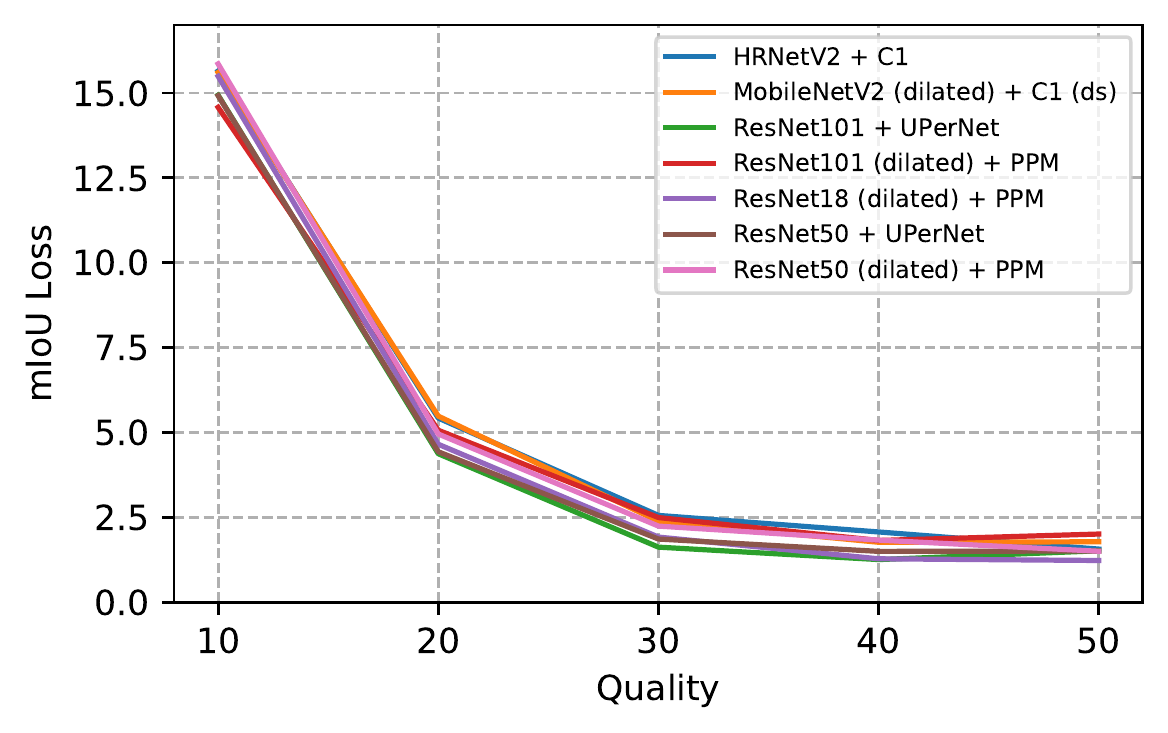}
    \caption[Performance Loss Due to JPEG Compression]{\textbf{Performance Loss Due to JPEG Compression} separated by task. Left: Classification, Middle: Detection, Right: Segmentation. The plots show all models from a single task with no mitigation applied. For segmentation tasks, the format of the model name is \texttt{Encoder Model} + \texttt{Decoder Model} and ``ds'' indicates that the model was trained with deep supervision. Note that methods which use a Pyramid Pooling Module (PPM) decoder always use deep supervision.}
    \labfig{ttac-nomitigation}
\end{figure*}
\begin{figure*}
    \centering

    \includegraphics[width=0.5\textwidth]{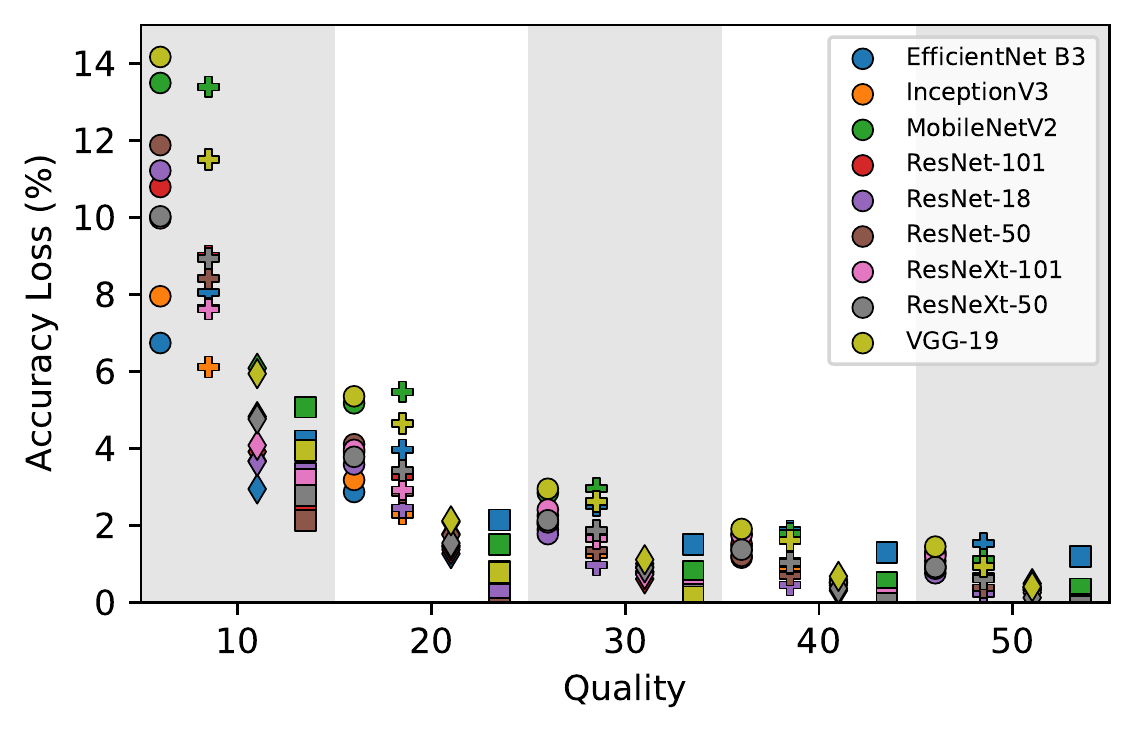}\hfill\includegraphics[width=0.5\textwidth]{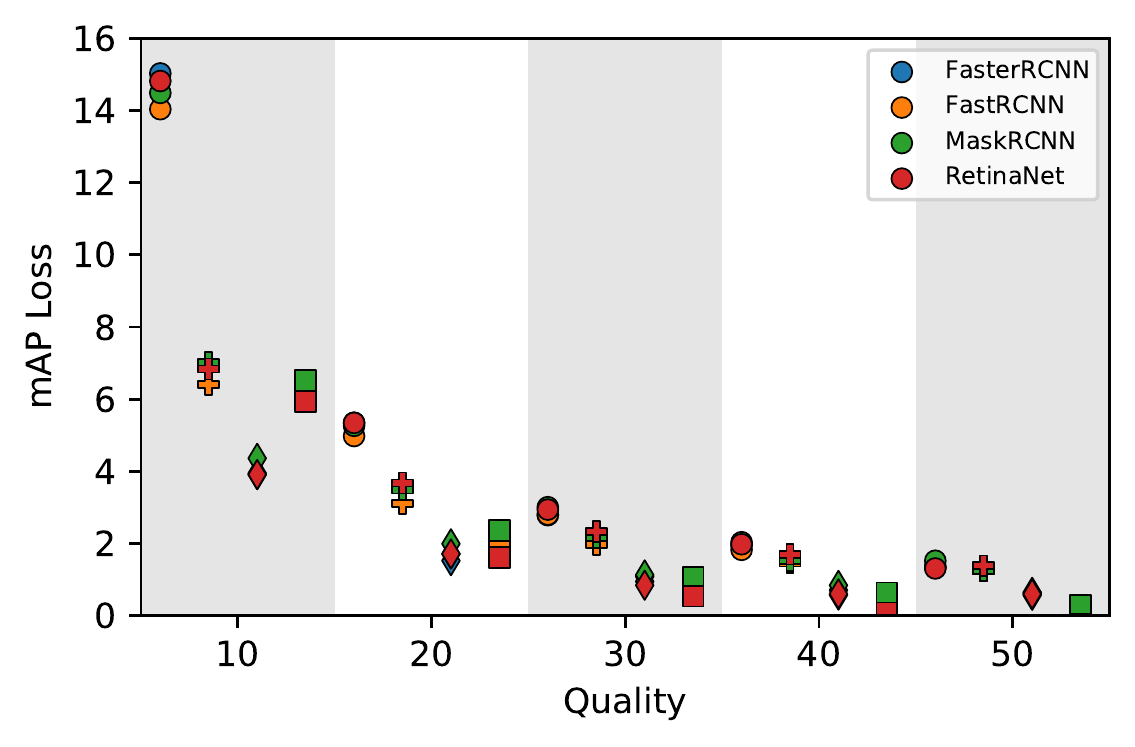}\hfill\includegraphics[width=0.5\textwidth]{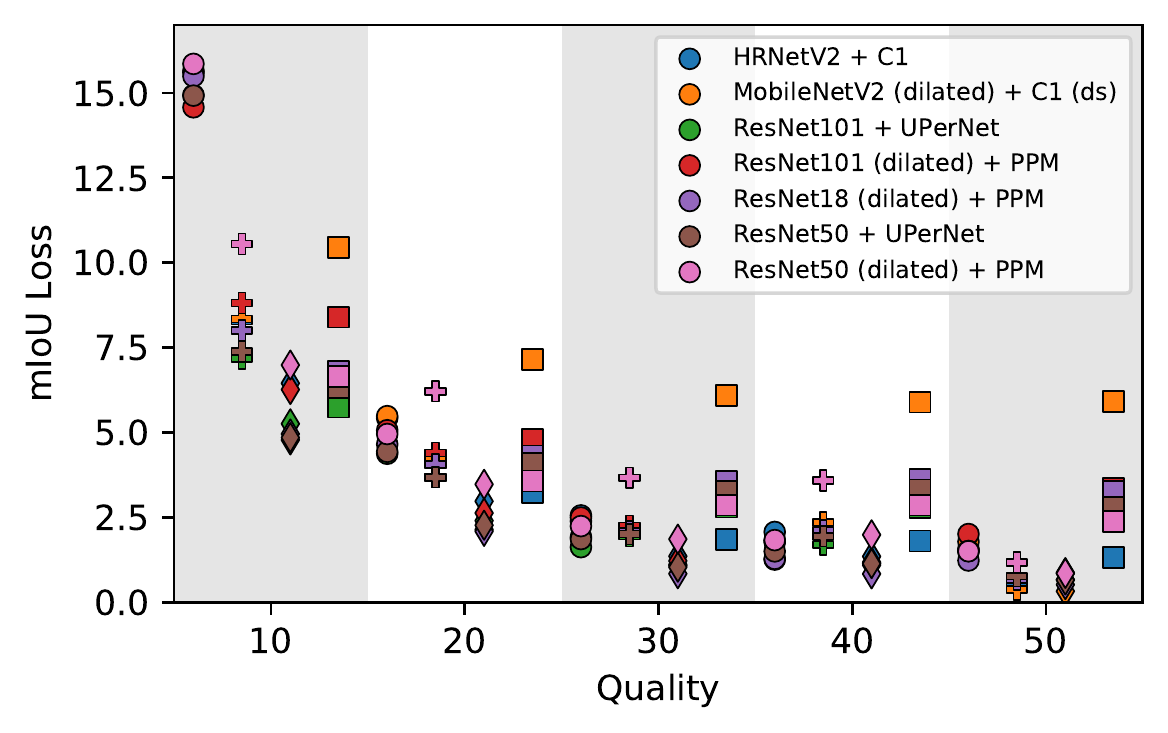}
    \caption[Performance Loss with Mitigations]{\textbf{Performance Loss with Mitigations} \textbf{$\pmb{\bullet}$ Circle}: No Mitigation, \textbf{$\pmb{+}$ Cross}: Off-the-Shelf Artifact Correction, \textbf{$\blacklozenge$ Diamond}: Task-Targeted Artifact Correction, \textbf{$\blacksquare$ Square}: Supervised Fine-Tuning.  The models in this figure correspond to those shown in \reffig{ttac-nomitigation}.}
    \labfig{ttac-withmitigation}
\end{figure*}

\begin{figure*}[t]
    \centering
    \includegraphics[width=0.5\textwidth]{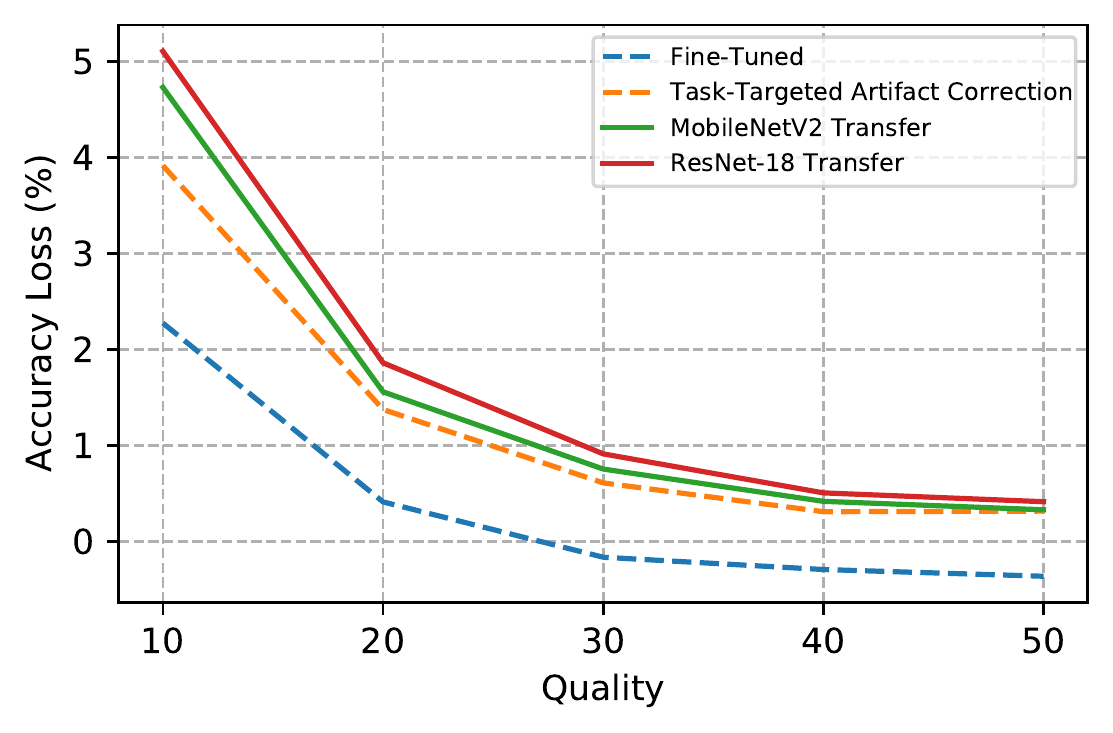}\hfill\includegraphics[width=0.5\textwidth]{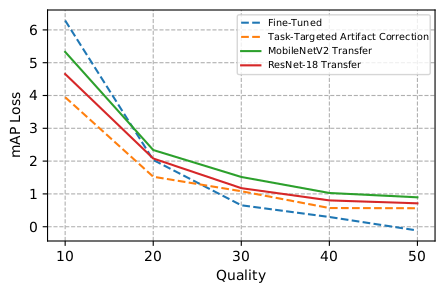}\hfill\includegraphics[width=0.5\textwidth]{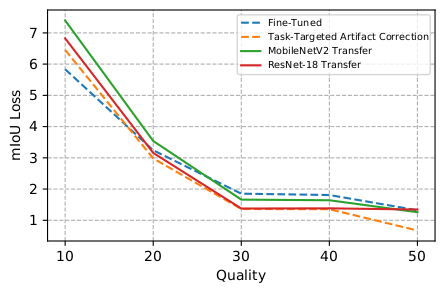}
    \caption[Transfer Results]{\textbf{Transfer Results.} Left: ResNet 101 (classification), Middle: Faster R-CNN (detection), Right: HRNetV2 encoder with C1 decoder (semantic segmentation). In all plots, we add an evaluation using artifact correction weights that were trained on ResNet-18 and MobileNetV2, our lightest weight models. Note that ``Fine-Tuned'' and ``Task-Targeted Artifact Correction'' methods are both trained using their respective task network directly \eg in (a) they use a ResNet 101. \textbf{- -} dashed lines indicate results shown in \refsec{ttac-effect}}
    \labfig{ttac-transfer}
\end{figure*}
\begin{figure*}[t]
    \centering
    \includegraphics[width=0.5\textwidth]{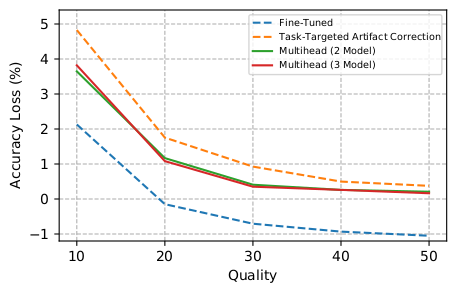}\hfill\includegraphics[width=0.5\textwidth]{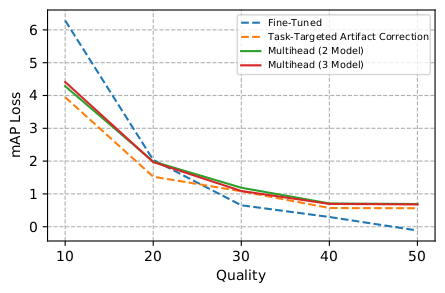}\hfill\includegraphics[width=0.5\textwidth]{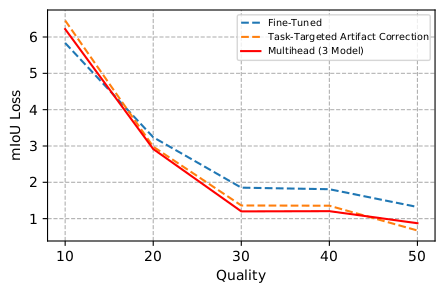}
    \caption[Multiple Task Heads]{\textbf{Multiple Task Heads.} Left: ResNet 50 (classification), Middle: Faster R-CNN (detection), Right: HRNetV2 encoder with C1 decoder (semantic segmentation). In all plots, we add an evaluation using artifact correction weights that were trained using multiple task networks. For the two task setup, we used ResNet-50 and FasterRCNN. For the three task setup, we used ResNet-50, FasterRCNN, and HRNetV2 + C1. Note that HRNetV2 + C1 has no two-task multihead model. \textbf{- -} dashed lines indicate results shown in \refsec{ttac-effect}.}
    \labfig{ttac-mh}
\end{figure*}

\section{Effect of JPEG Compression on Computer Vision Tasks}
\labsec{ttac-effect}

A legitimate question at this point is ``how much does JPEG actually affect computer vision tasks?''. We can answer this with a study, the conclusions of which are summarized in this section. The full results are relegated to Appendix \ref{app:study}.

For this study, we compressed images using quality in [10, 90] in steps of 10\sidenote{We only show [10, 50] in this section as these are the most interesting results.} using the test sets of the respective models we are evaluating. The input images are compressed, then restored, then they are transformed according to the requirements of the target model (\eg, cropping to $224 \times 224$). We evaluate supervised fine-tuning, off-the-shelf artifact correction, and task-targeted artifact correction, we do not evaluate stability training. For methods requiring fine-tuning, we train for 200 epochs varying the learning rate from $10^{-3}$ to $10^{-6}$ using cosine annealing \sidecite{loshchilov2016sgdr}. We compare all mitigation methods to a baseline of ``doing nothing'', \ie, accepting JPEG inputs with no modification. We evaluate the following tasks, datasets, and models:
\begin{description}
    \item[Classification] using ImageNet \parencite{deng2009imagenet} with MobileNetV2 \parencite{sandler2018mobilenetv2}, ResNet 18, 50, and 101 \parencite{he2016deep}, ResNeXt 50 and 101 \parencite{xie2017aggregated}, VGG-19 \parencite{simonyan2014very}, InceptionV3 \parencite{szegedy2016rethinking}, and EfficientNet B3 \parencite{tan2019efficientnet}
    \item[Detection] using MS-COCO \parencite{lin2014microsoft} with Fast R-CNN \parencite{girshick2015fast}, Faster R-CNN \parencite{ren2016faster}, and RetinaNet \parencite{lin2017focal}
    \item[Instance Segmentation] again using MS-COCO with Mask-RCNN \parencite{he2017mask}
    \item[Semantic Segmentation] using ADE-20k \parencite{zhou2017scene, zhou2016semantic} with encoding models MobileNetV2 \parencite{sandler2018mobilenetv2}, ResNet 18, 50, 101 \parencite{he2016deep}, and HRNet \parencite{sun2019high} and decoders C1 \parencite{zhao2017pyramid}, PSPNet \parencite{zhou2017scene}, and UPerNet \parencite{xiao2018unified}.
\end{description}

In \reffig{ttac-nomitigation} we see the result of these models for varying JPEG quality. All of the models face a steep penalty for the lowest quality settings which gradually abates as quality increases. This finding is intuitive and confirms our need for JPEG mitigation techniques. We follow this up with summary plots of the mitigation study in \reffig{ttac-withmitigation}. We can see some interesting behavior here, mainly that the different mitigations do not behave the same on different tasks. In particular, tasks that are very localization-heavy like segmentation do not benefit from supervised fine-tuning as much. These tasks are greatly aided by task-targeted artifact correction, however.

\begin{figure}[t]
    \centering
    \includegraphics{figures/mrcnn_tide.pdf}
    \caption[MaskRCNN TIDE Plots]{\textbf{MaskRCNN TIDE Plots.} From left to right the model was evaluated at quality 10, 50, 100 with no mitigations. Note that the bulk of the errors are missed detections at low quality. As quality increases, more objects are detected but they are not localized correctly.}
    \labfig{ttac:tide}
\end{figure}

\section{Transferability and Multiple Task Heads}
\labsec{ttac-trans}

One of the most intriguing properties of task-targeted artifact correction is the potential for transferability. Since the original task network is not changed in any way, and only the artifact correction network is fine-tuned, there is no reason that we cannot use the outputs of the artifact correction network for other tasks entirely. This opens up a range of new potential deployment scenarios. For example, a TTAC model could be targeted to MobileNetV2 \parencite{sandler2018mobilenetv2}, which is fast and lightweight to train, and then used for a much heavier semantic segmentation network which would have been impossible to train without significant compute power.

Of course this only works if the TTAC models can generalize. We examine this in \reffig{ttac-transfer}. For each plot, we take the supervised fine-tuning and task-targeted artifact correction results from \reffig{ttac-withmitigation}. These are shown in dashed lines. We compare this with a target-targeted artifact correction network which was trained with MobileNetV2 (green) and ResNet-18. As the plots show, the transfer works quite well even to different tasks. In the right hand plot, for example, the new results are almost indistinguishable from the task-targeted network which was fine tuned for segmentation and performs better than fine-tuning the segmentation network itself.

It is also worth noting that there is no reason a TTAC model needs to be trained with only one downstream task target. We can use as many downstream tasks as we have compute power for. We examine this in \reffig{ttac-mh}. In these plots we have added results from a TTAC model with 2 heads (classification and detection) and a TTAC model with 3 heads (classification, detection, and segmentation). Not only does this work perfectly well, but in many cases the additional model heads improved the generalizability of the TTAC model, leading to improved results.

\begin{figure*}[t]
    \centering
    \includegraphics{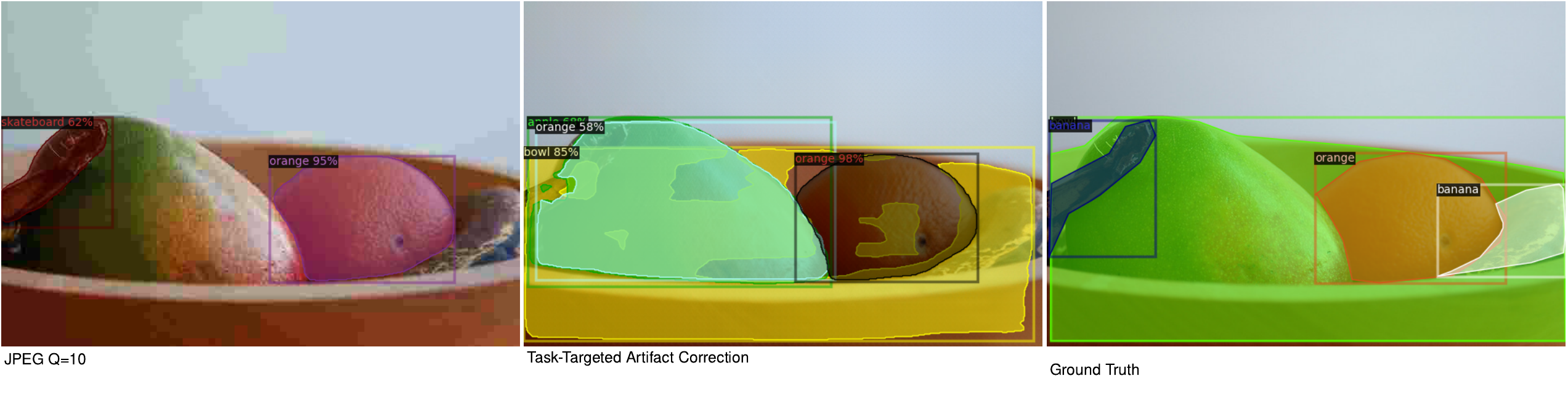}
    \caption[Mask R-CNN Qualitative Result]{\textbf{Mask R-CNN Qualitative Result.} Input was compressed at quality 10. This compares the JPEG result to TTAC and the ground truth.}
    \labfig{ttac:qual}
\end{figure*}

\section{Understanding Model Errors}

So far we have looked at model error in a very aggregate view. In other words, we are looking at overall accuracy and how it changes with increasing compression. In this section, using Mask R-CNN \sidecite{he2017mask} as a representative example, we examine the errors made by the model in more detail.

We start by using TIDE \sidecite{tide-eccv2020} to compute a breakdown of the exact errors that the network is making with increasing compression. These plots are shown in \reffig{ttac:tide} with no mitigations applied. The trend here is interesting. For low quality, the bulk of the errors are caused by missed detections. However, as the quality increases and the missed detections decrease, the localization error actually increases as the newly detected objects are not properly localized. This is sensible because it suggests that once enough information is present in the image for objects to be identifiable, the ``spread'' induced by the missing high frequency basis functions causes the exact boundaries of the objects to be obscured.

We can view this qualitatively as well. \reffig{ttac:qual} shows the result for a JPEG compressed at quality 10 both with and without TTAC (as well as the ground truth). In the uncorrected model, we can observe a significant number of missed detections as well as minor localization errors on the orange, although overall the orange is localized quite well given the significant blocking artifacts present on the boundaries. The TTAC output is also informative. Not only does the image appear significantly high quality (keep in mind it was the same as the left image before artifact correction) but there are also far fewer missed detections. What remains are some localization errors particularly on the bowl.

\begin{figure}[t]
    \centering
    \includegraphics{figures/throughput.pdf}
    \caption[Model Throughput]{\textbf{Model Throughput.} Throughput comparing TTAC FPS for training and inference. TTAC incurs a non-negligible throughput impact. }
\end{figure}
\section{Limitations and Future Directions}

There are two major limiting factors to TTAC in its current incarnation: speed and fidelity. Since TTAC requires placing an artifact correction network, specifically a QGAC network, before any task processing happens, it can severely limit performance. As long as this happens at the datacenter level, the impact is likely minimal but it is still a legitimate concern as GPU resources are still highly valuable and are currently required for artifact correction networks. This could be addressed by more efficient formulations for artifact correction or bespoke TTAC architectures that are intended to be lightweight.

Next, although TTAC has some marked advantages over data augmentation techniques, it currently struggles to outperform them in all cases. We expect that this can be addressed with deeper supervision on the task networks (\ie, matching more than just the final logits) however this is currently an open problem. There may be deep changes required to the scheduling for generation of suitable training data that would be required to see a clear numerical advantage.

Finally, the scope of TTAC is still somewhat restricted. Although JPEG artifacts are arguably the most important and prevalent type of degradation applied to images, we believe that TTAC is an invaluable tool for general degradations. This could mean corruptions like noise, masking, rotations, \etc or something more mundane like resampling.

\pagelayout{wide}
\addpart{Video Compression}
\pagelayout{margin}

\setchapterpreamble[u]{\margintoc}
\chapter{Modeling Time Redundancy: MPEG}
\labch{mpeg}

\lettrine{H}{aving} discussed image compression at length, we now move on to video compression. When considering uncompressed images, we modeled them as samples of a continuous 2D signal. We now allow those samples to vary over time to create a sort of ``flip-book''. Light intensity is captured in discrete steps in space to create ``frames''\index{frame}, and then, multiple frames are captured in discrete steps in time to create the video. By sampling frames at a sufficiently high frame rate, there is an illusion of smooth motion.

Naturally this significantly increases the size of the representation. Since each frame in the video is the size of a single image, videos increase in size quickly with increasing framerate and time. Since we classified images as big enough to warrant compression, videos also certainly need to be compressed. In fact, timely transmission of even short videos would be impossible without compression.

In this chapter we cover, at a high level, the first principles of video compression. There are many different video ``codecs''\index{codec}, or, different algorithms for compressing videos. Although most of the concepts we discuss here are applicable to all modern codecs in some form, when we need specific details, we will defer to MPEG and specifically, the AVC standard \sidecite{marpe2006h}. Readers may be familiar with AVC by other names like H.264 or MPEG4 part 10. We standardize on the AVC terminology to easily differentiate between HEVC/H.265 \sidecite{sullivan2012overview} and to align better with the naming of AOM codecs (like VP9, AV1, \etc). We focus on AVC because it is widely used \sidecite{vdr} and many of its key ideas are used as the foundation for continuing codec development.

As we will see, the important insight which makes video compression possible is that we can exploit time redundancy in the signal and remove information across time. This is in addition to the spatial manipulations we are to used in \gls{JPEG}, and the effect is synergistic. In other words, by exploiting temporal redundancy, we can remove additional spatial information which we would have needed to store if we only had a single image.

The dependence on the temporal dimension will create the need for three frames types:
\begin{description}
    \item[Intra Frames] or I-frames\index{intra frame} which are frames that can be decoded without information from any other frame, \ie, there is no temporal dependency.
    \item[Predicted Frames] or P-frames\index{predicted frame} are frames that requires at least one previous frame to decode We are said to predict the current frame based on the previous frame and any hints stored with the current frame.
    \item[Bipredicted Frames] or B-frames\index{bipredcted frame} are frames that requires at least one previous and future frame to decode. These frames are beyond the scope of this dissertation.
\end{description}
These frames together referred to as a Group of Pictures\index{group of pictures}, \ie, an I-frame and its associated P- or B- frames is a group of pictures.

\begin{figure*}[t]
    \centering
    \includegraphics[width=1.5\textwidth]{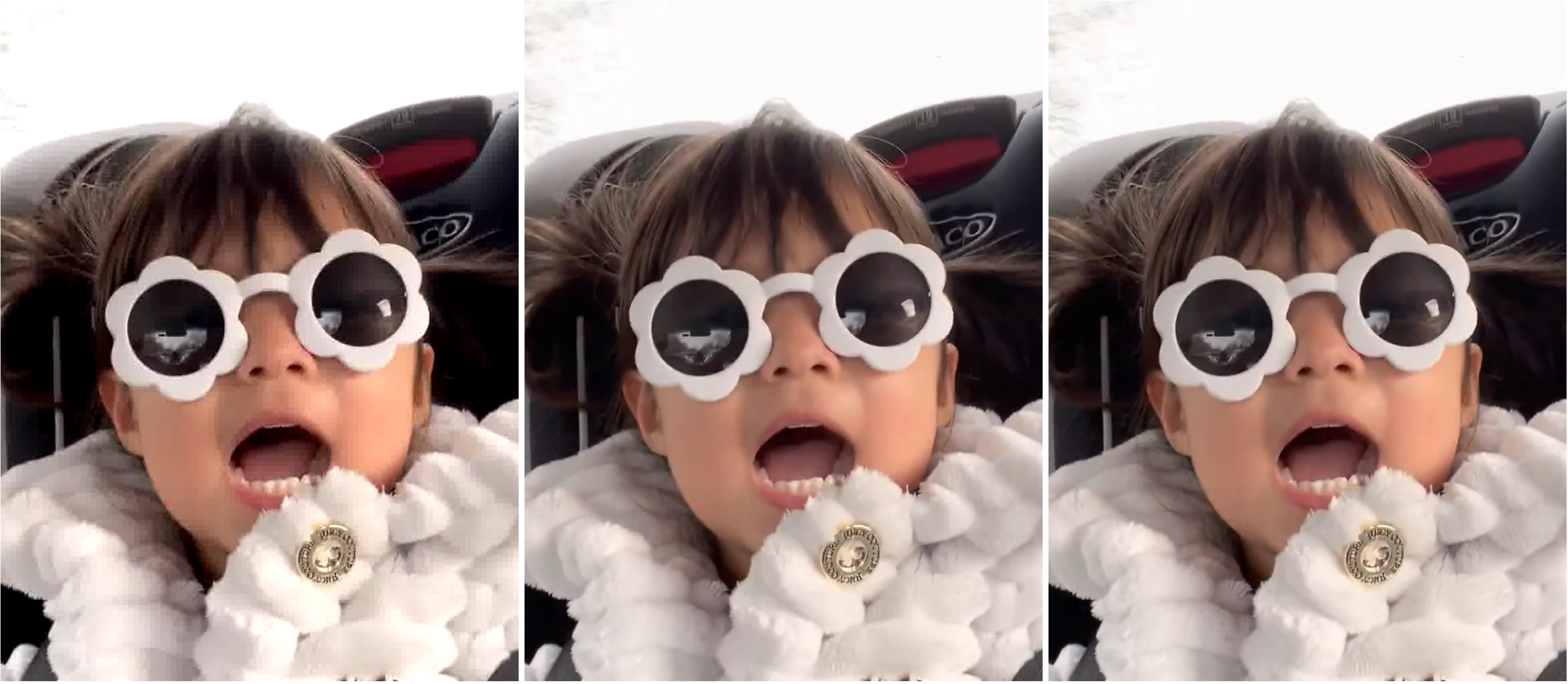}
    \caption[Motion JPEG Comparison]{\textbf{Motion JPEG Comparison.} Left: Motion JPEG frame, Center: AVC frame, Right: Original frame. The motion JPEG video was larger and poorer quality than the AVC frame, motivating compression in the time dimension as well as the spatial dimension.}
    \labfig{mpeg:mjpeg-compare}
\end{figure*}

\section{Motion JPEG}

Before we begin the discussion of ``true'' video codecs, it is worth discussing an obvious solution: \gls{mjpeg}. Motion JPEG can be thought of as a successor to MPEG \sidenote{Although historically MPEG-1 was technically standardized first, core MPEG-1 technology was based on work from the JPEG (committee).}. The idea is incredibly simple: each frame is compressed separately as a JPEG and stored in a file along with some kind of frame rate specification. This information is all that is needed to decode and play the video. Note that although Motion JPEG enjoys widespread use because of its simplicity, there is actually no standard which defines it and different software libraries will have different methods of specifying metadata.

As a quick example of this in action, we can take a raw  240frame, 24fps, 1080p video and try compressing with motion JPEG and with ffmpeg defaults for AVC. The original video in this case is 746,496,000 bytes ($1920 \times 1080 = 2,073,600$ bytes for luminance plane, 4:2:0 subsampling gives $2,073,600 / 4 + 2,073,600 / 4 = 1,036,800$ bytes for the chrominance planes so $3,110,400$ bytes per frame times 240 frames $ = 746,496,000$) or about 747MB. Pretty large for 10 seconds of 1080p video.

The Motion JPEG file generated from this video is 12.7MB, an impressive 62x compression ratio. This can be though of as a naive ``limit'' on how much compression is attainable with out considering the temporal dimension. The AVC file on the other hand is only 7.2MB, a 103x compression ratio. This is not the end of the story, however, as the AVC file is almost indistinguishable from the original frame, yet the Motion JPEG frames have significant blocking artifacts from compression (\reffig{mpeg:mjpeg-compare}). This example motivates our desire to study compression in the temporal dimension: the AVC frames are both high quality and smaller (in file size) than the Motion JPEG frames.

\begin{figure}[t]
    \centering
    \includegraphics{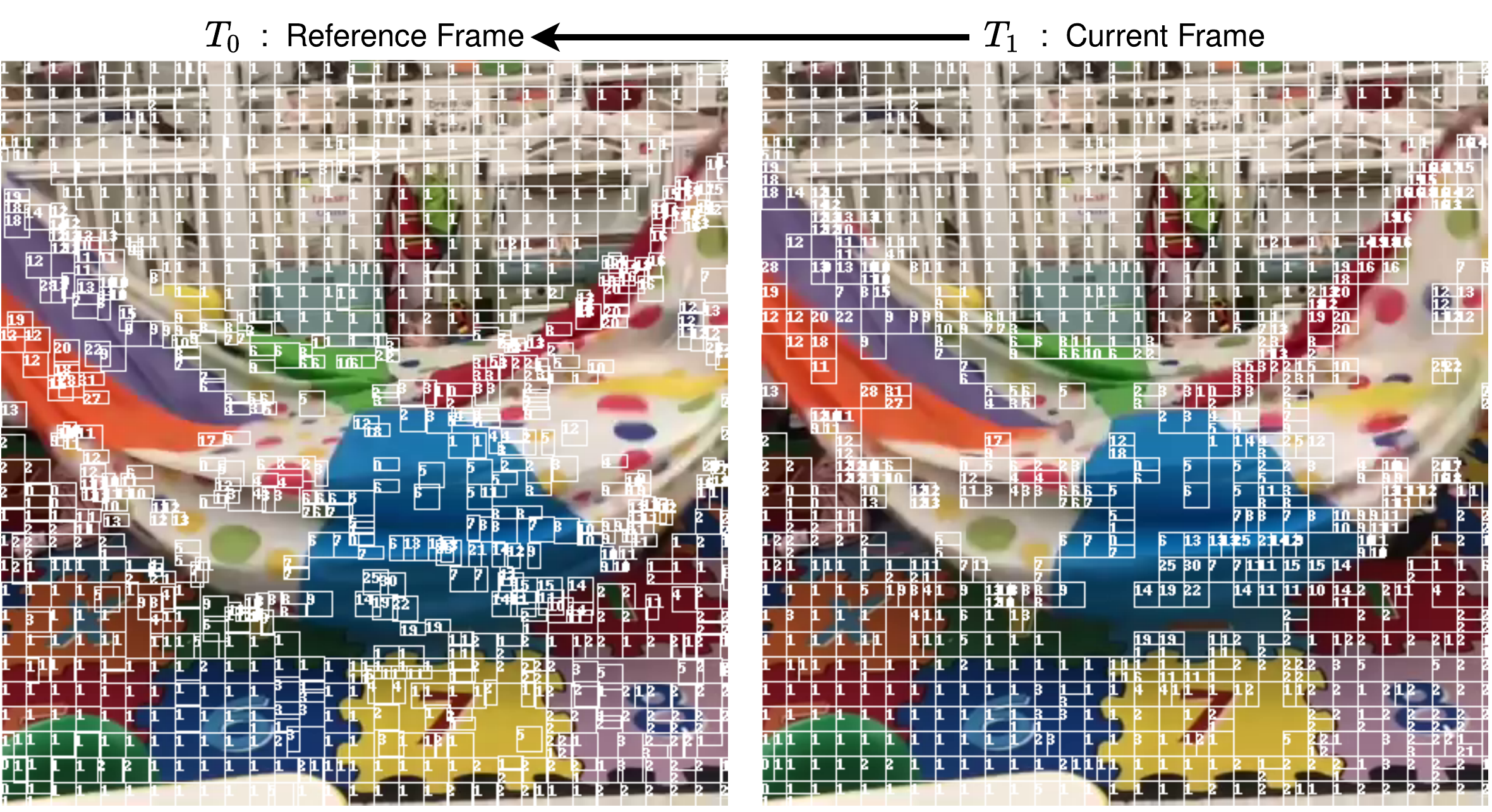}
    \caption[Motion Vector Grid]{\textbf{Motion Vector Grid.} The grid is defined based on the frame that is currently being decoded. Each motion vector indicates where in the previous frame a block of pixel moved from. The numbers on the grid cells indicate the motion strength.}
    \labfig{mpeg:mvgrid}
\end{figure}

\section{Motion Vectors and Error Residuals}

We will ``measure'' temporal information by modeling motion between neighboring frames. After modeling the motion, we can warp and subtract the frames giving a ``residual''. The motion modeling is designed to be compact and simple while still capturing some complex motions. Since at a high enough framerate, inter-frame motion is small\sidenote{Barring large cuts or scene changes}, the frames should share a significant amount of information after accounting for motion. Any additional information is stored in the residual, which is generally low entropy.

\begin{marginfigure}
    \centering
    \includegraphics{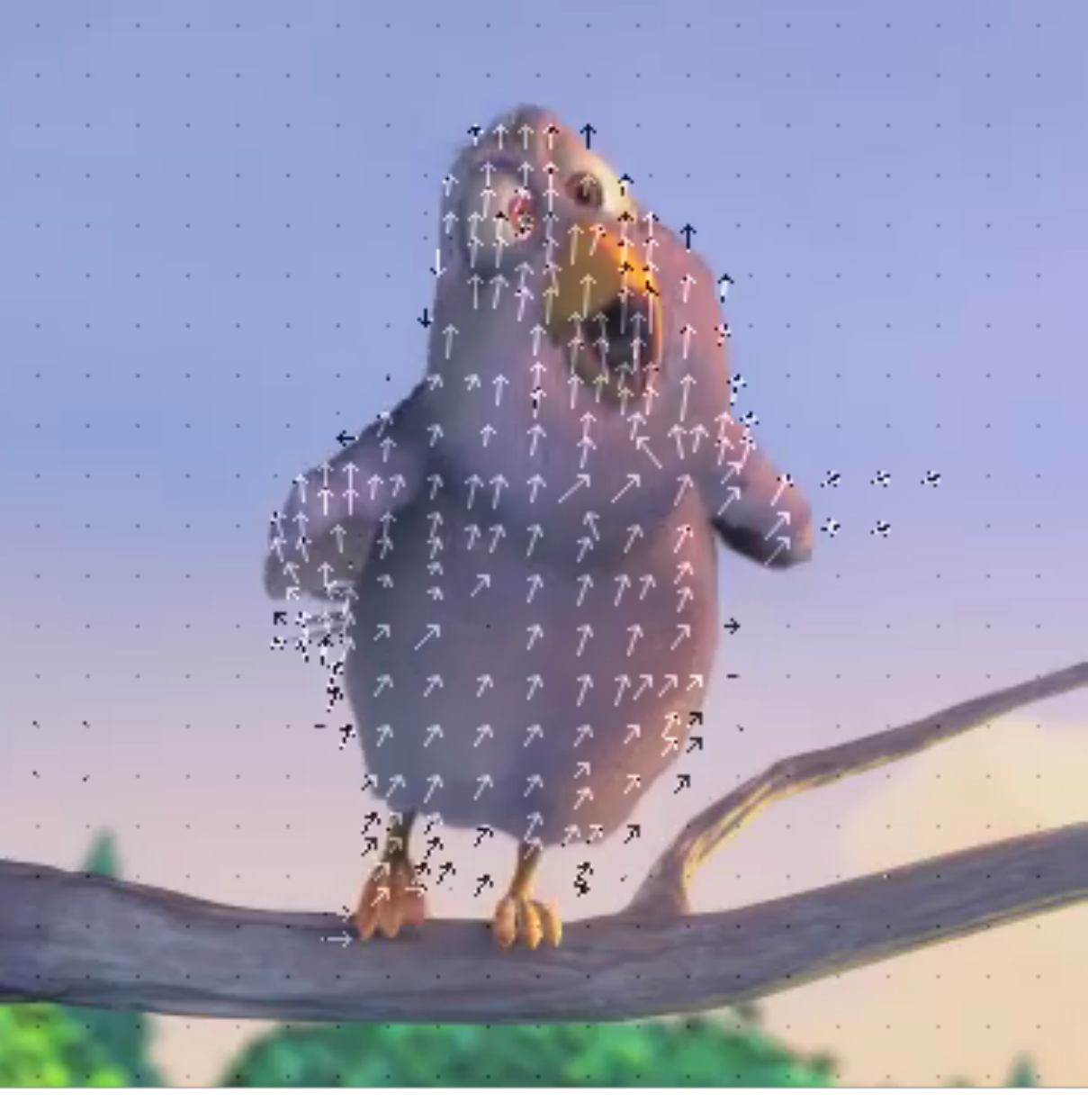}
    \caption[Motion Vector Arrows]{\textbf{Motion Vector Arrows.} Each vector is positioned in the center of the block, the arrow points to the position that block came from in the reference frame. \textit{Image credit: Big Buck Bunny \parencite{bbb}.}}
    \labfig{mpeg:mvarrow}
\end{marginfigure}

We call frames which are constructed from motion and residual information Predicted Frames (P-frames)\index{P-frame} since they must be predicted from a previous frame. We will call the frame which is being decoded the current frame and the previously decoded frame the reference frame. MPEG standards also define a Bipredicted Frame (B-Frame)\index{B-Frame} which is predicted from a previous and future frame. This takes advantage of the presentation timestamp and decoding timestamp features of video containers to store frames out-of-order and with a complex dependency graph. We will consider further discussion of B-frame and multi-reference decoding to be beyond the scope of this dissertation.

Motion information is stored in the form of \gls{motion vectors}\index{motion vector}. These vectors are computed by breaking the current frame into a regular grid and, for each grid cell, measuring where in the reference frame that grid cell moved from. For some cells there may be no motion and for others there may be large motions. For AVC, our example codec, the cells themselves can be $8 \times 8$, $16 \times 8$, $8 \times 16$, or $16 \times 16$ pixels. The blocks are stored as nine 16-bit integers\sidenote{reference frame, width, height, reference x, reference y, current x, current y, motion strength, and flags}. So this operation alone turns, at worst, 192 byte pixel blocks into 144 byte motions.

In \reffig{mpeg:mvgrid}, we show an example of the grid structure from a real video. Note that the grid is defined from the current frame which is broken into a clean regular grid. On the reference frame, these blocks may overlap. There are some grid cells missing from the current frame, for these blocks no motion was detected, so they are skipped. We can also visualize the vectors themselves as in \reffig{mpeg:mvarrow}. For each block in the image, we draw a vector starting from the center of that block and terminating at the position in the reference frame that the block came from.

The motion vectors are used in a process called \gls{motion compensation}\index{motion compensation}. This process simply copies blocks of pixels from their position in the reference frame to their position in the current frame, pasting over any content that was there previously. The resulting motion compensated frame represents a coarse warping of the reference frame to match the current frame. Of course there are still errors since the motion is only computed on blocks, so it is not usually a perfect representation of the current frame. See the left side of \reffig{mpeg:residual} for an example of this. The vectors are computed using a process called \gls{motion estimation}\index{motion estimation}. We do not cover this process here as it is not standardized\sidenote{The MPEG standards only define the decoder. It is up to the encoder to produce a standards-compliant bitstream however it wants.}.

In order to correct errors in the motion compensated frame, the encoder stores an \gls{error residual}\index{error residual}. One thing that is immediately noticeable in \reffig{mpeg:mvgrid} and \reffig{mpeg:mvarrow} is that not all blocks move. Indeed in both there are many blocks which are stationary. This means that the pixels in those blocks are exactly the same in the reference and current frames and therefore if we subtract the two frames those blocks will be filled with zeros. Motion compensation takes this a step further to try to match moving objects as closely as possible as well as stationary regions; this increases the likelihood of generating zero blocks. These zero blocks are extremely low entropy and aid in the compression process.

To compute this residual, we first compute the motion compensated current frame from the reference frame then subtract the true frame, yielding everything that was not accurately modeled by the motion estimation process. An example is shown on the right side of \reffig{mpeg:residual}. In effect we have told the decoder to reuse information it already had about those blocks without needing to store them. Small errors are then accumulated on edges and in rapidly moving objects which make up the bulk of the size of the compressed residual. Note also that the above discussion is appearance based. An object moving in the physical world may very well generate blocks in the frames that do not appear to move, like in the center regions of the parachute in \reffig{mpeg:mvgrid}. This information can still be freely used to fill in the current frame even though it is not an accurate reflection of the real-life parachute.

\begin{figure}[t]
    \centering
    \includegraphics{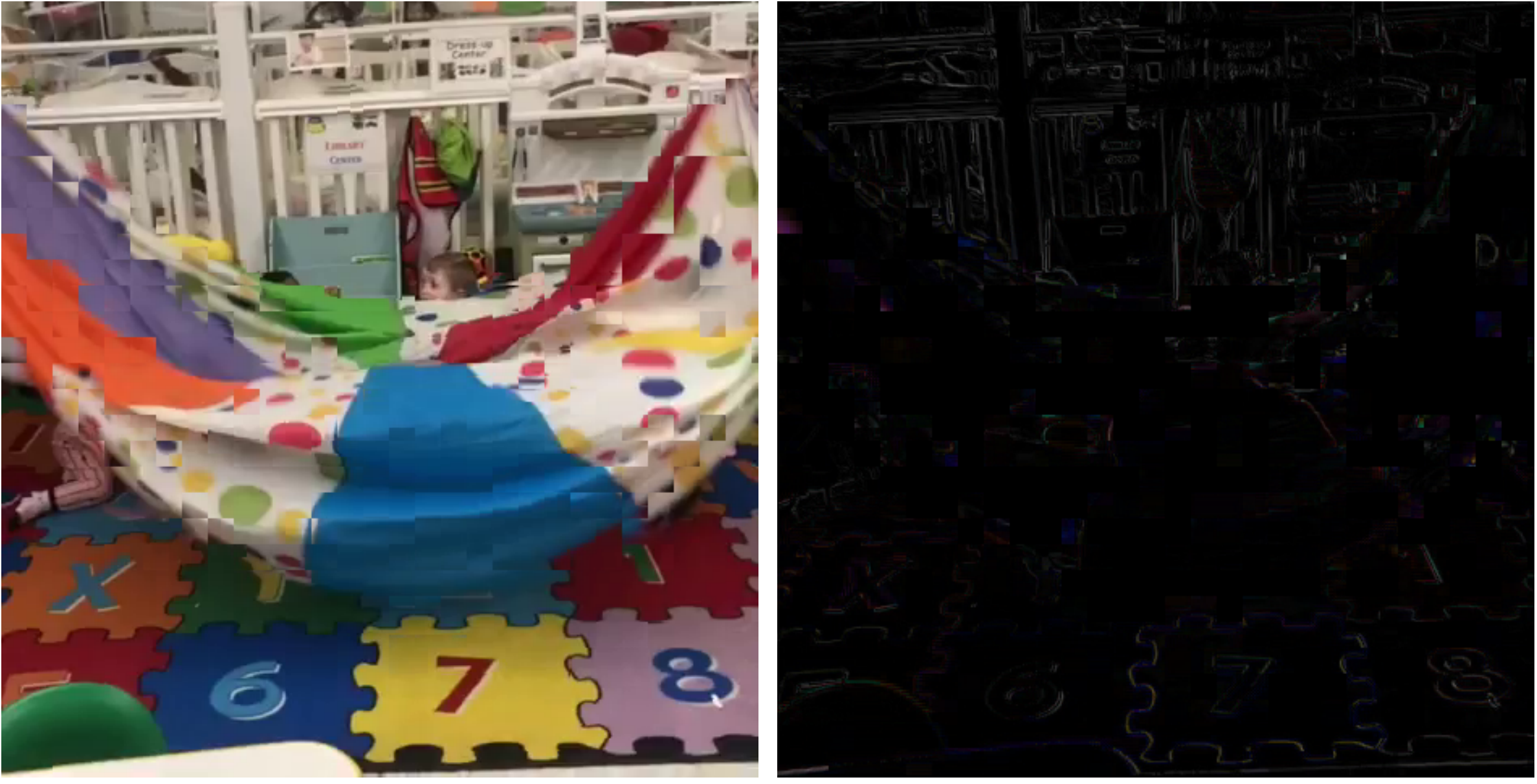}
    \caption[Motion Compensation and Error Residuals]{\textbf{Motion Compensation and Error Residuals.} Left: motion compensated frame, Right: error residual. Note the block artifacts in the motion compensated frame. corrections to these are stored in the error residual along with small edges. Note that the error residual is mostly zeros and therefore much easier to compress. To produce the residual, we subtracted the motion compensated frame (left) from the true current frame (\reffig{mpeg:mvgrid} right).}
    \labfig{mpeg:residual}
\end{figure}

To summarize: the encoder stores per-block motion. This motion is then combined with a low entropy residual and a previous frame to produce the current frame. This yields direct savings in that storing motion information is more compact than storing pixels, and indirect savings because the error residual is much more compressible than the original pixels.

\begin{figure*}[t]
    \centering
    \includegraphics[width=1.5\textwidth]{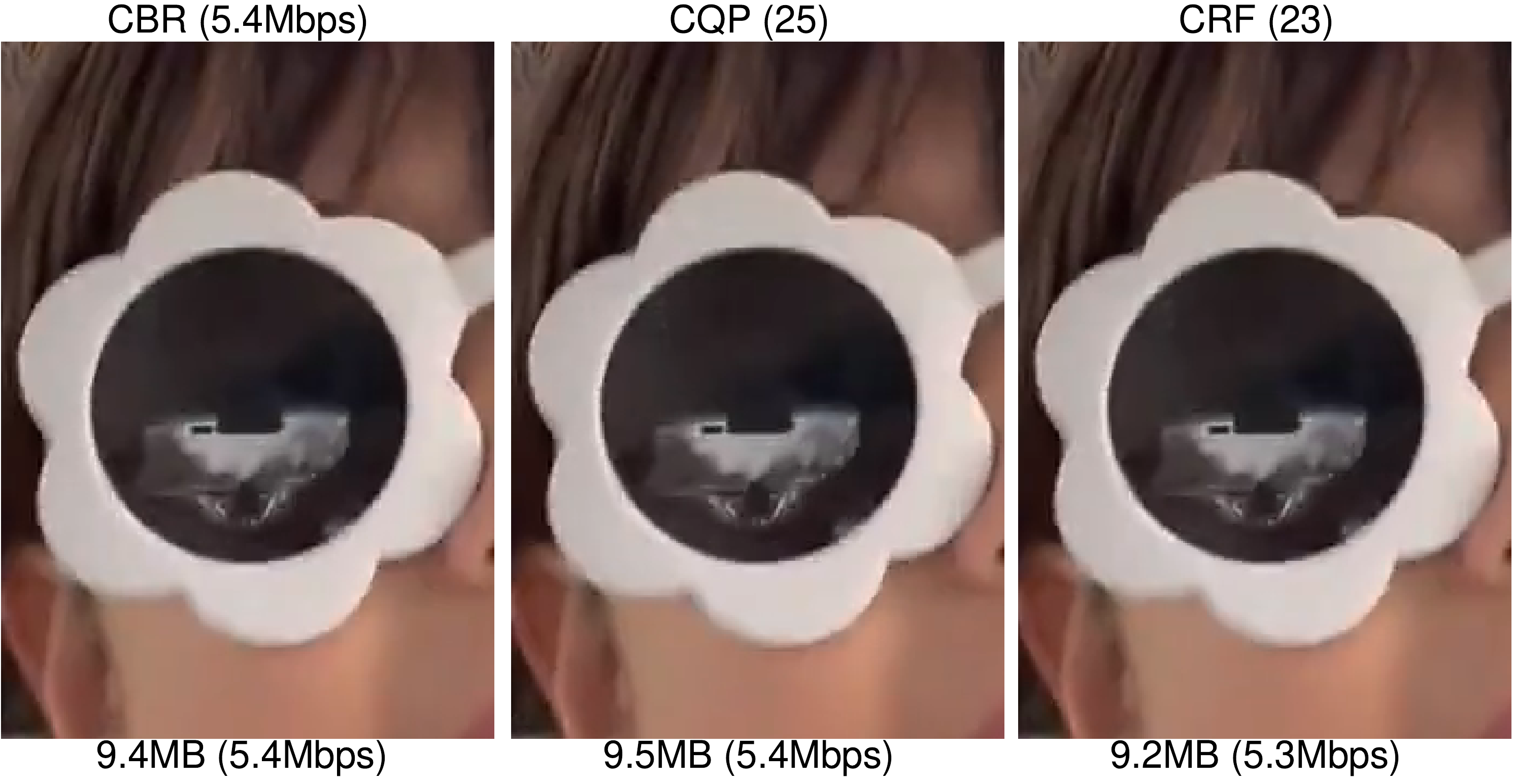}
    \caption[Rate Control Comparison]{\textbf{Rate Control Comparison.} The three rate control methods are tested targeting the same file size. Note the different artifacts produced by each method despite similar file sizes. Also note that for this video, CQP 26 undershoots the target to produces 6.6MB file but CQP 25 overshoots at 7.5MB. CBR and CRF are very close in file size.}
    \labfig{mpeg:rate-control}
\end{figure*}

\section{Slices and Quantization}
\labsec{mpeg:sq}

In addition to motion modeling, the AVC standard makes some notable changes to the way frames are structured. Also, similar to JPEG, AVC allows the use of quantization for rate control, although this is exposed to the user in several different ways. Unlike the last section, these ideas are applicable to both P-frames and I-frames.

\begin{marginfigure}
    \centering
    \includegraphics{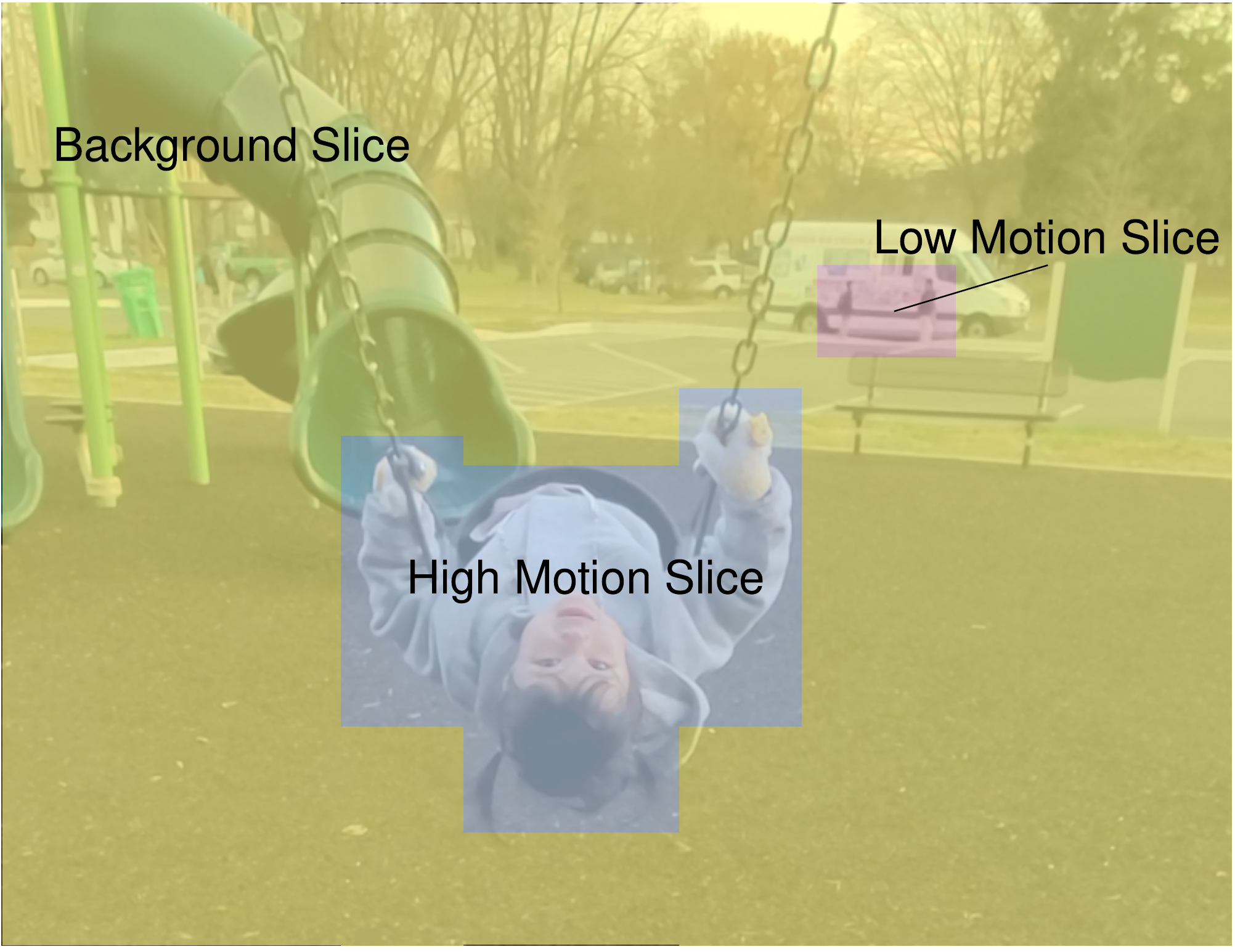}
    \caption[Slicing Example]{\textbf{Slicing Example.} The image has been broken up into three slices by the encoder: a background region, a high motion region, and a low motion region.}
    \labfig{mpeg:slicing}
\end{marginfigure}

The biggest departure from JPEG is the concept of \gls{slices}\index{slice}. A slice is a region of the frame made up of a whole number of \gls{macroblocks}\index{macroblocks} (usually $16 \times 16$ pixel blocks). Prediction (motion compensation) is only possible within a single slice. In the simplest case, the entire frame is one slice, but the general idea allows the encoder to break up the image into more meaningful groups. For example, the encoder might have two slices: a high detail slice with minimal quantization and a low detail slice with higher quantization. It may have high and low motion slices. Even more intriguing is the concept of I- and P-slices\index{intra slice}\index{predicted slice}. This means that a single frame can contain intra information and predicted information rather than the encoder making a blanket decision for the entire frame which may be sub-optimal. For example, highly detailed regions with low motion may be better stored as an I-slice but a high motion region may be more efficiently stored as a P-slice. An example of this is shown in \reffig{mpeg:slicing}.

In JPEG \gls{rate control}\index{rate control} was implemented by choosing a scalar quality in $[0 , 100]$ and mapping that scalar to a quantization matrix. For videos we have several options; these are compared visually in \reffig{mpeg:rate-control}. The most similar is Constant Quantization Parameter (CQP)\index{constant quantization parameter}. This is essentially the same idea but on a scale of $[0, 51]$ with 51 being the worst quality, this number is used to derive a quantization matrix for the coefficient blocks. This type of \gls{rate control} is not generally used because it is extremely simplistic to apply the same amount of quantization to every frame and generally produces sub-optimal results in both size and perceptual quality.

Instead the more common Constant Rate Factor (CRF)\index{constant rate factor} is used. This is also a number in $[0, 51]$ with 0 being truly lossless encoding (no quantization) and 51 being the worst quality. This method is tuned to hold perceptual quality constant and takes into account inter-frame motion and frame rate. Still objects will be less aggressively quantized and moving objects more so following a similar argument for removing high spatial frequencies: fast moving object are harder to perceive with detail.

The final \gls{rate control} method, which is also quite common, is Constant Bitrate (CBR)\index{constant bitrate}. In CBR encoding, the only thing the encoder is trying to optimize is the bitrate of the video, it should be as close as possible to a specified target without going over it. This is useful for maxing out a connection with a known bandwidth where it is desirable to get the maximum quality that the connection can support without dropping frames. However, there is no way for the encoder to know \textit{a priori} how to perform CBR encoding so it will regularly over- or undershoot the target unless two pass encoding is used.

\section{Recap}

We have now covered the high level ideas by which video codecs compress temporal data. The encoder computes and stores coarse motion between frames and use that motion to compute an error residual containing anything in the frames which cannot be modeled with motion from a previous frame. Internal to each frame, the encoder is free to slice the frame and make per-slice decisions. \Gls{rate control} is accomplished with the help of a target bitrate, a user defined CRF, or a user defined quantization parameter.

The spatial domain compression is similar to JPEG. Pixels are transformed into the frequency domain using a $4 \times 4$ \gls{dct}, \gls{dst}, or \gls{hadamard transform} (the encoder is free to choose this) and, depending on the rate control mechanism, a QP is computed or is given to the encoder. These QPs map directly to a standard set of quantization matrices for the blocks. Note that the QP is allowed to vary spatially so different blocks can receive different amounts of compression, as opposed to JPEG where one quantization matrix is used for every block in the image.

This yields a surprisingly straightforward compression algorithm. Given a set of frames, partition the frames into GOPs. This is usually a fixed number of frames per GOP but it can be based on the content. Encode the first frame of each GOP as an intra frame by quantizing the transform coefficients of its pixels. For each subsequent frame in the GOP, compute motion vectors from the previous frame, compute the motion compensated frame, and take the difference between the current frame and the motion compensated frame. Encode the predicted frame by storing the motion vectors and quantized coefficients of the error residual. Entropy code the frames. The decoder simply needs to loop through each frame and either decode it directly if it is an I-frame or warp the currently displayed frame using the motion vectors and add the error residual if it is a P-frame.

Unsurprisingly, an actual video codec is much more complex than this and is full of small details which make a large difference to the overall coding efficiency. As of this writing, the current revision of the AVC standard \sidecite{ITUTH264} was released on August 2021 and is 844 pages long. Aside from the core decompression algorithm it includes instructions for storing the resulting data in a stream\sidenote{This is not the same as an MP4 file, for example, but the MP4 file will contain AVC video streams.}, definitions of constants and other hard coded mappings, algorithms for scalable streams, algorithms for compressing multi-view/depth/3D videos, \etc. Although covering such details is beyond the scope of this dissertation, the high level intuition we developed in this chapter will be enough to guide us in the following chapters as we explore ways to improve video coding efficiency using deep learning.

\setchapterpreamble[u]{\margintoc}
\chapter{{\color{Plum}Metabit: Leveraging Bitstream Metadata}}
\labch{metabit}

\lettrine{U}{ntil} this point we have reviewed the basic principles of video compression including how to achieve compression over time by removing redundant motion information. We have also reviewed several methods for using deep learning to restore quality to compressed video frames. In order for video compression to function, \ie, in order to successfully decompress a compressed bitstream, we need additional information beyond simply transform coefficients. This information, such as QP values, GOP structure, and motion vectors among others, give a very strong prior for how the encoder has compressed the video stream and what information has been removed that should be restored.

We now turn our attention to developing a deep learning method which exploits this data to improve its reconstruction. This contribution of the dissertation is currently under submission for separate publishing and is available as a pre-print \sidecite{ehrlich2022leveraging}. If we closely examine the direction of prior works, there are some whispers of this idea. For example, MFQE \sidecite{Yang_Xu_Wang_Li_2018} contributed the idea of ``peak quality frames'' \index{peak quality frames} which were high quality frames that could be used to restore nearby (in time) low quality frames. STDF \sidecite{Deng_Wang_Pu_Zhuo_2020} does away with expensive motion compensation to rely on deformable convolutions.

However, both of these methods leave something to be desired, specifically by relying on outside computation for what is already stored by the encoder. While the concept of peak quality frames seems somewhat abstract, after all how can we predict the existence of such frames, their existence is grounded in first principles. These are I-frames. The encoder inserts them intentionally to create frames with high information content which improves the decoding fidelity. Recall that MFQE 1.0 scans the entire sequence to determine peak quality frames using an SVM and MFQE 2.0 \sidecite{Xing_Guan_Xu_Yang_Liu_Wang_2021} does the same using a Bi-LSTM \sidecite{schuster1997bidirectional}. These are computationally expensive algorithms which are essentially computing the I- and P-frame structure of the GOP, something which we can readily extract from a bitstream with no additional computation.

The MFQE family of networks also rely on optical flow to align nearby frames. While there are many methods for computing optical flow, they all vary in their speed and accuracy, although perfectly accurate optical flow may not be necessary in the first place \sidecite{Xue_Chen_Wu_Wei_Freeman_2019}. The major contribution of STDF was to move away from explicit motion estimation by using deformable convolutions to learn an implicit motion estimation. This is desirable because it reduces the computational burden of the algorithm: the deformable convolutions model both motion and mapping simultaneously. However, we can do better than both explicit and implicit motion estimation; we can do no motion estimation at all. Of course we still wish to align nearby frames and for this we can extract motion vector from the bitstream. This gives a coarse motion compensation which we show is not only good enough for accurate reconstruction but, taken with our other contributions, outperforms both MFQE and STDF as well as their later follow-up works.

The common theme among the contributions of our method is that we are removing things which were computed explicitly by prior algorithms and replacing them with things that are computed by the encoder and stored in the video. We view these computations as redundant. By reducing these redundant computations we are left with extra compute time per frame that we can re-invest in additional model parameters leading to an improved result. We take the additional step of moving away from the sliding window paradigm, where a block of seven frames produces a single frame output, and instead use a block based approach where all seven frames are produced in a single forward pass. The result of these efficiency improvements is a network which has almost twice the parameters of STDF and yet runs the same or faster than it depending on input resolution. It also outperforms STDF by a wide margin for many compression settings.

\begin{fpbox}
    \begin{itemize}
        \item Architecture captures GOP structure
        \item Explicit I- and P-frame representations based on expected information content
        \item Alignment using motion vectors
        \item High frequency restoration using targeted loss functions
    \end{itemize}
\end{fpbox}

\section{Capturing GOP Structure}
\labsec{mb:gop}

\begin{marginfigure}
    \centering
    \includegraphics{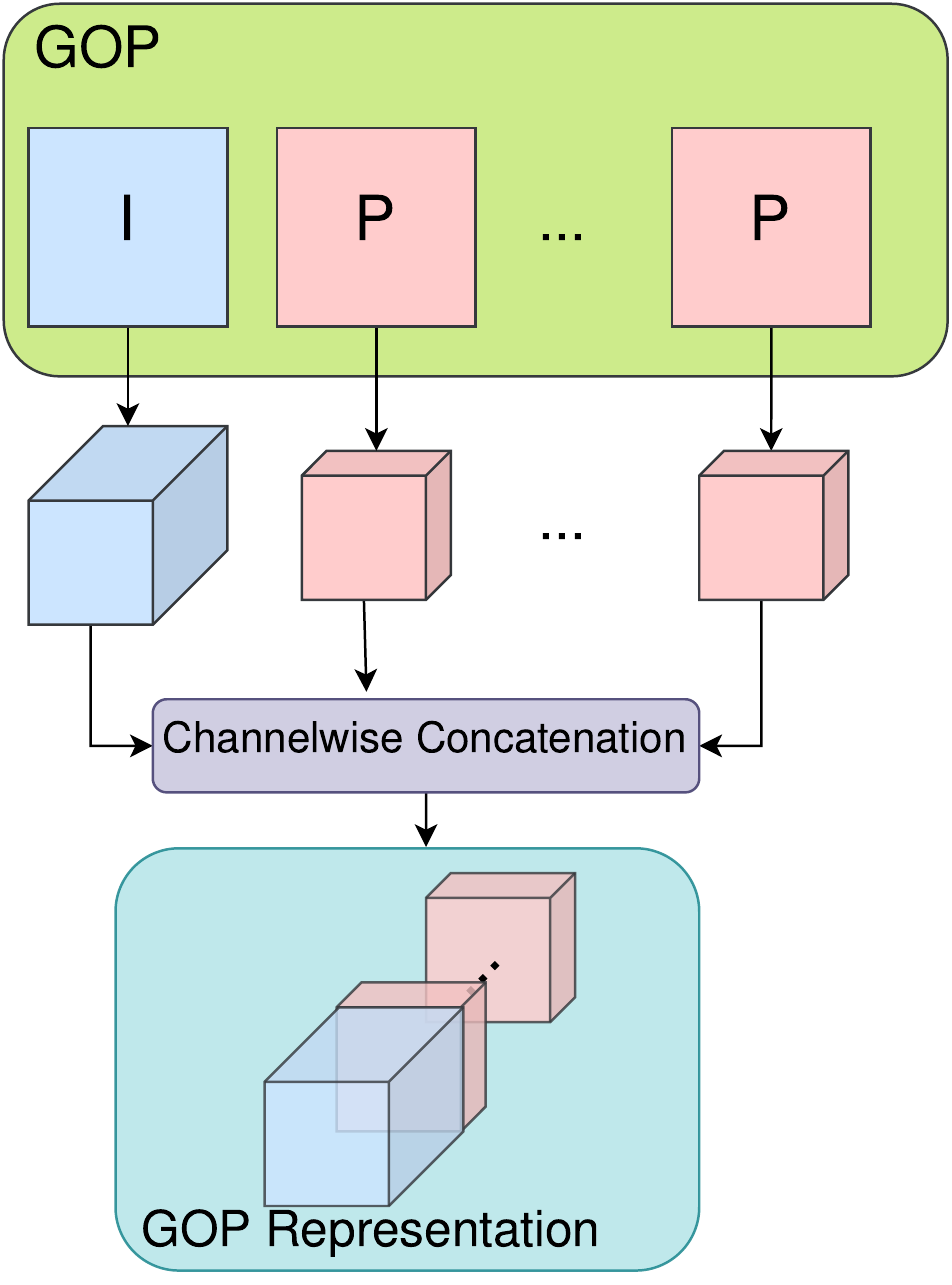}
    \caption[Capturing GOP Structure]{\textbf{Capturing GOP Structure.} The GOP representation is computed from wide I-frame feature extractors and narrow P-frame feature extractors.}
    \labfig{mb-gop}
\end{marginfigure}

One of the primary contributions of this work is the way in which our network takes into account GOP structure. Recall from \nrefch{mpeg} that (in the MPEG standards) frames can be either I-,P-,or B-frames where I-frames are ``intra frames''\index{intra frame} which can be reconstructed using only information in the frame itself, P-frames are ``predicted frames''\index{predicted frame} which require some previous frame to reconstruct, and B-frames are ``bipredicted frames''\index{bipredicted frame} which require a previous and future frame to reconstruct. The goal of using these different frames types is to rely more on information which is stored in other frames that would be redundant to store again. These frames are organized into a group-of-pictures (GOP)\index{group of pictures} which is an I-frame and its associated P-/B- frames. Without loss of generality we only consider P-frames in the following discussion.

Since the predicted frames intentionally do not store information which is stored in previous frames, we can observe that they contain less information and, due to prediction errors, generally have lower perceptual quality than their associated I-frame. When other models process video frames in a sliding window, they do not take this into account in any meaningful way and so the same network which processes I-frames is used to process P-frames.

We can view this as wasting compute resources. Since the bulk of the information is stored in the I-frame, we can process that with a wide representation. We can then use a narrower, and therefore faster to compute, representation for the P-frame to extract the additional information the P-frame contains. This is shown in \reffig{mb-gop}. Note that it is important to match the depth of the extraction networks so that the receptive fields are aligned. We view the resulting GOP representation as capturing the available information in the entire sequence and use it to reconstruct each frame in the GOP after warping.

This is a major gain in efficiency since the faster network is used for most frames in each sequence. Further, we will expend significant resources reconstructing the I-frame, which was already higher quality as it contains the most information in the frame. We will then use this restored I-frame as a base to compute the restored P-frames again using a lighter network. In other words, the GOP structure is encoded into our reconstruction algorithm in all stages.

\begin{figure}[t]
    \centering
    \includegraphics{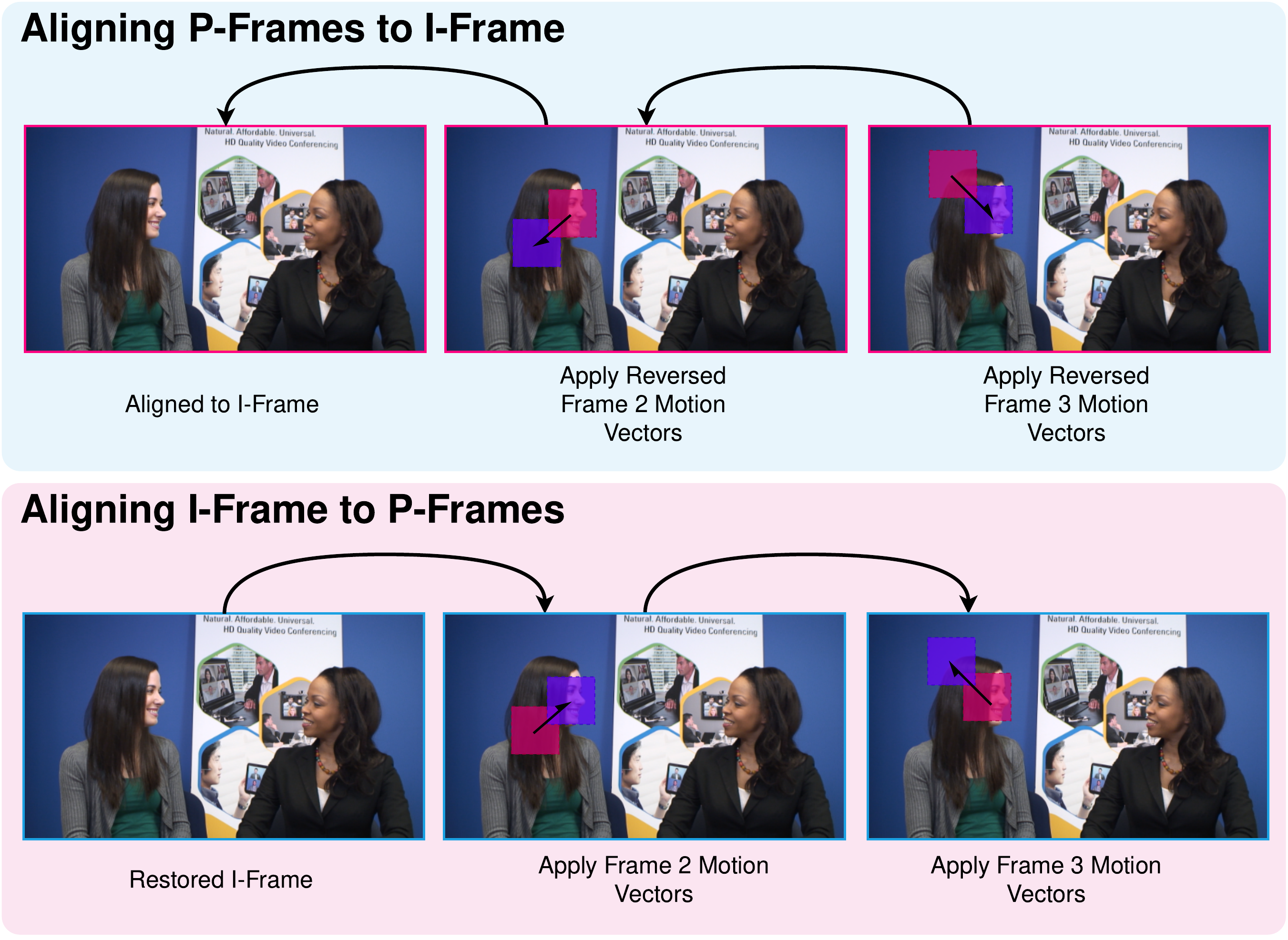}
    \caption[Motion Vector Alignment]{\textbf{Motion Vector Alignment.} P-frames are warped backwards to the I-frame during feature extract. The I-frame is warped forwards to align to the P-frames during frame generation.}
    \labfig{mb:alignment}
\end{figure}

\section{Motion Vector Alignment}
\labsec{mb:mv}

\begin{marginfigure}[-10\baselineskip]
    \centering
    \includegraphics{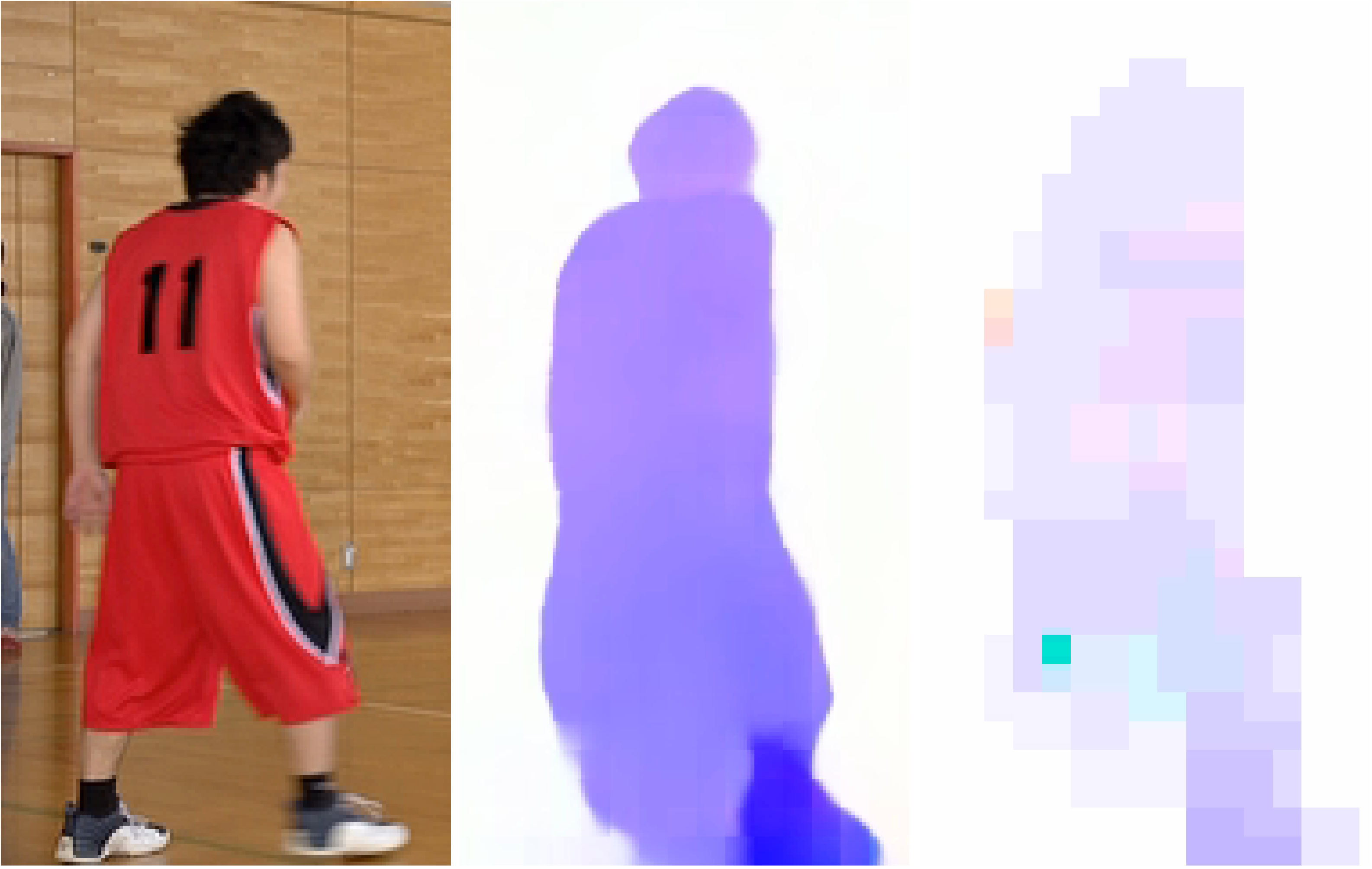}
    \caption[Motion Vectors \vs Optical Flow]{\textbf{Motion Vectors \vs Optical Flow.} The motion vectors resemble a coarse or downsampled version of the optical flow. Optical flow was computed with RAFT \parencite{teed2020raft}}
    \labfig{mb:flow}
\end{marginfigure}

In multiframe restoration problems, it is extremely common to align nearby frames, or features extracted from nearby frames, to ensure that various scene details are overlapping (see ToFlow \sidecite{Xue_Chen_Wu_Wei_Freeman_2019} and EDVR \sidecite{Wang_Chan_Yu_Dong_Loy_2019} among others). Conceptually, this should make the restoration task easier for the network since the additional information of nearby frames is in the correct location, ready to be exploited for additional reconstruction accuracy. The removal video compression defects is no different, and as discussed in the opening to the chapter, this is generally accomplished explicitly with optical flow as in MQFE and related networks \sidecite{Yang_Xu_Wang_Li_2018,Xing_Guan_Xu_Yang_Liu_Wang_2021} with STDF using deformable convolutions for an implicit alignment \sidecite{Deng_Wang_Pu_Zhuo_2020}.

While it is useful to compute high quality alignments between frames, it may not be necessary (this is discussed at length in ToFlow). Assuming that the constraint of high quality alignment can be relaxed, we have a convenient tool at our disposal: motion vectors which are compared with optical flow in \reffig{mb:flow}. The motion vectors relate nearby frames at the block level; for most resolutions the blocks are fine enough and the motion accurate enough that warping frames using motion vectors instead of optical flow works well \sidenote{where ``well'' is measured in terms of reconstruction accuracy.} The major advantage of using motion vectors is that they require no computation to produce since they are stored in the video bitstream. Compared with optical flow, they require no more computation to apply.

We will use the motion vector to align each P-frame to the I-frames during feature extract, and then to align the restored I-frame to each P-frame during frame generation. This is illustrated in \reffig{mb:alignment}. Since motion vector measures motion from the previous frame, we must reverse and warp each frame in sequence, \eg, frame 3 is warped by frame 2's motion vectors; the result is warped by frame 1's motion vectors, \etc. During frame generation we carry out the inverse process by warping the restored I-frames by each of the P-frame's motion vectors in sequence.

\begin{figure*}[t]
    \centering
    \includegraphics{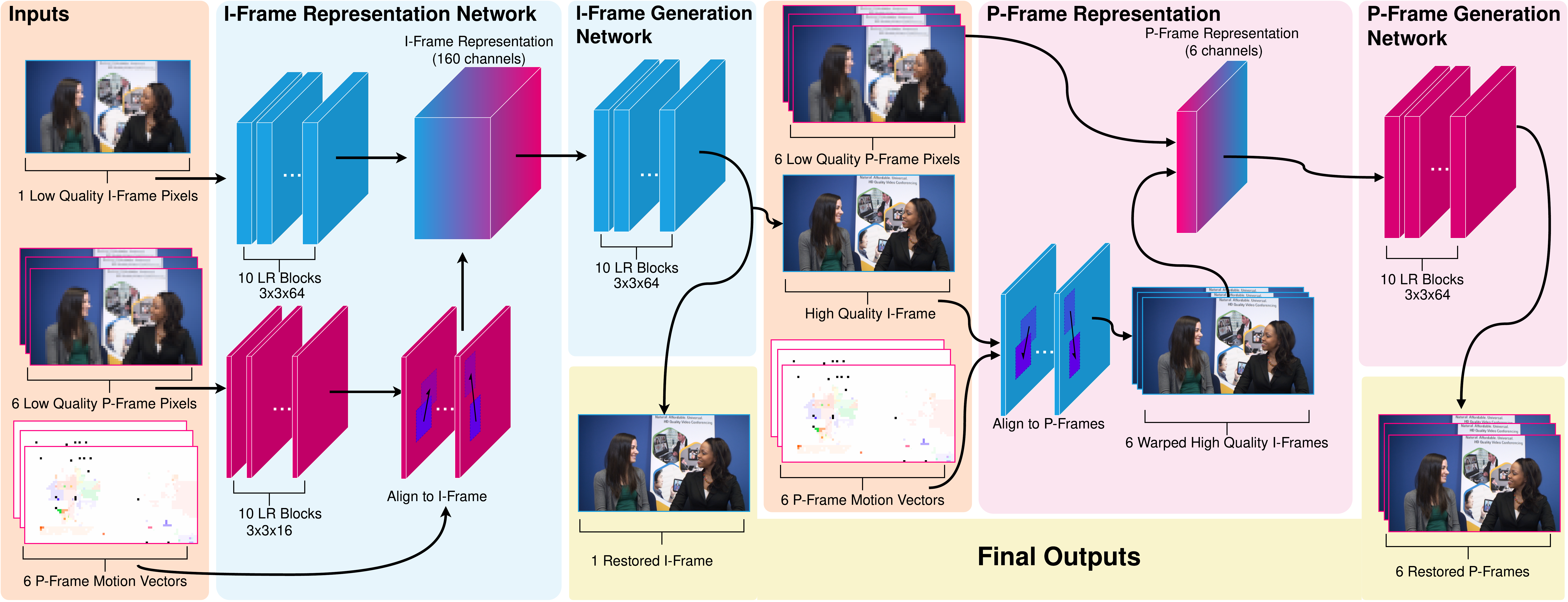}
    \caption[Metabit System Overview]{\textbf{MetaBit System Overview.} I-Frames are shown in {\color{ProcessBlue}\textbf{Blue}} and P-Frames are shown in {\color{RubineRed}\textbf{Pink}}. Our network takes an input ({\color{YellowOrange}\textbf{Orange}}) in the form of a low-quality Group-of-Pictures and first performs multi-frame correction on the {\color{ProcessBlue}\textbf{I-Frame}}. The resulting high-quality {\color{ProcessBlue}\textbf{I-Frame}} is used to guide correction of the low-quality {\color{RubineRed}\textbf{P-Frames}}. The final output of our network ({\color{Goldenrod}\textbf{Yellow}}) is the entire high-quality Group-of-Pictures.}
    \labfig{mb:overview}
\end{figure*}

\section{Network Architecture}

\begin{marginfigure}[-5\baselineskip]
    \centering
    \includegraphics{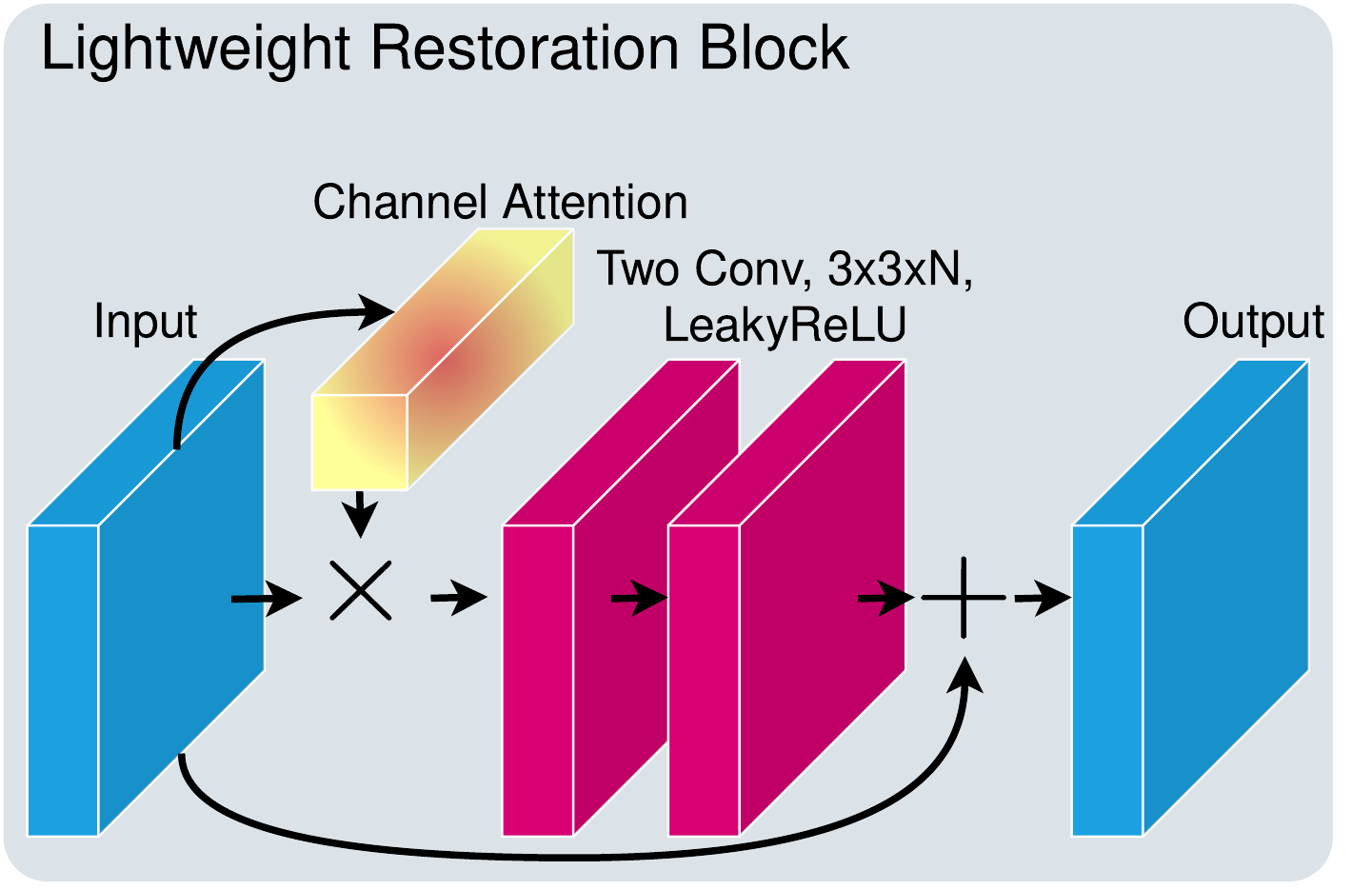}
    \caption[LR Block]{\textbf{LR Block.} The Lightweight Restoration block modifies the residual block to follow recent best practices in deep learning and image-to-image translation.}
    \labfig{mb:lr}
\end{marginfigure}

We are now in a position to develop a complete network architecture using the ideas in the previous two sections. Although we will develop a concrete architecture in this section, the high level idea to leverage specific bitstream metadata can actually be applied to many different architectures. First we need to develop a basic block to build the rest of our network with. We would like to base this on residual blocks (\nrefsec{resnets}) which are known to be effective at many tasks; however, residual blocks by themselves do not follow best practices for image-to-image problems. Conversely, the \gls{rrdb} layer (\sidecite{wang2018esrgan}) works well but is computationally inefficient. For videos, we require something which is lightweight and effective.

We make the following modifications (\reffig{mb:lr}) to the residual block which we call a lightweight restoration (LR) block. First, we remove batch normalization \sidecite{ioffe2015batch} which is known to perform poorly in image-to-image translation scenarios \sidecite{wang2018esrgan}. Next, we replace the ReLU layers with LeakyReLU. Finally, we add channel attention \sidecite{wang2020eca, vaswani2017attention} following recent best practices in deep learning methodologies.  To these residual blocks, we add our accounting for GOP structure and our motion vector alignment blocks. An overview of the Metabit system is shown in \reffig{mb:overview}.

The network is divided into several stages. The network takes a 7-frame GOP with no B-frames as input. In the first stage, the I- and P- frame representations are computed using separate feature extractors. As discussed in \nrefsec{mb:gop}, the I-frame representation is 64 dimensional while the P-frame representation is 16 dimensional. Each of the P-frames is warped using motion vectors as in \nrefsec{mb:mv} to align them to the I-frame. Given a 7 frame GOP this gives a final representation of 160 dimensions. This representation is then used as input to the I-frame generation network which produces the high quality I-frame. This high quality I-frame is then warped 6 times to generate 6 copies each aligned to the individual P-frames. Then, the aligned I-frame is concatenated with the low quality P-frames and the P-frame generation network generates the 6 high quality P-frames. This gives the final output: the high quality GOP consisting of 7 frames.

Note that this is quite different from the sliding window or even recurrent approaches used in other video restoration work. In sliding window, a new representation would be computed for every frame consisting of three preceding and three succeeding frames. In a recurrent formulation, an accumulated hidden state is used to condition the current frame on past frames. In contrast, the method we developed in this chapter compacts information for a block of frames into a compact representation and then projects that information forward in time such that each frame has some past and some future information. This mimics the process that the video decoder performs when it decodes a bitstream. The information is discarded when a new I-frame is encountered.

\section{Towards a Better Benchmark}

All video compression reduction work is tested on one primary dataset: the MFQE \sidecite{Yang_Xu_Wang_Li_2018} dataset. This dataset consists of a large training set of diverse sequences and an eighteen video test set. The test set contains diverse real-world scenes and a variety of resolutions. The videos are stored in raw (YUV) format. Overall, this dataset is satisfactory for the purposes of evaluating compression reduction.

The problem, however, comes in how the dataset is used. Prior works used only HEVC (H.265) compression with constant QP values in $\{27, 32, 37, 42\}$ or some subset of these depending on the paper. While numerical results should not generally be our primary concern when evaluating any proposed method, these compression settings do leave much to be desired. Firstly, HEVC compression incurs less degradation than other commonly encountered codecs. Although HEVC would no longer be considered ``new'', it is also much less frequently used with almost no browser support. Additionally, the constant QP compression method is simply not used in real videos and is mostly included as a debugging tool. The degradations that it causes are much simpler to model than CRF or CBR which are used frequently in real videos. See \nrefsec{mpeg:sq} for a deeper discussion of these terms.

Instead we propose to use AVC (H.264) compression for evaluation and we use CRF instead of CQP. This is a much better representation of real world video than the previous benchmark as AVC compression accounts for nearly 91\% of internet videos as of 2019 \sidecite{vdr} \sidenote{Although this has likely decreased since 2019 it would not be by much.}. We choose CRF values in $\{25, 35, 40, 50\}$ ranging from relatively little compression at 25 (this is the default for ffmpeg \sidecite{tomar2006converting}) to 50 which is only one less than the maximum. To reiterate, our goal with this benchmark is to ensure that compression reduction algorithms face tests which accurately represent videos in the real world. In the next section we will see that MFQE fails to converge for any CRF setting as it produces significantly more complex degradations than CQP, thus justifying our concern.

\pagelayout{wide}
\addpart{Concluding Remarks}

\lettrine{Y}{ou} have made it to the end of the dissertation, a journey developing the \gls{first principles} of deep learning and classical compression from the preliminaries through to the published research of the author. With the body of the dissertation behind us we can recap where we've been and where we're going. To reiterate, the overall goal of my dissertation was to present, explicitly, an approach that follows the \gls{first principles} of the compression problems. This is motivated by engineering as much as by science, and I have shown that incorporating engineering principles, both into the methodologies, \eg, considering how compression algorithms were developed, \textit{and the philosophy of the research}, \eg, approaching a scientific problem as an engineer and a scientist simultaneously, can be successful. While this is not an approach unique to myself, I find that it is rarely stated out loud. And yet, the engineers that developed the compression algorithms we all use on a daily basis were extremely smart, so is it not logical to follow in their example when studying deep learning? I hope that the implications of this thought extend far beyond this document.

In \nrefch{jdr} we developed a method for performing deep learning in the JPEG domain. This method operates on coefficients directly and requires no decompression. The goal of the method was to produce a result which is as close as possible to the pixel domain result. In this work we leveraged the first principles of JPEG compression by linearizing the JPEG transform and composing it with our pixel domain convolutions. We also leveraged this to produce closed form derivations of batch normalization and average pooling.However, we could not do this for ReLU and so we used an approximation technique. The primary issue with this work is that it uses substantially more memory to store feature maps and convolutions.

In \nrefch{qgac} we improved JPEG compression using deep learning. We noted that there were several issues preventing the widespread success of prior work in this field. Prior works were quality dependent models which trained a unique model for each JPEG quality factor. They did not handle color images, and they were focused only on regression. We solved each of these problems again leveraging first principles. By conditioning our network on the JPEG quantization matrix and processing DCT coefficients instead of pixels, we were able to encode quality information using a single network. By explicitly handling chroma subsampling and the additional quantization that color channels are subjected to, we improved results on color images. Finally, we incorporated GAN and texture losses to improve the visual result over a regression-only solution. While this method was highly successful, it had a distinct disadvantage when quantization data is either incorrect, as in multiple compression, or not available at all as in transcoding.

This was followed up in \nrefch{ttac} which extended the previous method to optimize the correction for machine consumption. This is somewhat of a novelty in artifact correction which is traditionally for humans only. By incorporating a loss based on a downstream task, we are able to greatly improve the performance of that task on JPEG compressed inputs, often outperforming data augmentation techniques that retrain the task with JPEG images. Our method had the added benefit that a network trained for one downstream task would also work well for other downstream tasks with no re-training required. The main drawback of this method is that it has increased running time (for multiple networks) and that it does not always out perform data augmentation.

Finally, in \nrefch{metabit}, we forayed into video compression by developing a correction based method for improving AVC and HEVC compression. We noted that prior works expended significant resources computing results which were already stored explicitly in the video bitstreams, such as high quality frame locations (I-frames) and motion data. We leveraged this data and used our increased compute budget to more than double the number of parameters in our network with almost no impact on throughput. We also incorporated scale-space loss along with the GAN and texture losses from the JPEG work, in order to improve high frequency reconstructions. While this method outperforms prior works and fully deep-learning based codecs at low birates, it does struggle with high bitrate reconstructions and it currently requires a unique model for each video compression setting. The method would also struggle to run in real time on any consumer hardware.

So where do we go from here? Aside from the multitude of related problems to work on in compression, from things as simple as improving the results to as tangential as data privacy, the number one focus for the next decade of compression and deep learning is going to be on making these techniques practical. One of the primary goals in writing this document is to instruct practitioners, \eg, professional engineers, developers, students, \etc, that while these techniques show extreme promise, actually getting them into the hands of consumers requires considerable effort.

Focusing solely on the techniques which improve compression performance, which would be of direct use to consumers who lack broadband internet, these methods currently require significant compute resources on the end user's side. This could be shifted to the data center side, \ie, a hybrid technique which extracts some deep representation during transmission. This could also be addressed simply by developing faster and lower-memory algorithms or by leveraging customized hardware. While the latter sounds like an expensive solution, consider that video decoders are almost exclusively implemented in hardware on modern processors both for desktop/laptop machines and mobile phones. This is also starting to happen for deep learning, \eg, the Google Tensor chips and edge TPUs, and the Apple A14 chips. In any case, I believe consumer applications for this technology are no more than two years off at the time of writing, and within the decade for fully deep learning based compression. The field of compression as a whole is progressing as fast as ever and it is an exciting time to be involved.

This is all in the midst of the global pandemic. In a world which was just beginning to address the inequality in internet access, suddenly in late 2019, we were forced to confront this issue as work and school became primarily remote. Remote work and school means communication with video and images which means compression. Those who did not have a strong internet connection were simply left behind as there were no suitable alternatives, and it remains to be seen what the long term ramifications of this will be. Now in early 2022, the world is quickly moving on from pandemic life. Yet it is important not to forget this lesson.

\noindent {\large Better compression has the ability to help people right now. }

\begin{kaobox}[frametitle=Author's Note]
    I invite the readers to now visit the appendices where they will find material which is just as interesting, and yet not directly related to, the dissertation proper. In particular we will review some additional qualitative results and briefly cover fully deep-learning based compression algorithms. Thank you for reading my dissertation!

    \includegraphics[width=0.5\textwidth]{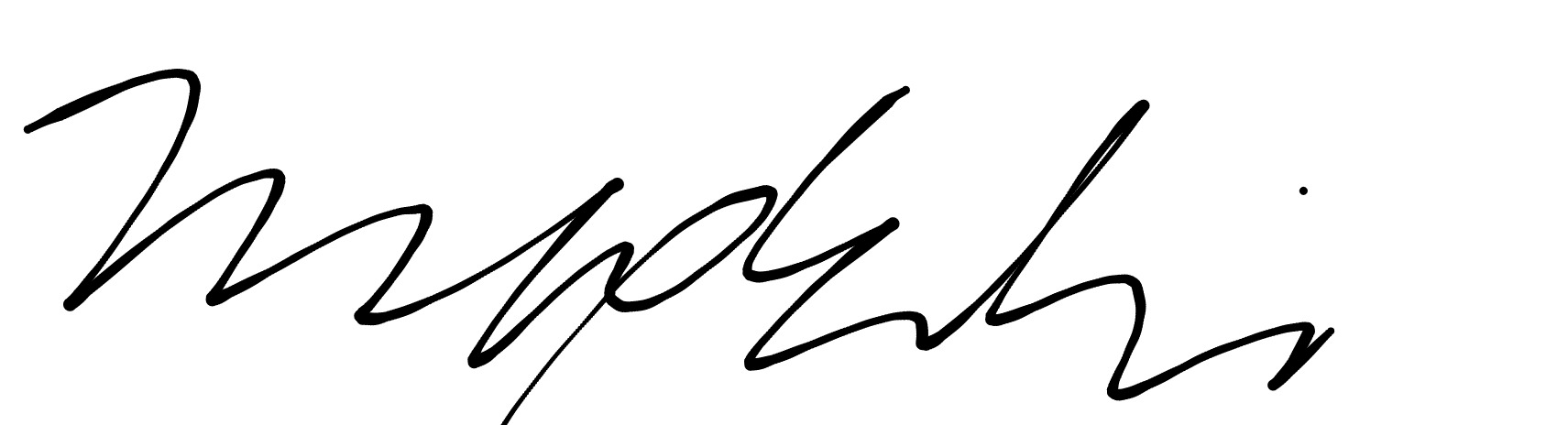} \\
    Max Ehrlich
\end{kaobox}

\appendix %

\pagelayout{wide}

\addpart{Appendix}

\setchapterstyle{lines}

\chapter{Study on JPEG Compression and Machine Learning}
\label{app:study}

This appendix reproduces the full plots and tables of the results of the study on JPEG compression and deep learning \parencite{ehrlich2021analyzing}. See \nrefch{ttac} for more details. These plots are for informational purposes only.

\section{Plots of Results}

\begin{figure}[H]
    \centering
    \includegraphics[width=0.8\textwidth]{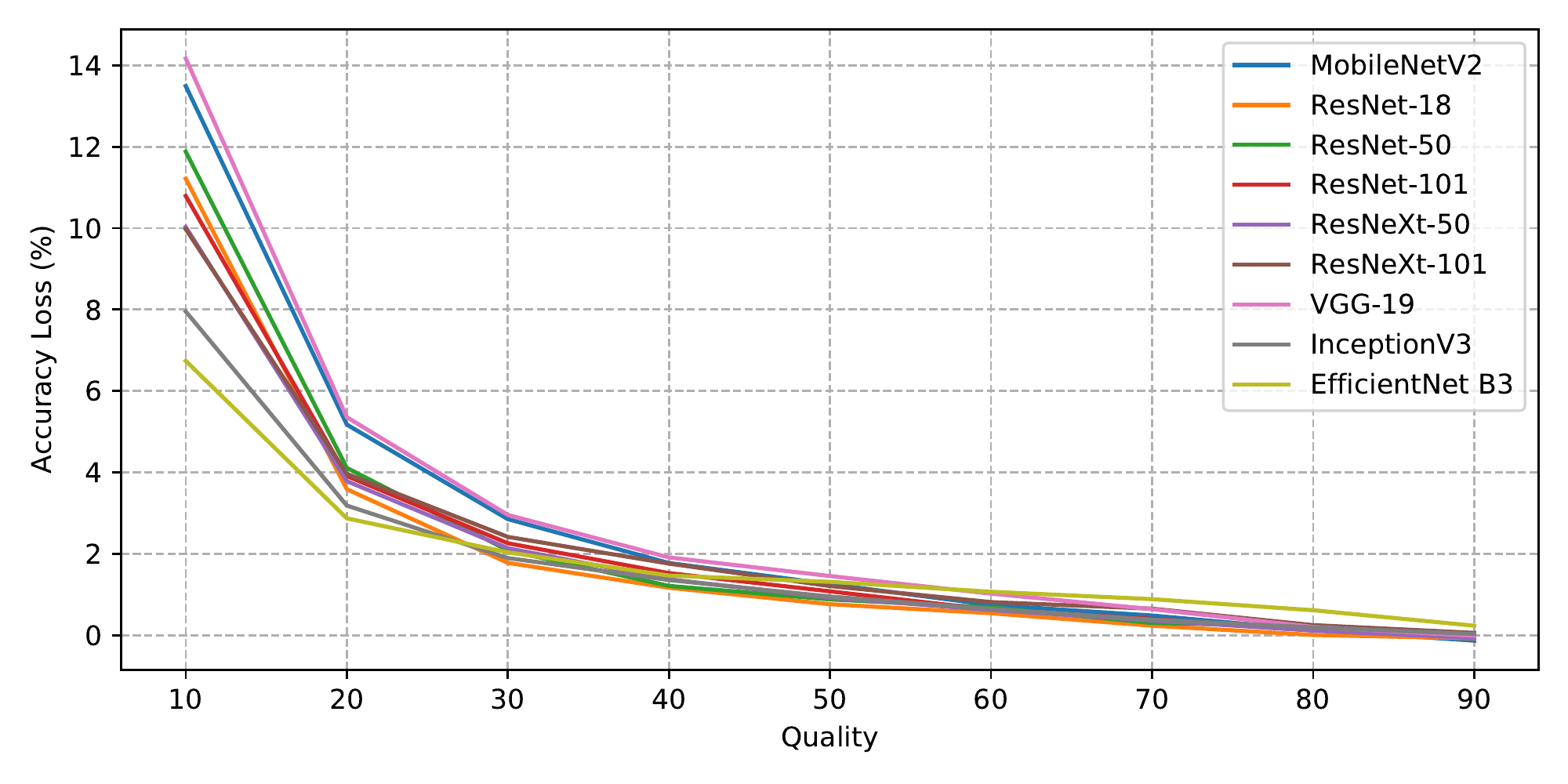}
    \caption[Overall Classification Results]{Overall Classification Results}
\end{figure}

\begin{figure}[H]
    \centering
    \includegraphics[width=0.8\textwidth]{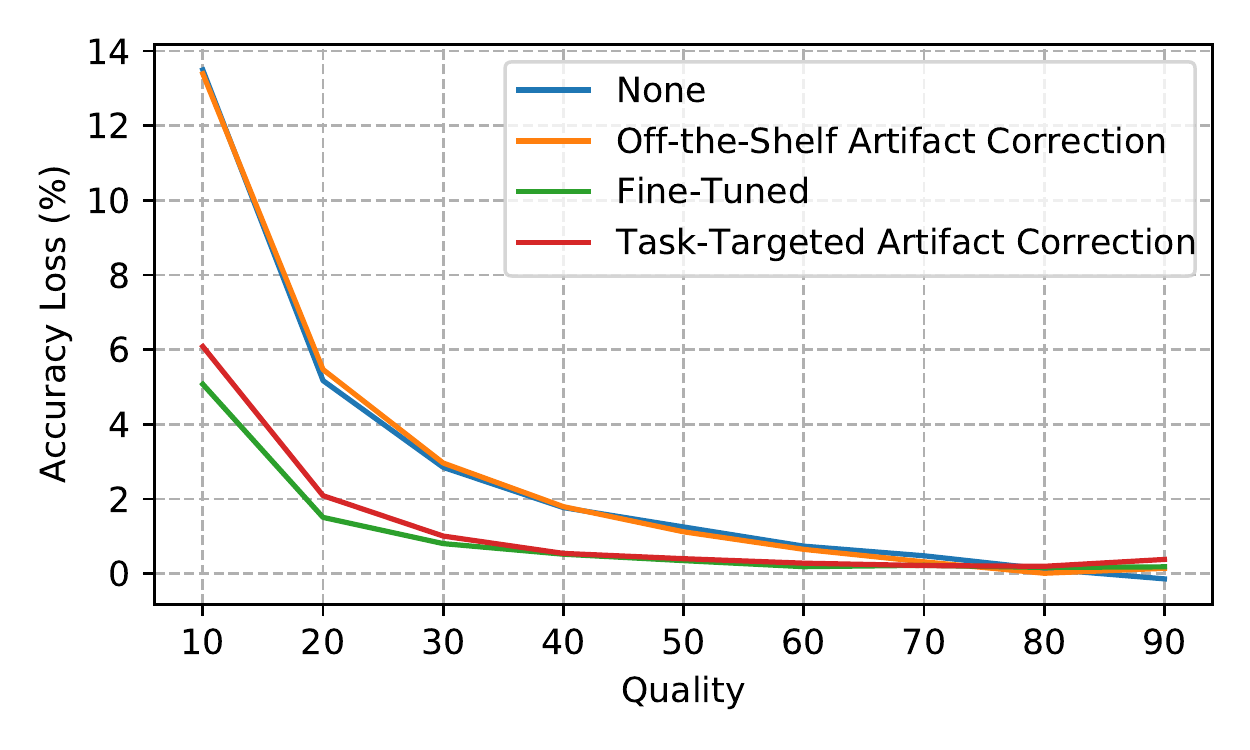}
    \caption[Classification Results: MobileNetV2]{Classification Results: MobileNetV2}
\end{figure}
\begin{figure}[H]
    \includegraphics[width=0.7\textwidth]{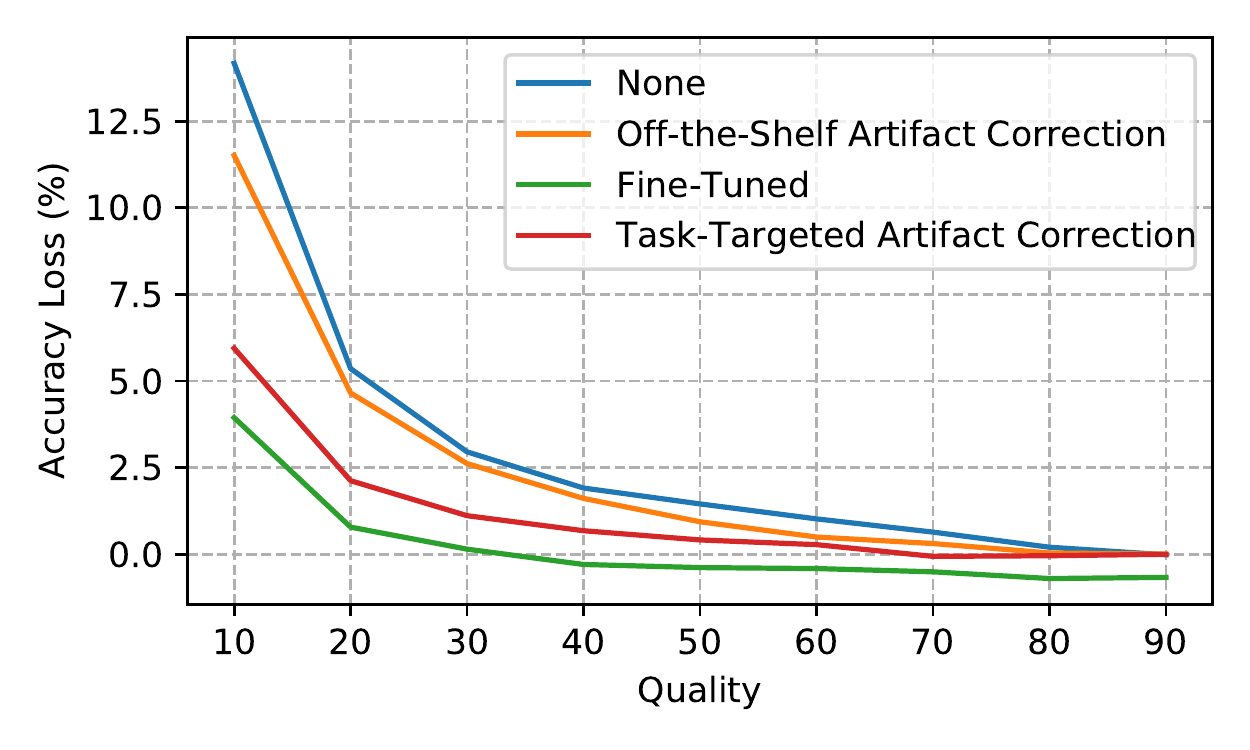}
    \caption[Classification Results: VGG-19]{Classification Results: VGG-19}
\end{figure}
\begin{figure}[H]
    \centering
    \includegraphics[width=0.7\textwidth]{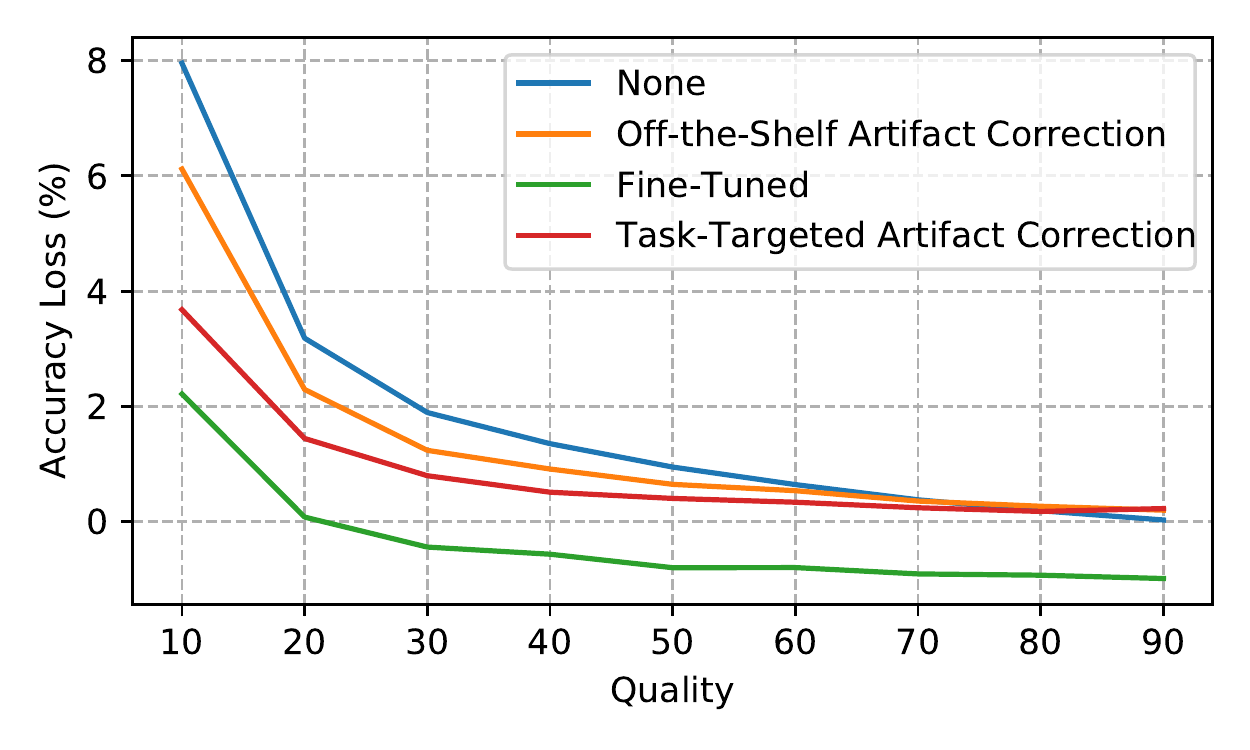}
    \caption[Classification Results: InceptionV3]{Classification Results: InceptionV3}
\end{figure}
\begin{figure}[H]
    \includegraphics[width=0.7\textwidth]{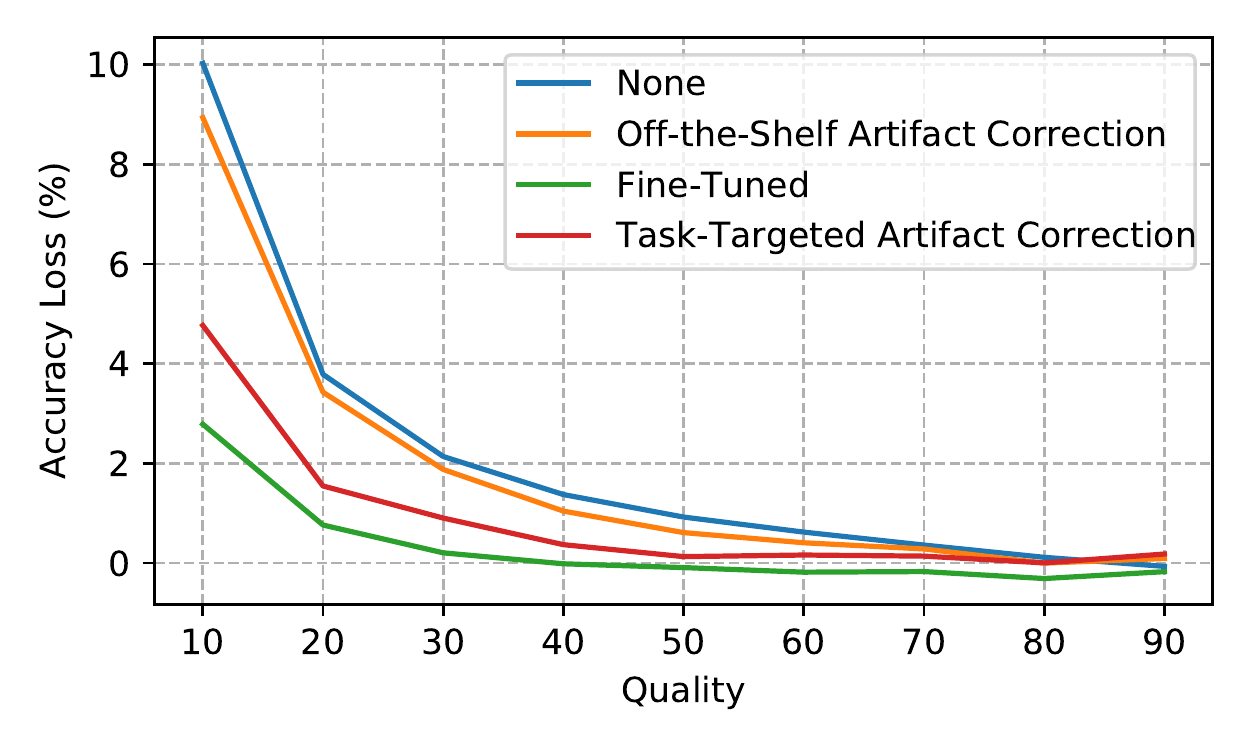}
    \caption[Classification Results: ResNeXt 50]{Classification Results: ResNeXt 50}
\end{figure}
\begin{figure}[H]
    \centering
    \includegraphics[width=0.7\textwidth]{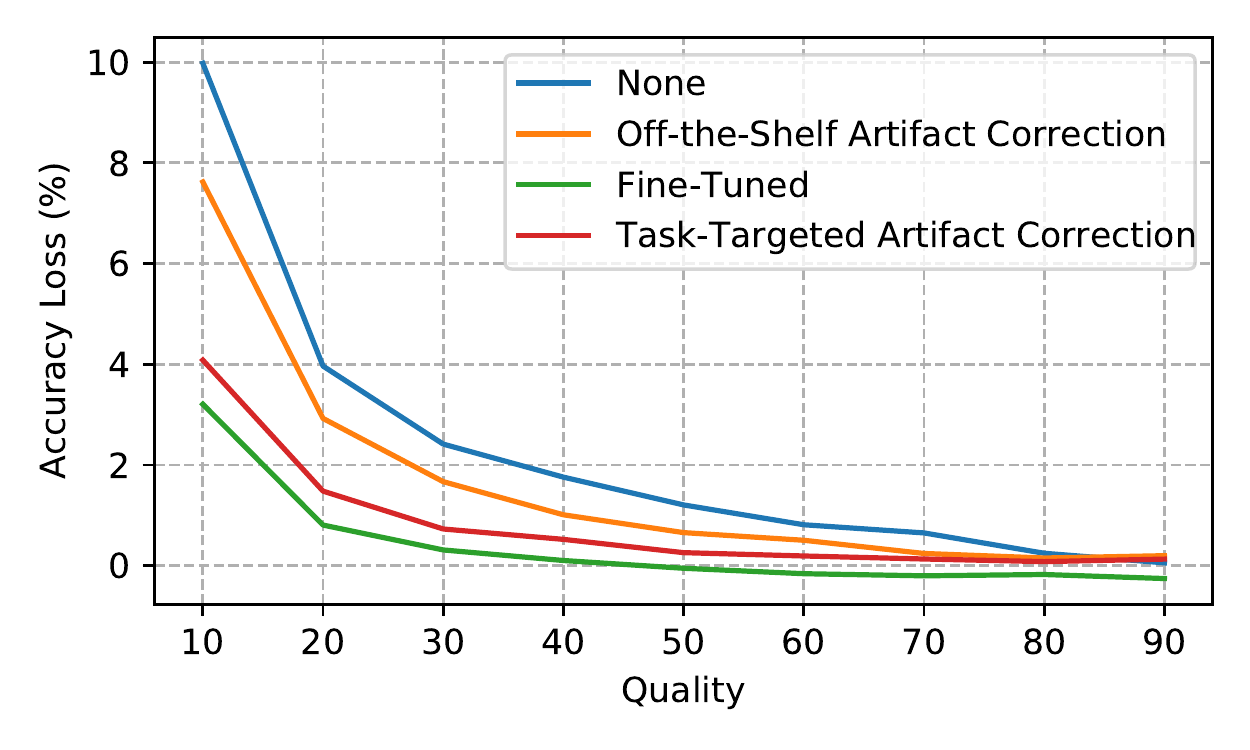}
    \caption[Classification Results: ResNeXt 101]{Classification Results: ResNeXt 101}
\end{figure}
\begin{figure}[H]
    \includegraphics[width=0.7\textwidth]{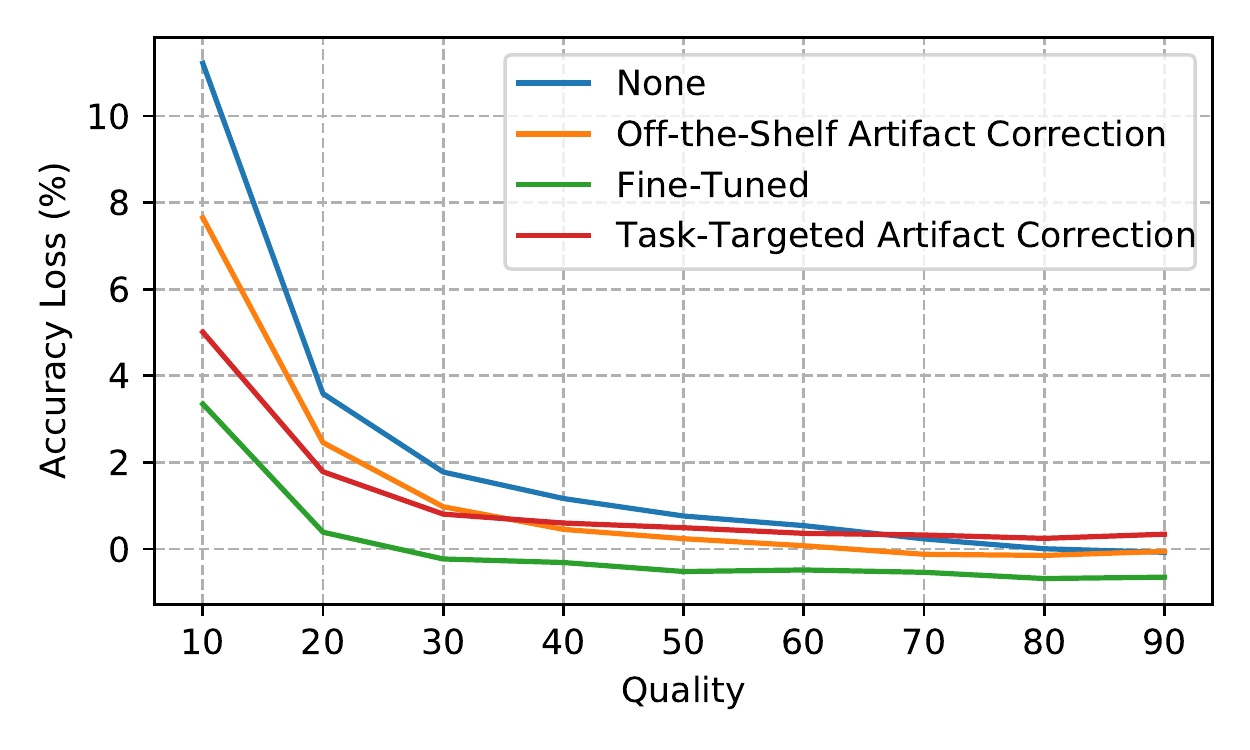}
    \caption[Classification Results: ResNet 18]{Classification Results: ResNet 18}
\end{figure}
\begin{figure}[H]
    \centering
    \includegraphics[width=0.7\textwidth]{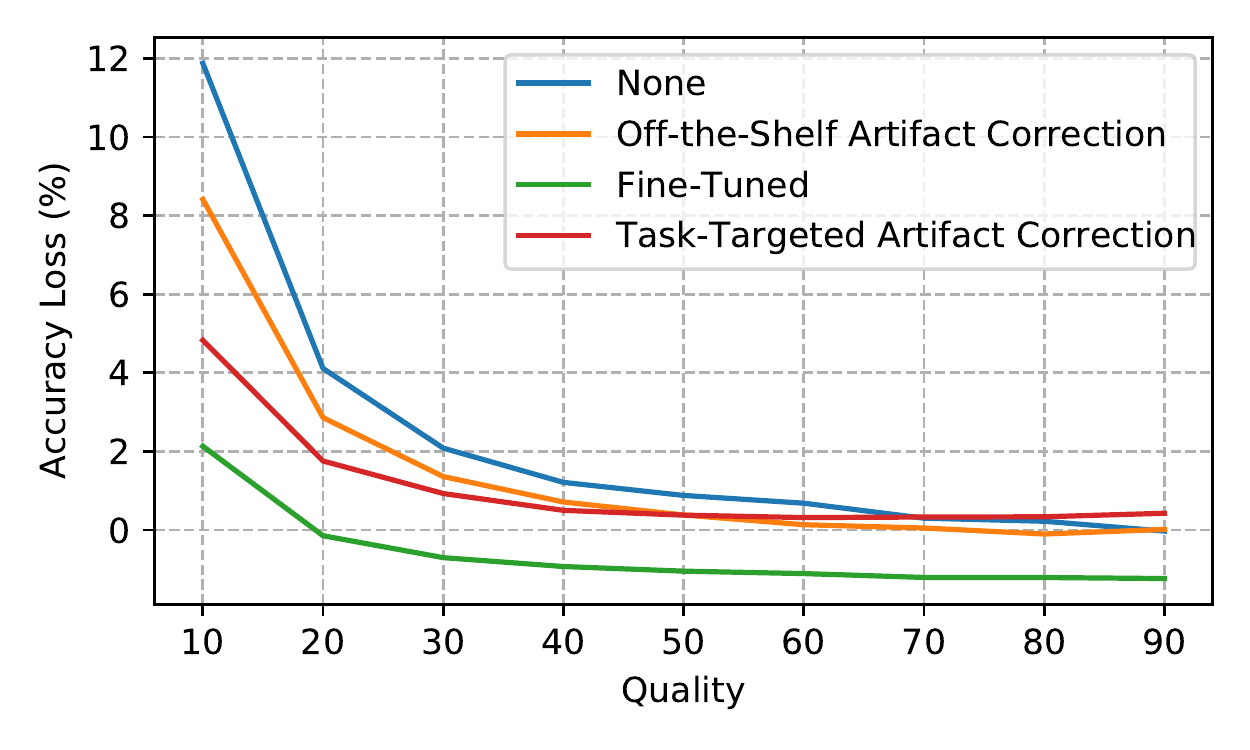}
    \caption[Classification Results: ResNet 50]{Classification Results: ResNet 50}
\end{figure}
\begin{figure}[H]
    \includegraphics[width=0.7\textwidth]{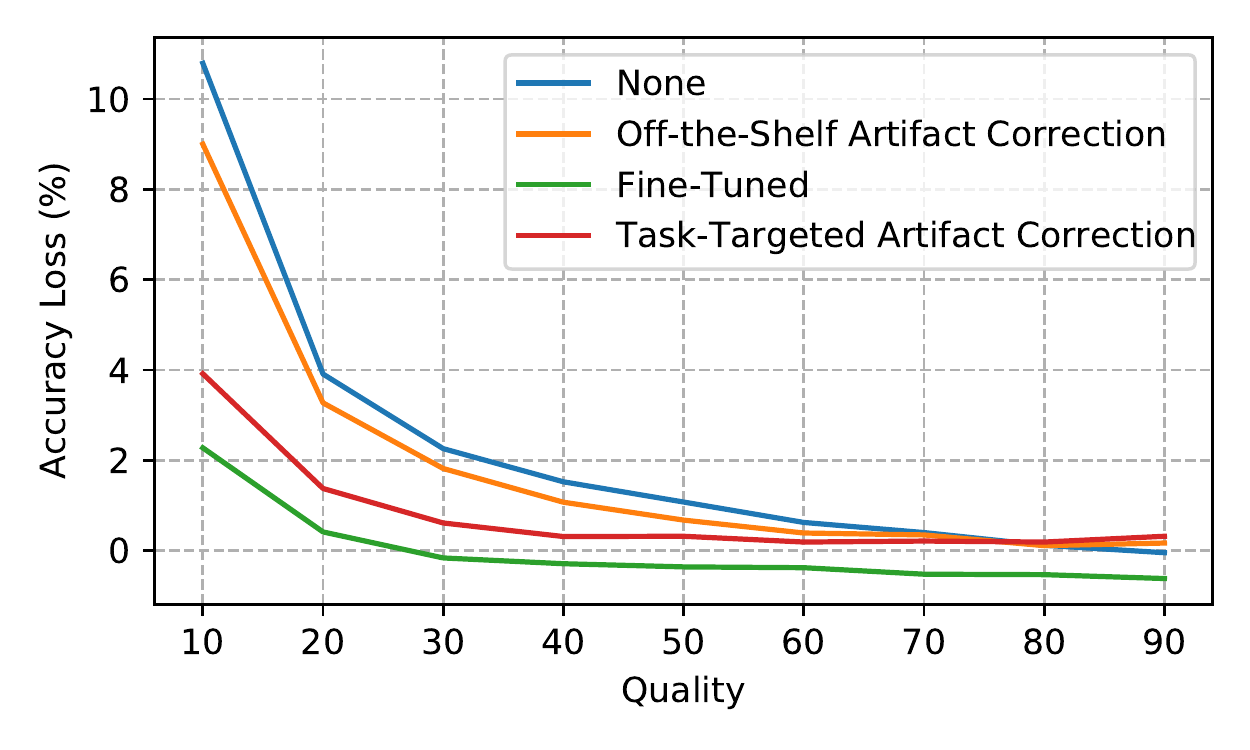}
    \caption[Classification Results: ResNet 101]{Classification Results: ResNet 101}
\end{figure}
\begin{figure}[H]
    \centering
    \includegraphics[width=0.7\textwidth]{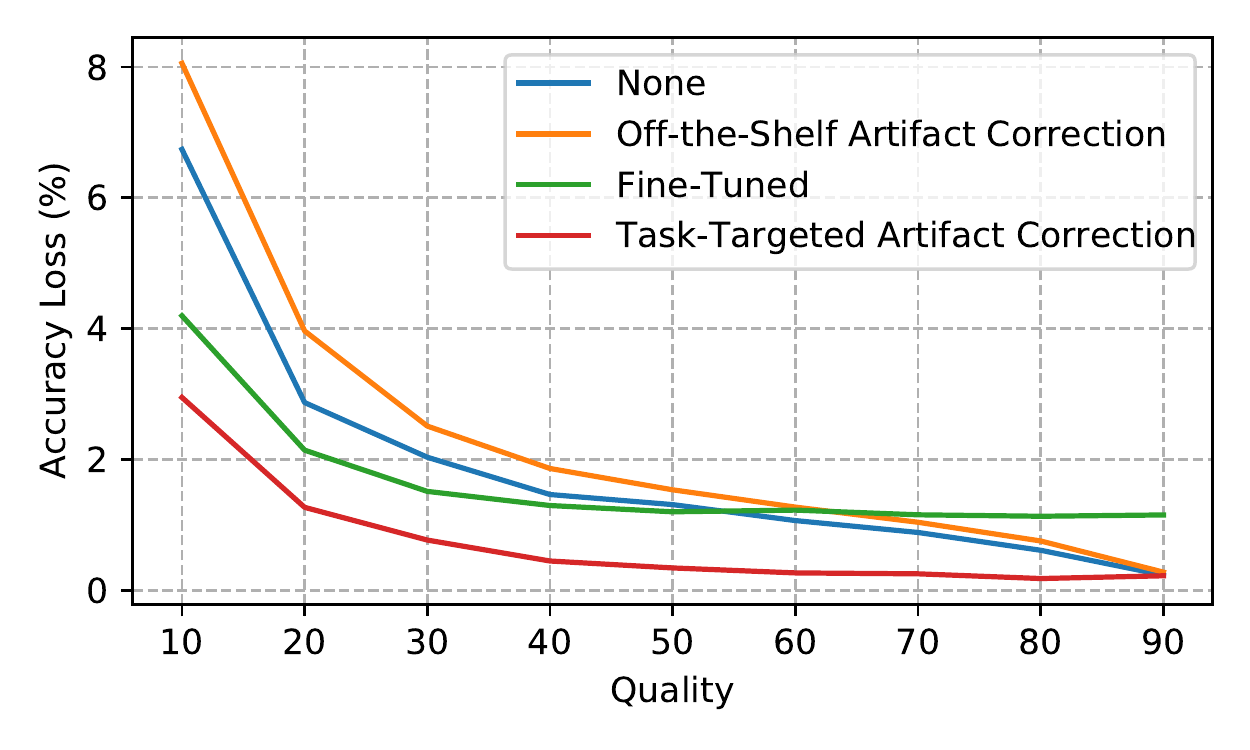}
    \caption[Classification Results: EfficientNet B3]{Classification Results: EfficientNet B3}
\end{figure}

\clearpage

\begin{figure}[H]
    \centering
    \includegraphics[width=0.7\textwidth]{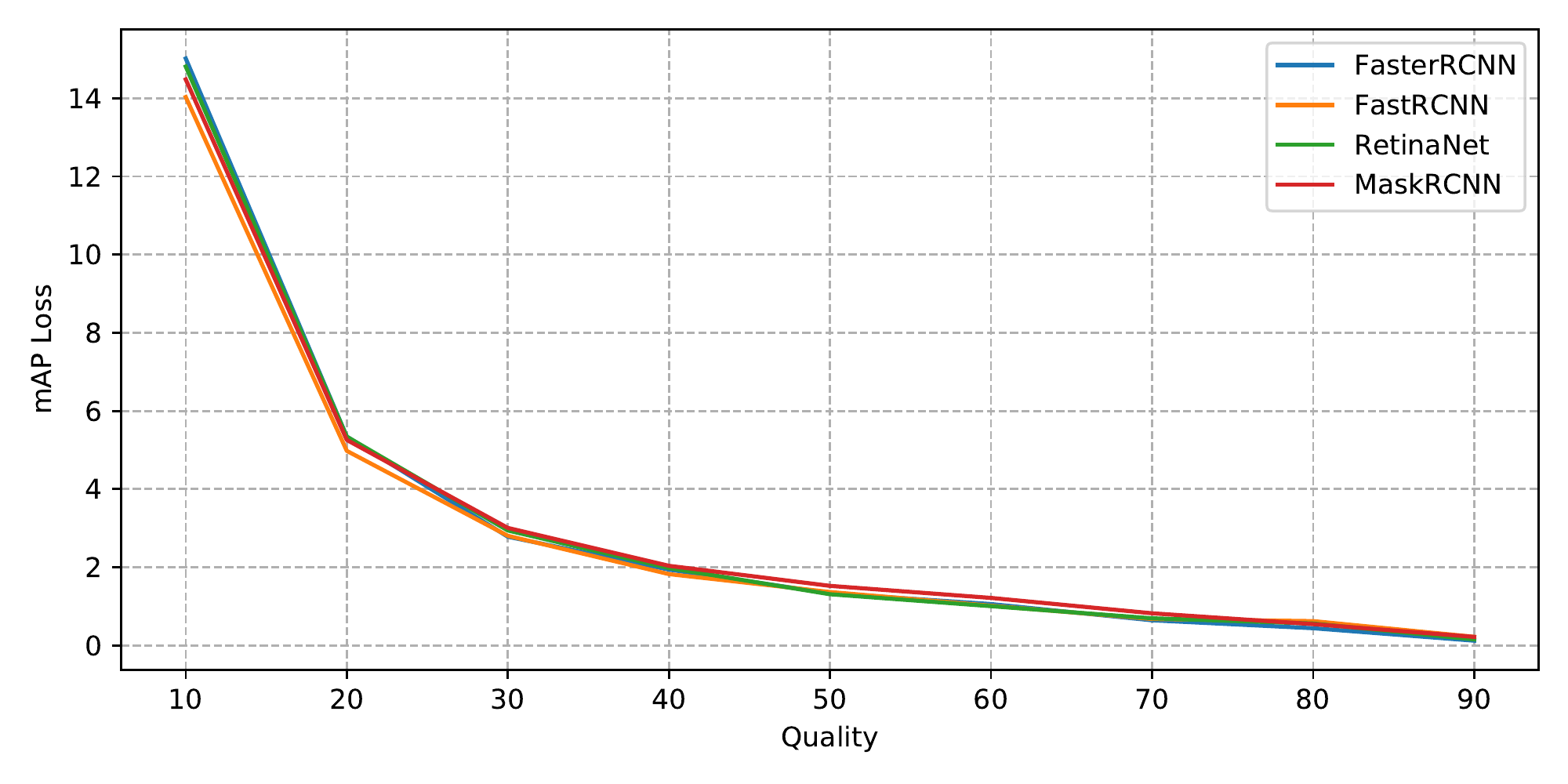}
    \caption[Overall Detection and Instance Segmentation Results]{Overall Detection and Instance Segmentation Results}
\end{figure}
\begin{figure}[H]
    \centering
    \includegraphics[width=0.7\textwidth]{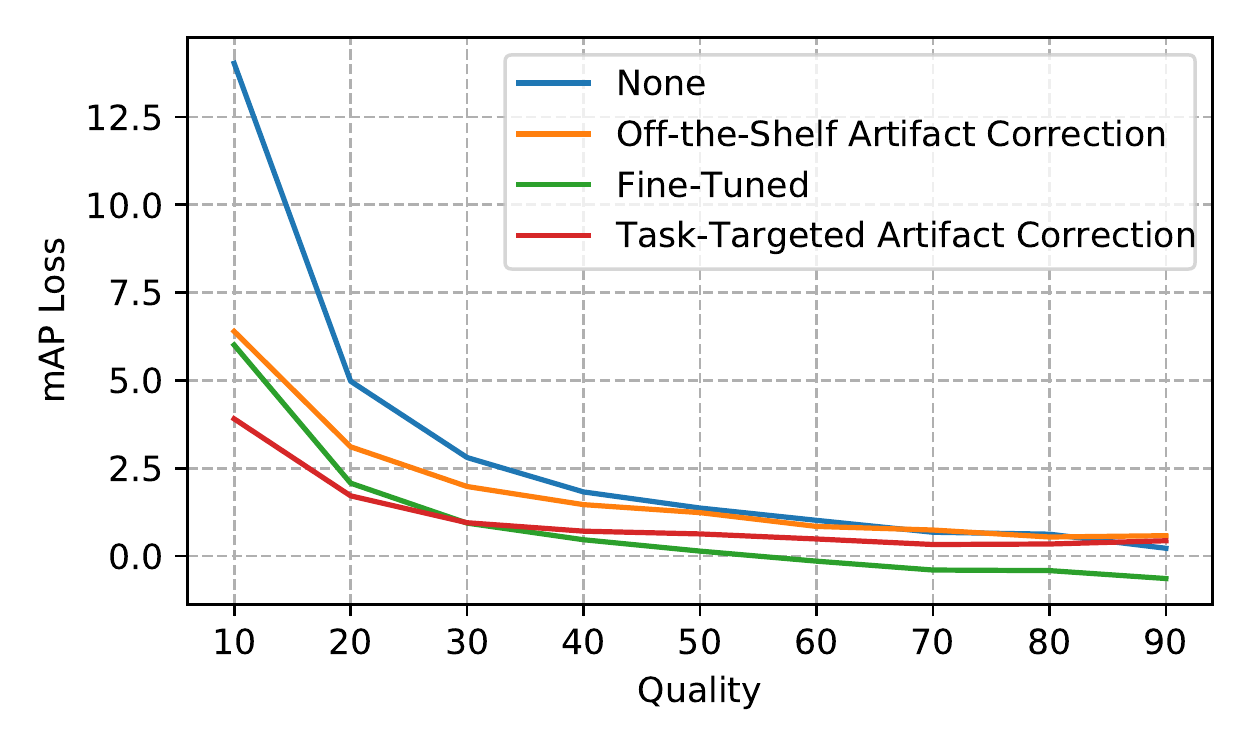}
    \caption[Detection Results: FastRCNN]{Detection Results: FastRCNN}
\end{figure}
\begin{figure}[H]
    \centering
    \includegraphics[width=0.7\textwidth]{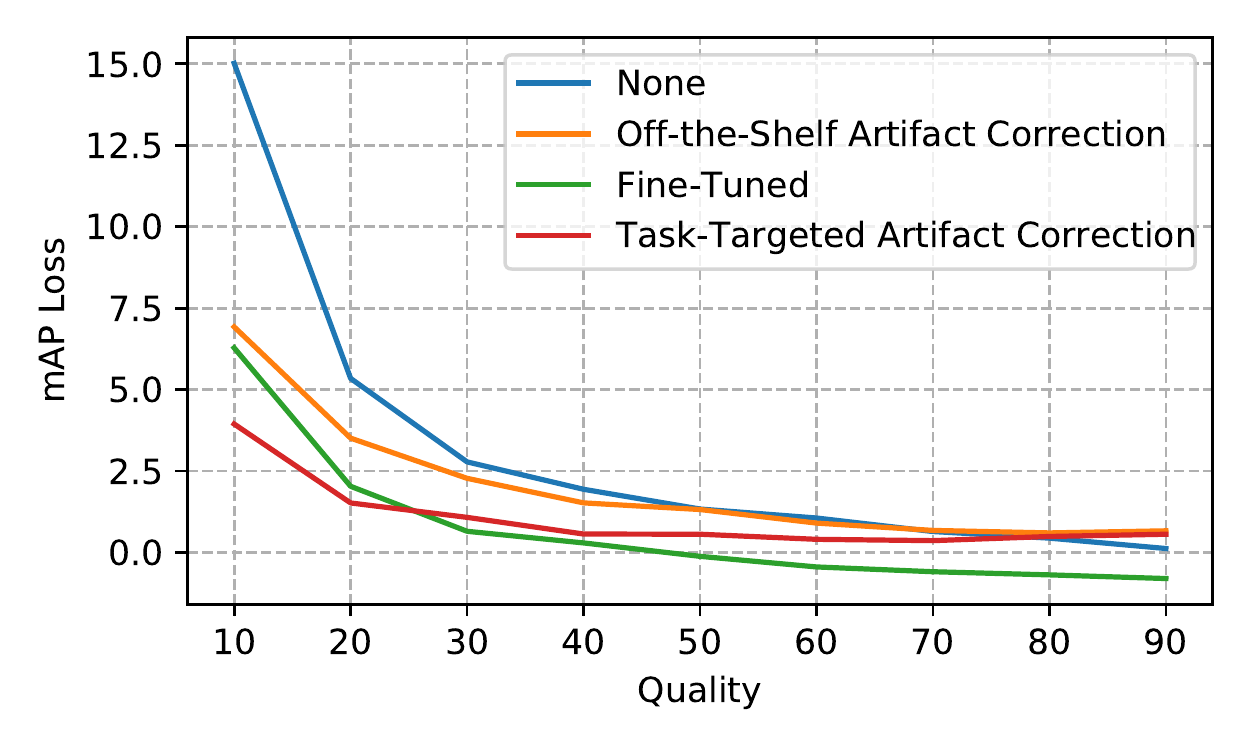}
    \caption[Detection Results: FasterRCNN]{Detection Results: FasterRCNN}
\end{figure}
\begin{figure}[H]
    \centering
    \includegraphics[width=0.7\textwidth]{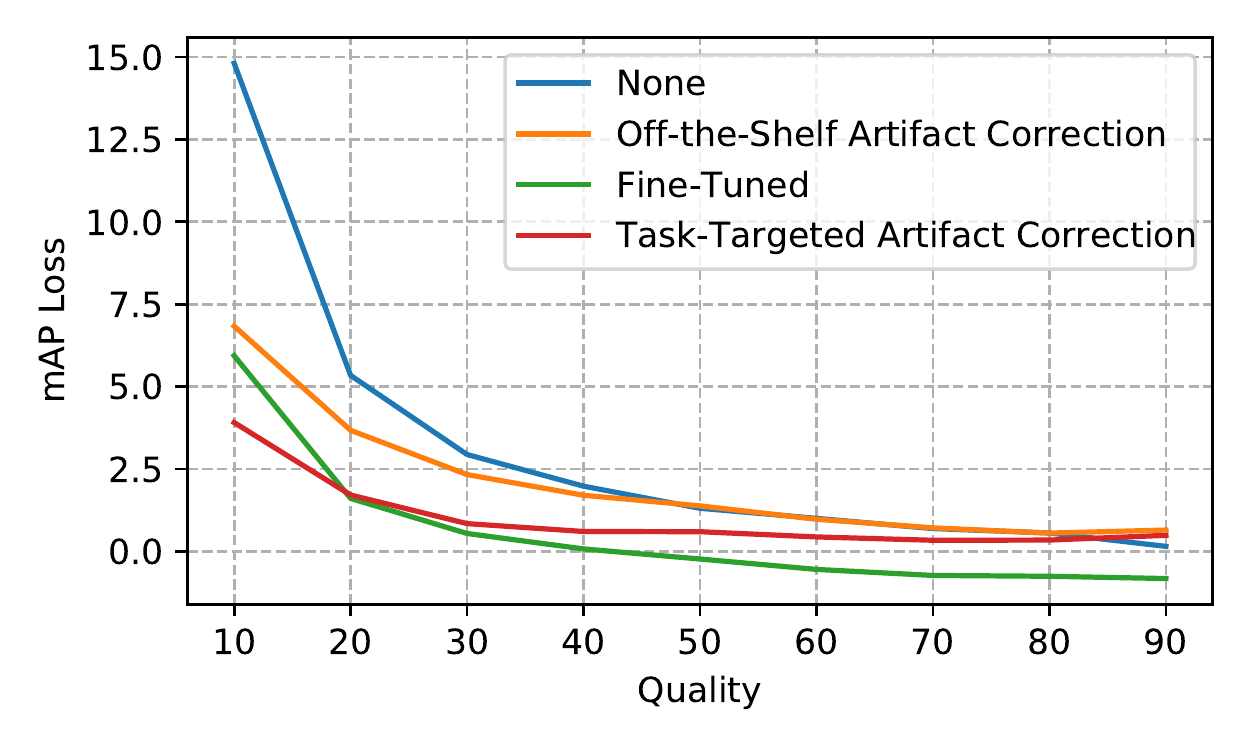}
    \caption[Detection Results: RetinaNet]{Detection Results: RetinaNet}
\end{figure}
\begin{figure}[H]
    \centering
    \includegraphics[width=0.7\textwidth]{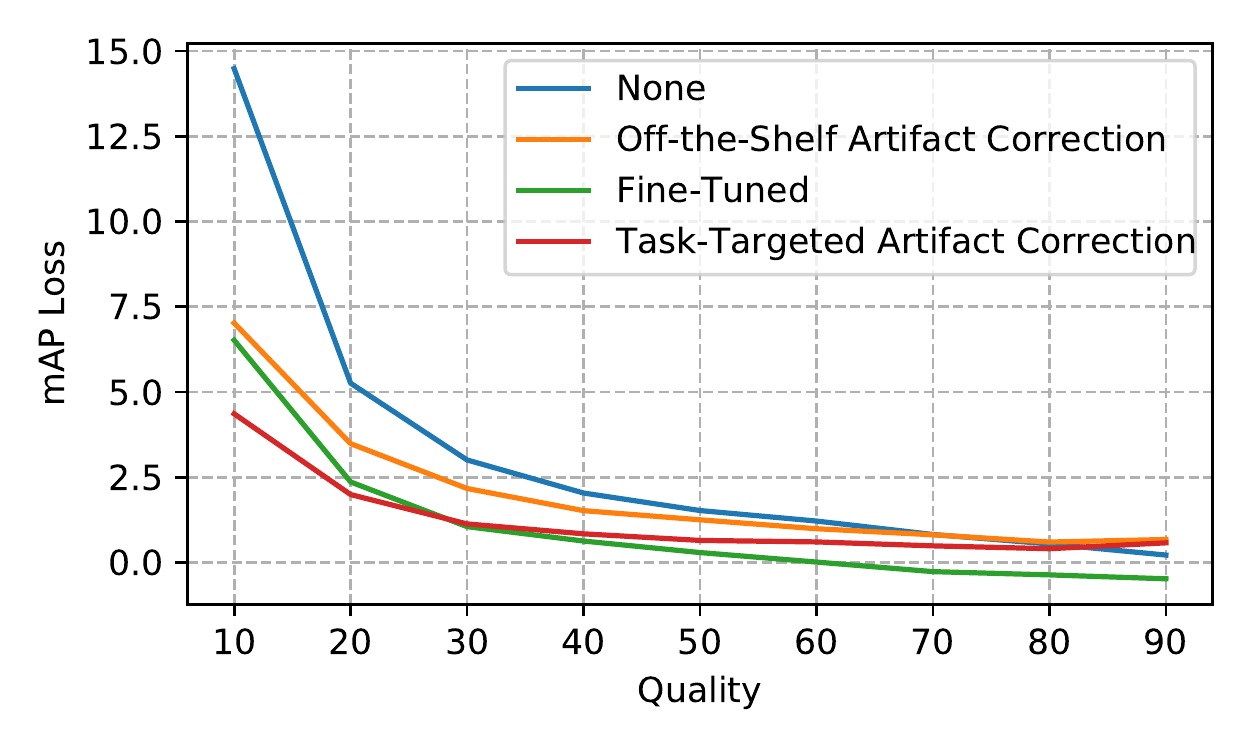}
    \caption[Instance Segmentation Results Results: MaskRCNN]{Detection Results: MaskRCNN}
\end{figure}

\clearpage

\begin{figure}[H]
    \centering
    \includegraphics[width=0.7\textwidth]{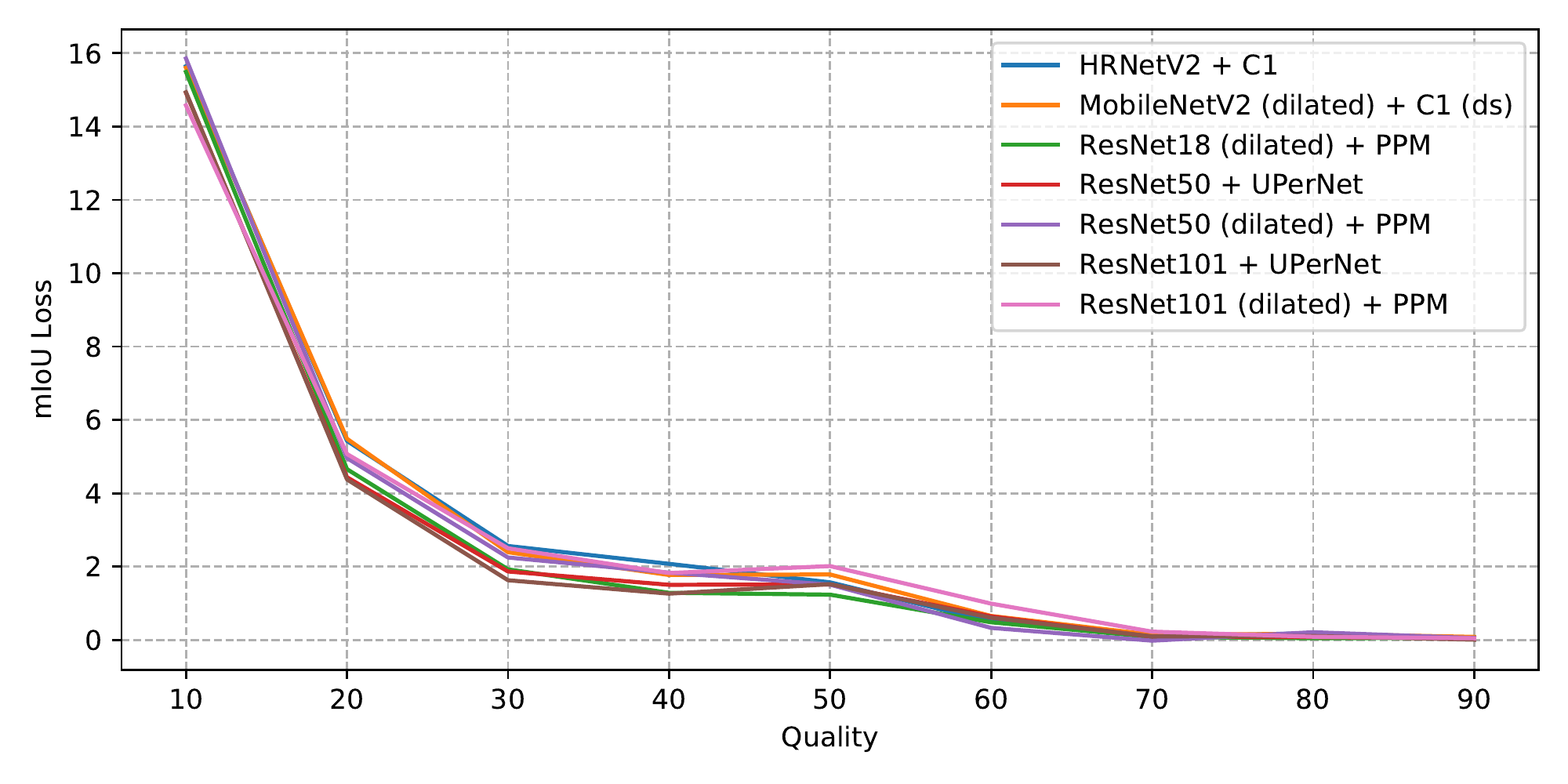}
    \caption[Overall Semantic Segmentation Results]{Overall Semantic Segmentation Results}
\end{figure}
\begin{figure}[H]
    \centering
    \includegraphics[width=0.7\textwidth]{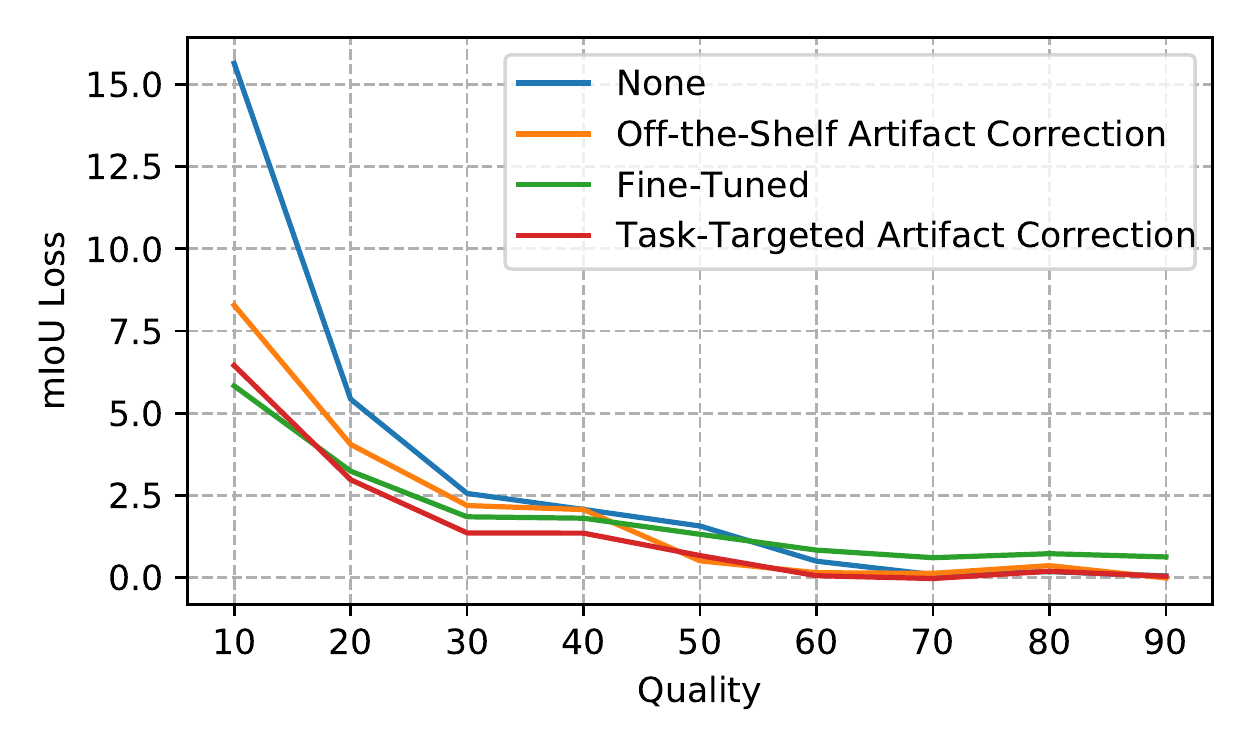}
    \caption[Semantic Segmentation Results: HRNetV2 + C1]{Semantic Segmentation Results: HRNetV2 + C1}
\end{figure}
\begin{figure}[H]
    \centering
    \includegraphics[width=0.7\textwidth]{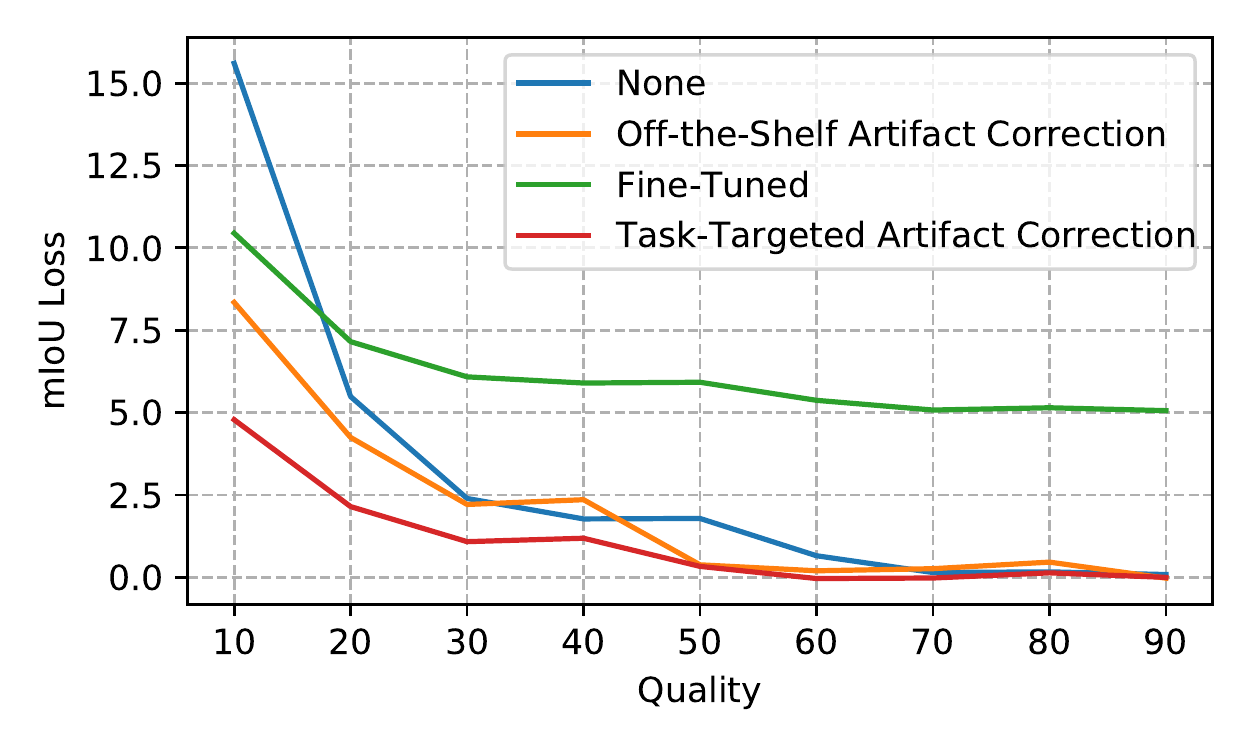}
    \caption[Semantic Segmentation Results: MobileNetV2 + C1]{Semantic Segmentation Results: MobileNetV2 + C1}
\end{figure}
\begin{figure}[H]
    \centering
    \includegraphics[width=0.7\textwidth]{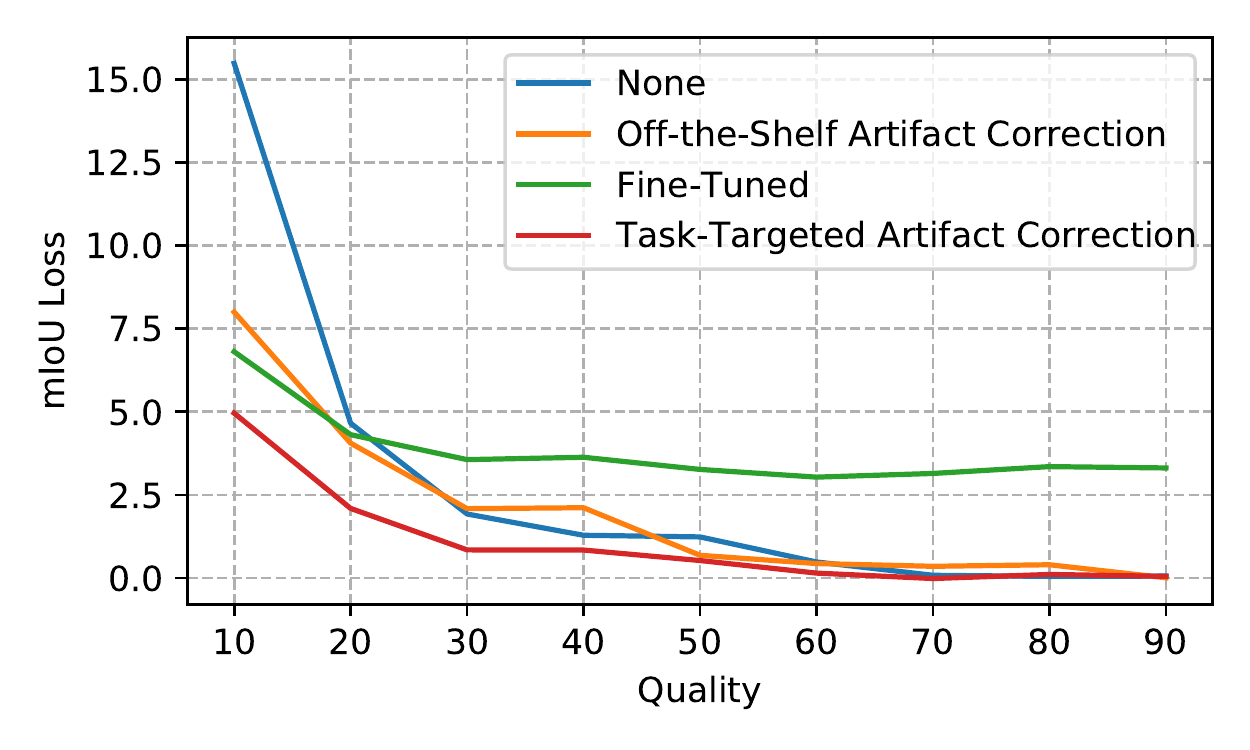}
    \caption[Semantic Segmentation Results: ResNet 18 + PPM]{Semantic Segmentation Results: ResNet 18 + PPM}
\end{figure}
\begin{figure}[H]
    \centering
    \includegraphics[width=0.7\textwidth]{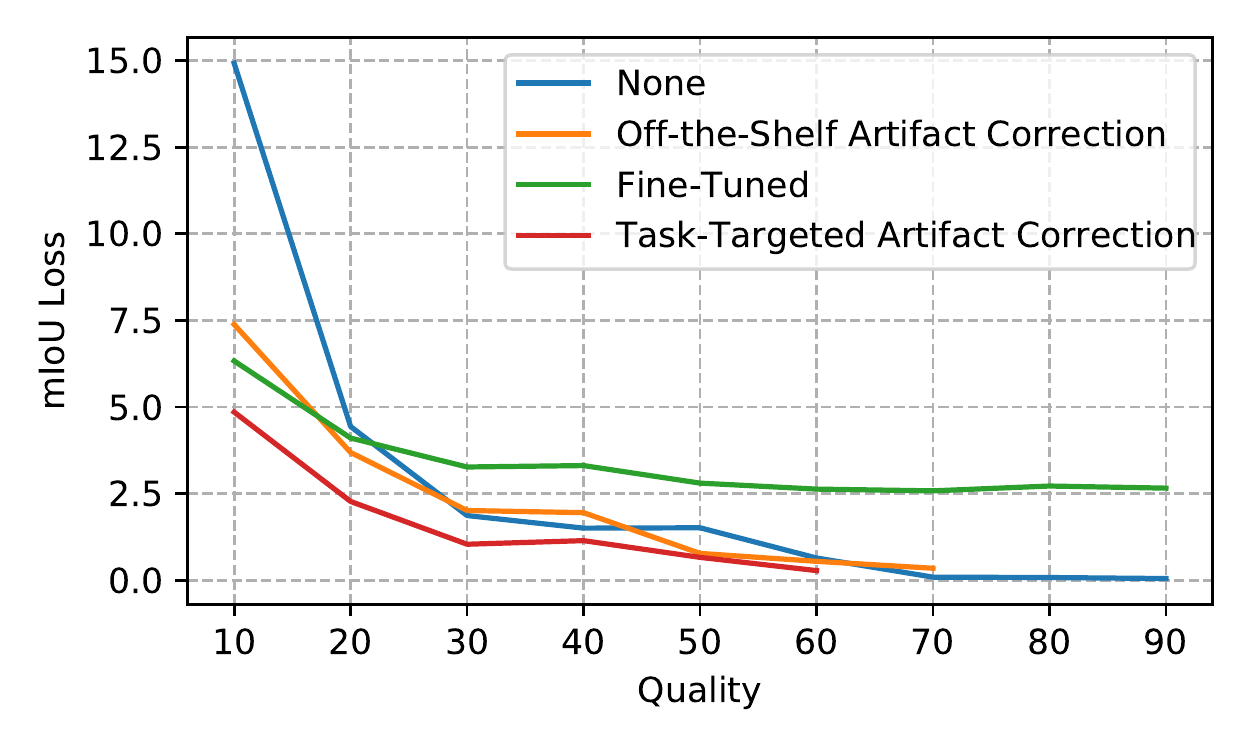}
    \caption[Semantic Segmentation Results: Resnet50 + UPerNet]{Semantic Segmentation Results: Resnet50 + UPerNet}
\end{figure}
\begin{figure}[H]
    \centering
    \includegraphics[width=0.7\textwidth]{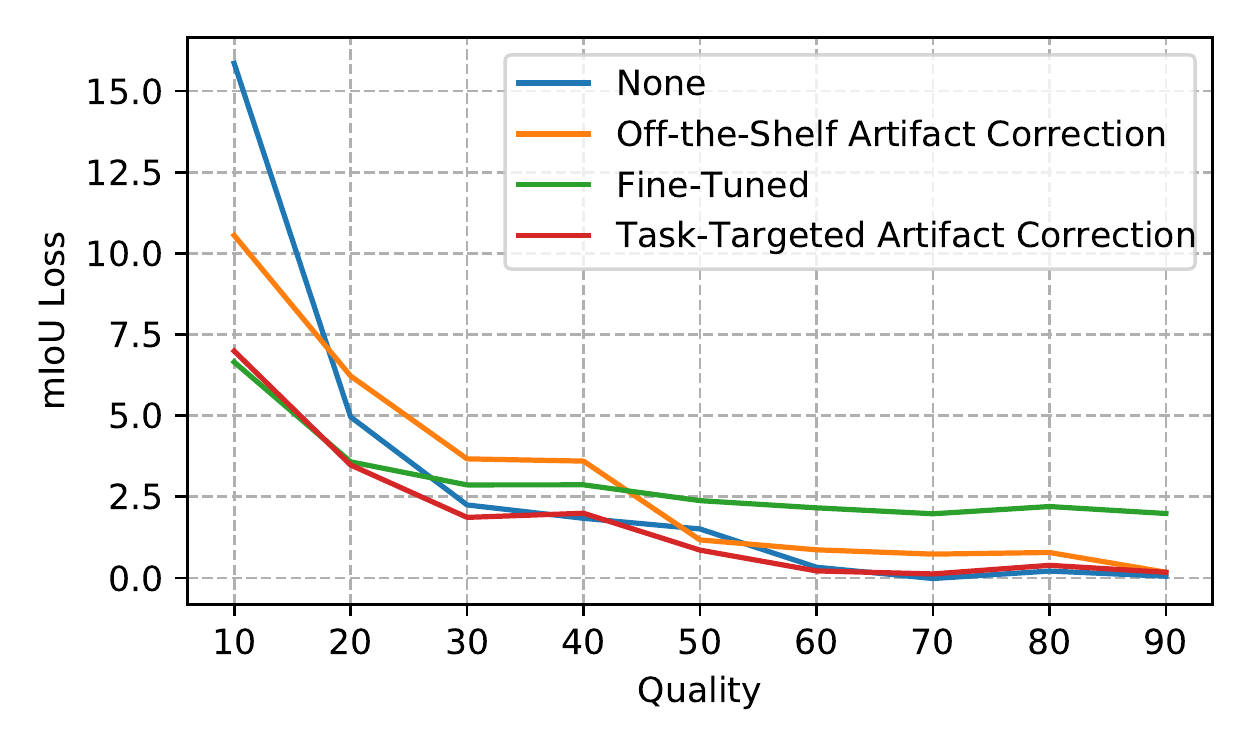}
    \caption[Semantic Segmentation Results: ResNet 50 + PPM]{Semantic Segmentation Results: ResNet 50 + PPM}
\end{figure}
\begin{figure}[H]
    \centering
    \includegraphics[width=0.7\textwidth]{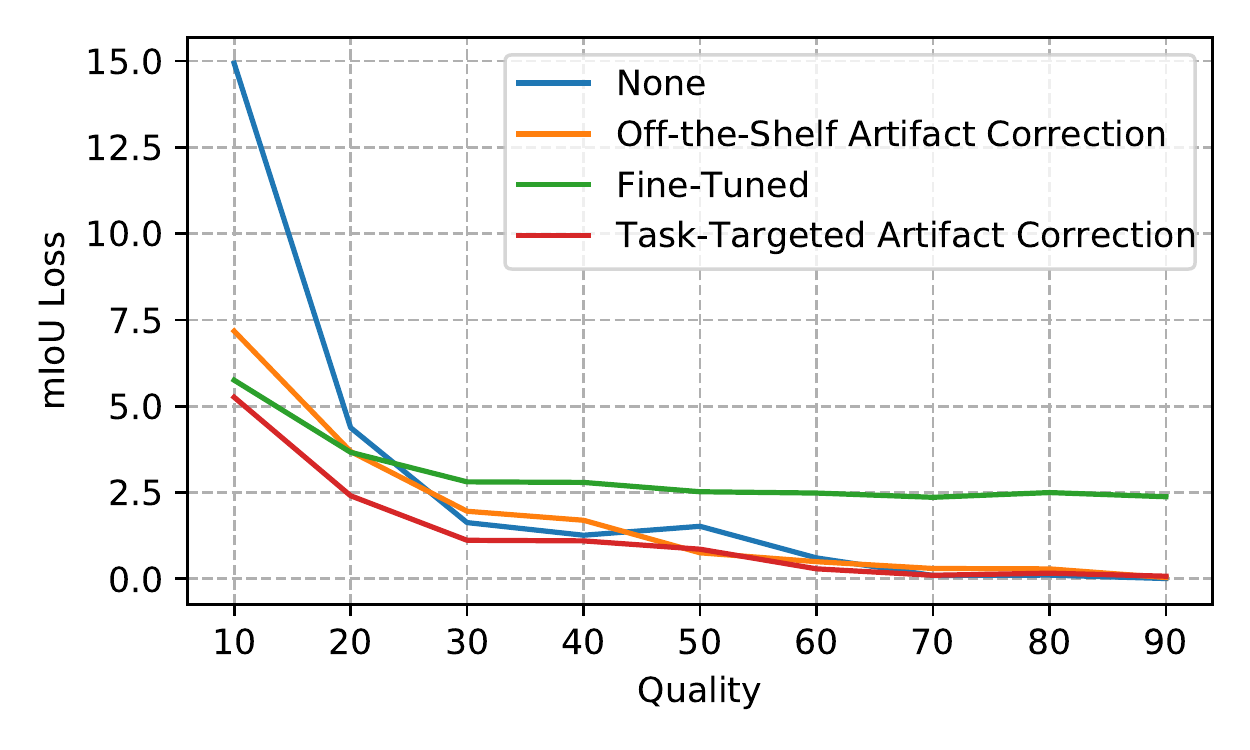}
    \caption[Semantic Segmentation Results: ResNet 101 + UPerNet]{Semantic Segmentation Results: ResNet 101 + UPerNet}
\end{figure}
\begin{figure}[H]
    \centering
    \includegraphics[width=0.7\textwidth]{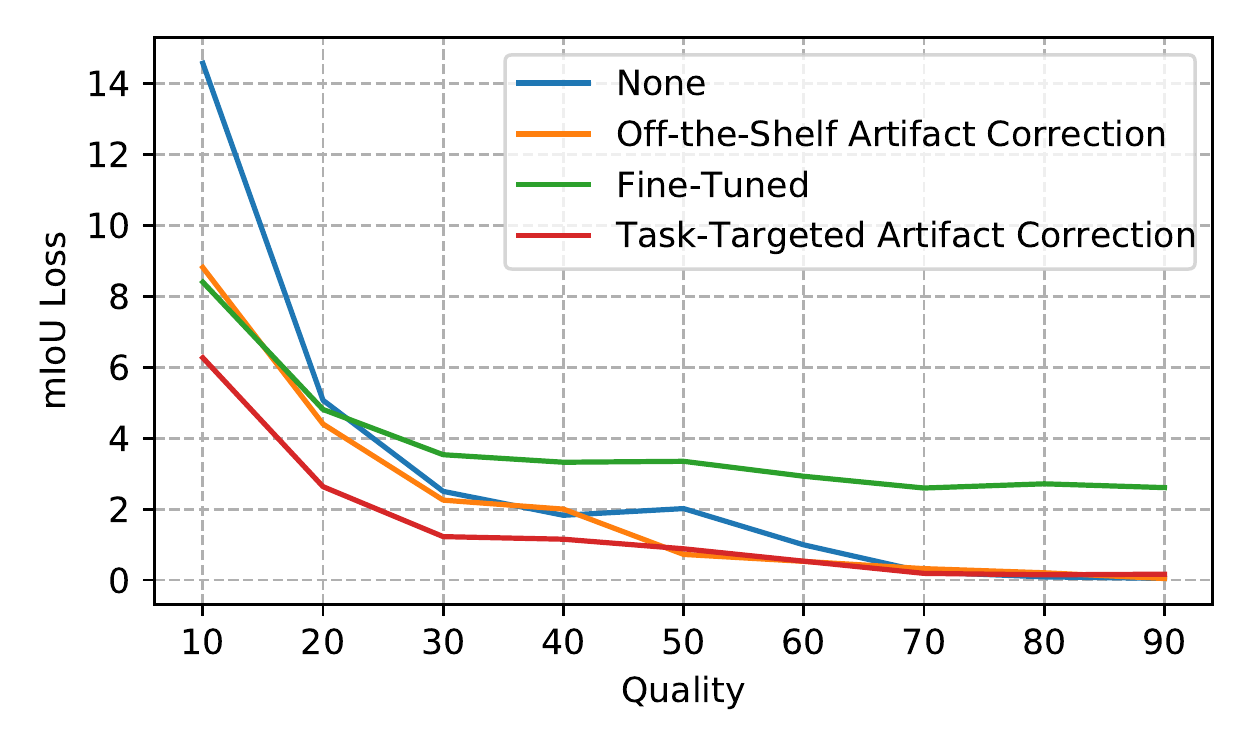}
    \caption[Semantic Segmentation Results: ResNet 101 + PPM]{Semantic Segmentation Results: ResNet 101 + PPM}
\end{figure}

\section{Tables of Results}

\resizebox{\textwidth}{!}{
    \begin{tabular}{llllrrrrrrrrr}
    \toprule
    Model                            & Metric                          & Reference              & Mitigation                        & Q=10  & Q=20  & Q=30  & Q=40  & Q=50  & Q=60  & Q=70  & Q=80  & Q=90  \\
    \midrule
    \multirow{4}{*}{EfficientNet B3} & \multirow{4}{*}{Top-1 Accuracy} & \multirow{4}{*}{83.98} & Supervised Fine-Tuning            & 79.78 & 81.84 & 82.47 & 82.68 & 82.78 & 82.75 & 82.83 & 82.85 & 82.83 \\
                                     &                                 &                        & None                              & 77.24 & 81.11 & 81.95 & 82.52 & 82.67 & 82.91 & 83.10 & 83.37 & 83.75 \\
                                     &                                 &                        & Off-the-Shelf Artifact Correction & 75.92 & 80.02 & 81.47 & 82.12 & 82.44 & 82.71 & 82.94 & 83.23 & 83.70 \\
                                     &                                 &                        & Task-Targeted Artifact Correction & 81.03 & 82.71 & 83.21 & 83.53 & 83.64 & 83.71 & 83.73 & 83.80 & 83.76 \\
    \midrule
    \multirow{4}{*}{InceptionV3}     & \multirow{4}{*}{Top-1 Accuracy} & \multirow{4}{*}{77.33} & Supervised Fine-Tuning            & 75.11 & 77.25 & 77.77 & 77.89 & 78.13 & 78.13 & 78.24 & 78.26 & 78.32 \\
                                     &                                 &                        & None                              & 69.38 & 74.15 & 75.44 & 75.98 & 76.38 & 76.69 & 76.95 & 77.14 & 77.30 \\
                                     &                                 &                        & Off-the-Shelf Artifact Correction & 71.21 & 75.04 & 76.09 & 76.42 & 76.68 & 76.79 & 76.97 & 77.06 & 77.13 \\
                                     &                                 &                        & Task-Targeted Artifact Correction & 73.65 & 75.89 & 76.53 & 76.82 & 76.93 & 76.99 & 77.09 & 77.15 & 77.10 \\
    \midrule
    \multirow{4}{*}{MobileNetV2}     & \multirow{4}{*}{Top-1 Accuracy} & \multirow{4}{*}{70.72} & Supervised Fine-Tuning            & 65.65 & 69.21 & 69.92 & 70.20 & 70.37 & 70.53 & 70.50 & 70.55 & 70.54 \\
                                     &                                 &                        & None                              & 57.23 & 65.55 & 67.87 & 68.95 & 69.47 & 69.98 & 70.24 & 70.60 & 70.86 \\
                                     &                                 &                        & Off-the-Shelf Artifact Correction & 57.33 & 65.25 & 67.76 & 68.93 & 69.60 & 70.07 & 70.40 & 70.71 & 70.58 \\
                                     &                                 &                        & Task-Targeted Artifact Correction & 64.64 & 68.63 & 69.71 & 70.18 & 70.32 & 70.44 & 70.50 & 70.52 & 70.34 \\
    \midrule
    \multirow{4}{*}{ResNet-101}      & \multirow{4}{*}{Top-1 Accuracy} & \multirow{4}{*}{76.91} & Supervised Fine-Tuning            & 74.63 & 76.50 & 77.07 & 77.20 & 77.27 & 77.29 & 77.43 & 77.44 & 77.53 \\
                                     &                                 &                        & None                              & 66.12 & 73.00 & 74.65 & 75.39 & 75.83 & 76.29 & 76.51 & 76.79 & 76.96 \\
                                     &                                 &                        & Off-the-Shelf Artifact Correction & 67.91 & 73.64 & 75.09 & 75.84 & 76.23 & 76.52 & 76.56 & 76.80 & 76.74 \\
                                     &                                 &                        & Task-Targeted Artifact Correction & 72.99 & 75.53 & 76.30 & 76.60 & 76.59 & 76.72 & 76.70 & 76.72 & 76.59 \\
    \midrule
    \multirow{4}{*}{ResNet-18}       & \multirow{4}{*}{Top-1 Accuracy} & \multirow{4}{*}{68.84} & Supervised Fine-Tuning            & 65.49 & 68.46 & 69.07 & 69.16 & 69.36 & 69.33 & 69.38 & 69.53 & 69.49 \\
                                     &                                 &                        & None                              & 57.62 & 65.26 & 67.07 & 67.68 & 68.08 & 68.30 & 68.61 & 68.84 & 68.92 \\
                                     &                                 &                        & Off-the-Shelf Artifact Correction & 61.19 & 66.39 & 67.87 & 68.39 & 68.61 & 68.77 & 68.97 & 68.99 & 68.90 \\
                                     &                                 &                        & Task-Targeted Artifact Correction & 63.83 & 67.06 & 68.04 & 68.24 & 68.35 & 68.48 & 68.52 & 68.60 & 68.50 \\
    \midrule
    \multirow{4}{*}{ResNet-50}       & \multirow{4}{*}{Top-1 Accuracy} & \multirow{4}{*}{75.31} & Supervised Fine-Tuning            & 73.18 & 75.46 & 76.02 & 76.24 & 76.36 & 76.42 & 76.52 & 76.52 & 76.55 \\
                                     &                                 &                        & None                              & 63.43 & 71.20 & 73.23 & 74.10 & 74.43 & 74.63 & 75.01 & 75.09 & 75.34 \\
                                     &                                 &                        & Off-the-Shelf Artifact Correction & 66.90 & 72.45 & 73.95 & 74.60 & 74.93 & 75.18 & 75.26 & 75.42 & 75.30 \\
                                     &                                 &                        & Task-Targeted Artifact Correction & 70.48 & 73.56 & 74.39 & 74.81 & 74.94 & 75.00 & 74.98 & 74.98 & 74.89 \\
    \midrule
    \multirow{4}{*}{ResNeXt-101}     & \multirow{4}{*}{Top-1 Accuracy} & \multirow{4}{*}{78.81} & Supervised Fine-Tuning            & 75.60 & 78.00 & 78.50 & 78.71 & 78.86 & 78.97 & 79.01 & 78.98 & 79.06 \\
                                     &                                 &                        & None                              & 68.83 & 74.84 & 76.39 & 77.05 & 77.60 & 78.00 & 78.16 & 78.56 & 78.75 \\
                                     &                                 &                        & Off-the-Shelf Artifact Correction & 71.19 & 75.88 & 77.14 & 77.80 & 78.15 & 78.30 & 78.57 & 78.66 & 78.61 \\
                                     &                                 &                        & Task-Targeted Artifact Correction & 74.73 & 77.33 & 78.08 & 78.29 & 78.55 & 78.62 & 78.68 & 78.73 & 78.68 \\
    \midrule
    \multirow{4}{*}{ResNeXt-50}      & \multirow{4}{*}{Top-1 Accuracy} & \multirow{4}{*}{76.99} & Supervised Fine-Tuning            & 74.21 & 76.23 & 76.79 & 77.01 & 77.08 & 77.18 & 77.16 & 77.30 & 77.17 \\
                                     &                                 &                        & None                              & 66.96 & 73.21 & 74.85 & 75.62 & 76.07 & 76.37 & 76.63 & 76.88 & 77.06 \\
                                     &                                 &                        & Off-the-Shelf Artifact Correction & 68.05 & 73.56 & 75.11 & 75.95 & 76.38 & 76.59 & 76.71 & 76.99 & 76.90 \\
                                     &                                 &                        & Task-Targeted Artifact Correction & 72.22 & 75.45 & 76.09 & 76.62 & 76.86 & 76.83 & 76.85 & 76.99 & 76.81 \\
    \midrule
    \multirow{4}{*}{VGG-19}          & \multirow{4}{*}{Top-1 Accuracy} & \multirow{4}{*}{73.44} & Supervised Fine-Tuning            & 69.50 & 72.66 & 73.29 & 73.74 & 73.83 & 73.85 & 73.95 & 74.14 & 74.11 \\
                                     &                                 &                        & None                              & 59.27 & 68.08 & 70.49 & 71.53 & 71.99 & 72.42 & 72.80 & 73.24 & 73.46 \\
                                     &                                 &                        & Off-the-Shelf Artifact Correction & 61.93 & 68.79 & 70.82 & 71.83 & 72.50 & 72.94 & 73.13 & 73.40 & 73.44 \\
                                     &                                 &                        & Task-Targeted Artifact Correction & 67.50 & 71.32 & 72.33 & 72.76 & 73.03 & 73.16 & 73.50 & 73.48 & 73.44 \\
    \bottomrule
\end{tabular}
}
\captionof{table}{Results for classification models.}
\medskip

\resizebox{\textwidth}{!}{
    \begin{tabular}{llllrrrrrrrrr}
    \toprule
    Model                       & Metric               & Reference              & Mitigation                        & Q=10  & Q=20  & Q=30  & Q=40  & Q=50  & Q=60  & Q=70  & Q=80  & Q=90  \\
    \midrule
    \multirow{4}{*}{FasterRCNN} & \multirow{4}{*}{mAP} & \multirow{4}{*}{35.37} & Supervised Fine-Tuning            & 29.09 & 33.34 & 34.72 & 35.08 & 35.49 & 35.82 & 35.96 & 36.06 & 36.17 \\
                                &                      &                        & None                              & 20.35 & 30.03 & 32.59 & 33.43 & 34.04 & 34.31 & 34.73 & 34.93 & 35.25 \\
                                &                      &                        & Off-the-Shelf Artifact Correction & 28.45 & 31.86 & 33.10 & 33.85 & 34.05 & 34.47 & 34.70 & 34.77 & 34.71 \\
                                &                      &                        & Task-Targeted Artifact Correction & 31.43 & 33.85 & 34.29 & 34.81 & 34.81 & 34.97 & 35.01 & 34.88 & 34.81 \\
    \midrule
    \multirow{4}{*}{FastRCNN}   & \multirow{4}{*}{mAP} & \multirow{4}{*}{34.02} & Supervised Fine-Tuning            & 28.01 & 31.94 & 33.08 & 33.56 & 33.88 & 34.17 & 34.42 & 34.44 & 34.66 \\
                                &                      &                        & None                              & 19.99 & 29.04 & 31.22 & 32.19 & 32.65 & 33.00 & 33.34 & 33.40 & 33.80 \\
                                &                      &                        & Off-the-Shelf Artifact Correction & 27.62 & 30.91 & 32.04 & 32.56 & 32.78 & 33.18 & 33.28 & 33.48 & 33.44 \\
                                &                      &                        & Task-Targeted Artifact Correction & 30.11 & 32.31 & 33.07 & 33.31 & 33.39 & 33.53 & 33.69 & 33.68 & 33.59 \\
    \midrule
    \multirow{4}{*}{MaskRCNN}   & \multirow{4}{*}{mAP} & \multirow{4}{*}{32.84} & Supervised Fine-Tuning            & 26.32 & 30.48 & 31.79 & 32.21 & 32.55 & 32.83 & 33.11 & 33.20 & 33.32 \\
                                &                      &                        & None                              & 18.35 & 27.58 & 29.83 & 30.80 & 31.32 & 31.62 & 32.02 & 32.29 & 32.62 \\
                                &                      &                        & Off-the-Shelf Artifact Correction & 25.82 & 29.35 & 30.67 & 31.32 & 31.59 & 31.85 & 32.03 & 32.24 & 32.16 \\
                                &                      &                        & Task-Targeted Artifact Correction & 28.48 & 30.85 & 31.71 & 32.00 & 32.19 & 32.24 & 32.35 & 32.43 & 32.26 \\
    \midrule
    \multirow{4}{*}{RetinaNet}  & \multirow{4}{*}{mAP} & \multirow{4}{*}{33.57} & Supervised Fine-Tuning            & 27.64 & 31.97 & 33.03 & 33.50 & 33.80 & 34.12 & 34.30 & 34.33 & 34.40 \\
                                &                      &                        & None                              & 18.76 & 28.23 & 30.63 & 31.59 & 32.27 & 32.57 & 32.88 & 33.02 & 33.42 \\
                                &                      &                        & Off-the-Shelf Artifact Correction & 26.74 & 29.90 & 31.24 & 31.87 & 32.19 & 32.60 & 32.86 & 33.02 & 32.93 \\
                                &                      &                        & Task-Targeted Artifact Correction & 29.66 & 31.86 & 32.73 & 32.97 & 32.98 & 33.13 & 33.24 & 33.23 & 33.09 \\
    \bottomrule
\end{tabular}
}
\captionof{table}{Results for detection models.}
\medskip

\resizebox{\textwidth}{!}{
    \begin{tabular}{llllrrrrrrrrr}
    \toprule
    Model                                            & Metric                & Reference              & Mitigation                        & Q=10  & Q=20  & Q=30  & Q=40  & Q=50  & Q=60  & Q=70  & Q=80  & Q=90  \\
    \midrule
    \multirow{4}{*}{HRNetV2 + C1}                    & \multirow{4}{*}{mIoU} & \multirow{4}{*}{40.59} & Supervised Fine-Tuning            & 34.76 & 37.35 & 38.74 & 38.78 & 39.27 & 39.75 & 39.98 & 39.86 & 39.96 \\
                                                     &                       &                        & None                              & 24.95 & 35.16 & 38.03 & 38.52 & 39.02 & 40.09 & 40.50 & 40.41 & 40.54 \\
                                                     &                       &                        & Off-the-Shelf Artifact Correction & 32.30 & 36.54 & 38.40 & 38.52 & 40.08 & 40.44 & 40.46 & 40.22 & 40.60 \\
                                                     &                       &                        & Task-Targeted Artifact Correction & 34.14 & 37.61 & 39.23 & 39.24 & 39.92 & 40.53 & 40.62 & 40.39 & 40.55 \\
    \midrule
    \multirow{4}{*}{MobileNetV2 (dilated) + C1 (ds)} & \multirow{4}{*}{mIoU} & \multirow{4}{*}{29.52} & Supervised Fine-Tuning            & 19.07 & 22.37 & 23.43 & 23.62 & 23.60 & 24.15 & 24.44 & 24.37 & 24.46 \\
                                                     &                       &                        & None                              & 13.92 & 24.03 & 27.13 & 27.75 & 27.73 & 28.86 & 29.37 & 29.35 & 29.43 \\
                                                     &                       &                        & Off-the-Shelf Artifact Correction & 21.17 & 25.27 & 27.31 & 27.16 & 29.14 & 29.32 & 29.26 & 29.06 & 29.54 \\
                                                     &                       &                        & Task-Targeted Artifact Correction & 24.74 & 27.37 & 28.44 & 28.33 & 29.19 & 29.56 & 29.54 & 29.38 & 29.52 \\
    \midrule
    \multirow{4}{*}{ResNet101 + UPerNet}             & \multirow{4}{*}{mIoU} & \multirow{4}{*}{41.08} & Supervised Fine-Tuning            & 35.32 & 37.41 & 38.27 & 38.28 & 38.55 & 38.59 & 38.72 & 38.58 & 38.70 \\
                                                     &                       &                        & None                              & 26.14 & 36.70 & 39.45 & 39.81 & 39.55 & 40.47 & 40.98 & 40.97 & 41.07 \\
                                                     &                       &                        & Off-the-Shelf Artifact Correction & 33.90 & 37.39 & 39.12 & 39.38 & 40.32 & 40.58 & 40.78 & 40.79 & 41.04 \\
                                                     &                       &                        & Task-Targeted Artifact Correction & 35.82 & 38.67 & 39.96 & 39.98 & 40.22 & 40.79 & 40.97 & 40.91 & 41.00 \\
    \midrule
    \multirow{4}{*}{ResNet101 (dilated) + PPM}       & \multirow{4}{*}{mIoU} & \multirow{4}{*}{40.26} & Supervised Fine-Tuning            & 31.86 & 35.45 & 36.73 & 36.94 & 36.91 & 37.33 & 37.67 & 37.55 & 37.65 \\
                                                     &                       &                        & None                              & 25.68 & 35.19 & 37.76 & 38.43 & 38.24 & 39.27 & 40.03 & 40.17 & 40.21 \\
                                                     &                       &                        & Off-the-Shelf Artifact Correction & 31.44 & 35.86 & 38.01 & 38.26 & 39.54 & 39.73 & 39.94 & 40.06 & 40.22 \\
                                                     &                       &                        & Task-Targeted Artifact Correction & 33.99 & 37.63 & 39.04 & 39.11 & 39.38 & 39.73 & 40.07 & 40.11 & 40.10 \\
    \midrule
    \multirow{4}{*}{ResNet18 (dilated) + PPM}        & \multirow{4}{*}{mIoU} & \multirow{4}{*}{36.65} & Supervised Fine-Tuning            & 29.84 & 32.33 & 33.08 & 33.01 & 33.38 & 33.61 & 33.50 & 33.29 & 33.33 \\
                                                     &                       &                        & None                              & 21.16 & 31.99 & 34.72 & 35.36 & 35.41 & 36.16 & 36.56 & 36.60 & 36.59 \\
                                                     &                       &                        & Off-the-Shelf Artifact Correction & 28.64 & 32.59 & 34.56 & 34.53 & 35.96 & 36.21 & 36.29 & 36.25 & 36.64 \\
                                                     &                       &                        & Task-Targeted Artifact Correction & 31.69 & 34.55 & 35.80 & 35.80 & 36.12 & 36.50 & 36.66 & 36.54 & 36.60 \\
    \midrule
    \multirow{4}{*}{ResNet50 + UPerNet}              & \multirow{4}{*}{mIoU} & \multirow{4}{*}{39.21} & Supervised Fine-Tuning            & 32.88 & 35.11 & 35.94 & 35.90 & 36.41 & 36.58 & 36.63 & 36.49 & 36.55 \\
                                                     &                       &                        & None                              & 24.29 & 34.78 & 37.34 & 37.71 & 37.70 & 38.57 & 39.12 & 39.13 & 39.16 \\
                                                     &                       &                        & Off-the-Shelf Artifact Correction & 31.83 & 35.52 & 37.20 & 37.26 & 38.44 & 38.67 & 38.87 & 38.86 & 39.12 \\
                                                     &                       &                        & Task-Targeted Artifact Correction & 34.36 & 36.94 & 38.17 & 38.07 & 38.55 & 38.93 & 39.14 & 39.06 & 39.09 \\
    \midrule
    \multirow{4}{*}{ResNet50 (dilated) + PPM}        & \multirow{4}{*}{mIoU} & \multirow{4}{*}{38.91} & Supervised Fine-Tuning            & 32.26 & 35.33 & 36.04 & 36.04 & 36.53 & 36.75 & 36.93 & 36.71 & 36.92 \\
                                                     &                       &                        & None                              & 23.05 & 33.95 & 36.66 & 37.07 & 37.40 & 38.58 & 38.93 & 38.70 & 38.86 \\
                                                     &                       &                        & Off-the-Shelf Artifact Correction & 28.36 & 32.69 & 35.24 & 35.31 & 37.74 & 38.04 & 38.18 & 38.13 & 38.73 \\
                                                     &                       &                        & Task-Targeted Artifact Correction & 31.92 & 35.43 & 37.04 & 36.92 & 38.05 & 38.69 & 38.79 & 38.52 & 38.74 \\
    \bottomrule
\end{tabular}
}
\captionof{table}{Results for segmentation models.}
\medskip

\begin{center}
    \begin{tabular}{lr}
    \toprule
    Model                      & Value                                                             \\
    \midrule
    \multicolumn{2}{@{}l}{ImageNet Classification, Metric: Top-1 Accuracy}                         \\
    \cmidrule[\cmidrulewidth]{1-2}
    ResNet 18                  & 68.84                                                             \\
    ResNet 50                  & 75.31                                                             \\
    ResNet 101                 & 76.91                                                             \\
    ResNeXt 50                 & 76.99                                                             \\
    ResNeXt 101                & 78.81                                                             \\
    VGG 19                     & 73.44                                                             \\
    MobileNetV2                & 70.72                                                             \\
    InceptionV3                & 77.33                                                             \\
    EfficientNet B3            & 83.98                                                             \\
    \midrule
    \multicolumn{2}{@{}l}{COCO Object Detection and Instance Segmentation, Metric: mAP}            \\
    \cmidrule[\cmidrulewidth]{1-2}
    FastRCNN                   & 34.02                                                             \\
    FasterRCNN                 & 35.38                                                             \\
    RetinaNet                  & 33.57                                                             \\
    MaskRCNN                   & 32.84                                                             \\
    \midrule
    \multicolumn{2}{@{}l}{ADE20k Semantic Segmentation, Metric: mIoU}                              \\
    \cmidrule[\cmidrulewidth]{1-2}
    HRNetV2 + C1               & 40.59                                                             \\
    MobileNetV2 (dilated) + C1 & 29.52                                                             \\
    ResNet 18 (dilated) + PPM  & 36.65                                                             \\
    ResNet 50 (dilated) + PPM  & 38.91                                                             \\
    ResNet 101                 & 41.08                                                             \\
    ResNet 101 (dilated) + PPM & 40.26                                                             \\
    \bottomrule
\end{tabular}
\end{center}
\captionof{table}{Reference results (results with no compression).}
\chapter{Additional Results}
\label{app:ar}

In this appendix we examine more interesting outputs from various methods discussed in the body of the dissertation. These are mostly qualitative results. While these images are not critical to understanding the methods, everyone likes looking at pictures!

\begin{warningbox}
    The results presented here are intended to be reproductions from the published papers, so there may be some repeats from the body of the dissertation.
\end{warningbox}

\section{Quantization Guided JPEG Artifact Correction}

These results are from the method presented in \nrefch{qgac}.

We first show more equivalent quality examples next, recall that equivalent quality performs restoration on an image then uses SSIM to find the matching JPEG quality to the restored image which can give an indication of how much space is saved by using QGAC.

\begin{figure}[H]
    \centering
    \includegraphics[width=0.7\textwidth]{figures/eq_qual.pdf}
    \caption{Equivalent quality visualizations. For each image we show the input JPEG, the JPEG with equivalent SSIM
        to our model output, and our model output.}
    \label{fig:eq_qual_res}
\end{figure}

Next we show the full frequency domain results. Recall that these results show the frequency domain content of the images comparing JPEG compression, regression restoration, and GAN restoration.

\begin{figure}[H]
    \centering
    \includegraphics{figures/freq_woman.pdf}
    \caption{Frequency domain results 1/4}
\end{figure}
\begin{figure}[H]
    \centering
    \includegraphics{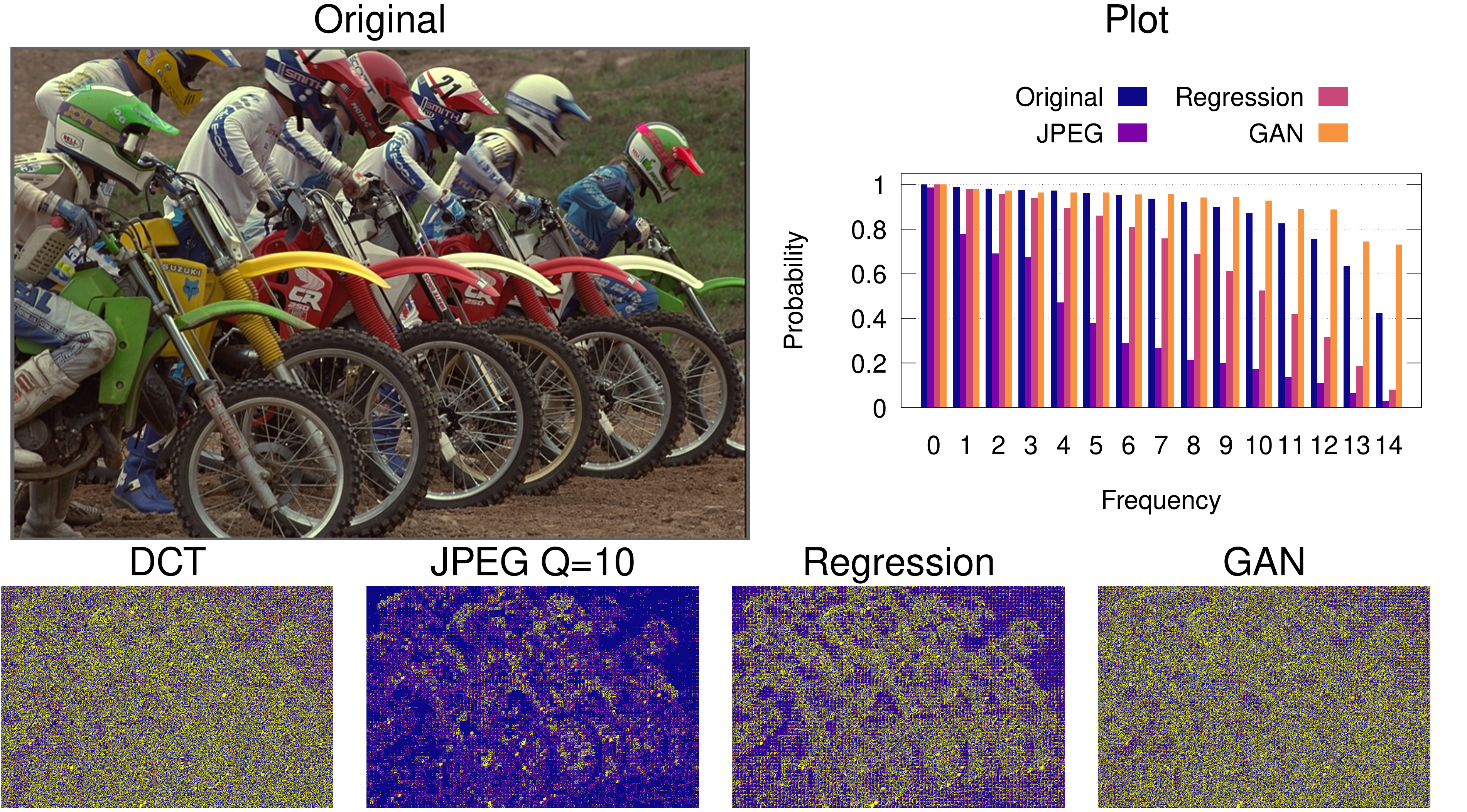}
    \caption{Frequency domain results 2/4}
\end{figure}
\begin{figure}[H]
    \centering
    \includegraphics{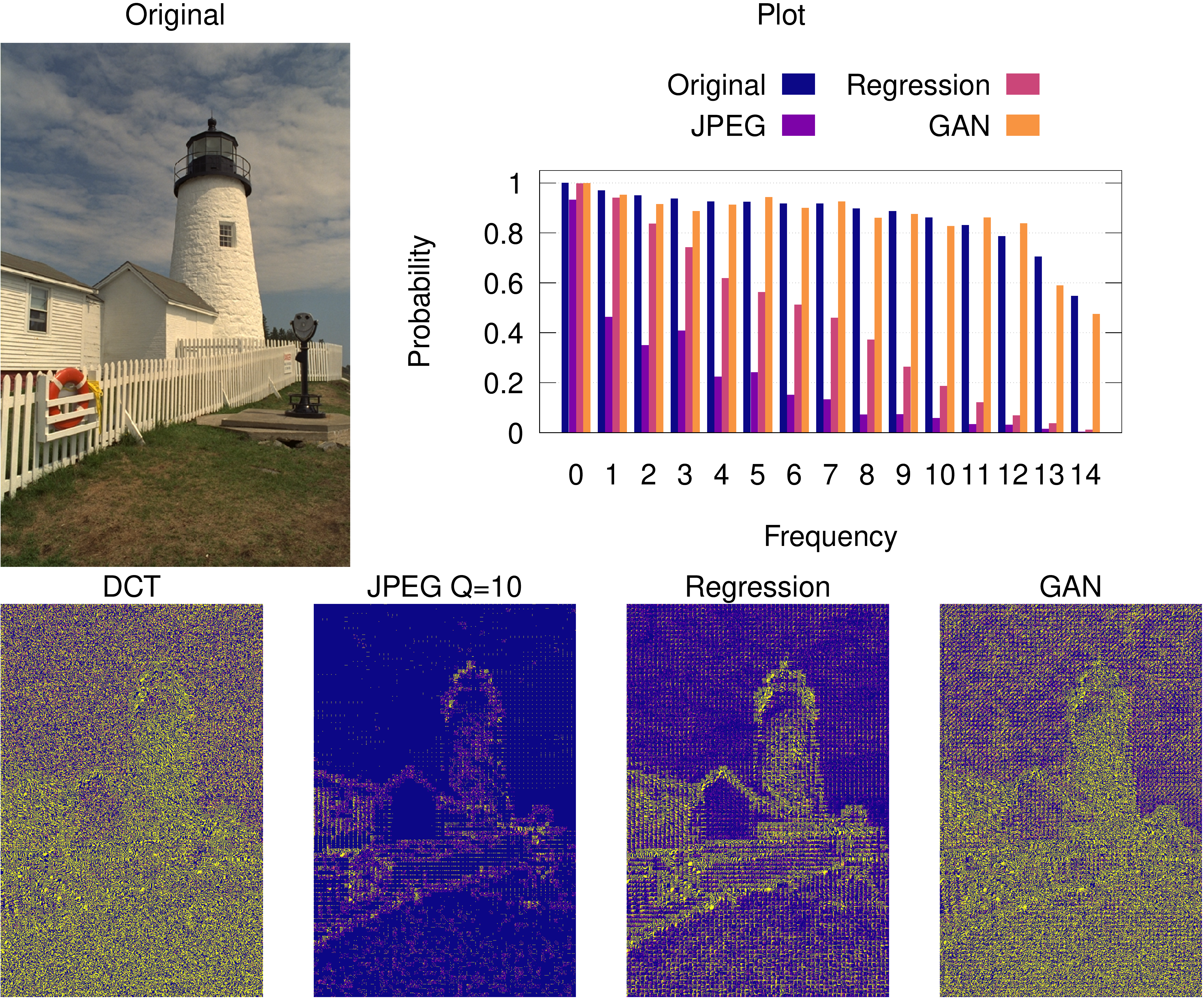}
    \caption{Frequency domain results 3/4.}
\end{figure}
\begin{figure}[H]
    \centering
    \includegraphics{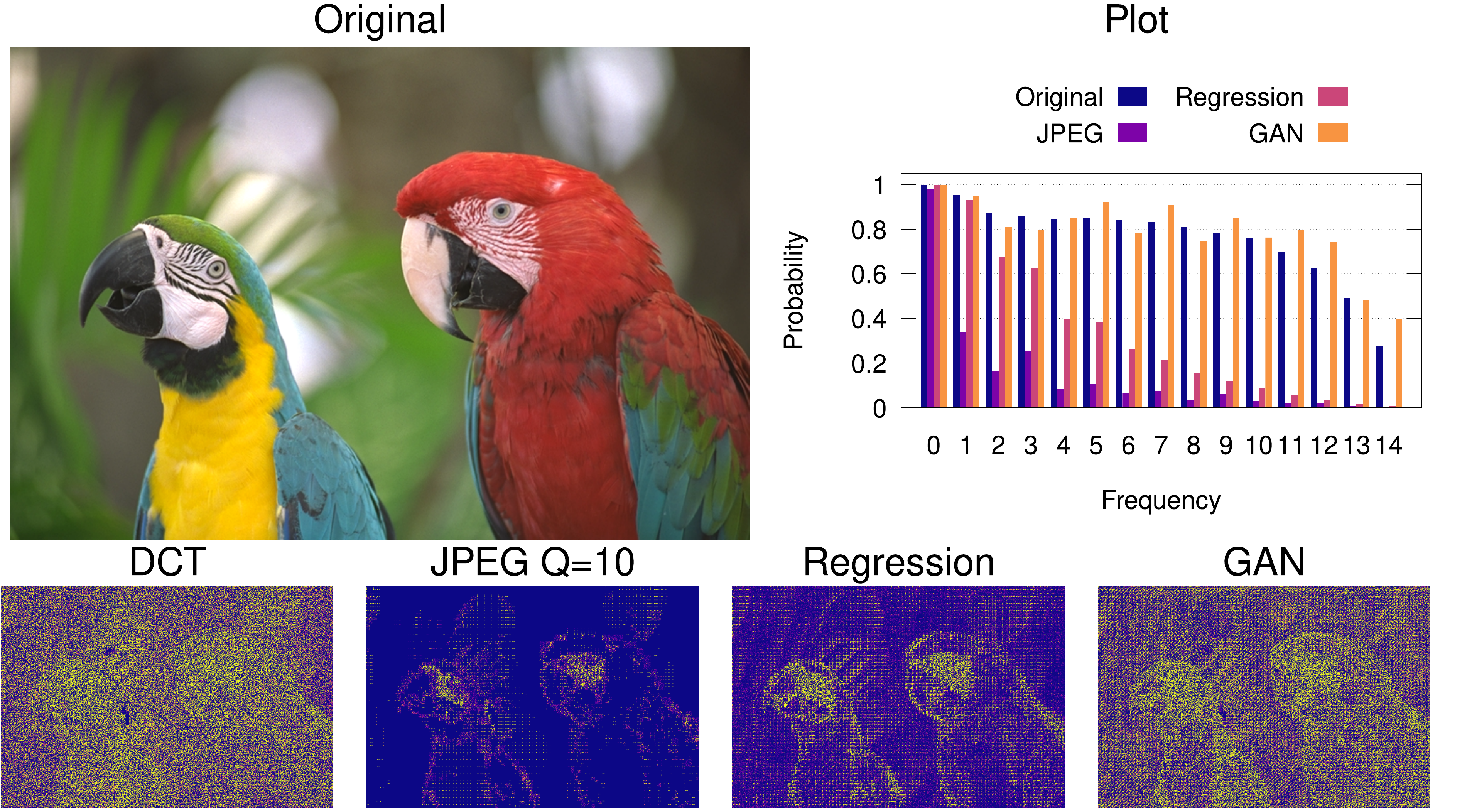}
    \caption{Frequency domain results 4/4.}
\end{figure}

One way to reduce any artifacts caused by divergent GAN training is to use model interpolation \parencite{wang2018esrgan}. Model interpolation simply takes the regression weights $W_R$ and the GAN weights $W_G$ along with a scalar $\alpha$ and computes new model parameters
\begin{align}
    W_I = (1 - \alpha)W_R + \alpha W_G
\end{align}
We show close up views of different textured regions for different choices of $\alpha$.

\begin{figure}[H]
    \centering
    \includegraphics[width=0.6\textwidth]{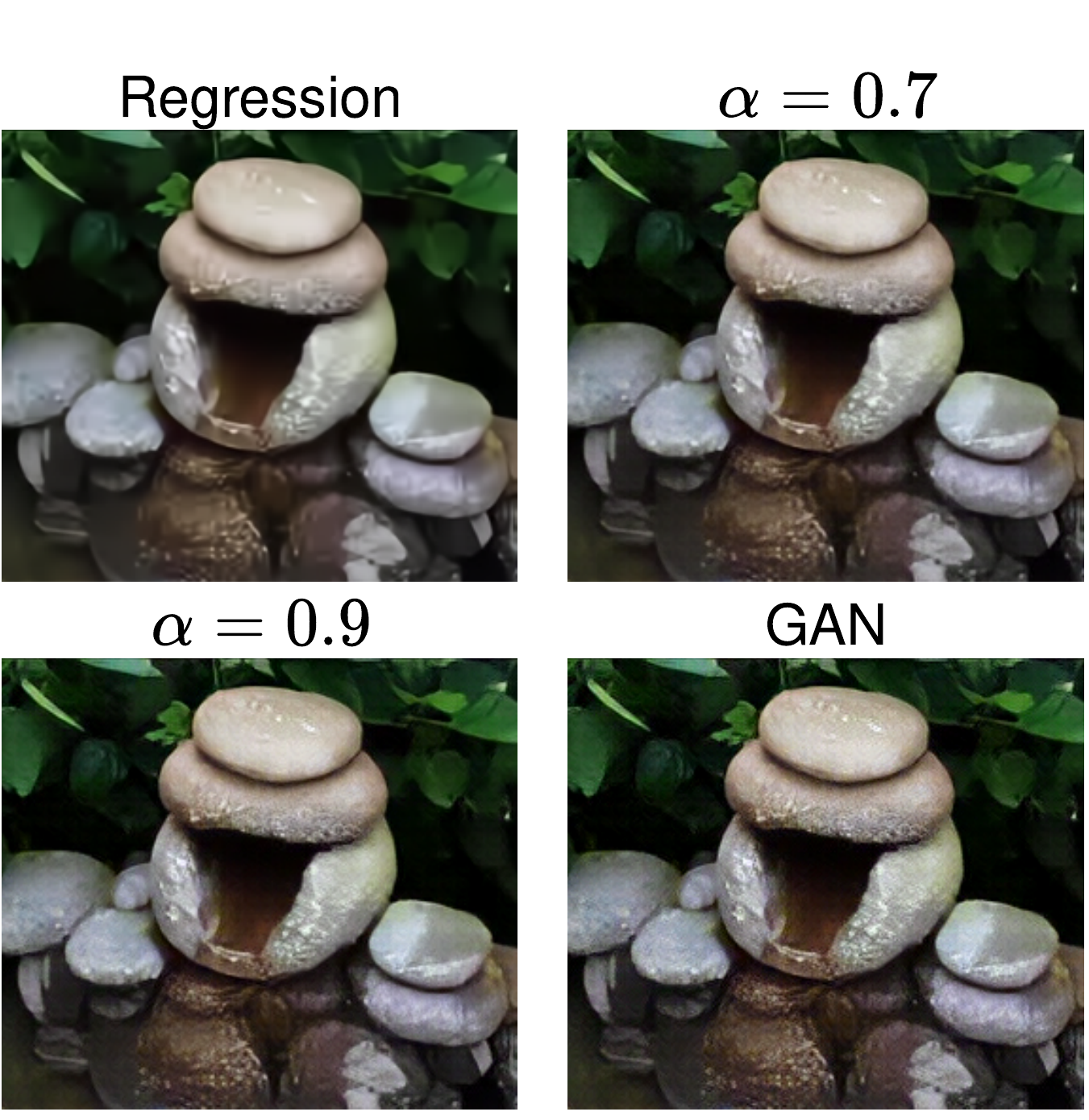}
    \caption{Model interpolation results 1/4}
\end{figure}

\begin{figure}[H]
    \centering
    \includegraphics[width=0.75\textwidth]{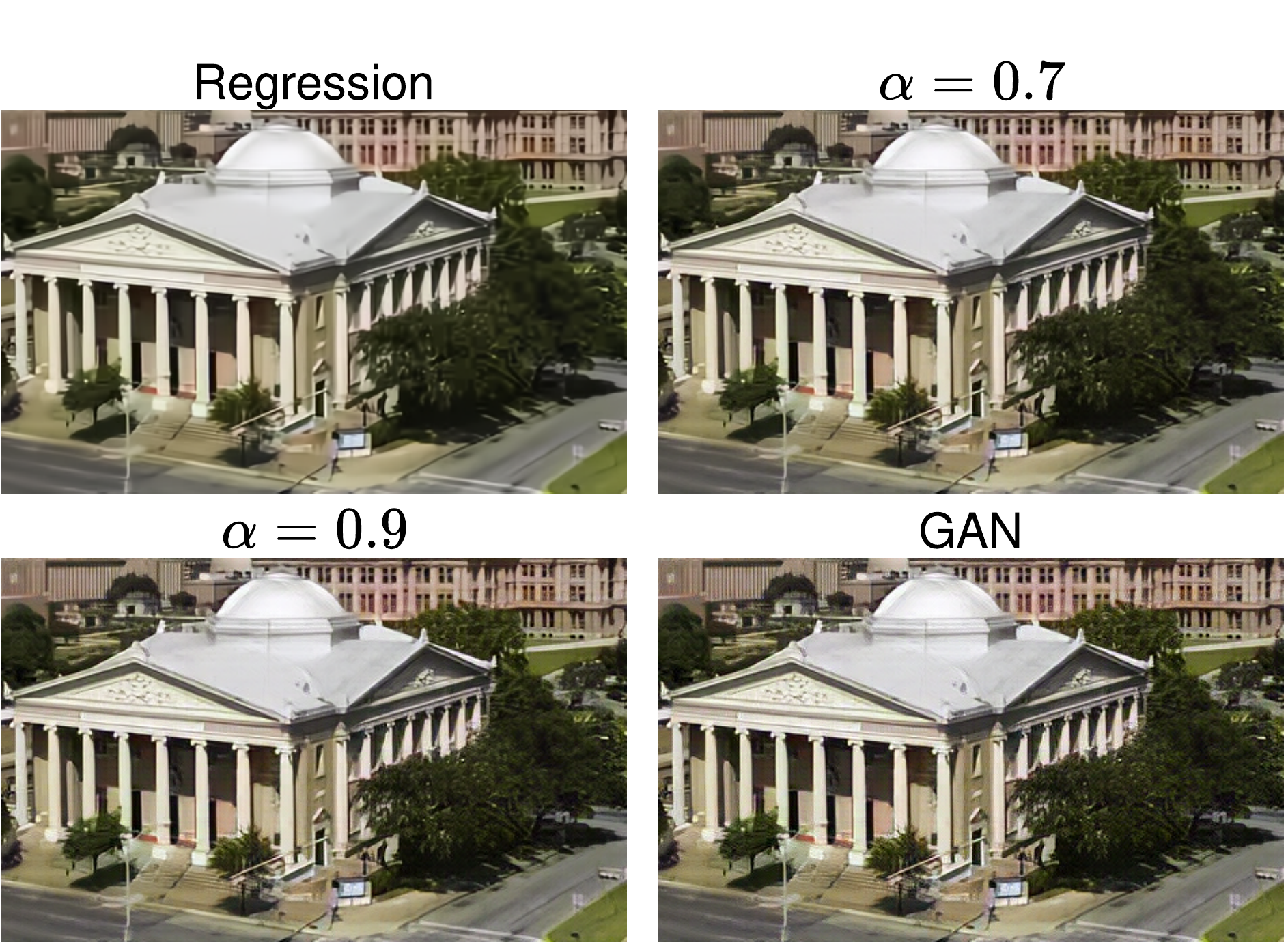}
    \caption{Model interpolation results 2/4}
\end{figure}

\begin{figure}[H]
    \centering
    \includegraphics[width=0.8\textwidth]{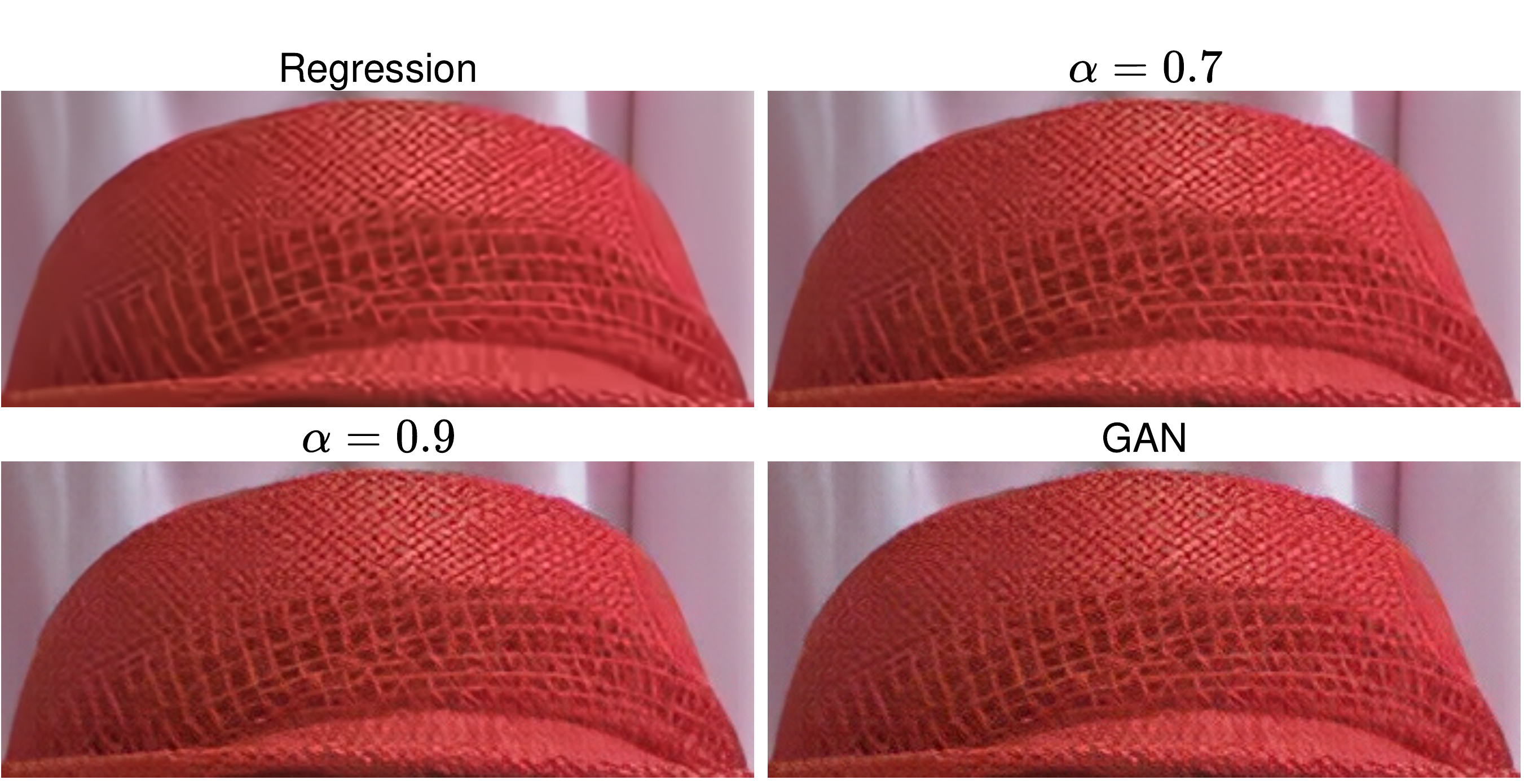}
    \caption{Model interpolation results 3/4}
\end{figure}

\begin{figure}[H]
    \centering
    \includegraphics[width=0.8\textwidth]{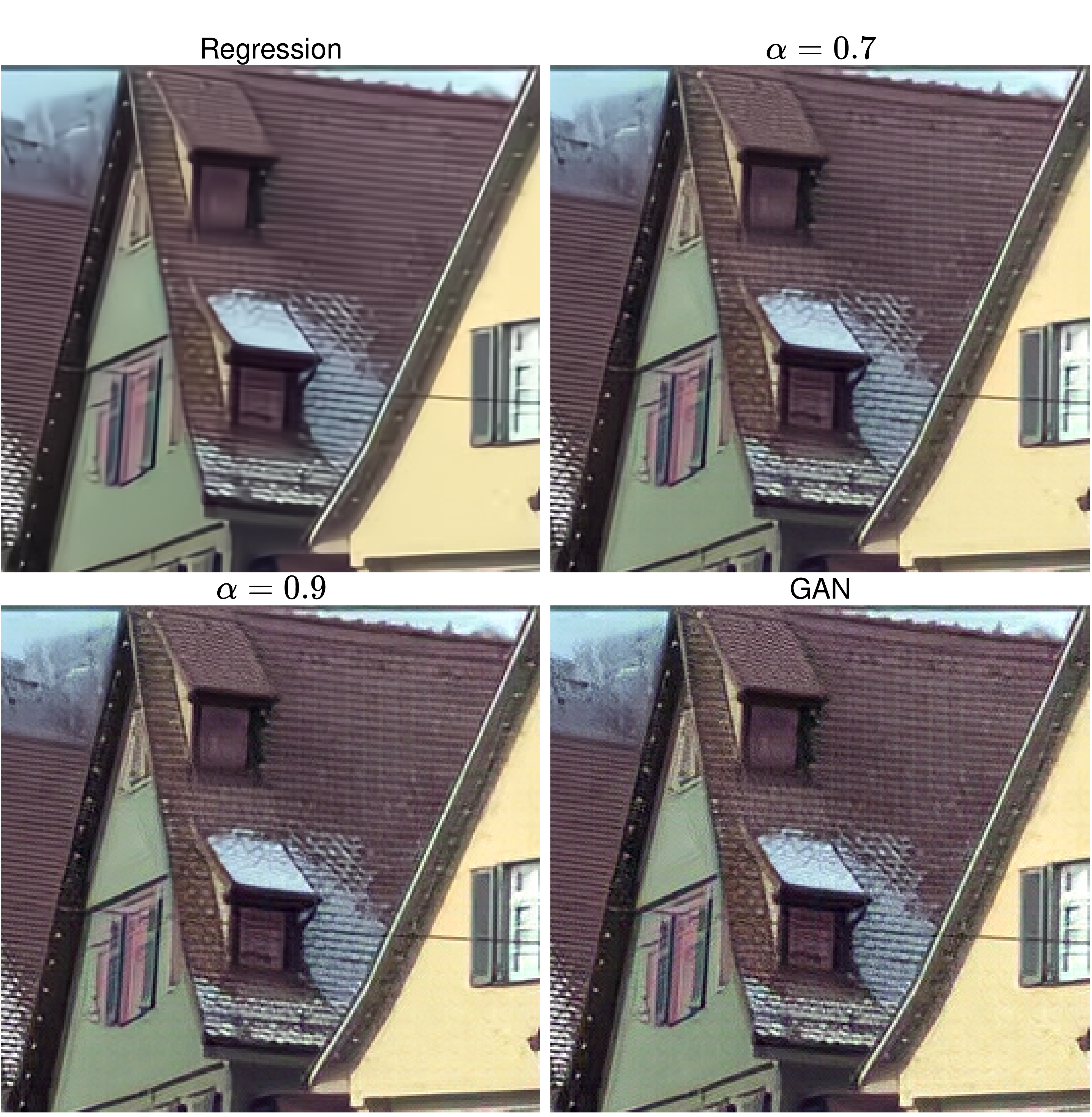}
    \caption{Model interpolation results 4/4}
    \label{fig:model_interp}
\end{figure}

We close with purely qualitative results. These are for quality 10 as in \nrefch{qgac} and quality 20 which was not shown there to save space.

\begin{figure}[H]
    \centering
    \includegraphics{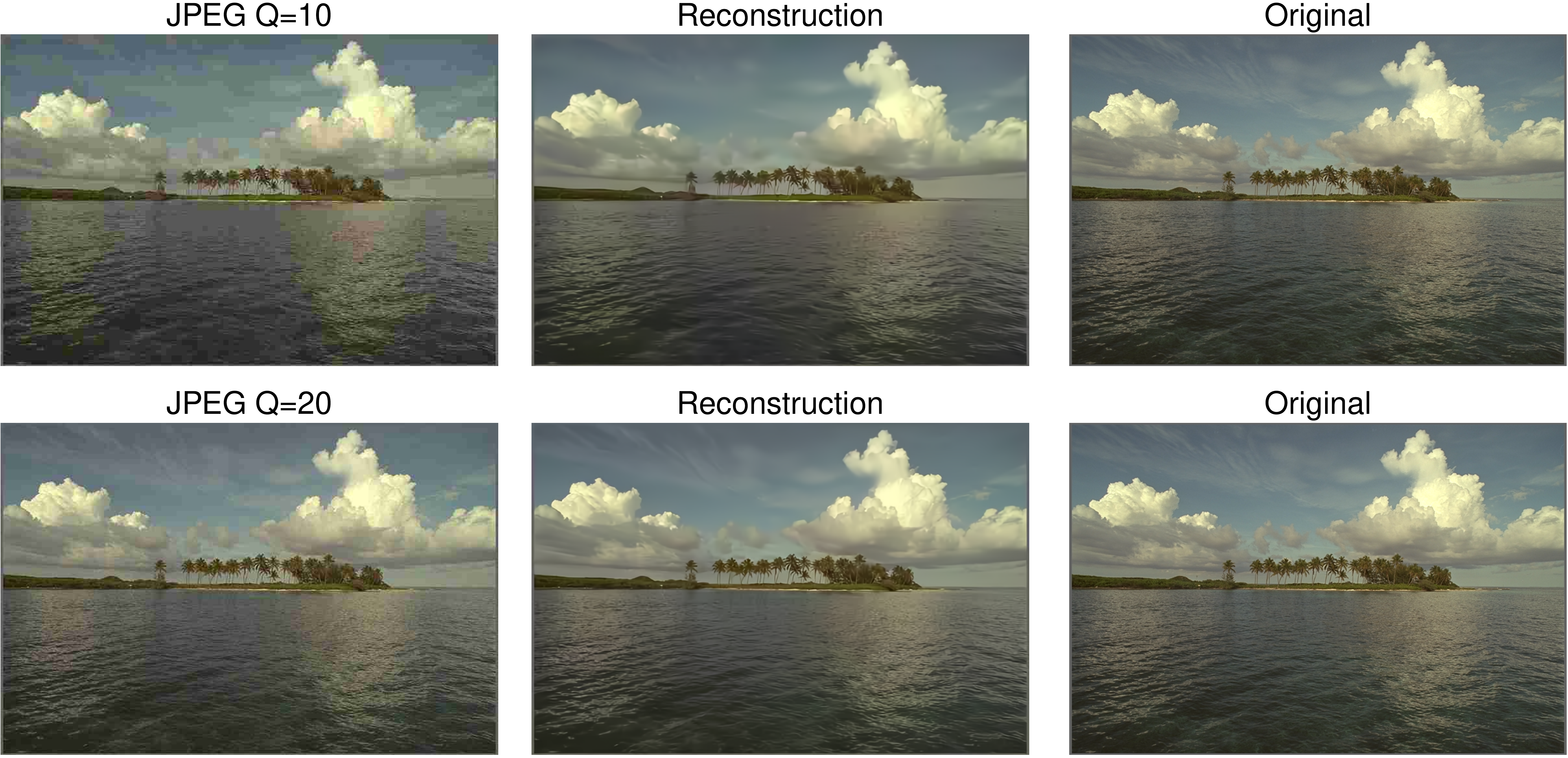}
    \caption{Qualitative results 1/4. Live-1 images.}
\end{figure}
\begin{figure}[H]
    \centering
    \includegraphics{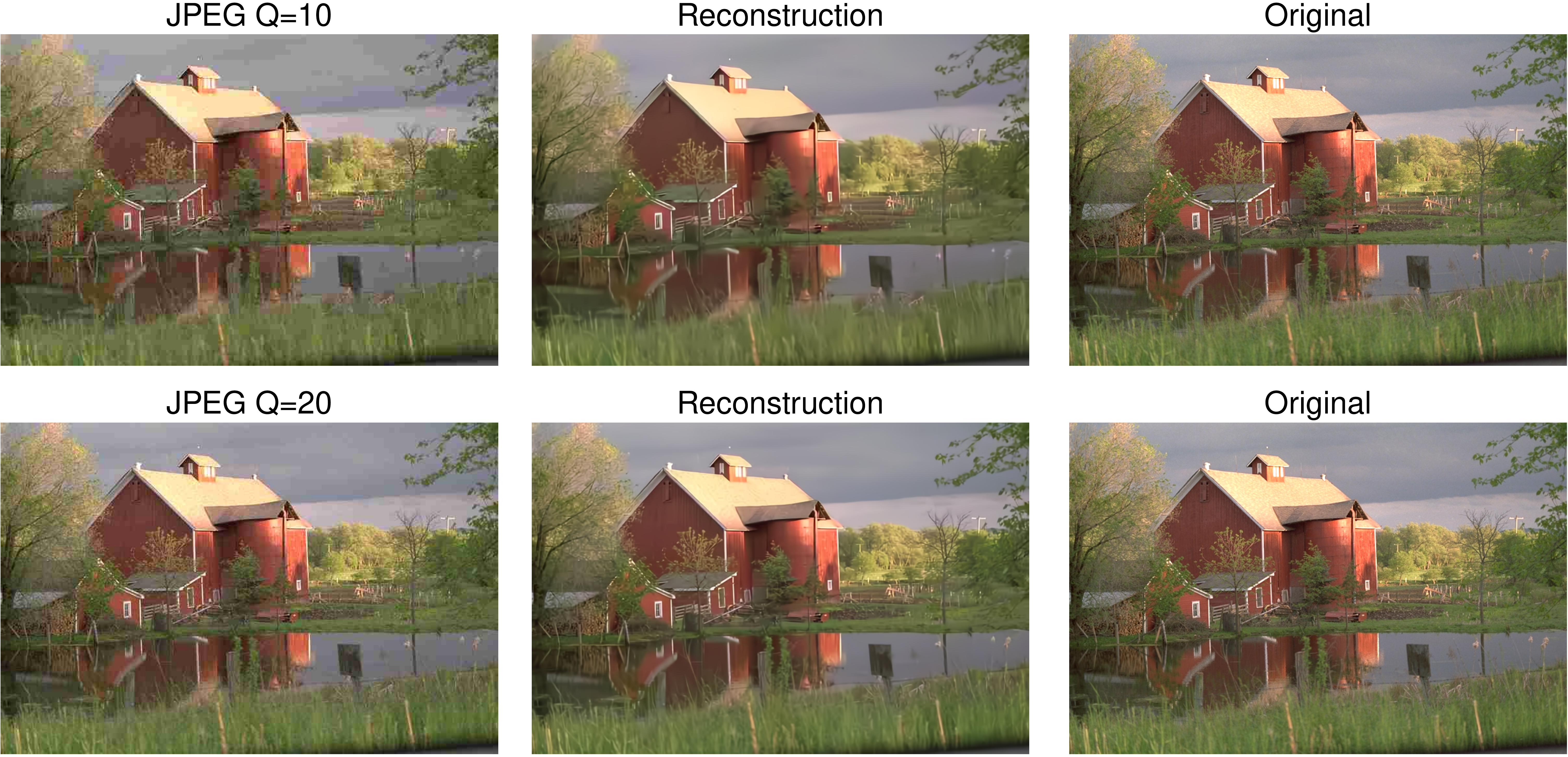}
    \caption{Qualitative results 2/4. Live-1 images.}
\end{figure}

\begin{figure}[H]
    \centering
    \includegraphics{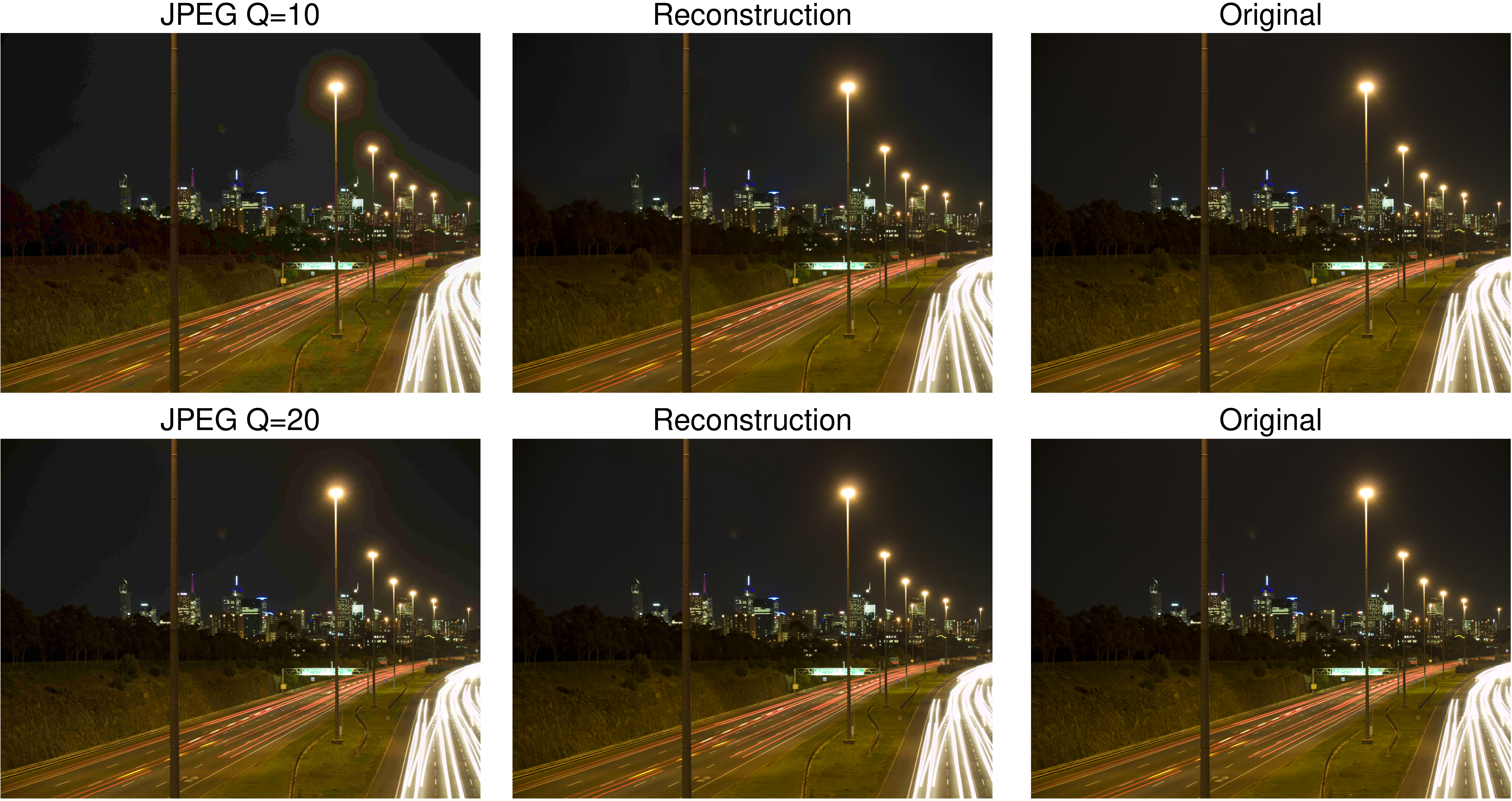}
    \caption{Qualitative results 3/4. Live-1 images.}
\end{figure}
\begin{figure}[H]
    \centering
    \includegraphics[width=0.9\textwidth]{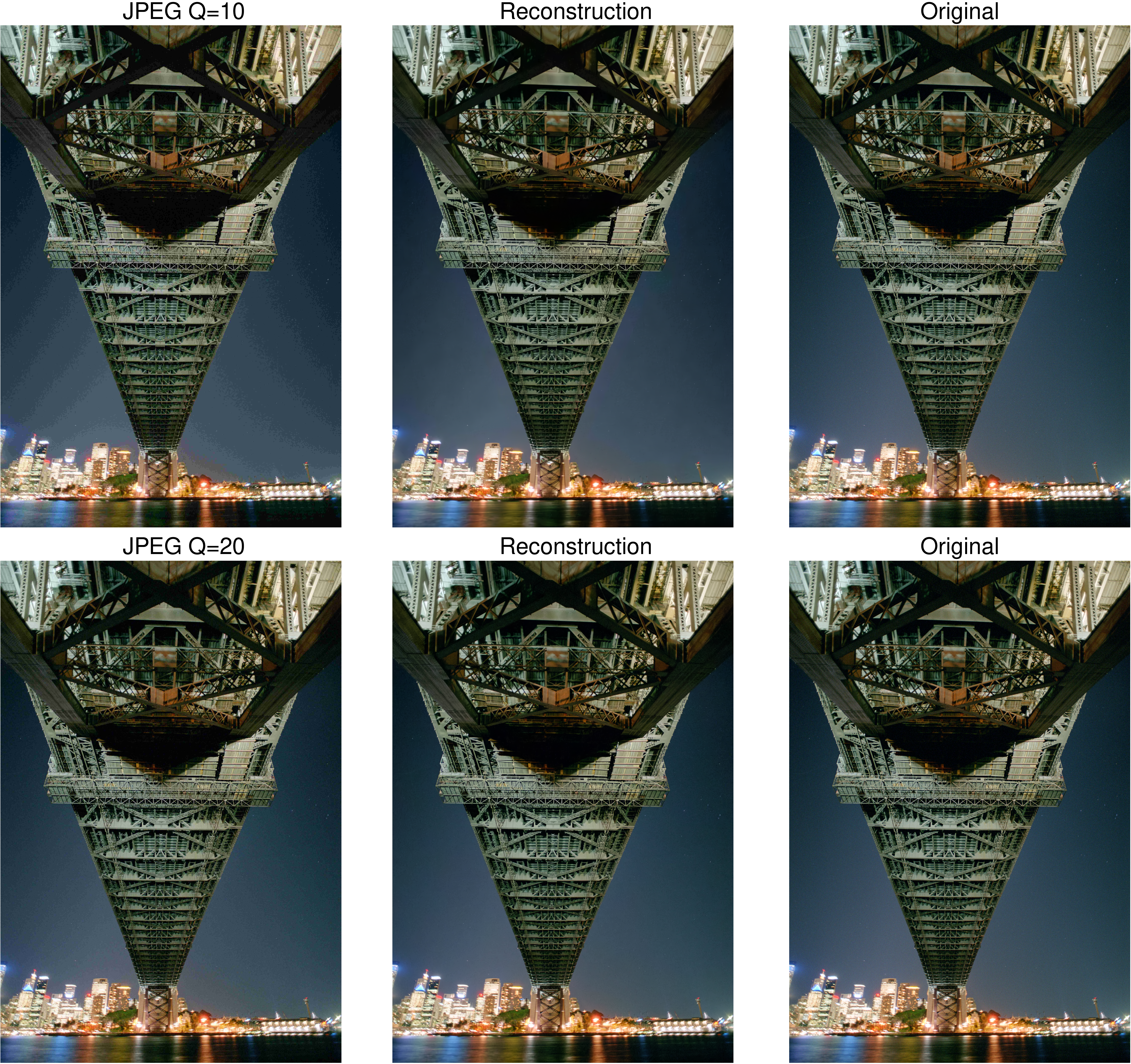}
    \caption{Qualitative results 4/4. ICB images.}
\end{figure}

\section{Task Targeted Artifact Correction}

These results are from the method of \nrefch{ttac}. We start with visualizations of model errors, first using GradCam \parencite{selvaraju2017grad}. This shows how the model focus is impacted by JPEG compression and how it can be corrected using the various mitigation techniques we studied. The figures show some interesting behavior. In terms of localization, the JPEG compressed input actually does well, and the localization is in fact more accurate than the original model with an uncompressed input. The problem with the JPEG compressed input seems to be with the gradient, which is extremely noisy. Mitigation seems to help with this, with the supervised method providing the cleanest gradient although there is a loss of localization accuracy.

\begin{figure}[H]
    \includegraphics{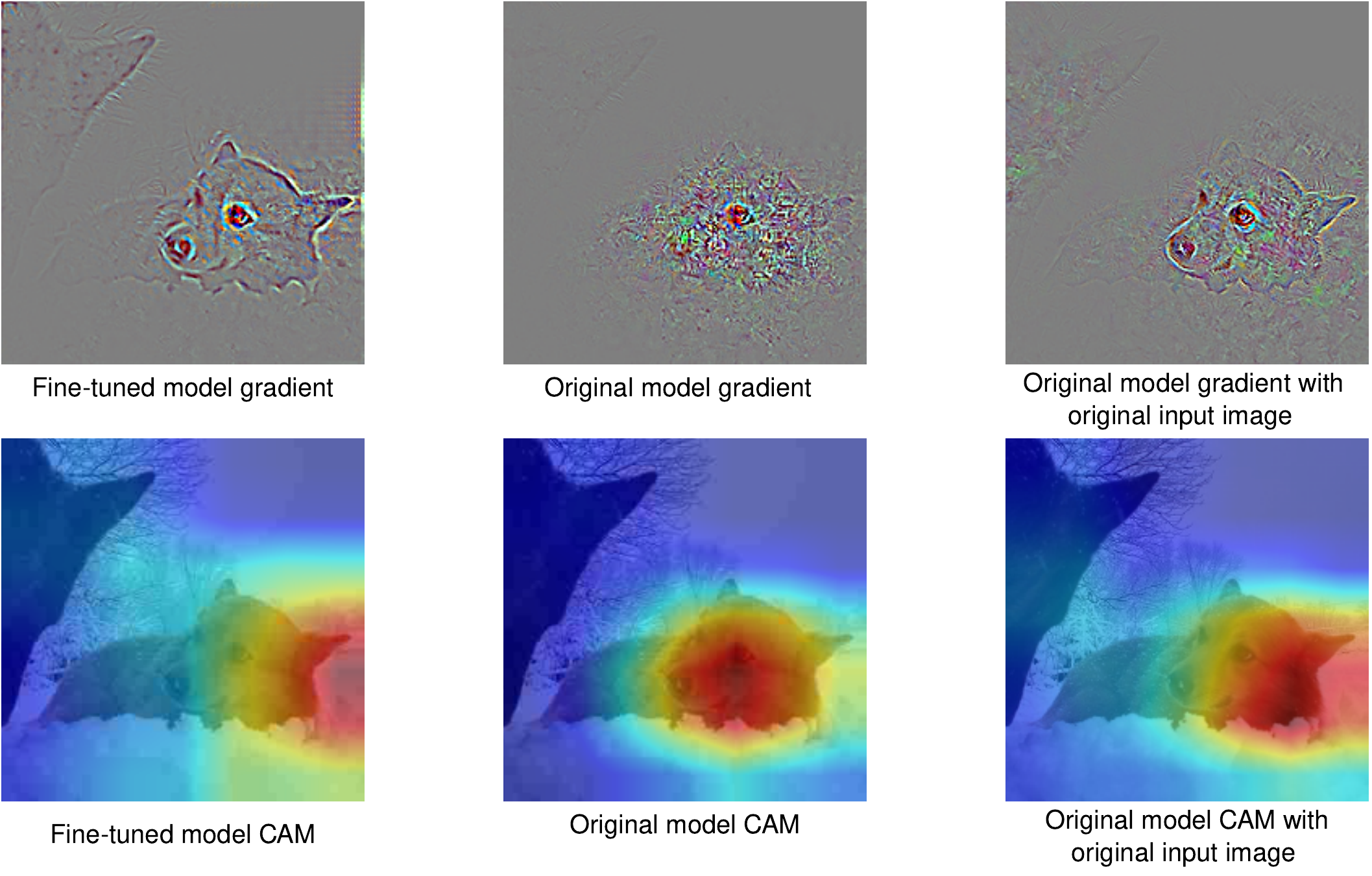}
    \caption{Fine Tuned Model Comparison}
\end{figure}

\begin{figure}[H]
    \includegraphics{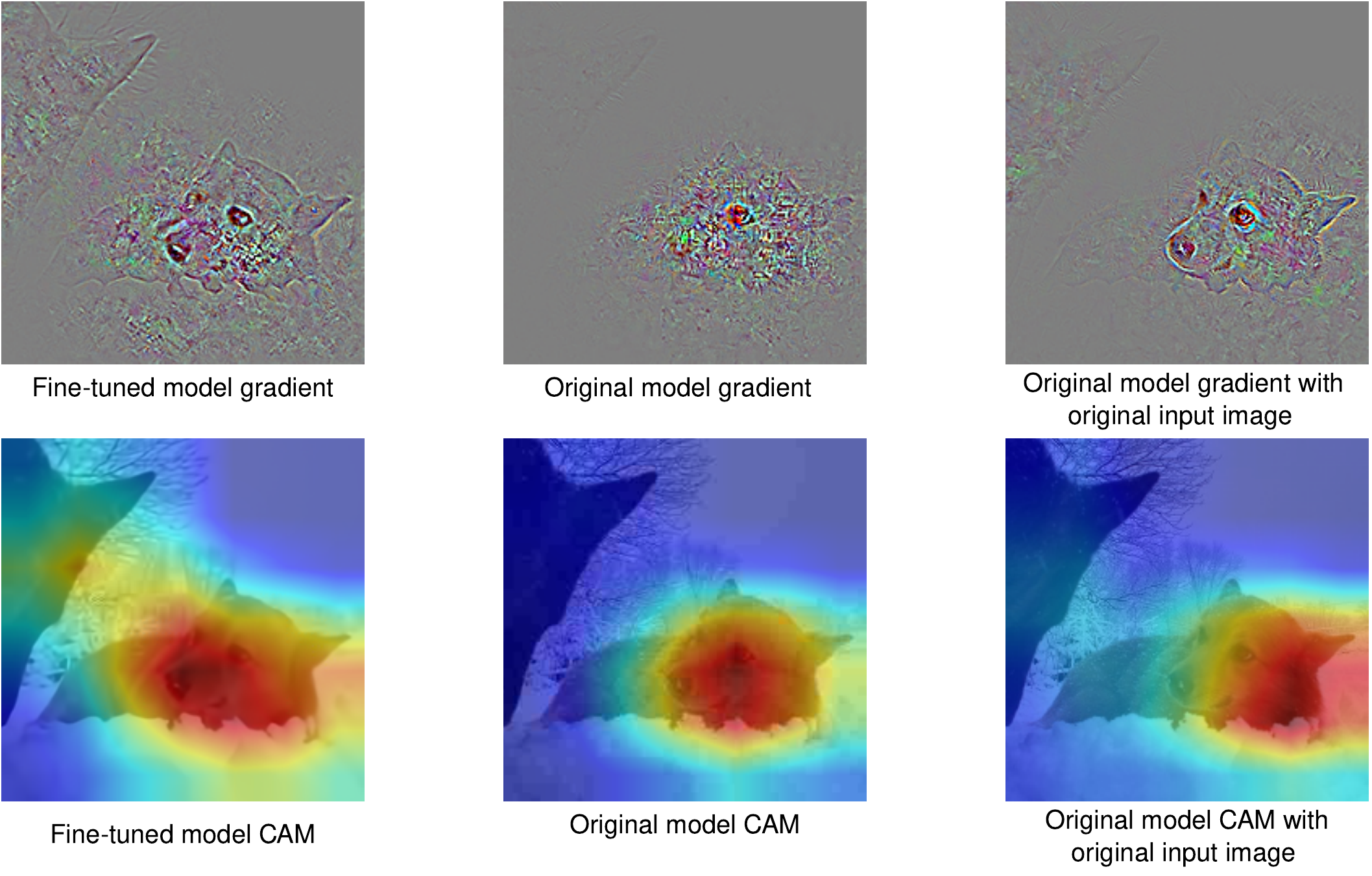}
    \caption{Off-the-Shelf Artifact Correction Comparison}
\end{figure}

\begin{figure}[H]
    \includegraphics{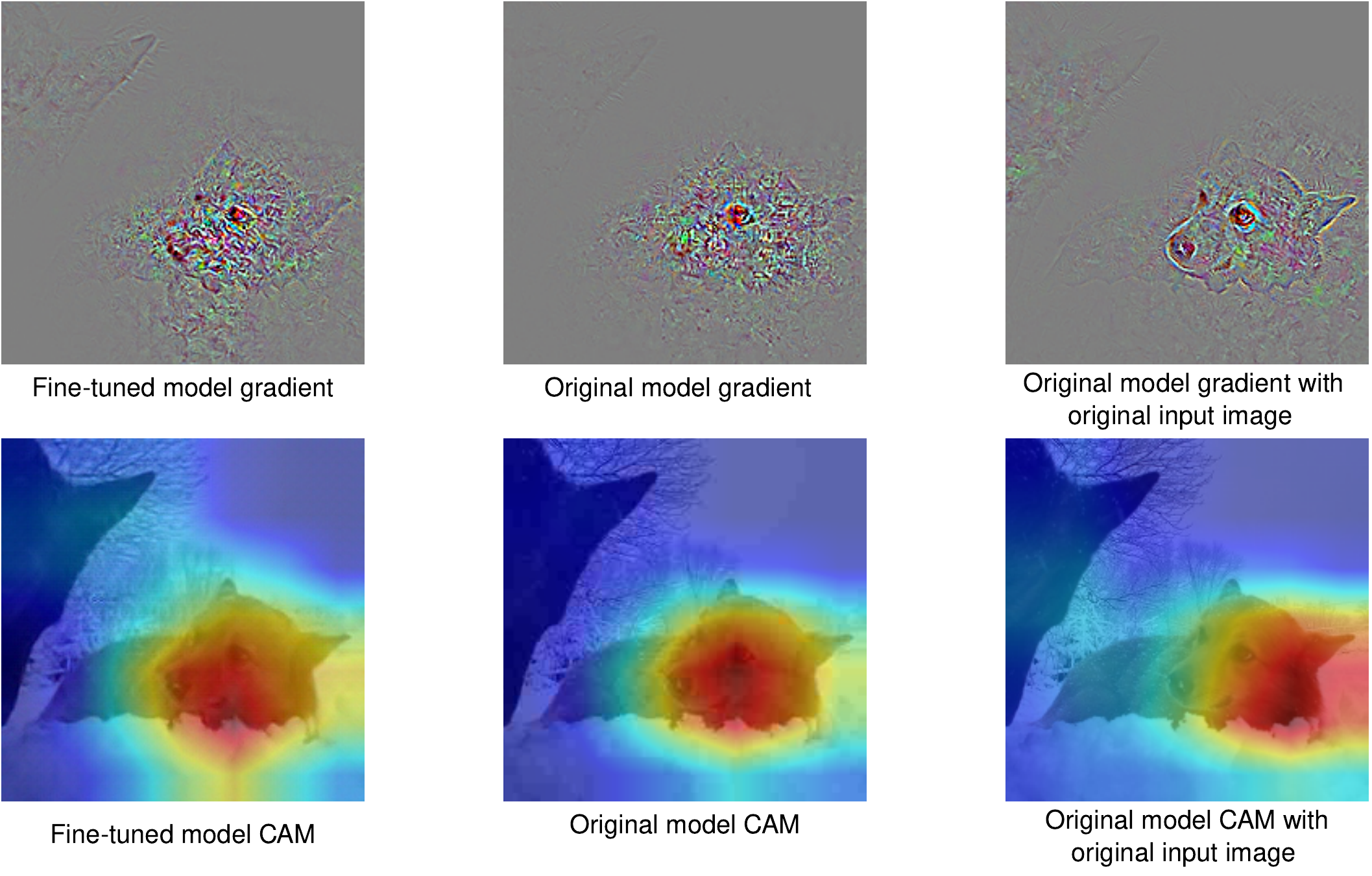}
    \caption{Task-Targeted Artifact Correction Comparison}
\end{figure}

For visualizing detection results we provide plots generated using TIDE \parencite{tide-eccv2020}. We show these for FasterRCNN \parencite{ren2016faster} and MaskRCNN \parencite{he2017mask}. The results show a significant number of missed detection for low quality inputs. This is overtaken by localization errors as quality increases.

\begin{figure}[H]
    \includegraphics{figures/frcnn_tide.pdf}
    \caption{FasterRCNN TIDE Plots. Left: quality 10, Middle: quality 50, Right: quality 100.}
\end{figure}

\begin{figure}[H]
    \includegraphics{figures/mrcnn_tide.pdf}
    \caption{MaskRCNN TIDE Plots. Left: quality 10, Middle: quality 50, Right: quality 100.}
\end{figure}

We close the section with qualitative results including visualizations of the results where appropriate.

\begin{figure}[H]
    \centering
    \includegraphics{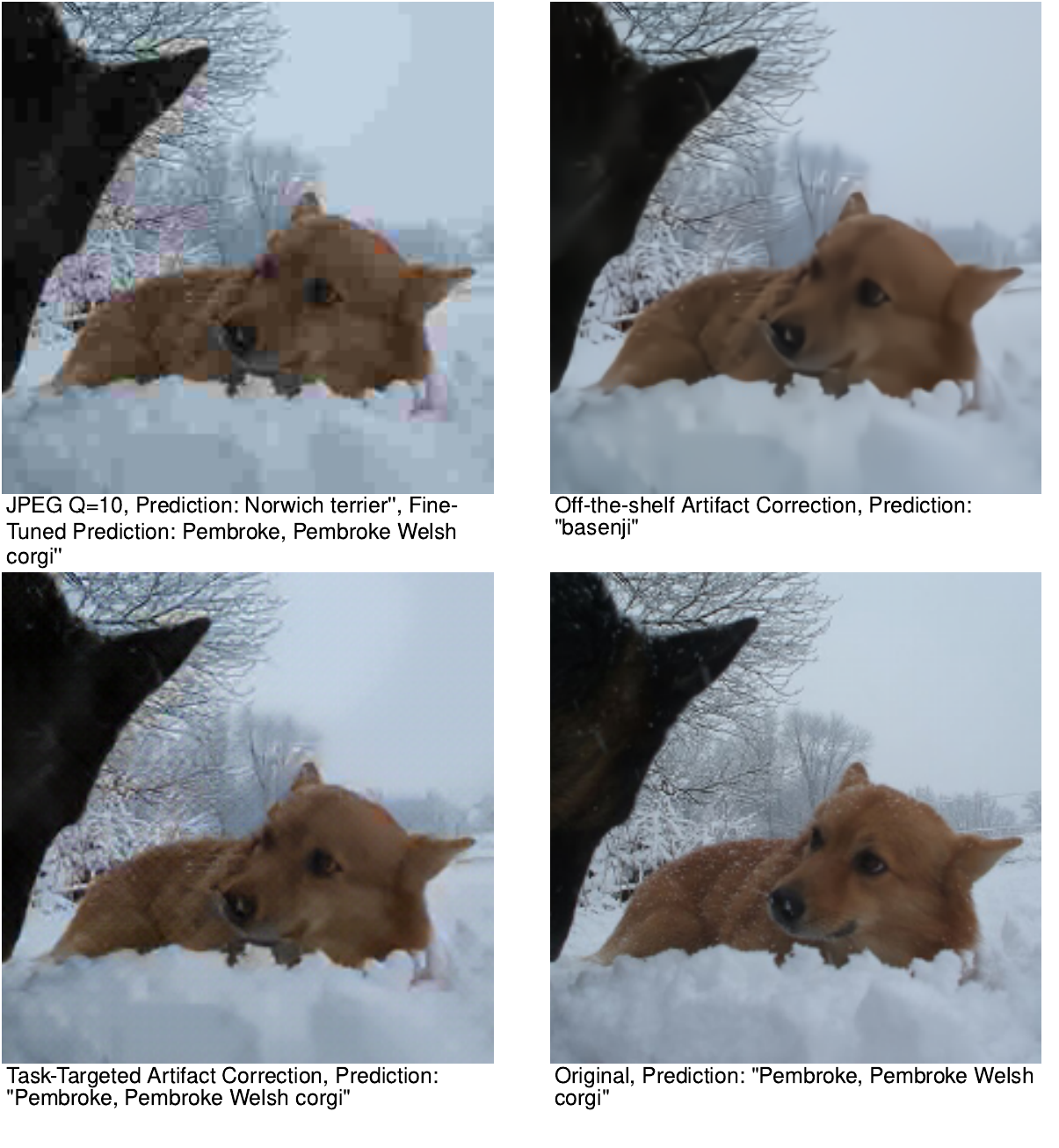}
    \caption{MobileNetV2, Ground Truth: ``Pembroke, Pembroke Welsh corgi''}
    \label{fig:mnv2_qual}
\end{figure}

\begin{figure}[H]
    \centering
    \includegraphics{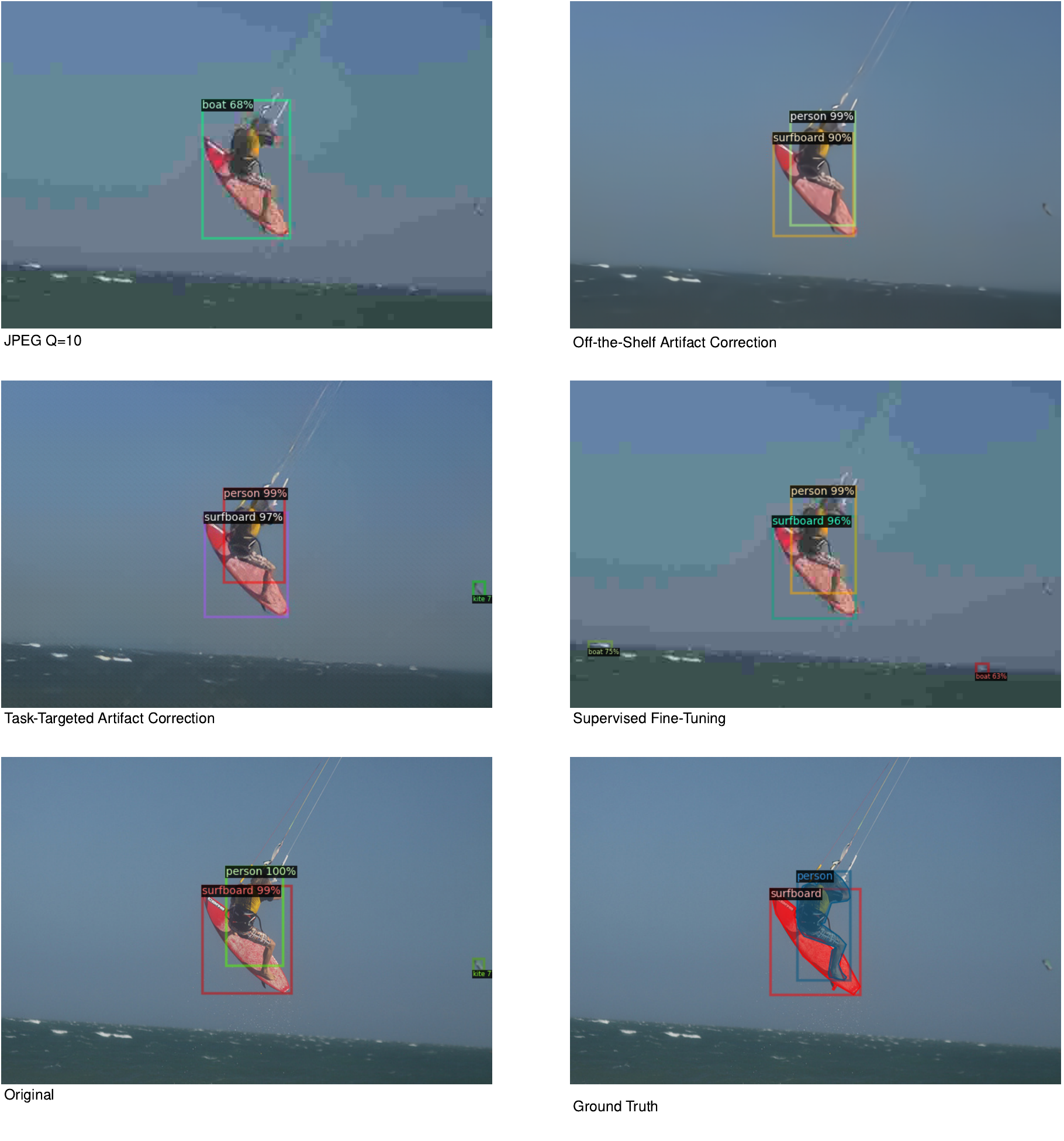}
    \caption{FasterRCNN}
\end{figure}

\begin{figure}[H]
    \centering
    \includegraphics{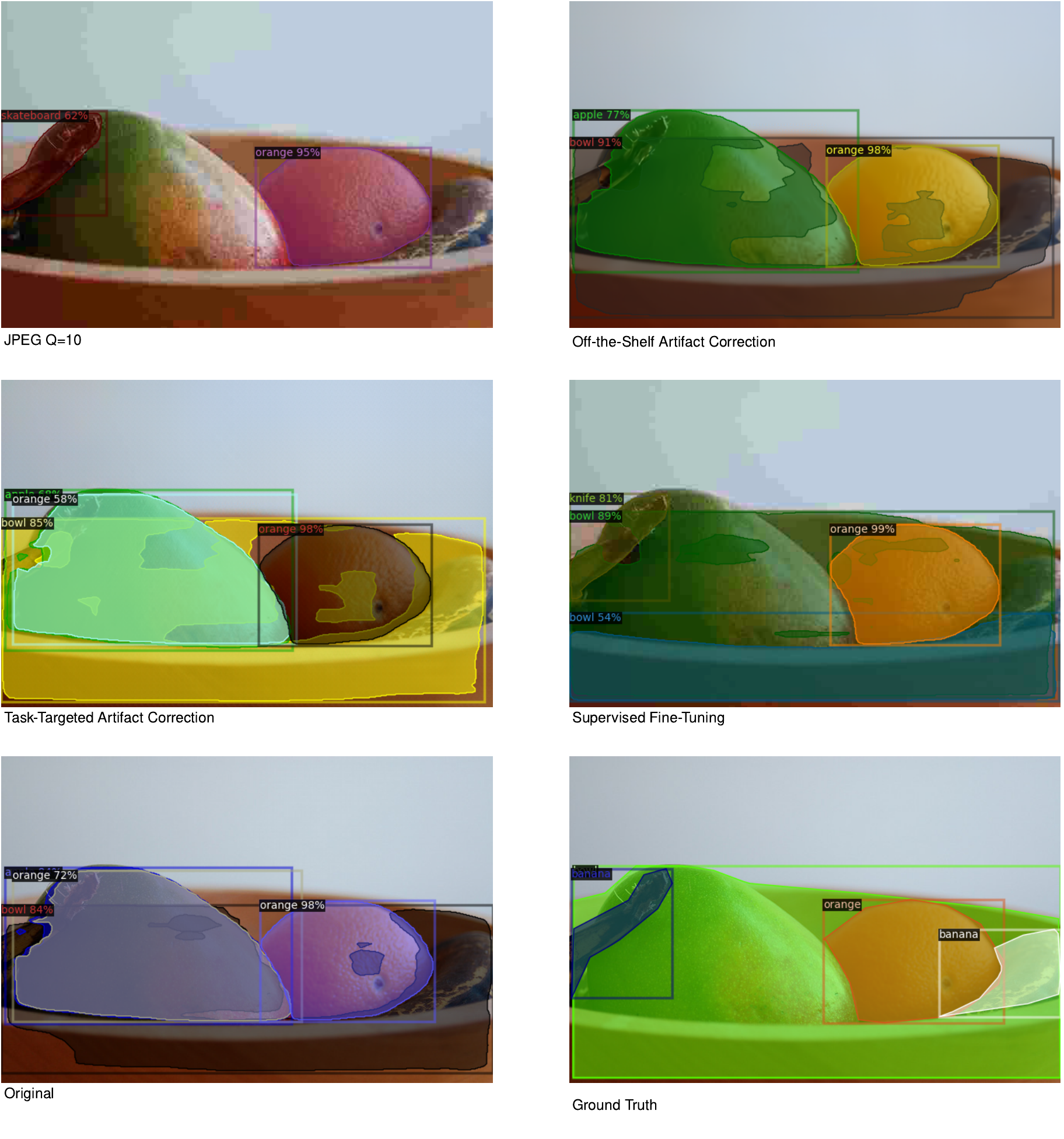}
    \captionof{figure}{MaskRCNN}
\end{figure}
\begin{figure}[H]
    \centering
    \includegraphics{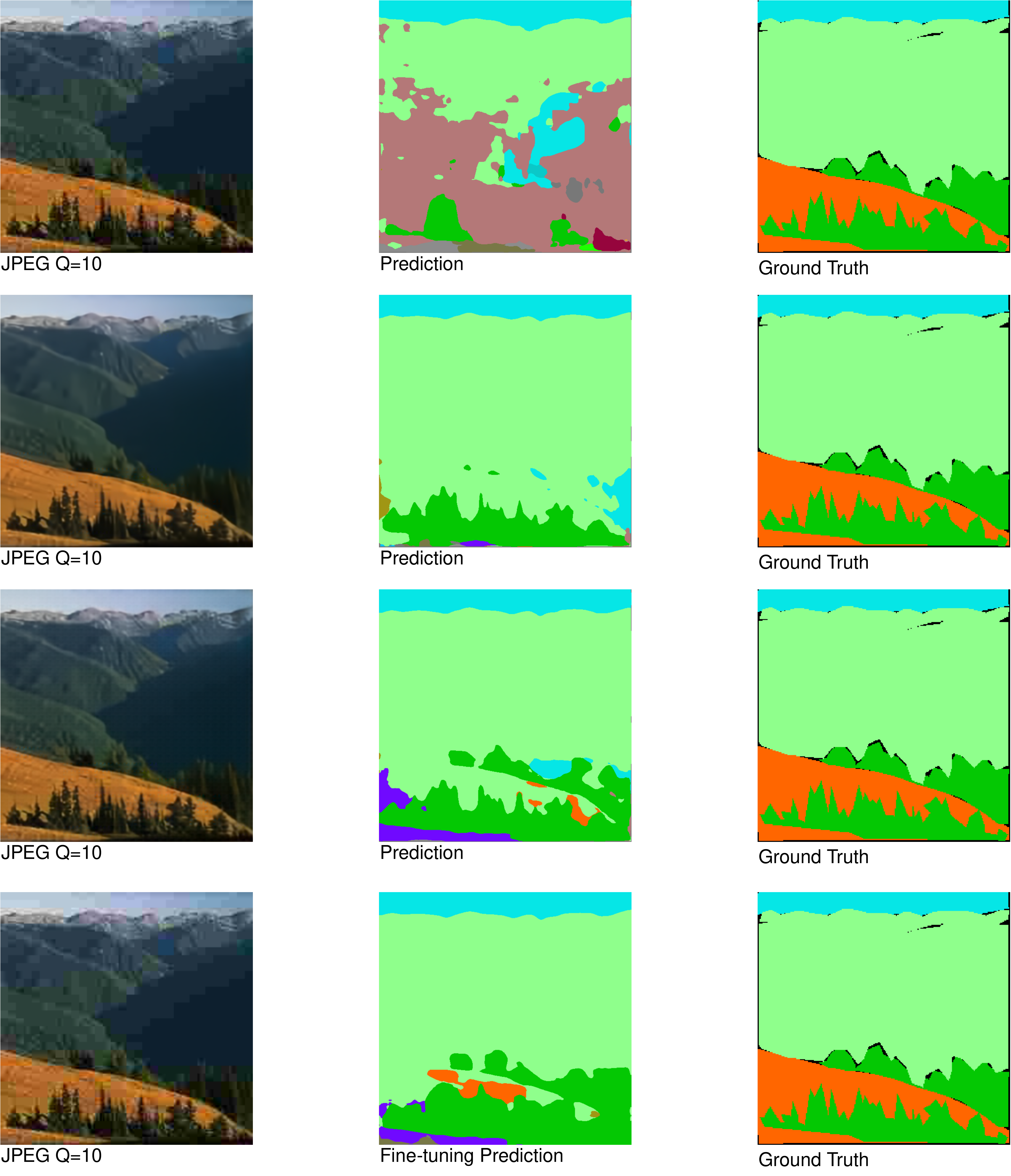}
    \caption{HRNetV2 + C1}
\end{figure}

\clearpage

\section{Metabit}

The results in this section are from the method of \nrefch{metabit}. These are purely qualitative results but they do highlight specific successes and failures of the method.

\begin{figure}[H]
    \centering
    \includegraphics[width=\textwidth]{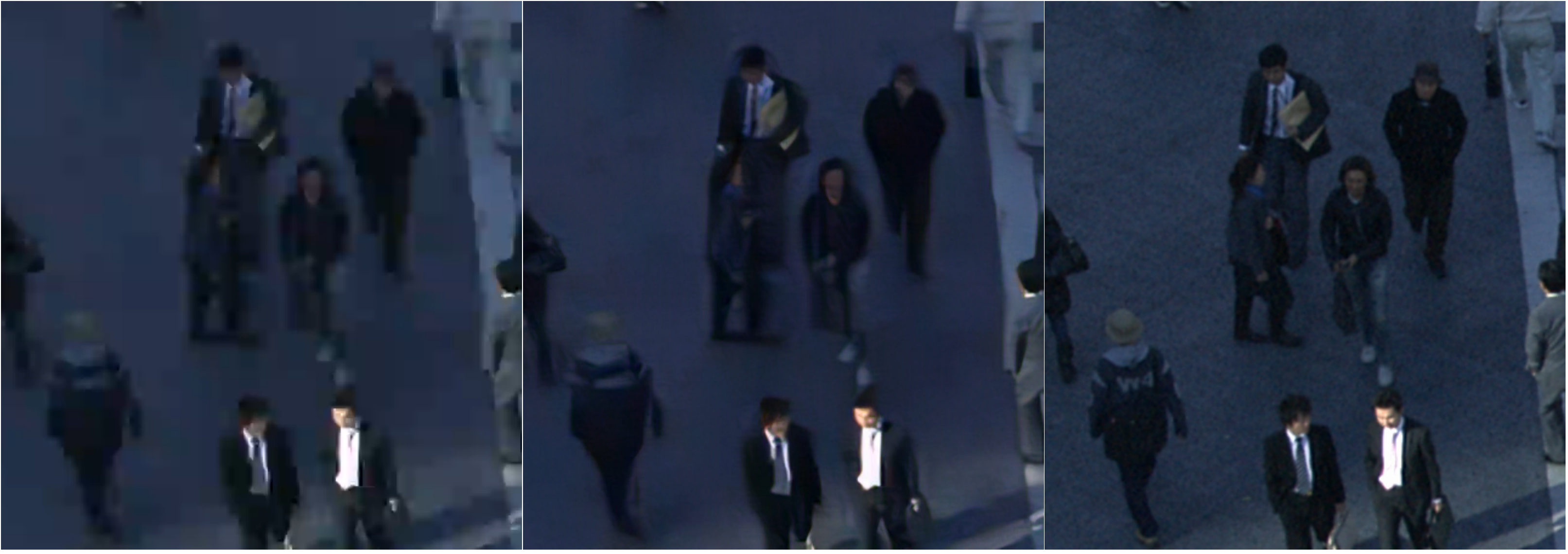}
    \caption[Dark Region]{\textbf{Dark Region.} Crop from $2560 \times 1600$ ``People on Street''. The dark region, is poorly preserved by compression. Our GAN restoration struggles to cope with the massive information loss in this region.}
    \label{fig:dark}
\end{figure}

\begin{figure}[H]
    \centering
    \includegraphics[width=\textwidth]{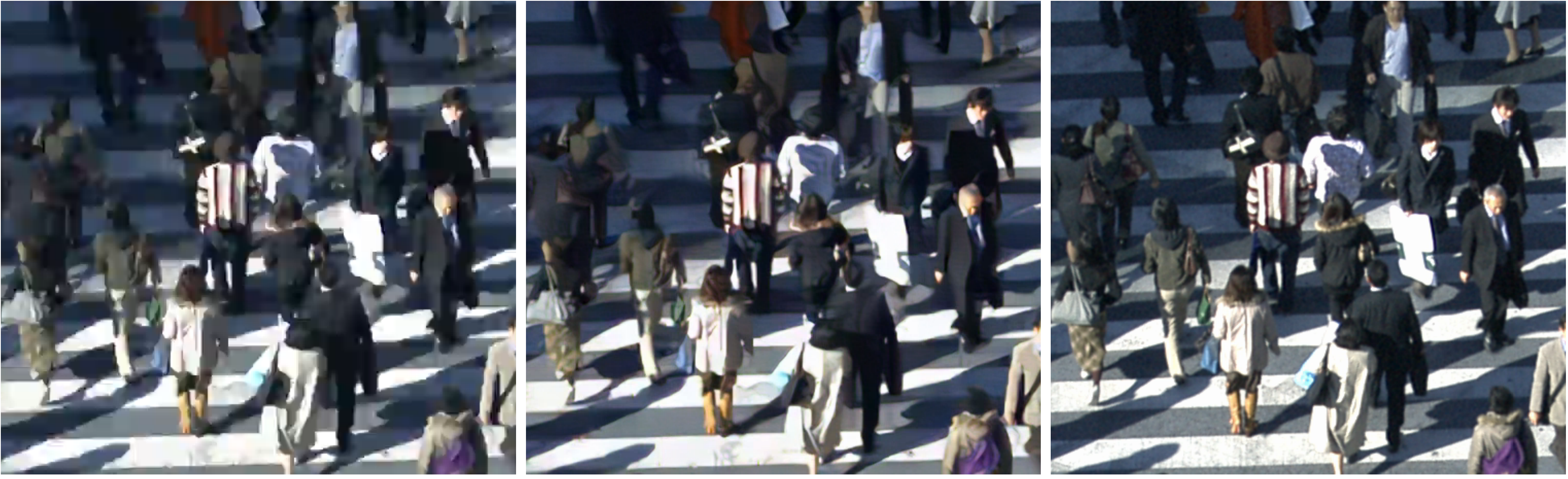}
    \caption[Crowd]{\textbf{Crowd.} Crop from $2560 \times 1600$ ``People on Street''. The image shows an extremely dense crowd. Despite the chaotic nature, our GAN is able to produce a good restoration although there is detail missing.}
    \label{fig:crowd}
\end{figure}

\begin{figure}[H]
    \centering
    \includegraphics[width=\textwidth]{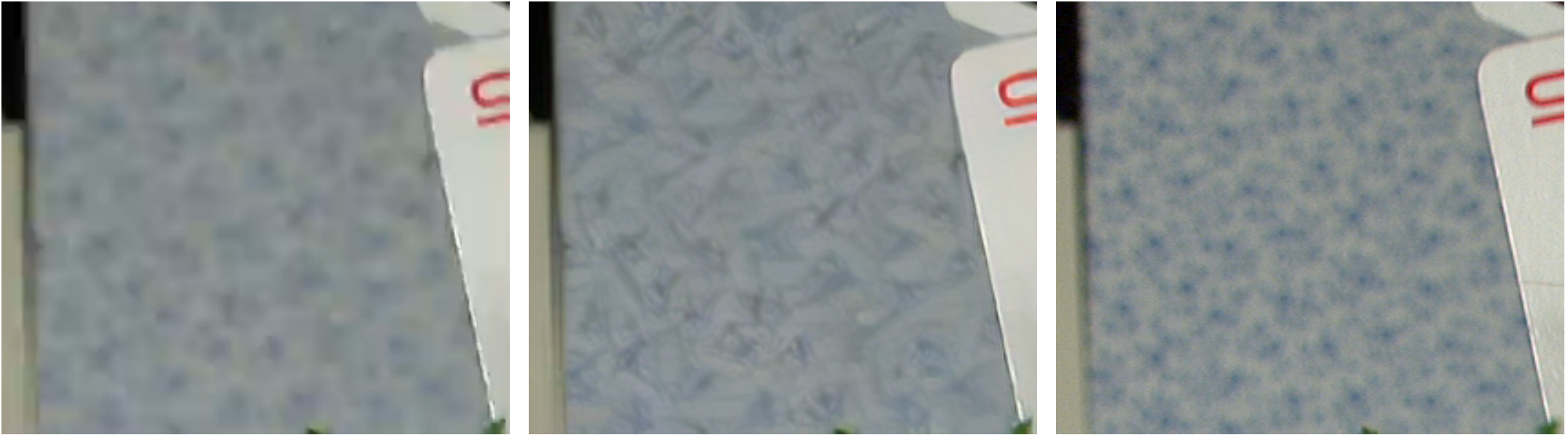}
    \caption[Texture Restoration]{\textbf{Texture Restoration.} Crop from $1920 \times 1080$ ``Cactus''. The texture on the background is destroyed by compression. Our GAN reconstructs a reasonable approximation to the true texture.}
    \label{fig:texture}
\end{figure}

\begin{figure}[H]
    \centering
    \includegraphics[width=0.9\textwidth]{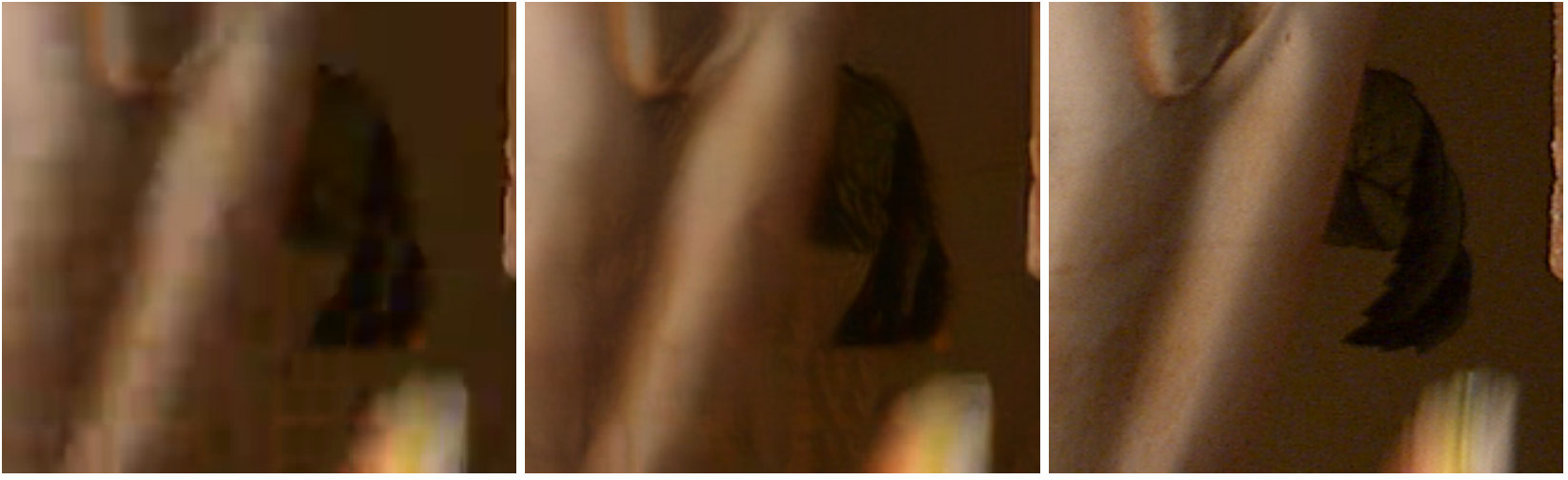}
    \caption[Compression Artifacts Mistaken for Texture]{\textbf{Compression Artifacts Mistaken for Texture.} Crop from $1920 \times 1080$ ``Cactus''. The compressed image exhibits strong chroma subsampling artifacts (lower right corner). These are mistaken by the GAN is a texture and restored as such.}
    \label{fig:art}
\end{figure}

\begin{figure}[H]
    \centering
    \includegraphics[width=0.9\textwidth]{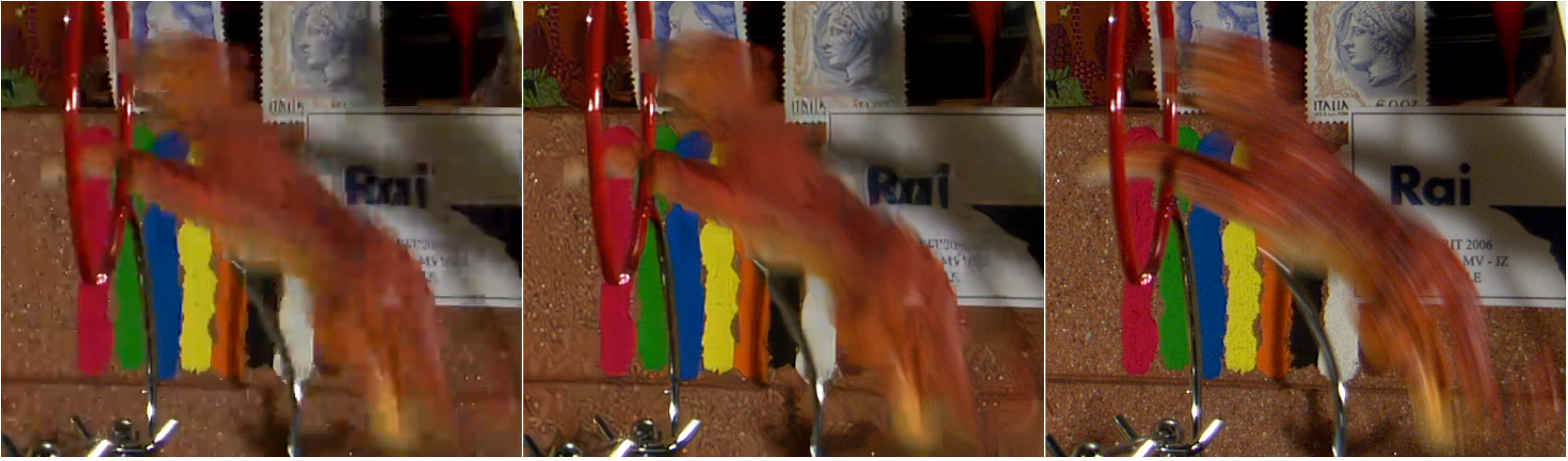}
    \caption[Motion Blur]{\textbf{Motion Blur.} Crop from $1920 \times 1080$ ``Cactus''. The tiger exhibits high motion which presents itself in the target frame as motion blur. This blur is destroyed by compression and is not able to be restored by the GAN loss. The GAN loss is also ``rewarded'' for sharp edges which would make reconstructing blurry objects difficult. As an aside, note the additional detail on the background objects in the GAN image when compared to the compressed image.}
    \label{fig:motion}
\end{figure}

\begin{figure}[H]
    \centering
    \includegraphics{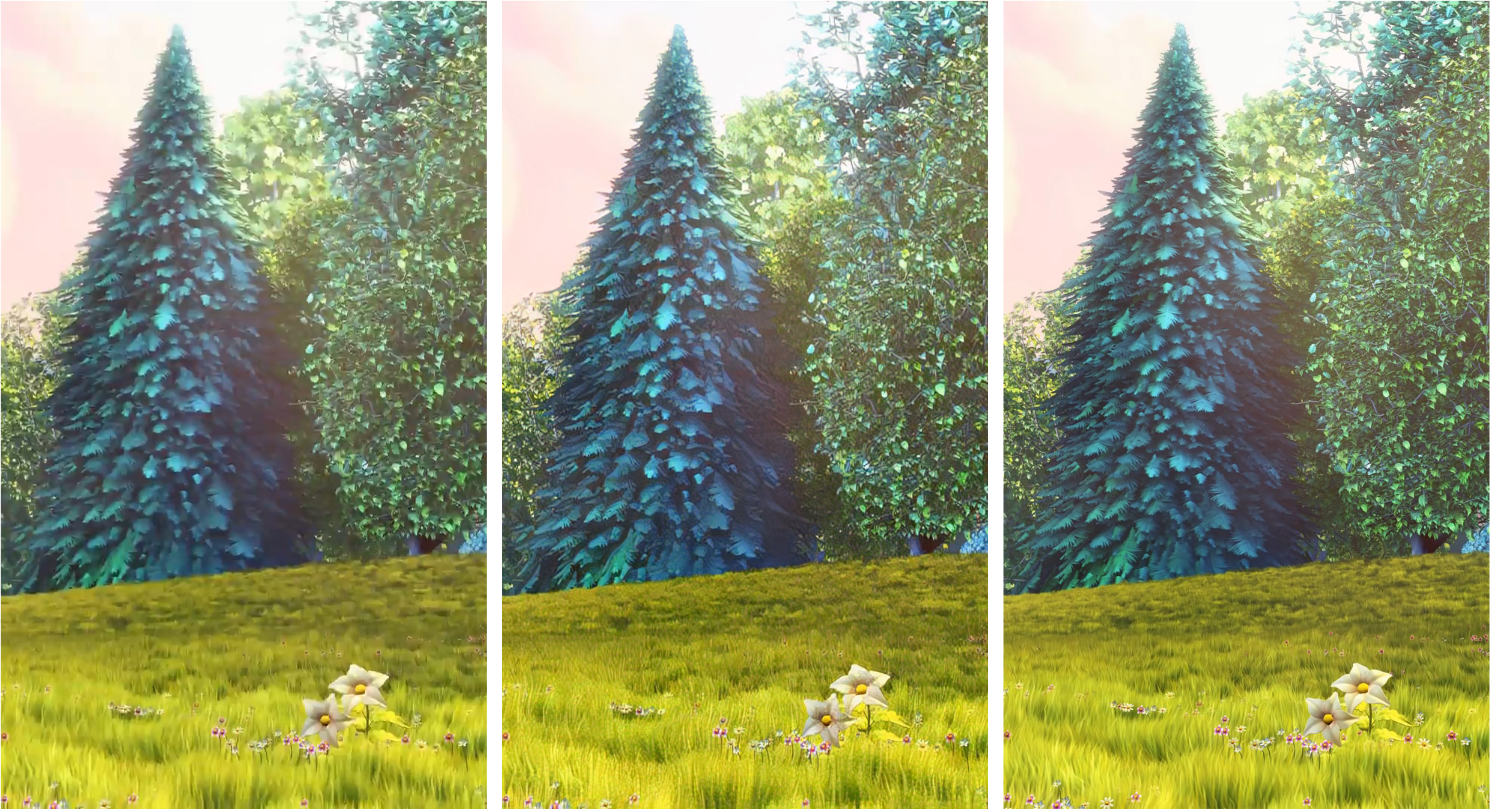}
    \caption[Artificial]{\textbf{Artificial.} Crop from $1920 \times 1080$ ``Big Buck Bunny''. This artificial scene is restored accurately despite a lack of artificial training data. Note the grass and tree textures, sharp edges, removal of blocking on the flower, and preservation of the smooth sky region.}
    \label{fig:bbb}
\end{figure}

\chapter{Survey of Fully Deep-Learning Based Compression}
\label{app:fulldl}

\setlength{\parindent}{1em} %

Although fully deep-learning based compression methods are generally considered out of the scope of this dissertation, there is general interest in these technologies and they are certainly related to the work presented in the body of the document. Therefore, in this appendix, we conduct a brief survey of the major points of image and video compression that depends entirely on deep learning to produce the encodings. While deep learning based compression shows extreme promise, it is still a very academic problem. Models currently require expensive hardware to train and to compress new media in a timely manner. This also leads to high memory usage. In general, important compression concepts like rate control are still largely missing.

In terms of objective performance, the most recent methods at the time of writing are on par with classical compression on some benchmarks. This is not always easy to evaluate, however, as methods depending on generation like GANs \parencite{goodfellow2014generative} often do not produce meaningful rate-distortion curves in the traditional sense.

In a rare personal opinion, based on my observation of the state of the art, I believe that machine learning, wherever it may end up, is the future of compression. Within the decade (\ie, before 2030) we will begin to see machine learning techniques used in consumer application. In contrast, the techniques presented in the body of the dissertation will likely be seen in consumer application in the next one or two years. There are currently a number of companies competing for deep-learning based compression market share, \eg, Google and Wave One. While these companies are delivering important research contributions it is unlikely that their proprietary solutions will win out in the long term given the compression community's reliance on standardization. Although Google was able to gain traction in classical compression with its VP codecs, even these were eventually standardized into the Alliance for Open Media (AOM) and development continued with the AV codecs. Notable standardization efforts include JPEG-AI and MPEG-AI which are much more likely to see success, meaning that any new players in this field would do well to work with the standards bodies.

\section{Image Compression}

We start with image compression. The goal of these models is to train a CNN to encode pixels into a feature vector with another CNN trained simultaneously to decode the feature vectors to an image, essentially a fancy autoencoder. The feature vectors are quantized and losslessly compressed before ``transmission'', or in this case, evaluating their size in bytes. The networks will be trained to minimize both the size of the feature vectors when stored on disk and the error of the reconstruction. There are three obvious problems here which drive the works we will consider in this section
\begin{itemize}
    \item The size on disk is not differentiable and therefore not suitable for use in a loss function.
    \item Classical compression algorithms incorporate rate-control to make their use more flexible. It is not trivial to incorporate such a side parameter into a CNN.
    \item Minimizing the error term does not necessarily produce a visually pleasing result.
\end{itemize}

Likely the first modern work in image compression with deep learning was the work by \cite{toderici2015variable} for thumbnails with a follow up for full resolution images \parencite{toderici2017full}. Todereci's work is based on recurrent networks specifically Long-Short-Term Memory networks\index{long-short-term memory} (LSTMs). The output of the LSTM at time $t$ is subtracted from the input and this residual used as the input to the LSTM at time $t + 1$ after starting the process with the input patch generating a fixed length code for a given bitrate setting. The network is only trained to minimize the $l_2$ error. Considering how early these architectures were developed they have some nice properties, including reasonable results compared to JPEG and a rudimentary attempt at variable rate encoding.

Next, \cite{theis2017lossy} proposed compressive autoencoders to generate a compressed representation. The idea is to produce a deep encoding of the input image which is then quantized for transmission and decoded by another deep network. The objective can be written as
\begin{align}
    -\log(Q(f(x))) + \beta d(x, g(f(x)))
\end{align}
where $Q()$ is the quantization function, $f()$ is the encoder, $g()$ is the decoder, and $d()$ is a measure of distortion (\ie, error). The left term here is measuring the size of the representation (number of bits) and the right term is measuring the error. Of course this objective cannot be minimized directly since $Q()$ is not differentiable. To get around this they define a differentiable approximation to the rounding step. By adjusting this approximation, they are able to produce a much more accurate variable rate encoder, although the empirical results show that training for a single rate naturally works better.

Also in 2017, \cite{NIPS2017_86b122d4} developed ``soft-to-hard vector quantization''. They start with the same problem of \cite{theis2017lossy}, that the quantization step is not differentiable. They solve the problem by using a soft assignment of the features to symbols, \ie, instead of a hard rounding they compute
\begin{align}
    \phi(z) = \text{softmax}(-\sigma(\|z - c_1\|^2, \ldots, \|z - c_n\|^2))
\end{align}
for $n$ symbols $c$. This equation is fully differentiable. However, this alone would be a poor approximation so during training the ``hardness'' is ``annealed'' from some initial condition to infinity which produces a more and more accurate approximation of the hard assignment which is used at test time. This allows the network to quickly converge on the easier soft solution in early training while increasing the problem difficulty to match the real scenario in late training.

\cite{balle2016end} propose yet another solution to this problem. Their solution is motivated by Shannon's information theory and, although somewhat questionable in the motivation, has become a staple technique for approximating quantization. Bell\'e \etal observe that the discrete quantization processes is essentially introducing noise into the signal which is output by the deep encoder. Of course, entropy of a noisy channel is something that Shannon studied quite extensively \parencite{shannon1948mathematical}. Therefore, the solution is to simply add Gaussian noise to the signal which is a simple and differentiable process. Of course the issue with this is that Gaussian noise is very different in appearance from quantization noise and CNNs are very sensitive to the actual appearance even if the entropy analysis is the same (entropy is essentially giving an aggregate view of the information loss). Nevertheless the method does work well.

\cite{mentzer2018conditional} specifically focus on designing a method for variable rate encoding. Although this was a feature of prior works, their primary focus was on overcoming the non-differentiable quantization. Mentzer \etal use both the soft-to-hard technique \parencite{NIPS2017_86b122d4} and the compressive autoencoders technique \parencite{theis2017lossy} to deal with quantization. To model the rate term in the loss, Mentzer \etal treat the feature vectors as a conditional distribution, \ie,
\begin{align}
    P(z) = \prod_{i = 1}^N P(z_i | z_{i - 1}, \ldots z_1)
\end{align}
in raster order. So each feature vector is considered to have its own probability which is conditioned on all previous features. They then model $P(z)$ and the conditional distributions using another deep network (which is differentiable). Specifically they use a 3D convolution since this is efficient and respects the ``causality constraint.'' In other words, the previous feature vectors cause the current feature vector since they are conditional distributions. This formulation for $P(z)$ allows them to compute an approximation for the entropy and therefore the rate which they use as a loss term.

In something of a departure from prior works, \cite{agustsson2019generative} formulate a compression algorithm based on GANs \parencite{goodfellow2014generative}. The advantage here is that these algorithms can produce striking, faithful, images in extreme settings. In this case the distortion term is replaced with a GAN loss instead of the traditional $l_2$ loss of prior works. This is based on the correct observation that $l_2$ loss does not capture human perception well. Although the paper offers limited further insight into the mathematics of deep learning based compression systems, the imagery their method produces is truly remarkable often outperforming JPEG by a wide margin while saving considerably more space.

This work is continued by \cite{mentzer2020high} and they make a number of advancements over \cite{agustsson2019generative}. Where Aggustsson \etal showed visually realistic results there were major deviations from the true outputs. The preservation of the output seemed to be more semantic than visual which makes sense given that the GAN training uses deep networks classifying real/fake. Mentzer \etal by comparison is extremely faithful to the original images often deviating in ways that are indistinguishable to the human eye. Aggustsson \etal is also quite limited in the size of the input that it can accept (an efficiency concern), whereas the Mentzer \etal formulation works efficiently on sizes up to $2000 \times 2000$ which is a respectable size for a modern image. An interesting avenue of analysis in this work is in the effectiveness, or lack thereof, of metrics. After conducting an extensive user study, they found that no metric was adequate for matching the human's responses. This is not at all surprising.

Although we end in 2020 image compression continues to be an active area of research, although it remains to be seen which works of 2021 will emerge as the most influential. In the interest of space, we conclude the discussion of image compression here. Although the advance of \cite{mentzer2020high} was extremely promising, there is still no deep learning algorithm that is suitable for deployment in a consumer application. This is partly an efficiency concern but it is also a flexibility concern. JPEG was extremely well thought out to work for the widest range of situations which is part of the reason it has persisted for 30 years. Deep learning methods are only just scratching the surface of this kind of long term thinking.

\section{Video Compression}

We now turn to video compression. Similar to the previous section, the goal will be to train encoder and decoder CNNs with some kind of quantization of the encoded feature vectors. Unlike the last section, however, we now have a temporal component in everything we do. In addition to the challenges of image compression, dealing with the temporal component is a problem by itself. Some methods will treat the time component as independent, essentially image compression with different features over time. Some will attempt to incorporate the temporal component into the prediction itself either in a recurrent or motion based solution. Still others will use an implicit representation, essentially over-fitting a network for each video.

We start with the method of \cite{wu2018video}. The key insight is that ``keyframes'' can be defined which are then encoded using off-the-shelf image encoders. Then, the intermediate frames are produced using image interpolation networks that take the two keyframes as input and produce the intermediate frames. Naturally, the longer the interval between the keyframes the higher the error of the predictions. While the method certainly works, it does not clearly outperform H.264 in the same way that image compression algorithms were clearly outperforming JPEG. This was still a major advancement, however, since prior to this no one had tried to produce a video codec using deep learning. By focusing on keyframe compression and interpolation, the method is efficient which was a major concern with video compression.

Next, DVC \parencite{lu2019dvc} proposes an end-to-end technique which encodes motion and residual information for predicted frames. This is intentionally designed to mimic the classical compression loop which stores intra-frames and then predicts intermediate frames using motion warping and low-entropy error residuals. In this case, each component is modeled separately with a CNN. The method has several moving parts
\begin{description}
    \item[Motion Estimation] which uses a task-specific optical flow network to produce per-pixel motion. This motion is then encoded using another CNN for compression and quantized. The decoder performs the inverse process to produce the flows
    \item[Motion Compensation] Also uses a deep network. First the decoded optical flow is used to warp the reference image, then the reference image, warped images, and optical flow are all used as input to another deep network to predict the true frame.
    \item[Transform] The residual between the predicted and true frame is taken and encoded using yet another CNN to produce the quantized encoding. This is similar to image compression techniques.
\end{description}
Overall the method is fairly complex and heavy consisting of several convolutional networks. While all of this does pay off in terms of the overall result compared with \cite{wu2018video}, the actual codec itself struggles to match H.265. Furthermore, per-pixel flow is likely wasteful, at least we know that classical video codecs do not make use of dense motion information.  That being said the end-to-end nature and the idea of replacing each part of a traditional video encoder with a CNN are major advances to the state-of-the-art.

In 2020, we finally had a technique capable of outperforming H.265. The method of \cite{liu2020conditional} proposed a simple but effective technique. Use a standard deep learning based image compression algorithm to generate initial codes for each frame. Then perform internal learning to generate an ``optimal'' code for that frame and use a conditional entropy model to produce a final code for the current frame that is conditioned on the previous frame. Note that the internal learning method is actually learning a small CNN just for that particular frame, so the encoding time is increased but the decoding is still fast since the decoder only needs to perform inference on the resulting network to obtain the code. The conditional entropy model also helps the encoder reuse information from prior frames to reduce the final code length. The contribution is very straightforward. Aside from these two ideas there are no special formulations (this is a  good thing). The result is impressive, with results that consistently outperform the classical codecs for higher bitrates. The method does struggle at low bitrates, however.

Continuing with internal learning is NeRV \parencite{chen2021nerv}. This method is entirely an internal learning technique, which means that for each video, the compression process is to train a neural network which predicts only that video (over-fitting it) and then the neural network weights are compressed using a model compression technique like pruning. To decode, the transmitted model weights are used in an inferencing pass to retrieve the frames. In particular, NeRV proposes a frame-based implicit representation \vs the pixel based approach in something like SIREN \parencite{sitzmann2020implicit}. What this means practically is the NeRV takes a time $t$ as input and produces the frame of the video at time $t$ instead of taking the triple $x, y, t$ and producing the pixel at position $x, y$ at time $t$. Not only is the NeRV formulation simpler for the network to learn (leading to better results) it is also significantly faster, requiring $T$ forward passes to produce a video of length $T$ instead of $H \times W \times T$ forward passes. While the overall idea here is interesting the results do leave something to be desired, as NeRV struggles to outperform even H.264. Furthermore, although the network is small making decoding time fast, encoding (\ie, training the network) is on the order of hours.

We close the section with ELF-VC \parencite{rippel2021elf}, a method which is groundbreaking both in its results and in its methodical design. The approach is fast, provides a well motivated method for I- and P-frame encoding with deep networks, supports variable rate encoding, and compares well to other classical and deep learning codecs. For the I-frame model, standard image deep learning compression is used. For P-frames, the method is more interesting. Motion is predicted using a flow model as in \cite{lu2019dvc} and the residual and flows are both stored. The decoder uses a prior frame as an initial estimate of the warped frame before incorporating the flow vectors and residual. Variable rate encoding is achieved using a level map\index{level map} where the rate-distortion curve is discretized in levels. The level is tiled spatially and used as input to the encoder and decoder with the loss encouraging the network to hit the specified bitrate target. This provides a simple way to tune the bitrate. In terms of results the ELF-VC method largely outperforms other work on all benchmarks with the exception of AV1, the latest in classical compression.

Although ELF-VC hits on a number of ideas that would be required for a commercial video codec, that goal is still very far off. With optimizations ELF-VC can decode a 1080p video at 18fps, which is not fast enough, and requires a GPU with a large amount of memory. As new methods are developed which are more efficient (this needs to be a continuing focus, however) and hardware speed increases, the likelihood of deep learning compression finding its way to consumer applications increases. These reasons contribute to the 10 year estimate.

\section{Lossless Techniques}

The previous two section dealt exclusively with lossy compression\index{lossy compression}. We expect that the networks will remove information from images during encoding, and even if they do not the quantization\index{quantization} process will. But we can also use machine learning for lossless compression\index{lossless compression}. This takes a number of forms which we discuss in this section. Importantly this is a fairly interesting use case and could potentially see practical application sooner than the lossy methods although it would be in niche scenarios. For example, these techniques have uses in lossless transcoding\index{transcode} of classically compressed images. This particular application is important because it would allow large datacenters, which have the resources to run deep learning models at scale, to save on storage costs by transcoding images and videos to a smaller deep learning based file. The media are then transcoded back to their consumer format before being transmitted so that the consumer does not need special hardware or software to view the media. When we discussed entropy coding in \nrefch{entropy}, we noted that entropy coders work by assigning shorter codes to probable symbols and longer codes to improbable symbols. In order to work well the encoder needs an accurate probability distribution which is difficult to come up with particularly for image data. The techniques in this section are primarily focused on learning such distributions.

PixelRNNs \parencite{van2016pixel, van2016conditional} are generative models that predict each pixel in an image as a discrete conditional distribution. This has an advantage over other generative methods like GANs \parencite{goodfellow2014generative} because the model predicts the distribution explicitly instead of simply producing samples from the distribution. In standard fashion, each pixel is treated as a distribution conditioned on all previous pixels
\begin{align}
    p(z) = \prod_{i=i}^N p(z_i | z_{i - 1}, \ldots, z_1)
\end{align}
Each pixel can then be generated by sampling from the learned distribution pixel by pixel. So how is this relevant to compression? With the likelihood of each pixel, we can use these distributions to produce probabilities for entropy coders \parencite{huffman1952method,rissanen1979arithmetic}.

Integer discrete flows\index{integer discrete flows}(IDFs) \parencite{hoogeboom2019integer, berg2020idf++} are similar in spirit. The idea again is to learn an explicit distribution of the image data and produce a latent code from the distribution with the advantage of much faster sampling. IDFs in particular are designed to overcome an explicit problem with flows in general: that they assume a continuous random variable. Images are discrete random variables so quantization of the resulting model to fit the discrete distribution may introduce loss. By formulating an integer discrete flow, the authors can provably reproduce exactly the given input from a code. The flow itself is based on a change of variables formula
\begin{align}
    P_X(x) = P_z(f(x))\left|\frac{\partial z}{\partial x}\right|
\end{align}
The flow is then reformulated to be in integer form where the Jacobian is one. The method was extended more recently in iVPF \parencite{zhang2021ivpf} which used volume preserving flows instead of integer discrete flows (they are quite similar in operation however). In either case, the learned flow can then be used directly as a probability distribution for entropy coding.

While Bits-back encoding \parencite{wallace1990classification, hinton93keeping, frey1996free, frey1998bayesian} has been around for some time, the Bits-Back ANS \parencite{townsend2019practical} method was the first algorithm using neural networks for the learning component and which was shown to be efficient on large datasets. Without going into too much detail, the idea of bits-back encoding is to assume that the given symbol $s$ has some latent variable $y$ associated with it and that we have a way of measuring $p(y)$, $p(s | y)$, and $p(y | s)$. Bits-back encoding allows us to leverage this knowledge of the latent distribution to store $s$ with fewer bits. Bits-back ANS uses a variational autoencoder (VAE)\index{kingma2013auto} for the latent model. Bit-swap \parencite{kingma2019bit} and Hilloc \parencite{townsend2019hilloc} extend this with hierarchical latent variables, and LBB \parencite{ho2019compression} merges flows with bits-back encoding.

We close with a very different approach, \cite{mentzer2020learning}. This method actually leverages lossy compression in order to improve the lossless compression rate. The idea is to start with BPG \parencite{bpg} and use a network to predict an optimal quantization parameter controlling how aggressive BPG should behave. BPG of course loses information so the residual of the true frame and the compressed frame is taken and another network predicts the probability of the residual given the input image. The residual is then encoded using an entropy coder with the learned distribution. Since the encoded residual is stored with the BPG compressed image, there is no information loss.

Overall, this field is full of interesting and practical ideas. Although somewhat niche in their application, these are highly developed techniques that could already be useful engineering applications. However, these ideas are by definition not suitable for consumers as they are really one part of a more complex whole and their performance can not match lossy algorithms. Their use is more suited to specialized applications in medical imaging, datacenters, or remote sensing where loss of data many not be acceptable.

\backmatter %
\setchapterstyle{plain}

\printbibliography[heading=bibintoc]

\setglossarystyle{listgroup}
\printglossary[title=Terminology, toctitle=Terminology]

\printglossary[type=\acronymtype]

\printindex

\chapter*{Figure Credits}
\addcontentsline{toc}{chapter}{Figure Credits}

\noindent Unless listed here, figures are either generated by the author, in the public domain. The original authors of these works do not endorse any changes made for this document.

\begin{description}
    \item[\vreffig{ha:morlet}] Wikipedia. User JonMcloone \\
        \url{https://commons.wikimedia.org/wiki/File:MorletWaveletMathematica.svg}. \\
        \href{https://creativecommons.org/licenses/by-sa/3.0/deed.en}{CC-BY-SA 3.0}. \\
        \textit{Removed axes.}
    \item[\vreffig{ha:uncertainty}] Wikipedia. User JonMcloone \\
        \url{https://commons.wikimedia.org/wiki/File:MorletWaveletMathematica.svg}. \\
        \href{https://creativecommons.org/licenses/by-sa/3.0/deed.en}{CC-BY-SA 3.0}. \\
        \textit{Removed axes, added scaled version to show hierarchy.}
    \item[\vreffig{ha:haar}] Wikipedia. User Omegatron \\
        \url{https://commons.wikimedia.org/wiki/File:Haar_wavelet.svg}. \\
        \href{https://creativecommons.org/licenses/by-sa/3.0/deed.en}{CC-BY-SA 3.0}. \\
        \textit{Removed axes, added scaled version to show hierarchy.}
    \item[\vreffig{ent:comms}] \sdcite{shannon1948mathematical}
    \item[\vreffig{dl:mlp}] Wikipedia. User Glosser.ca \\
        \url{https://commons.wikimedia.org/wiki/File:Colored_neural_network.svg}. \\
        \href{https://creativecommons.org/licenses/by-sa/3.0/deed.en}{CC-BY-SA 3.0}.
    \item[\vreffig{dl:hog}] \sdcite{dalal2005histograms}
    \item[\vreffig{dl:ss}] \sdcite{lowe1999object}
    \item[\vreffig{dl:cnn}] \sdcite{lecun1998gradient}
    \item[\vreffig{dl:unet}] \sdcite{ronneberger2015u}
    \item[\vreffig{dl:resblock}] \sdcite{he2016deep}
    \item[\vreffig{mpeg:mvarrow}] Big Buck Bunny. \sdcite{bbb} \\
        \url{https://peach.blender.org/} \\
        \href{https://creativecommons.org/licenses/by-sa/3.0/deed.en}{CC-BY-SA 3.0}. \\
        \textit{Motion vector arrows added to frame.}

\end{description}

\end{document}